\documentclass[aps,rmp,reprint,amsmath,amssymb,graphicx]{revtex4-1}
\usepackage{natbib}
\usepackage{color}
\usepackage{amsmath}
\usepackage{amssymb}
\usepackage{graphicx}
\usepackage{mdframed}
\usepackage{collectbox}

\begin{document}

\title{Colloquium: Criticality and dynamical scaling in living systems}

\author{Miguel A. Mu\~noz}
\address{Instituto Carlos I de
  F{\'\i}sica Te\'orica y Computacional and Departamento de
  Electromagnetismo y F{\'\i}sica de la Materia, Facultad de Ciencias
  Universidad de Granada 18071 Granada Spain}

\begin{abstract}
  A celebrated and controversial hypothesis suggests that some
  biological systems --parts, aspects, or groups of them-- may extract
  important functional benefits from operating at the edge of
  instability, halfway between order and disorder, i.e. in the
  vicinity of the critical point of a phase transition. Criticality
  has been argued to provide biological systems with an optimal
  balance between robustness against perturbations and flexibility to
  adapt to changing conditions, as well as to confer on them optimal
  computational capabilities, huge dynamical repertoires, unparalleled
  sensitivity to stimuli, etc.  Criticality, with its concomitant
  scale invariance, can be conjectured to emerge in living systems as
  the result of adaptive and evolutionary processes that, for reasons
  to be fully elucidated, select for it as a template upon which
  further layers of complexity can rest.  This hypothesis is very
  suggestive as it proposes that criticality could constitute a
  general and common organizing strategy in biology stemming from the
  physics of phase transitions.  However, despite its thrilling
  implications, this is still in its embryonic state as a well-founded
  theory and, as such, it has elicited some healthy skepticism.  From
  the experimental side, the advent of high-throughput technologies
  has created new prospects in the exploration of biological systems,
  and empirical evidence in favor of criticality has proliferated,
  with examples ranging from endogenous brain activity and
  gene-expression patterns, to flocks of birds and insect-colony
  foraging, to name but a few.  Some pieces of evidence are quite
  remarkable, while in some other cases empirical data are limited,
  incomplete, or not fully convincing. More stringent experimental
  set-ups and theoretical analyses are certainly needed to fully
  clarify the picture. In any case, the time seems ripe for bridging
  the gap between this theoretical conjecture and its empirical
  validation.  Given the profound implications of shedding light on
  this issue, we believe that it is both pertinent and timely to
  review the state of the art and to discuss future strategies and
  perspectives.
\end{abstract}

\date{\today}


\maketitle
\tableofcontents

\section{Introduction: Statistical physics of biological systems}
\label{Intro}

One of the greatest challenges of Science is to shed light on the
essence of the phenomenon that we call ``life'', with all its
astonishing diversity and complexity.  Cells --the basic
building-blocks of life-- are intricate dynamical systems consisting
of thousand types of interacting molecules, being created, used and
destroyed every minute; multicellular organisms rely on the perfectly
orchestrated motion of up to trillions of interacting cells, and
communities assemble dozens of individuals, interacting in
countless ways, forming entangled ecosystems, and giving rise to a
mind-blowing hierarchy of \emph{``complexity''}.

The standard viewpoint in biology, stemming from the reductionist
tradition, is that each molecular component (protein, nucleic acid,
metabolite...) is specific and requires individualized scrutiny. This
one-at-the-time approach has successfully identified and quantified
most of the components and many of the basic interactions of life as
we know it, as stressed by the rapid advance of the ``omics'' sciences
(genomics, proteomics, metabolomics...).  Still, unfortunately, it
offers no convincing explanation of how systemic properties emerge
\cite{Sauer}. Questions such as ``how are those myriads of elements
and interactions coordinated together in complex living creatures?''
or ``how does coherent behavior emerge out of such a soup of highly
heterogeneous components?'' \cite{Schrodinger} remain largely
unanswered.

A complementary strategy consists in looking at complex biological
problems from a global perspective, shifting the focus from specific
details of the molecular machinery to integral aspects
\cite{Kaneko-book,Bialek-book,Goldenfeld,Stan1999,Sauer,Alon-book}.
System approaches to biology rely on the evidence that some of the
most fascinating phenomena of living systems --such as memory and the
ability to solve problems-- are collective ones, stemming from the
interactions of many basic units, and might not be reducible to the
understanding of elementary components on an individual basis
\cite{Bialek-perspectives}. Theoreticians have long struggled to
elucidate whether simple and general principles --such as those in
physics-- could be of any help in tackling biological complexity. More
specifically, they have long been seduced by the idea of adapting
concepts and methods from statistical mechanics to shed light onto the
large-scale organization of biological systems\footnote{The
  possibility that biological problems may stretch the frontiers of
  physics by uncovering phenomena and mechanisms unknown in purely
  physical systems is also inspiring \cite{Goldenfeld,Ask}.}
\cite{Schrodinger,Bialek-book,Alon-book,Anderson,Hopfield1982,Amit-book2,Parisi,Sneppen-book,Kelso,Kelso1985}.

One of the most striking consequences of interactions among elementary
constituents of matter (atoms, molecules, electrons...) is the
emergence of diverse \emph{phases} whose behavior bears little
resemblance with that of their basic components or small groups of
them \cite{Stanley-book,Chaikin,Anderson}.  Systems consisting of very
many (microscopic) components may exhibit rather diverse types of
(macroscopic) collective behavior, i.e. phases, with different levels of
internal order. Moreover, slight changes in external conditions
(e.g. temperature, pression...)  or in the strength of interactions
may induce dramatic structural rearrangements, i.e. \emph{phase transitions}.

It is thus tempting to hypothesize that biological states might be
manifestations of similar collective phases and that shifts between
them could correspond to phase transitions
\cite{Anderson,Hopfield1994}.  As a matter of fact, phase transitions
are a common theme in biology \cite{Pollack-book,Sole-book}, as
illustrated by the following
non-exhaustive list of examples: (i) synchronization phase
transitions in collective biological oscillators such as circadian
clocks \cite{PNAS-Ojalvo}; (ii) percolation transitions of fibers in
connective tissues such as collagen \cite{Forgacs1,Forgacs2,McIntosh},
(iii) melting phase transition in DNA strands \cite{DNA,Poland}; and (iv)
transitions between different dynamical regimes (oscillations,
bursting,...) in neuronal networks
\cite{Freeman2005,Freeman2013,Rabinovich,Werner,Haken2013,Kelso},
etc.  

Life --guided by evolution-- has found its way to exploit very diverse
types of order: crystalline structures (seashells, skeletons...),
liquid states (blood, lymph, sap...), gels (vitreous humor, cell
cytoplasm), etc.  However, some aspects of biological systems --think
e.g. of neural networks or flocks of birds-- exhibit intermediate
levels of organization, half way between order and disorder, less
regular than perfect crystals but more structured than random
gases. Remarkably, it has been conjectured that, under some
circumstances, living systems --i.e. parts, aspects, or groups of
them-- could draw important functional advantages from operating right
at the borderline between ordered and disordered phases, i.e. at the
very edge of a (continuous) phase transition or critical
point\footnote{Phase transitions may occur in either a
  discontinuous/abrupt fashion \cite{Binney} --with associate
  bistability of the two different phases and an abrupt/discontinuous
  jump at the transition point-- or in continuous/progressive way with
  an associated critical point. Our main focus here is on continuous
  ones, but we will also encounter discontinuous transitions, which
  may also play a relevant role in biology.}
\cite{Origins,Bak,Chialvo2010,Plenz-Physics,Beggs-criticality,Chialvo2008,Schuster}.
For instance, rather generically, living systems need to achieve a
tradeoff between robustness (resilience of the system state to
external perturbations; which is a property of ordered phase), and
flexibility (responsiveness to environmental stimuli, which is a
feature of disordered phases). An optimal balance between these two
conflicting tendencies can be accomplished by keeping the system
dynamical state at the borderline of an order-disorder phase
transition, i.e. at criticality. Signatures of criticality, such as
the spontaneous emergence of long-range spatio-temporal correlations
and the exquisite sensitivity to stimuli are also susceptible to be
exploited for functional purposes, e.g. to create coordinated global
behavior, as we shall discuss in what follows.  The idea that --in
some special circumstances-- evolution might have favored states close
to the edge of a phase transition is certainly tantalizing, as it
suggests that operating near criticality could be an overarching
strategy in biological organization
\cite{Origins,Bak,Mora-Bialek,Plenz-Physics,Beggs-criticality,Schuster,Chialvo2010}.

Critical points have long been appreciated to exhibit striking
features. Still, given the need of careful fine tuning for them to be
observed, they were long treated as rarities. The development of some
of the most remarkable intelectual achievements of the second half of
the $20th$ century, such as the scaling hypothesis and the
renormalization group theory \cite{Wilson,Fisher}, changed this view
and led to an elegant and precise theory of criticality, with
unsuspected implications in many fields, from particle physics to
polymer science\footnote{See,
  e.g. \citet{Stanley-book,Binney,Henkel,Sethna-book,DeGennes,Tauber2017,Delamotte}.}.
A chief conclusion is that many features at critical points are quite
robust and largely independent of small-scale details, giving rise to
\emph{universality} in the large-scale behavior. This has very
important consequences for e.g. studies in biology, as criticality and
its concomitant scale-invariance can be understood through simple
stylized models --neglecting many irrelevant details of individual
components and putting the emphasis on how they interact-- paving the
road to the understanding of collective aspects of living systems in
relatively simple terms.

From the experimental side, the advent of high-throughput techniques
and big-data analyses have created new prospects in the exploration of
biological systems.  This is true, for example, in neuroscience
--where it is now possible to record activity from individual spiking
neurons to entire brains with previously-unthinkable resolution
\cite{Techniques}-- and, similarly, in genomics
\cite{Lesk-book} or in collective motion analyses \cite{Starflag}. As
a result, recent years have witnessed an upsurge of empirical works
reporting on putative scale-invariance and/or criticality in diverse
biological systems, supporting the above theoretical speculations. In
some cases the evidence appears to be robust, while in others it is
marginal, incomplete, or, to say the least, doubtful.  In any case,
time seems to be ripe for bridging the gap between the theoretical
hypothesis and its empirical validation.

The purpose of the present Colloquium is to briefly review the main ideas
and motivation behind the criticality hypothesis as a possible guiding
principle in the collective organization of living systems and to
scrutinize and discuss in a critical way the existing empirical
evidence and prospects. It also aims at providing the reader with a
self-consistent view of what is criticality and what it is not, as
well as an overview of the literature on this active and fascinating
research field with countless ramifications.

Let us remark that there exist excellent articles reviewing some of
these topics to different extents; the list includes the very
influential paper by \textcite{Mora-Bialek} which popularized the
subject, and other focused on neural dynamics
\cite{Schuster,Healthy,Gross2014,Plenz-Functional,Beggs-criticality,Chialvo2008,Chialvo2010,Breakspear-review},
gene networks \cite{Serra-review}, and collective motion
\cite{Vicsek2012}, respectively.  The present paper aims at
overviewing and complementing them, putting the emphasis
on dynamical aspects, and discussing together empirical evidence and
theoretical approaches.

\section{Criticality and scale invariance}
\label{criticality}

Many discussions about ``criticality'' are semantic ones. Depending on
authors and fields rather diverse contents are assigned to terms such
as ``critical'', ``quasi-critical'', ``dynamically critical'',
``generically critical'', or ``self-organized critical''. Given the
broad audience this paper is aimed at, we esteem that a section
devoted to present a synthetic overview of basic concepts and to fix
ideas and notation is necessary.\footnote{For a more exhaustive
  introduction to critical phenomena we refer to the standard
  literature; e.g.
  \textcite{Stanley-book,Binney,Moloney-book,Marro,Henkel,Sethna-book}.} Readers
familiar with these concepts can skip it.

\subsubsection{Scale-invariance and power laws}
In a seminal paper entitled ``Problems in Physics with many scales of
length'' K. Wilson emphasized that ``one of the more conspicuous
properties of nature is the great diversity of size or length
scales'', and cited oceans as an example where phenomena at vastly
disparate wavelengths coexist \cite{Wilson1979}. Different scales are
usually decoupled and the ``physics'' at each one can be separately
studied. However, there are situations --known as scale-invariant or
scale-free-- where broadly diverse scales make contributions of equal
importance. A remarkable instance of this --but just an example-- are
the critical points of continuous phase transitions where the
microscopic, mesoscopic and macroscopic scales are all alike.

Power-law (or Pareto) distributions such as $P(x) = A x^{-\alpha}$,
where $\alpha$ is a positive real number and $A$ a normalization
constant, are the statistical trademark of \emph{scale-invariance} or
``\emph{scaling}''\footnote{A well-known example is the
  Guttenberg-Richter equation for the probability distribution of
  observing an earthquake of dissipated energy $E$,
  $P(E) \propto E^{-\alpha}$, \cite{Corral-quakes}.}.  Actually, they
are the only probability distribution functions for which a change of
scale from $x$ to $\Lambda x$, for some constant $\Lambda$, leaves the
functional form of $P(x)$ unaltered, i.e.
$P(\Lambda x) = A(\Lambda x)^{−\alpha} = A
\Lambda^{−\alpha}x^{−\alpha} = \Lambda^{−\alpha}P(x)$, in such a way
that the ratio $P(\Lambda x)/P(x)=\Lambda^{-\alpha}$ does not depend
on the variable $x$, i.e. it is scale invariant
\cite{Newman-powerlaws,Sornette-book}.  As opposed to e.g. exponential
distributions, power-laws lack a relevant characteristic scale,
besides natural cut-offs.

Distributions with power-law tails appear in countless scenarios,
including the statistics of earthquakes, solar flares, epidemic
outbreaks, etc.
\cite{Mandelbrot,Newman-powerlaws,Sornette-book,West-Scale}. They are
also a common theme in biology
\cite{Stanley2002,heart,Meijer,Gisiger,West-physio}. For example,
physiological and clinical time-series data have typically a spectrum
that decays as a power of the frequency \cite{Mandelbrot-1/f} and
mobility patterns often exhibit scale-free features
\cite{Bara-human,Geisel-human,Proekt}.  Moreover, a number of
commonly-observed statistical patterns of natural-world data --such as
Zipf's law\footnote{This states that the frequency with which a given
  pattern is observed declines as a negative power law of its
  \emph{rank}, i.e. its position in the list of possible patterns
  ordered from the most frequent to the rarest one
  \cite{Zipf}.},\footnote{A very elegant and illuminating approach
  allowed Mora and Bialek to map the Zipf's law to underlying
  statistical criticality in a very precise way \cite{Mora-Bialek}.
  Within this setting, it was observed, however, that the Zipf's law
  (and its concomitant statistical criticality) may emerge rather
  generically if there is a fluctuating unobserved (hidden) variable
  that affects the system, such as e.g a common input, even in systems
  not tuned to criticality \cite{Mehta,Latham}; see also
  \textcite{Bialek-thermo2} for a discussion of these issues and how
  can they influence the conclusions about statistical criticality of
  empirical data.}
\cite{Sornette-book,Mora-Bialek,Visser,Minnhagen,Marsili-Zipf},
Bendford's law \cite{Benford,Pietronero}, and Taylor's law
\cite{Taylor,Cohen,Cohen+Amos}-- stem from underlying power-law
distributions.

Disputes on the validity and possible significance of power-laws have
a long history in diverse research fields. For some authors they reveal
fundamental mechanisms, while some others perceive them as largely
un-informative \cite{Stumpf,Linken-cognitive} or even
``more-normal-than-normal'' distributions \cite{More-normal}.  Still,
in some cases, there is very robust evidence of scale invariance and
it certainly provides valuable insight\footnote{An important example
  are allometric scaling laws, which are power-law relationships
  between different measures of anatomy/physiology
  \cite{Kleiber1932,West,Amos-form,Amos-form2}. These have been
  elegantly shown to stem from the constraint that living systems have
  an underlying optimal (e.g nutrient) transportation network
  \cite{Amos-networks,Simini2010}.}.

The detection and statistical characterization of power-law
distributions in real-world data is often hindered by sampling
problems since very rare but large events control the
statistics. Accordingly, the quality of power-law fits to empirical
data has been recently scrutinized, showing that many claims of
scale-invariance actually lack statistical significance and,
presently, more stringent statistical tests have become a must
\cite{Clauset}.

\begin{figure}
  \includegraphics[scale=0.41]{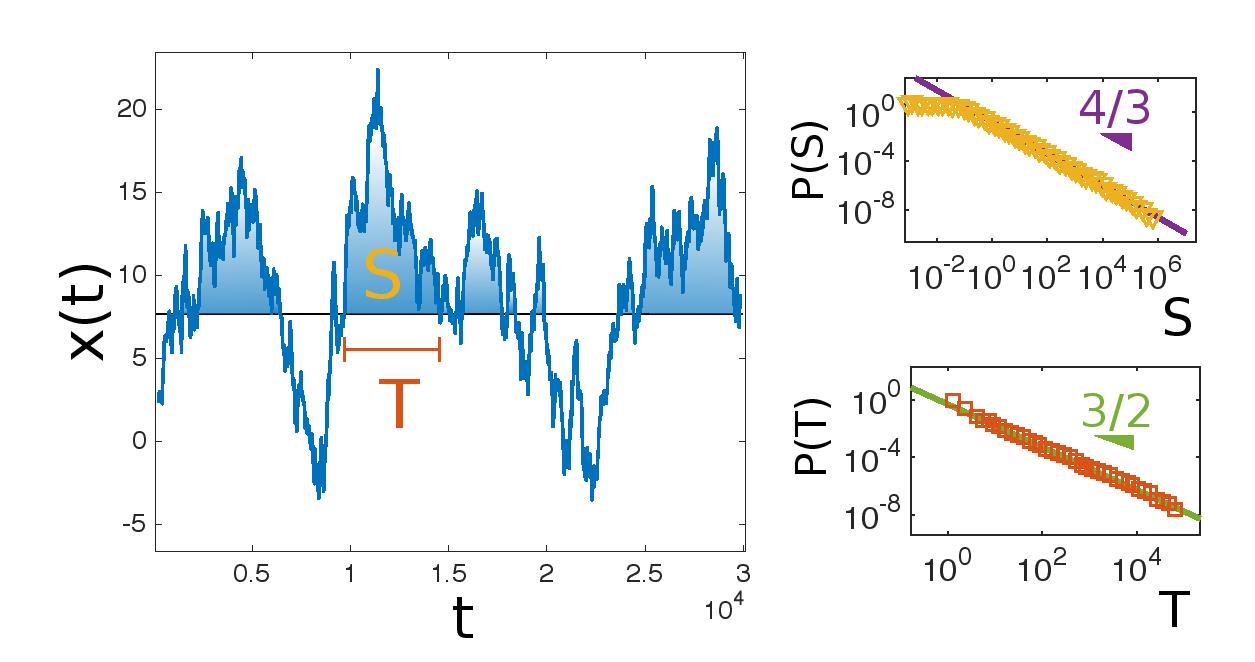}
  \caption{Random walks, such as the one illustrated in the left
    panel, lack a characteristic scale. As a consequence, the
    distribution of return times to the origin, $T$, of the
    one-dimensional (unbiased) random walk obeys
    $P(T)\sim T^{-\alpha}$ with $\alpha=3/2$ and the areas/sizes, $S$,
    covered by their excursions before returning to the origin
    (i.e. ``avalanches'') obey $P(S) \sim S^{-\tau}$ with $\tau=4/3$
    (right panels) \cite{Redner-book,Serena-BP}. Some biological
    systems exhibit scaling as a consequence of an underlying
    random-walk process; see e.g. \textcite{Mandelbrot-RW,Berg}.}
\label{RW}
\end{figure}

From the mathematical side, very diverse explanatory mechanisms for
the emergence of scaling in empirical data have been put forward
\cite{Mitzenmacher,Newman-powerlaws,Sornette-powerlaws,Gros-powerlaws,Willis}.
For example, random walks give rise to power laws in the distribution
of return times and ``avalanche'' sizes as illustrated in
Fig.\ref{RW}.
Other examples are: (i) Underlying multiplicative processes
\cite{Sornette+Cont,Sornette-multiplicative,Solomon,Reed}, (ii)
Preferential attachment processes \cite{Yule,Simon,BA}. (iii)
Optimization and constrained optimization \cite{HOT,Seoane}.

Even if --as the previous enumeration illustrates-- empirical
power-law distributions can in principle be ascribed to a handful of
possible different generative mechanisms, in the forthcoming sections
we discuss the most prominent and general mechanism, able to account
for scale invariance both in space and time in a rather robust,
powerful, and universal way: \emph{criticality}.

\subsubsection{Criticality in equilibrium systems and beyond}
The concept of criticality was born in the context of systems at
\emph{thermodynamic equilibrium}. A paradigmatic example are
ferromagnets. These exhibit a continuous/second-order phase transition
at a critical temperature, $T_c$, below which the orientational
symmetry of spins is spontaneously broken --i.e. a preferred direction
emerges-- and, progressively, more ordered/magnetized states emerge as
the temperature is lowered. On the other hand, above $T_c$ thermal
fluctuations dominate and the system remains disordered. This change
in the collective state is usually encoded in an \emph{order parameter}
(e.g. the overall magnetization) which measures the degree of order as
the phase transition proceeds.

The described \emph{symmetry-breaking} is a collective phenomenon that
requires a system-wide coordination for the global re-organization to
emerge. This implies that the correlation length among individual
components needs to span the whole systems at criticality. Similarly,
 when the system is becoming incipiently ordered, it is
highly fluctuating in the orientation to be chosen. For example,
classical experiment with liquid-gas transitions (e.g. with $CO_2$)
shows that, right at criticality, light of many different wavelengths
scatters with internal structures of the mixture (i.e. there are
density fluctuations of all possible length scales), causing the
normally transparent liquid to appear cloudy in a phenomenon called
critical opalescence \cite{Binney,Stanley-book}.

Importantly, the concepts and methods developed in the context of
equilibrium systems were soon extended to time-dependent and
non-equilibrium problems
\cite{HH,Marro,Hinrichsen,Tauber,Henkel,Kamenev,Tauber2017}.  All
along this paper, we adopt a view of criticality and phase transitions
focused mostly on dynamical and non-equilibrium aspects. This seems to
be the most natural choice to analyze living systems, which are
dynamical entities kept away from thermal equilibrium by permanently
exchanging energy and matter with their surroundings. It is important
to underline that there exists an important alternative
``statistical-criticality'' approach to the analysis of biological
data.  It focuses on the statistics of existing configurations
(without regard to the temporal order in which they appear, much as in
equilibrium statistical mechanics) rather than on possible underlying
dynamical processes, and it is only briefly discussed here where, as
said above, we choose to focus on dynamical aspects.

\subsubsection{Non-equilibrium phase transitions: an example}

In order to turn the foregoing wordy explanations into a more formal
approach, we describe in detail --as a guiding example-- one of the
simplest possible dynamical models exhibiting a non-equilibrium phase
transition.
\begin{figure}
 \includegraphics[scale=0.35]{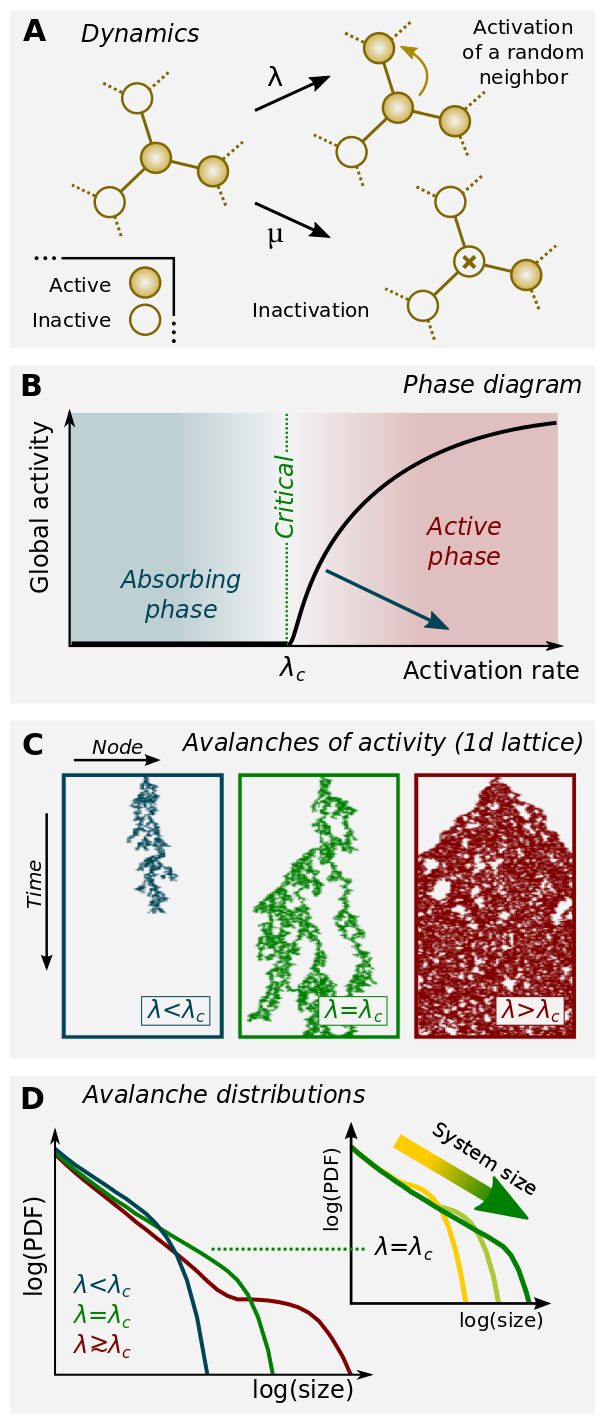}
 \caption{Sketch of the main aspects of the contact process. (A)
   Dynamical rules. (B) Phase diagram, including a critical point. (C)
   Temporal raster plots of activity (avalanches) in the different
   regimes, illustrating the complex patterns emerging at criticality,
   which involve many different scales. (D) Avalanche size
   distributions in the different phases (main) and, right at the
   critical point for different system sizes (inset), illustrating
   finite-size scaling, i.e. the emergence at criticality, of a
   straight line in a double-logarithmic plot, as corresponds to scale
   invariance (see also Fig.3).}
\label{CP}
\end{figure}
The \emph{contact process} (CP) is a prototypical toy model to study
the dynamics of propagation of some type of ``activity'' (as
e.g. infections in epidemic spreading; see Fig.2)
\cite{Harris,Hinrichsen,Marro,Henkel}. At any given time, each of the
nodes $i=1,2...N$ of a given network (which in particular can be a
lattice, a fully connected network, or one with a more complex
architecture, describing the pattern of connections among units/nodes)
is in a state $s_i$ that can be either occupied/active ($s_i=1$) or
empty/quiescent ($s_i=0$).  Occupied sites are emptied at rate $\mu=1$
and new active nodes are created at (empty) randomly-selected nearest
neighbors of active ones at rate $\lambda$.  Considering, for the sake
of simplicity, a fully connected network with $N$ nodes and performing
a large-$N$ expansion of the corresponding Master equation
\cite{vanKampen}, one readily obtains a 
``mean-field'' or deterministic equation:
\begin{equation} \dot{\rho}(t) = \lambda \rho(t) (1- \rho(t)) - \rho(t)=
  (\lambda-1) \rho(t)-\lambda \rho^2(t)
\label{rho}
\end{equation}
where the dot stands for time derivative of the activity density
$\rho=\sum_{i=1}^{N}s_i/N$.  This simple one-variable approximation
already illustrates some of the essential features of
criticality. Eq.(\ref{rho}) reveals the presence of a bifurcation at a
value $\lambda_c=1$, separating a subcritical (also called
``absorbing'' or ``quiescent'') phase ($\lambda < 1$) in which
transient activity decays to the only possible steady-state,
$\rho_{st}=0$, from a supercritical (or ``active'') one ($\lambda>1$)
with a sustained activity $\rho_{st}=1-1/\lambda$ (see
Fig.\ref{CP}). Thus, defining $\delta=|\lambda-1|$ as the distance to
criticality, $\rho_{st}\sim \delta$ for small $\delta$. In the
quiescent (or absorbing) phase\footnote{A similar argument holds in
  the active phase.}, an initial density decays exponentially,
$\rho(t) =\rho(0) \exp(-\delta t)$, implying that there is a
characteristic time scale proportional to $\delta^{-1}$. Note that
such time diverges at criticality, i.e. it takes a huge time for the
system to ``forget'' its initial state, reflecting a generic feature
of criticality: the so-called ``critical slowing down''. Indeed, right
at the critical point, the activity decays asymptotically as a
power-law, $\rho(t) \sim t^{-1}$.

Introducing an external field that creates activity at empty sites at
rate $h$, the overall response or ``susceptibility'', defined as
$\Xi =\frac{\partial \rho_{st}}{\partial h}|_{h \rightarrow 0}$, is
$\Xi \propto\delta^{-1}$ that, again, diverges right at $\delta=0$,
(i.e. $\lambda=1$), illustrating the diverging response to
infinitesimal perturbations, another important generic feature of
criticality.

A useful tool to analyze this type of transitions consists in
performing ``spreading experiments'' in which the evolution of a
single localized seed of activity in an otherwise absorbing/quiescent
state is monitored (see Fig.\ref{CP}C).  In this case, given the small
number of active sites, the dynamics is chiefly driven by fluctuations
and cannot be analyzed within the deterministic approximation above.
Stochastic cascades of spatio-temporal activity, or ``avalanches'' of
variable sizes and durations can be generated from the initial seed
before the system returns to the quiescent state (extinction). In this
framework the critical point separates a regime of sure extinction
(absorbing phase) from one of non-sure extinction (active phase).
Right at the critical point, the sizes and durations of avalanches are
distributed as power-laws with anomalously large (formally infinite)
variance (Fig.\ref{CP}C) \footnote{The large variability of possible
  patterns is a generic key feature of criticality.  In particular, in
  systems at equilibrium, the divergence at criticality of the
  specific heat reflects the huge variability of possible internal
  states \cite{Binney}.}.  To understand this mathematically, one needs
the next-to-leading correction to Eq.(\ref{rho}) in the large-$N$
expansion to include the effect of ``demographic'' fluctuations. This
leads to an additional term $+\sqrt{\rho} \eta(t)$, where $\eta(t)$ is
a Gaussian white noise of variance $\sigma^2 =(\lambda+1)/N$.
\footnote{The square-root noise stems from the central limit theorem
  \cite{vanKampen}.}

A simple analysis of the resulting stochastic equation\footnote{See
  \textcite{Serena-BP} for a pedagogical derivation of this.} shows
that right at the critical point, the time required to return to the
quiescent state, i.e. the avalanche-durations $T$ are distributed as
power laws: $F(T)\sim T^{-\alpha}$ with $\alpha=2$; similarly,
avalanches sizes $s$ obey $P(S)\sim S^{-\tau}$, with $\tau=3/2$ These
mean-field exponents coincide with those of the (Galton-Watson)
unbiased branching process \cite{Watson,Harris,Liggett}, introduced to
describe the statistics of family-names, and often employed to
illustrate the statistics of critical avalanches. Away from
criticality, as well as in finite systems, cut-offs appear in the
avalanche distributions (see Fig.\ref{CP}). In particular, as a
reflection of the underlying scale-invariance at criticality, the
finite-size cut-offs obey scaling laws such as
\begin{equation}
P(S,N) \sim S^{-\tau} {\cal{G}}(S/N)
\label{FSS}
\end{equation}
where the power-law $S^{-\tau}$ is cut-off by an unspecified function,
${\cal{G}}$, at an $N$-dependent scale
\cite{Binder,Stanley-book,Binney}.  This enforces that plotting
$P(S,N) S^{\tau}$ as a function of the rescaled variable $S/N$ should
give a unique curve into which all individual curves for different
sizes $N$ collapse. This \emph{finite-size scaling} method constitutes
an important tool for analyzing critical phenomena (both in computer
simulations and in experiments) as perfect power-laws/divergences can
only appear in the infinite-size limit, not reachable in biological
problems. Indeed, while in finite systems true criticality does not
exist, still, these may exhibit a progressive transition between order
and disorder. This can be characterized by the existence of a peak in
some quantity such as the susceptibility or the correlation length
that usually diverge at (true) criticality; this is used as a proxy
for ``approximate'' criticality in finite systems\footnote{Similarly,
  systems in the presence of an external driving force are not truly
  critical; in these cases, the Widom line --signaling e.g. the
  position of maximal susceptibility or correlation-- can be taken as
  a surrogate of criticality \cite{Rashid}.}.

As a result of universality, all models exhibiting a phase transition
to an absorbing/quiescent phase (without any additional symmetry or
conservation law) share the same set of critical exponents and scaling
functions --i.e. the same type of scale-invariant organization-- with
the contact process \cite{Henkel}\footnote{To study
  spatial effects one needs to replace $\rho(t)$ in Eq.(\ref{rho}) by
  a field $\rho(x,t)$ and to introduce a diffusive coupling term
\cite{Henkel,Hinrichsen,Odor}. }

Even if the simple propagation model discussed above is not intended
as a faithful description of the actual dynamics of any specific
biological system, in some cases --such as neural and gene regulatory
networks-- it can constitute an adequate effective representation of
``damage spreading'' experiments, in which two identical replicas of
the same system are considered; a localized perturbation in the state
of one unit/node is introduced in one of the two, and the difference
between both replicas is monitored as a function of time
\cite{Derrida}. Depending on the the system dynamical state, such
perturbations may grow (active phase), shrink (quiescent phase), or
fluctuate marginally (critical point), providing a practical tool to
gauge the level of internal order\footnote{The precise relationship
  between the damage spreading threshold and the system's actual
  critical point is an important and subtle issue
  \cite{damage1,damage3,damage4}.}.  Even if the actual dynamics might
be much more complicated, the resulting damage spreading process is
susceptible to be described in simple terms if local effective error
``propagation'' and error ``healing'' rates can be estimated.

\subsubsection{Self-organization to criticality}
As we have seen criticality requires of parameter fine tuning to a
precise point to be observed.  How it is possible that natural systems
(such as earthquakes, Barkhaussen noise, etc.)  exhibit signatures of
criticality, but without any apparent need for parameter tuning to
settle them in at the edge of a phase transition?  To answer this
question P. Bak and collaborators introduced the important concept of
``self-organized criticality'' (SOC) through a series of archetypical
models \cite{Bak,BT,Dhar,Frette,OFC,FF1,FF2,Corral}, including its
most famous representative: the \emph{sandpile} model \cite{BTW}.

In the sandpile model a type of ``stress'' or ``energy'' (sandgrains)
accumulates at a slow timescale at the sites of a (two-dimensional)
lattice, and when the accumulated stress overcomes a local instability
threshold, it is instantaneously redistributed among nearest neighbor
sites --and, possibly, released/dissipated at the system boundaries.
This can create a cascade or ``avalanche'' of further
instabilities. Remarkably, the durations and sizes of such avalanches
turn out to be distributed as power laws, i.e. the system becomes
critical without any apparent need for fine tuning\footnote{Stochastic
  variants of the original (deterministic) sandpile model
  \cite{Manna,Oslo} show much cleaner scaling behavior than it
  \cite{anomalies1,anomalies3}.}
\cite{BTW,Bak,Jensen,Moloney-book,Pruessner-book,SOC-25,BJP,Turcotte}.
\begin{figure}
\includegraphics[scale=0.45]{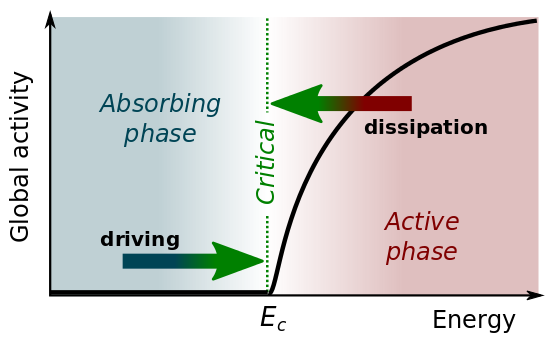}
\caption{The self-organization-to-criticality (SOC) mechanism works by
  establishing a feedback loop between the dynamics of the activity
  and that of the control-parameter (total accumulated
  energy/stress/sandgrains) at separated timescales. In particular,
  the control parameter itself becomes a dynamical variable that
  operates in opposite ways depending on the system's state:
  \emph{fast dissipation} (negative force) dominates while the control
  parameter lies within the active phase and by \emph{slow driving}
  dynamics (positive force) dominates in the absorbing/quiescent
  phase.  This feedback self-organizes the system to the critical
  point of its second-order phase transition if the separation between
  slow and fast timescales is infinitely large and the dynamics is
  conservative \cite{BP,SOBP,FES-PRL,FES-PRE,JABO1}.  Otherwise, the
  system is just self-organized to the neighborhood of the critical
  point with excursions around it, i.e.  ``self-organized
  quasi-criticality'' \cite{BJP,JABO1}.}
\label{SOC}
\end{figure}
The mechanism for self-organization to criticality in sandpile models
is described in Fig.\ref{SOC}.  It can be seen that it is
characterized by a dynamical feedback that acts differentially
depending on the actual system state. This is just an example of a
broader class that has been extensively analyzed in the context of
\emph{control theory}
\cite{Sontag,Magnasco-anti,Sornette-sweeping,SOB}, which is very
likely to emerge in biological systems, as we shall discuss. Two
important variants of this mechanism are as follows:

(a) \emph{Self-organized quasi-criticality} is analogous to SOC but
occurs when the dynamics is non-conservative and/or when the
separation of timescales is not perfect (relevant for biological
problems). This self-organization mechanism drags the system back and
forth around the critical point without sitting exactly at it, and is
able to generate effective scale-invariance across quite a few scales
\cite{JABO1,Kinouchi2018}.

(b) \emph{Adaptive criticality} is a variant of SOC from a network
perspective, in which connections among nodes in a network are
susceptible to be added, removed, or rewired depending on the system's
dynamical state, creating a feedback loop between network architecture
and dynamics in a sort of co-adaptive process.\footnote{Different
  variants of this idea have been proposed in the literature
  \cite{BR2000,Bornholdt2012,Rohlf2008,Meisel-Gross,Saito,Sole-adaptive,Anne-Ly,Perotti,Kuehn,Macarthur,Doro-evol,Gros-book,Blasius}.}
This mechanism can drive the dynamics to criticality
\cite{Doro-critical,Bassler,Bianconi2004} and, in parallel, the
network architecture develops a highly non-random structure, thus
capturing the feedback between dynamics and architecture in
actual biological networks.

\subsubsection{Classes of criticality}
Not all dynamical phase transitions of relevance in biology occur
between quiescent and active phases, nor can be described by an
associated activity-propagation process, such as the contact
process. Other important classes of phase transitions to be found
across this paper are: (i) synchronization transitions, at which
coherent behavior of oscillators emerges, as described by the
prototypical \emph{Kuramoto model}
\cite{Kuramoto,Kuramoto-review,synchro-book}. (ii) transitions to
collective ordered motion, as represented for instance by the
\emph{Vicsek model} \cite{Vicsek,Vicsek2012}) and its variants; and
(iii) percolation transitions \cite{Moloney-book}, and (iv) even
(thermodynamic) transitions such as that of the Ising model
\cite{Binney}, to name but a few. Each of these classes has its own
type of emerging ordering and its own scaling features. However, all
of them share the basic features that constitute the fingerprints of
criticality, such as diverging correlations and response, large
variability, scale invariance, etc.

\subsubsection{Criticality on complex networks}

Thus far we have discussed criticality in homogeneous
systems. However, in many biological problems the substrates on top of
which dynamical processes run are highly heterogeneous
\cite{Barabasi-Review,Newman-Review,Newman-book,Caldarelli-book}. In
particular, complex systems, including biological ones, can be
described as networks, where nodes represent units (neurons, genes,
proteins, ...)  and links stand for allowed pairwise interactions
among them. Such complex networks have been found to exhibit one or
more of the following important architectural features: (i) large
heterogeneity with a few highly connected nodes and many loosely
connected ones; actually the distribution of connection can be scale
free \cite{BA}, (ii) the small-world property \cite{SW}, (iii) modular
organization\footnote{Biology is ``modular'' in many aspects
  \cite{Bara-meta,Alon-book}, meaning that some components in
  biological networks (nodes) are connected among themselves more
  often or more strongly that they do with others
  \cite{Alon-bionetworks}.}, (iv) hierarchical organization,
etc. \cite{Corominas1}.  These structural features usually entail
profound implications on the dynamics of processes running on top of
them \cite{Doro-critical,Romu-Review,Yamir-review,Vespi-book}. For
instance, synchronization transitions proceed in a stepped way on
modular networks \cite{Arenas-synchro}, and broad critical-like phases
can emerge in hierarchical modular networks (as discussed e.g. in
\textcite{GPCN,Moretti}, Appendix A and in what follows).

\subsubsection{Generic scale invariance}
\label{generic1}
We have discussed the paradigm of a critical point --with its
concomitant spatio-temporal scale-invariance-- separating two
alternative phases. However, in some systems with peculiar symmetries,
conservation laws or structural disorder, critical-like features may
appear in extended regions in the phase space and not just at a
critical point. This is called \emph{generic scale invariance}
\cite{GG91} and can account for empirically reported scale-invariance
in some biological problems without the need to invoke precise tuning
to criticality.  Mechanisms for the emergence of generic scale
invariance are discussed in Appendix A.

\subsubsection{Statistical criticality}
To end this introductory section, we briefly discuss an (already
mentioned above) alternative perspective to criticality, particularly
useful to analyze the wealth of high-quality data now available for
living systems \cite{Mora-Bialek}. It relies on the idea that some
fundamental questions in biology can be tackled within a probabilistic
setting (for instance, analyzing the statistics of spiking patterns
may help deciphering the way in which neurons encode information)
\cite{Bialek1995}.  Bialek and coworkers developed a data-driven
maximum entropy (statistical physics) approach to biological problems,
that consists in approximating the probability distribution of
different patterns in a given dataset by a probabilistic model that
consistently reproduces its main statistical features (e.g. mean
values and pairwise correlations; see Appendix B). The resulting
models are akin to the Ising models.\footnote{And since the inferred
  interactions among ``spins'' have both signs, they are a sort of
  spin glasses \cite{Bialek-thermo1,spin-glass}.}  Rather remarkably,
Bialek and collaborators observed that the emerging probabilistic
models for a number of high-dimensional problems --including
biological ones, from retinal neural populations
\cite{Bialek-weak,Bialek-thermo1,Bialek-thermo2,Bialek-searching} to
flocks of birds \cite{Bialek-birds} and the immune system
\cite{Bialek-antibody}, for which excellent empirical data sets are
available-- have parameter values sitting close to the edge of a phase
transition, i.e the emerging probabilistic models seem to be critical
in a very precise sense \cite{Mora-Bialek} (see Appendix B).

%%%%%%%%%%%%%%%%%%%%%%%%%%%%%%%%%%%%%%%%%%

\section{Functional advantages of criticality}
\label{s-edge}
Having discussed basic aspects of criticality and scale invariance, we
move on to ask: what are the potential virtues of them susceptible to
be exploited by living systems to enhance their functionality?  To
shed light onto this, we first describe a well-understood
case in which both theoretical and empirical evidence match, and where
the essential and beneficial role played by criticality in a
biological system is clear and illuminating. Later on we discuss a
set of possible functional advantages of criticality from a general
perspective.

\subsection{Criticality in the auditory and other sensory systems}
The inner ear of vertebrates is able to detect acoustic stimuli with
extraordinary sensitivity and exquisite frequency selectivity across
many scales \cite{Hud-Review}.  At the basis of these exceptional
features there are \emph{hair cells}, the ear's sensory receptors,
which oscillate spontaneously even in the absence of stimuli, being
able to resonate with acoustic inputs
\cite{Gold,Hudspeth1,Julicher2001}. Intrinsic oscillations are either
damped or self-sustained depending on the concentration of Calcium
ions, with a Hopf bifurcation separating these two regimes.
Empirical
evidence reveals that the ion concentration is regulated in such a way
that hair cells operate in a regime very close to the Hopf bifurcation
\cite{Eguiluz2,Camalet}. This has been argued to entail important
consequences for signal processing
\cite{Eguiluz1,Hudspeth1,Julicher2001,Julicher2010}, as we discuss
now.

In the simplest possible setting, a hair cell can be effectively
described as Hopf oscillator \cite{Strogatz-book}: 
\begin{equation}
  \dot{\phi}(t) = (a + i \tilde{\omega}) \phi(t) - |\phi|^2 \phi(t) 
\label{Hopf}
\end{equation}
where the $\phi$ is a complex number, $\tilde{\omega}$ the
resonance frequency, and $a$ is the control parameter (ion concentration)
setting the dynamical regime. Eq.(\ref{Hopf}) exhibits self-sustained
oscillations of the form $\phi(t)= \sqrt{a} e^{i \tilde{\omega} t}$ if
$a>0$, while if $a<0$ oscillations are damped.\footnote{See
  \textcite{Stoop2003} from where this discussion is adapted.}
Introducing stimuli of the characteristic frequency
$\omega=\tilde{\omega}$ and small amplitude $F$ (i.e. adding
$+F e^{i \tilde{\omega} t}$ to Eq.(\ref{Hopf})), and writing
$\phi (t)=R(t) e^{i \omega t}$, one finds
\begin{equation}
\dot{R}(t) = R(t) [a- R^2(t)] + F.
\label{bifurcation}
\end{equation}
In the oscillatory regime, $a>0$, the response $R$ is proportional to
the input amplitude $F$.  However, at the bifurcation (or critical)
point, $a=0$, the response $R$ is strongly non-linear, as $R=F^{1/3}$
and, consequently, the ratio response-to-signal $R/F = F^{-2/3}$
diverges at $F\rightarrow 0$, leading to a huge response to tiny
signals of the characteristic frequency.  On the other hand, if the
input has some other frequency $\omega \neq \tilde{\omega}$ the
response is much smaller.  This entails an extremely efficient
frequency-selection and amplification mechanism, vividly illustrating
the advantage of working close to the instability point.

The described phenomenon involves a single hair-cell with a specific
intrinsic frequency and it is thus not a collective critical
phenomenon. However, the Cochlea is arranged in such a way that it
involves an (almost uni-dimensional) array of diverse and coupled hair
cells.  When coupling many different Hopf oscillators results in the
emergence of a true phase transition --i.e. a critical point with
scale-free avalanches-- which entails sharpened frequency response
\cite{Magnasco2003,Duke2003} and enhanced input sensitivity
\cite{Stoop2003,Stoop2015,Stoop2016}.

Summing up, woking at criticality has been shown to be essential to
generate the extraordinary features of vertebrate hearing, even the
most intricate ones \cite{Stoop2016}.  Similar virtues of criticality
have been explored in the olfactory system \cite{Olfactory} and the
visual cortex \cite{Shew2015b} (see also
\textcite{Chialvo-senses,Deco2012}).

\subsection{Exploiting criticality} 

\subsubsection{Maximal sensitivity and dynamic range}
\begin{figure}
  \includegraphics[scale=0.31]{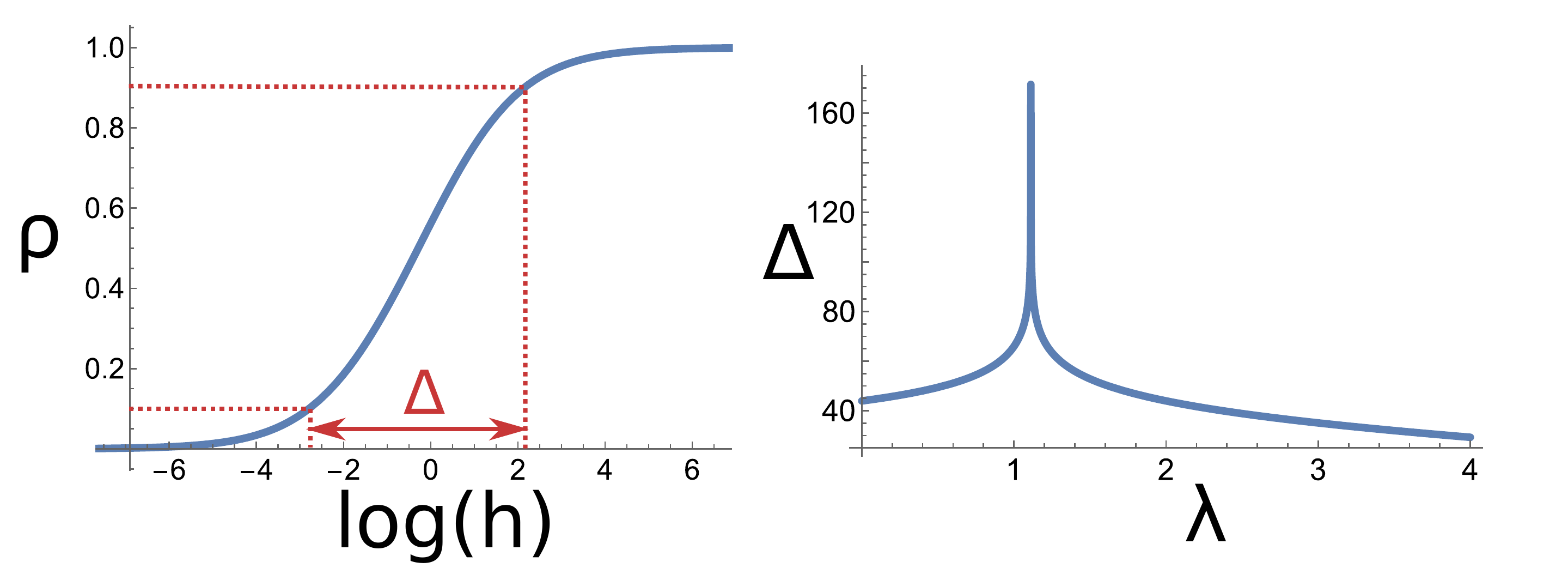}
  \caption{Sketch of the behavior of the dynamic range, defined as ,
    near a critical point. (Left) Steady state density $\rho$ as a
    function of the driving force $h$ (in log scale) for a given value
    of the control parameter $\lambda$; the dynamic range, $\Delta$,
    defined as $\Delta = 10 \log[h(\rho=0.9)/h(\rho=0.1)]$, signals
    the interval where distinguishable responses (i.e. values of
    $\rho$) can be measured. (Right) $\Delta$ exhibits a pronounced
    peak at criticality.}
\label{DR}
\label{peaks}
\end{figure}

As discussed above, an important trademark of critical points is the
divergence of the response (or susceptibility) which is likely to be
exploited in biological sensing systems, needing to optimize their
response to environmental cues.  To better quantify this, a related
quantity, dubbed \emph{dynamic range}, was introduced in
\textcite{Kinouchi-Copelli}. Consider a model for activity propagation
(similar to the contact process) with a critical point ($\lambda_c=1$)
running on a random network, under the action of an external stimulus,
$h$, able to create activity at empty nodes. The dynamic range,
$\Delta$ (see Fig.\ref{DR}) gauges the range of diverse stimuli
intensities where variations in input $h$ can be robustly coded by
variations in the response, discarding stimuli that give almost
indistinguishable outputs. $\Delta$ turns out to exhibit a marked peak
at $\lambda_c=1$, indicating that, at criticality, discriminative
outputs can be associated to a very large variety of inputs, with
obvious functional advantages for signal detection and processing.

\subsubsection{Large correlations }
The emergence of arbitrarily large correlation lengths at criticality
is an important feature susceptible to be exploited by living systems
in order to induce coordinated behavior of individual units across
space and time. This can be relevant for coordination purposes in
e.g. neural systems where coherent behavior across extended areas is
observed \cite{Taglia}, in flocks of birds \cite{Cavagna2010} and in
micro-organism colonies \cite{Endres}.  Similarly, the emergence of
very large correlation times and critical slowing down may provide
biological systems with a useful mechanism for the generation of
long-lasting and/or slow-decaying memories at multiple timescales (see
e.g. \textcite{Deco-Jirsa}).

\subsubsection{Statistical complexity and large repertoires}
The variability of possible spatio-temporal patterns is maximal at
criticality (as illustrated in Fig.2); this may allow biological
systems to exhibit a very wide spectrum of possible responses,
sometimes called ``dynamical repertoire'' \cite{Ramo,Ramo2,
  Shew2012}. This is consistent with the finding that e.g.  models for
brain activity reach highest signal complexity, with a variety of
attractors and multistability when operating near criticality
\cite{Deco-Jirsa,Haimovici}. Similarly, (i) the number of metastable
states \cite{Haldeman}, (ii) the variability of attractors to support
memories \cite{Lucilla2010,Basin}, and (ii) the diversity in
structure-dynamics relationships \cite{Nykter2} have been predicted to
be maximized at criticality. All this suggests that in order to
spontaneously generate complex patterns --required e.g. to store
highly diverse tokens of information-- operating near criticality can
be an excellent solution for living systems. As a consequence of this,
the capacity to store and process information is optimal at
criticality, as we discuss in more depth in what follows.

\subsubsection{Computation exploiting criticality}
It was long-ago conjectured that the extraordinary ``computational
power'' of living systems could be the result of collective behavior,
emerging out of a large number of simple components
\cite{Hopfield1982,Amari,Grossberg1,Grossberg2}.  By ``computation''
it is usually meant an algorithm or system that --with the aim of
performing some task-- assigns outputs to inputs following some
internal logic. Thus, the computational power of a given device is
quantified by estimating the amount and diversity of associations of
inputs to outputs that it can support.  As first suggested in
\cite{Turing1950,Ashby} and much further developed in the context of
machine learning \cite{Packard1988,Crutchfield1988,Langton1990,Li1990}
networked systems operating at criticality can have exceptionally high
computational capabilities. In particular, \citeauthor{Langton1990}
formulated the question: \emph{under what conditions will physical
  systems support the basic operations of information transmission,
  storage, and modification, required to support computation?'} His
answer was that systems\footnote{Cellular automata in this case
  \cite{Wolfram2002}.} operating at the ``edge of chaos'' are
especially suitable to perform complex computations\footnote{This
  proposal triggered a heated debate; see, e.g.
  \cite{Crutchfield1988,Mitchell,Crutchfield2012}. }. The ``edge of
chaos'' or critical point (as we rather call it here) is the
borderline between two distinct phases or regimes: the
chaotic/disordered one in which perturbations and noise propagate
unboundedly (thereby corrupting information storage) and the
frozen/ordered phase whereas changes are rapidly erased (hindering the
capacity to react and transmit information). Therefore, the critical
point confers on computing devices composed of equivalent units an
optimal tradeoff between information storage and information
transmission, two of the key ingredients proposed by Turing as
indispensable for universal computing machines \cite{Turing1950}.

In artificial intelligence, criticality is exploited in so-called
``\emph{reservoir computing}''
\cite{Reservoir-review} that was developed
independently in the fields of machine learning (``echo state
networks'' of \textcite{Jaeger}) and computational neuroscience
(``liquid state machine'' in \textcite{Maass2002}). These machines
consist of a network of nodes and links, ``the reservoir'', where each
node represents an abstract ``neuron'' and links between them mimic
the connectivity of actual biological circuits.  A series of seminal
works showed that such machines can perform real-time computations
--responding rapidly to time varying input signals-- in a coherent yet
flexible way if they operate near a critical point
\cite{Maass2002,Ber-Nat,Ber-Nat2,Legenstein2007,Asada}.

These ideas are corroborated by information-theoretic analyses
\cite{Cover-Thomas}, which have unveiled that the overall transmission
of information between units in a network --as measured by diverse
indicators\footnote{Such as the transfer entropy
  \cite{Sole1995,Lizier-transfer,Shriki2016}, Fisher information
  \cite{FI} and, more in general, statistical complexity (as discussed
  above) \cite{Ramo,Lizier,Basin}.}-- is maximal if the underlying
dynamical process is critical\footnote{See
  \textcite{Proko,Li1990,Beggs-criticality,Ribeiro2008,Luque}; and
  \textcite{Abbott,Barnett} for a discrepant view.}.

Let us also mention that (i) state-of-the-art deep learning machines
\cite{Deep} may rely on some form of intrinsic scale invariance or
even criticality
\cite{Mehta-RG,Tegmark,Deep-Marsili,Toth1,Toth2,Ringel}, opening
exciting research avenues to understand how artificial-intelligence
machines achieve their extraordinary performance, and (ii) from the
empirical side, recent work has revealed that a mechanism akin to
reservoir computing enables neuronal networks of the cerebellum to
perform highly complex tasks in an efficient way by operating at
criticality \cite{cerebellum}.

\section{Alleged criticality and scaling in biological systems}

Having discussed putative virtues of critical dynamics, susceptible in
principle to be exploited by biological systems, we now start a trip
through some of the most-remarkable existing empirical evidence
revealing signatures of criticality in such systems.  We warn the
reader that --even if the aim is to present a collection as extensive
and exhaustive as possible-- the selection of topics as well as the
extent in which they are discussed might be biased by our own
experience. Also, importantly, even if some of the experiments and
findings to be discussed are very appealing, evidence in many cases is
not complete and conclusions should be always taken with caution.
Indeed, for many of the forthcoming examples, we also discuss existing
criticisms and potential technical or interpretative problems.

\subsection{Neural activity and brain networks}
\label{s-neuro}

\subsubsection{Spontaneous cortical activity}
The cerebral-cortex of mammalians is never silent, not even under
resting conditions nor in the absence of stimuli; instead, it exhibits
a state of ceaseless spontaneous electro-chemical activity with very
high variability and sensitivity\cite{Arieli,Yuste,Raichle,Fox}. 
Understanding the genesis and functionality of
spontaneous cortical activity -- which accounts for about $20\%$ of
the total oxygen consumption of a person at rest-- is key to shedding
light onto how the cortex processes information and computes
\cite{Arieli,He,Deco-RSN1,Deco-RSN2}.  Criticality might play a key
role to generate such a variable and sensitive activity as diverse
empirical results suggest.

An adult human brain consists of almost $10^{11}$ neurons and up to
$10^{15}$ synaptic connections among them, forming an amazingly
complex network through which electric signals propagate
\cite{Numbers}.  Neurons integrate presynaptic excitatory and
inhibitory inputs from other neurons, and fire an action potential
when a given threshold is overcome, stimulating further activity.
This generates irregular cascades or outbursts of activity
interspersed by quiescent periods, as empirically observed both
\emph{in vitro} \cite{Vives-network,Segev1,Segev2,Latham2003,Eytan}
and \emph{in vivo} \cite{Meister1991,Steriade1993} (see
Fig.\ref{neuro}). Is this activity related to inherent critical
behavior? In what follows we discuss empirical pieces of evidence
suggesting diverse possible connections with different types of phase
transitions.

\subsubsection{The edge of activity propagation: avalanches}
\begin{figure}
  \includegraphics[scale=0.35]{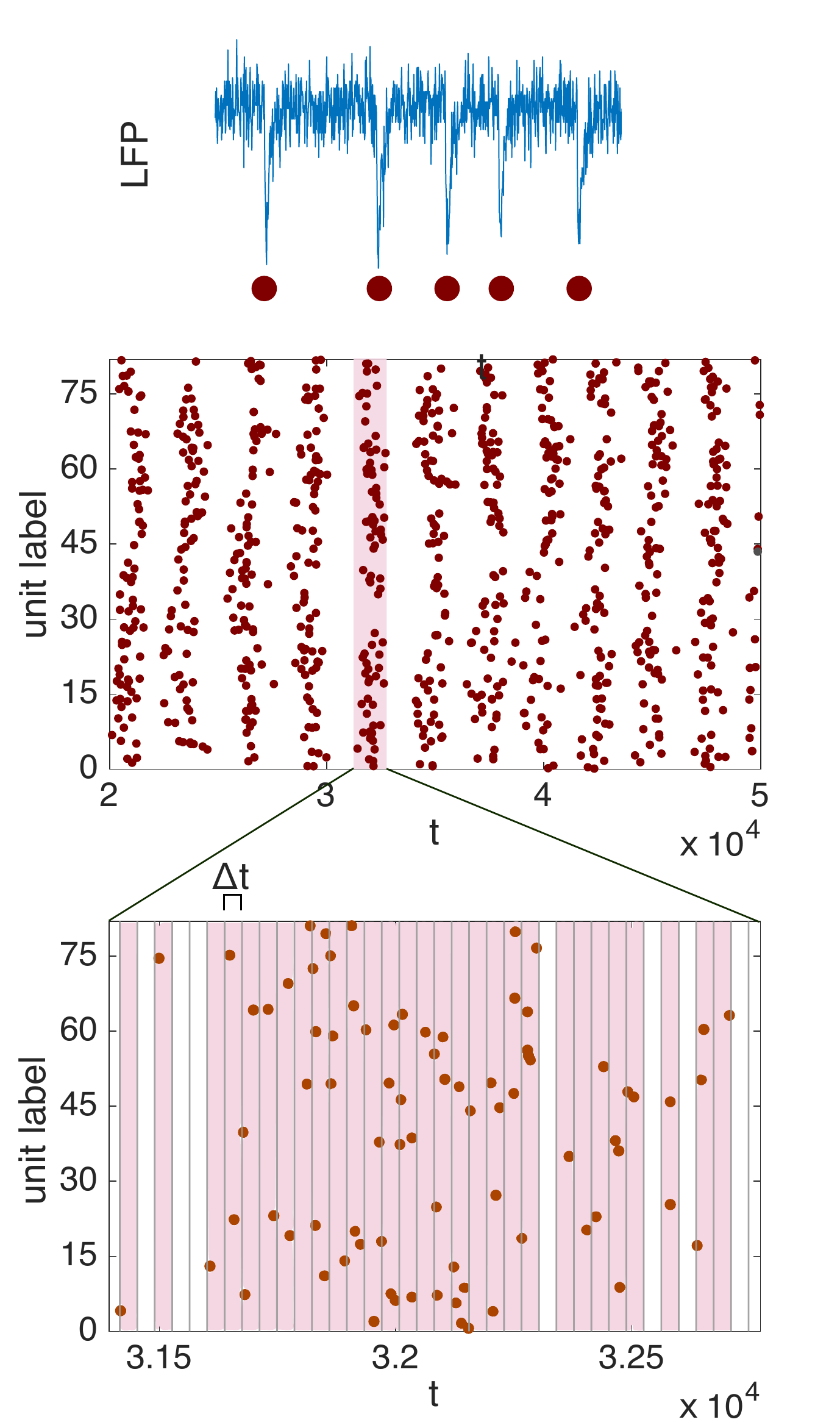}
  \caption{Sketch illustrating how neuronal avalanches are measured
    from Local field potential (LFP).  (Top) LFPs are measured at
    different locations; negative peaks of the time series correlate
    with large population spikes of the underlying neurons within each
    local region. (Middle) Raster plot illustrating the times at which
    peaks of the LPF occurs for different sites, revealing a high
    degree of temporal clustering. (Bottom) Enhancing the temporal
    resolution, it is possible to resolve the spatio-temporal
    organization within apparently coherent large-scale events; is
    occurs occur in the form of ``neuronal avalanches'' (shaded
    columns) interspersed by periods of quiescence (white columns). }
\label{neuro}
\end{figure}

In a remarkable breakthrough, \textcite{BP2003} succeeded at resolving
the internal spatio temporal organization of the above-mentioned
outbursts of neuronal activity. They analyzed mature cultures as well
as acute slices of rat cortex, and recorded spontaneous local field
potentials (LFP) --which provide coarse-grained measurements of
electrochemical activity-- at different locations and times.  Local
events of activity are defined as (negative) peaks of the LFP signals,
which are indicative of local population spikes \cite{BP2003}. As
illustrated in Fig.\ref{neuro}, events at different sites have a
tendency to cluster in time, producing network spikes of
activity. Each of these outbursts of activity when temporally
resolved, consists in a cascade of succesive local events, organized
as \emph{neuronal avalanches} interspersed by periods of quiescence
\cite{BP2003,BP2004}.  The avalanche sizes (i.e. number of local
events each one includes) and durations were found to be distributed
as power-laws with exponents $\tau \approx 3/2$ and
$\alpha \approx 2$, respectively, with cut-offs that increase with
system size in a scale-invariant way (i.e. the distributions obey
finite-size scaling\footnote{Instead, if data are
  temporally reshuffled the distributions become exponential ones,
  meaning that large coherent events disappear
  \cite{BP2003,Plenz2007}.}; see
\textcite{BP2003,Torre2007,Peterman2009}).  The observed exponents
coincide with those of the (mean-field) critical contact/branching
processes as described above and, thus, seem to describe a marginal
activity-propagation process. Moreover, the mean temporal profile of
neuronal avalanches of widely varying durations is quantitatively
described by a single universal scaling function \cite{Dahmen,Sethna},
and scaling relationships between the measured exponents are fulfilled
\cite{Dahmen}.
Similar avalanches have been observed \emph{in vitro}
\cite{Torre2007,Pasquale2008} and \emph{in vivo} for different species
\cite{Gireesh,Hahn2010,Peterman2009,Ribeiro2010,Yu2011} and across
resolution scales, from single neuron spikes to rather coarse-grained
measurements.\footnote{This includes single unit recordings
  \cite{2-photon}, local field potentials (LFP)
  \cite{BP2003,Peterman2009}, electroencephalography (EEG)
  \cite{EEG-Freeman,EEG-Allegrini,Meisel2013}, electrocorticography
  (ECoG) \cite{Magnasco1}, magnetoencephalography (MEG)
  \cite{Novikov,Linken2012,Palva2013,Plenz-Shriki2013}, and functional
  magnetic resonance imaging (fMRI) \cite{Taglia,Haimovici}.}  The
fact that at quite different resolution scales similar results are
reported is, by itself, strongly supportive of the existence of
underlying scale-invariant dynamical processes.\footnote{Some studies
  suggest that even single neurons can be intrinsically critical to
  optimize their inherent excitability
  \cite{single-Marom,single-Gollo}.}

All this evidence regarding neuronal avalanches seems to make a strong
case in favor of criticality. However, some caveats need to be made:

(i) Thresholding: A source of ambiguities in extracting (discrete)
events from (continuous) time-series analyses comes from thresholding;
i.e. signals at any given spatio-temporal location need to overcome
some threshold to be declared an ``event''of
activity. \textcite{Peterman2009} compared results for different
thresholds in LFPs time series and found that exponent values remain
unchanged, suggesting the existence of a truly scale-invariant
organization of events. However, a word of caution is still required
as recent works have underlined the ``perils'' associated with
thresholding, which in some controlled cases has been shown to
generate spurious effects such as effective exponent values and
correlations in the timings of consecutive avalanches
\cite{Perils1,Perils2,Perils3}. Further clarifying this issue is key
to make solid progress in the empirical analysis of avalanching
systems.

(ii) Time binning: Avalanches can only be defined by employing a
criterion to establish when an avalanche starts and when it ends. This
requires setting a discrete time binning to be applied to the data: an
avalanche starts when a time-bin with some activity within it follows
a series of preceding consecutive quiescent ones, and ends when a new
quiescent time-bin appears \cite{BP2003} (see Fig.\ref{neuro}). This
introduces some ambiguity, and the measured avalanche exponents have
been shown to be sensitive to the choice of the time-bin. However,
taking the time bin to coincide with the mean inter-event interval, the
mean-field branching process exponents seem to be systematically
recovered \cite{BP2003,Peterman2009,Taglia,Haimovici}. As above,
further work is needed to mathematically clarify this important issue.

(iii) Sub-sampling: A related problem is that of sub-sampling as a
result of observational and resolution limitations. Owing to these
factors the statistic is not complete, and this might affect the shape
of the observed distributions.  \textcite{Viola-1,Viola-2,Viola-3}
argued that --taking into consideration sub-sampling effects--
empirical data are best characterized by a slightly sub-critical
dynamics (additionally driven my external forces) rather than by a
critical one.

(iv) Limited scales: In general, no more than two, at most three,
orders of magnitude in avalanche statistics have been reported which
is somehow unsatisfactory. Obtaining much broader regimes of scale
invariance is technically challenging, but would make a stronger
case for actual scale-invariance \cite{Yu2014}.

(v) Some authors support different interpretations of the observed
power-laws, which are unrelated to criticality
\cite{Destexhe2006,Destexhe2009,Touboul1,Touboul2}.
  
These series of observations, taken together, seem to shed some doubts
on evidence in favor of criticality relying on avalanches. To further
stregthen it, we now discuss other complementary experimental
signatures of criticality from different perspectives.

\subsubsection{The edge of neural synchronization}
Much attention has been historically devoted to brain rhythms observed
in EEG, MEG, and LFP measurements \cite{Buzsaki}. Such rhythms emerge
owing to the transient synchronization between different neural
regions/circuits, and they play a key role in neural function
\cite{Steriade1996}. Clusters of neurons with coherent neural activity
have a much stronger coordinated effect on other neuronal assemblies
than asynchronous neurons do \cite{Kelso1986,Kelso1987,Brunel2}. Thus,
phase synchrony is essential for large-scale integration of
information \cite{Varela}, and abnormalities in the level of
synchronization --either by excess or by defect-- are a signature of
pathologies such as epilepsy, Parkinson's disease, schizophrenia, or
autism \cite{Shew2012}.  Empirically, the measured level of
synchronization across (resting) brain regions and across time has
been found to be highly variable and with strong long-range
correlations. Such spatio-temporal variability can be interpreted as a
template to codify vastly different tasks and to allow for a large
dynamical repertoire \cite{Arieli}, and has been observed to diminish
when the subject is engaged in a specific task
\cite{synchro-task}. 

The role that criticality might play in keeping intermediate and
variable levels of synchrony --which could for example be essential to
achieve a good balance between integration and segregation
\cite{Tononi}-- has been empirically analyzed as we discuss
now. Analyzing spontaneous bursts of coordinated activity (as in
Fig.\ref{neuro}; \cite{Segev1,Segev2}) the overall level of phase
synchrony between different electrodes has been recorded under
different pharmacological conditions, ranging from
excitation-dominated to inhibition-dominated regimes. It was observed
that the is a critical point at which excitation and inhibition
balance \cite{Shew2012}. At such a point --i.e. at ``the edge of
synchrony'' \cite{Brunel1,Deco-balanced,Battaglia-synchro}-- the level
of synchronization variability is maximal and scale-free avalanches of
activity can be concomitantly observed
\cite{Shew2012,Gireesh}. Actually, a recent theoretical work
emphasizes that if the cortex operates at a critical point, it should
be a synchronization critical point, where marginal synchronization
and scale-free avalanches emerge together \cite{Serena-LG}.  Last but
not least, the amazingly detailed computational model built within the
large-scale collaborative Blue brain project \cite{BBP} also suggests
that the cortical dynamics operates at the edge of a synchronization
phase transition \cite{Cell-Markram}.

\subsubsection{The edge of global stability}  
High temporal-resolution electrocorticography data from human reveal
time-varying levels of activity across different spatial locations
\cite{Magnasco1,Magnasco-anti}. Representing the system's state at a
given time as a vector, its time evolution can be approximated as a
series of linear (matricial) transformations between successive
time-discrete vector states \cite{AR}. By employing an eigenvector
decomposition of each of such matrices, it is possible to monitor the
temporal dynamics of the leading eigenvalues (Lyapunov exponents). In
awake individuals, the leading eigenvalue turns out to oscillate
closely around the threshold of instability, indicating that the
dynamics is self-regulated at the edge of a phase transition between
stable and unstable regimes. Quite remarkably, in anesthetized
subjects eigenvalues become much more stabilized, suggesting that
operating at the edge of stability is a property of functional brain
and that deviations from such point can be used as a measure of loss
of consciousness \cite{Magnasco2}.

\subsubsection{The edge of percolation}
Cortical dynamics can be viewed as a sort of percolation phenomenon.
\textcite{Taglia} analyzed functional magnetic resonance imaging
(fMRI) time series at different regions of (resting) humans. By
thresholding them they obtained discrete spatio-temporal maps of
activity (much as in Fig.5). They found that --using the density of
``active'' sites at a given time as a control parameter, and the size
of the largest connected cluster at each time as a percolation order
parameter-- there is a value of the control parameter nearby which the
dynamics spends most of the time and, remarkably, it corresponds to
the value for which the total number of different connected clusters
as well as their size variability are maximal, as happens at the
threshold of percolation transitions.  These empirical data reveal
that the dynamics is close to the critical percolation density value,
but with broad excursions to both, sub- and super-critical phases,
suggesting that regulatory mechanisms keep the system hovering around
a percolation transition (much as suggested by the mechanism of
``self-organized quasi-criticality'' discussed in Sect.I). In other
words, the resting brain spends most of the time near the point of
marginal percolation of activity, neither too inactive nor exceedingly
active.

\subsubsection{The edge of a thermodynamic transition}
The state of a neural network at a given small time window can be
represented by a binary vector encoding whether each individual neuron
has spiked or remained silent within it \cite{Bialek-thermo1}.
Questions of interest are, how often does a given simultaneous
(i.e. within a given time bin) spike pattern appear? What is the
simplest probabilistic model (in the sense of equilibrium statistical
mechanics) able to reproduce such statistical patterns?

Pioneering empirical studies obtained data from large-scale
multielectrode array recordings to determine the statistics of
patterns of neural activity in large populations of retinal (ganglion)
cells of salamander \cite{Marre2012}.  Employing such high-resolution
data and inferring from them maximum-entropy probabilistic
(Ising-like) models (as briefly described in Appendix B),
\textcite{Bialek-thermo1} observed that the associated specific heat
diverges as a function of sample size as occurs in thermodynamic
critical points. Furthermore, introducing an effective temperature
they observed that empirical data are poised near the critical point
of the (temperature-dependent) generalized model suggesting that the
visual cortex might operate in a close-to-critical regime
\cite{Bialek-thermo2,Mora-Bialek,Mora2015-retina,Bialek-searching}.

A possible interpretation of these results --backed by recent
empirical evidence \cite{Shew2015a,Shew2015b}-- is that adaptation to
sensory inputs has tuned the visual cortex to statistical criticality,
thus optimizing its performance. A competing view is that the observed
signatures of criticality could reflect an effective averaging over
un-observed variables (such as common external inputs in the case of
retinal populations), lacking thus any relationship with possible
functional advantages \cite{Macke2017,Latham} (see Appendix B).  We
refer to \textcite{Bialek-thermo2} and \textcite{Macke2017} for a more
thorough discussions on these alternative viewpoints.

\subsubsection{Large-scale cortical dynamics}
\label{resting}
Large research initiatives have allowed for the measurement of network
of physical (neuro-anatomical) connections between different regions
of the human brain, i.e. the ``\emph{human connectome
  network}''\footnote{The resulting \emph{human connectome} turns out
  to be a network organized in hierarchical
  modular way 
  \cite{Hagmann,Sporns2005,Sporns-book,Review-Kaiser,Review-Bullmore,Sporns2014,SCKH,Breakspear2017}.}.

On the other hand, functional magnetic resonance imaging (fMRI)
studies performed in the resting-state --i.e., while the subject is
awake not performing any specific task-- reveal the emergence of
spatio-temporal patterns of strongly coherent fluctuations in the
level of activity at the large scale. This allows for the
determination of so-called ``resting state networks'', encoding
pairwise correlations between different brain regions, or in words,
brain regions that become active or inactive together, and that are
consistently found in healthy individuals\footnote{See the vast
  literature on this, e.g. \textcite{RSN1,RSN2,RSN3,RSN4,Deco-RSN1,Deco-RSN2,Ibai}.}.

Diverse studies of simple dynamical models on top of the empirically
determined human connectome network it was found that spatio-temporal
correlations similar to those of the empirically-measured in the
resting state are reproduced only if the models operate close to
criticality \cite{Haimovici,Fraiman,Plenz-Physics,Cabral}, suggesting
that resting-state spatio-temporal patterns of activity emerge from
the interplay between critical dynamics and the large-scale underlying
architecture of the brain. Thus, resting state networks reflect
structured/critical fluctuations among a set of possible attractors
suggestive of a state of alertness facilitating rapid task-dependent
shifts \cite{Deco-Jirsa,Ghosh,Deco2013}.

On the other hand, one could expect that scale-invariance emerges in
broad regions of parameter space and not just at critical points (see
Appendix A), owing to the modular and highly heterogeneous
architecture of structural brain networks.  This has indeed been
verified to be true for models of neural activity propagation
\cite{Moretti} as well as for synchronization dynamics
\cite{Villegas1,Shanahan2010,Thurner-Kuramoto}, and implies that
cortical dynamics might not be required to be exactly critical to
reproduce empirical findings, but just to be located in a broad region
in parameter space exhibiting generic scale invariance (e.g. in a
\emph{Griffiths phase}; see Appendix A).

\subsubsection{Disruptions of criticality in pathological conditions}
Important pieces of evidence that scale invariance and criticality
might be specific of awake and healthy brain activity emerge from
experimental analyses of neural activity under pathology or modified
physiological conditions. For example, signatures of criticality have
been reported to fade away during epileptic seizures
\cite{epilepsy,Meisel2012} as well as during anomalously large periods
of wakefulness \cite{Meisel2013} or while performing simple tasks
\cite{Deco2017,Tomen}. Also, long-range temporal correlations
--characteristic of the awake state \cite{Expert2011,fMRI-He}-- break
down during anesthesia \cite{Ribeiro2010,2-photon}, unconsciousness
\cite{Taglia-unconcious} or under deep sleep \cite{Taglia-breakdown},
suggesting that critical dynamics is specific to the state of
wakefulness. Interestingly, sleep has been interpreted as a mechanism
to restore the overall dynamics to a critical state \cite{sleep}.

By pharmacologically altering the ratio of excitation to inhibition,
i.e. breaking the balance condition that characterizes functional
neural networks \cite{Sompo,Doiron,Reyes}-- induces a tendency to
super-critical propagation of activity, including many large
system-spanning avalanches, clearly disrupting scale-invariant
behavior \cite{BP2003,Torre2007}. Similarly, only during naturally
balanced conditions the dynamic range (as defined above) is
empirically observed to be maximal \cite{Shew2009,Shew2015a}.

There is also experimental evidence supporting the idea that
developing cortical networks go through different stages in the
process of maturating: they shift from being supercritical, to
subcritical, and then finally, converge towards criticality only when
they become mature \cite{development2,development1}.

Taken together, these observations suggest that criticality is the
baseline state of mature, healthy, and awake neural networks and that
deviations from criticality have profound functional consequences
\cite{Healthy}.

\subsubsection{Mathematical models of neuro-criticality}
Since the idea that the computational power of the brain could emerge
out of collective properties of neuronal assemblies
\cite{Hopfield1982,Hertz-book}, a large and disparate number of
modeling approaches have been proposed to scrutinize neural dynamics
\cite{Kandel,Dayan,Amit-book2,Amit-Brunel,Izhi1,Izhi2,Vives-exploring}.
These models uncovered a large variety of phases and possible
dynamical regimes such as up and down states
\cite{Holcman,Mejias0,Parga,Vives-exploring,Hidalgo2012}, synchronous
and asynchronous phases \cite{Brunel1,Brunel2,Abbott-asynchronous}, as
well as phase transitions separating them. Our aim here is not to
review them exhaustively but, rather, to discuss those aimed at
justifying the possible emergence of criticality in actual neural
networks.

P. Bak and collaborators are to be acknowledged for first
proposing that concepts of self-organization to criticality could play
a role in neural dynamics\footnote{See
  e.g. \textcite{Bak-brain,Bak,Chialvo-Bak1,Chialvo-Bak2,Chialvo2004}. Also,
  early work by Haken, Kelso and coworkers brought about the role that
  critical fluctuations and critical slowing-down might play in neural
  dynamics
  \cite{Haken1977,Kelso1986,Kelso1987,Haken2013}.}. \textcite{Herz}
realized that stylized (integrate-and-fire) models of neuronal
networks were mathematically equivalent to SOC archetypes. 

Short-time synaptic depression \cite{TM96,TM97,Sussillo} was
introduced in SOC-like neural-network models (in which some form of
neural ``stress'' is accumulated and then released to connected units
in a conserved way) as a mechanism to regulatory mechanism able to
auto-organize them to the edge of a phase transition
\cite{Levina2007,Levina2009,Lucilla2006,Kappen}\footnote{This opened
  the door to studies of the interplay between critical dynamics,
  memory and learning
  \cite{Lucilla2010,Lucilla2011,Lucilla2012,Lucilla2014,Lucilla-dragon}.};
synaptic resources become depleted owing to network activity and
remain so for a characteristic recovery period, while they slowly
recover to their baseline level. The alternation of these
activity-dependent mechanisms (i.e. slow charging and fast
dissipation) generates a feedback loop that, allegedly, guides the
networks to criticality, much as in SOC (Fig.3).

Alternative regulatory (\emph{homeostatic}) mechanisms such as
spike-timing dependent plasticity
\cite{Rubinov,Shin-Kim,Meisel-Gross,Levina2015}, retro-synaptic
signals \cite{Herrmann2017}, and Hebbian plasticity
\cite{Lucilla2010,Levina2013}, have been proposed to explain
self-organization to criticality \cite{Bienenstock}.

However, these SOC-like approaches might not be biologically
plausible, as they rely on conservative or almost-conservative
dynamics (while neurons and synapses are leaky) and, even more
importantly, they require of an unrealistically large (infinite)
separation of timescales between dissipation and recovering to
actually self-tune the dynamics to a critical state
\cite{JABO2,Copelli2015}. If the separation of timescales in these
models is fixed to moderate (finite) values, critical
self-organization is not achieved; instead, the system hovers around
the critical point with excursions to both sides of it --as in the
above-discussed self-organized quasi-criticality
\cite{JABO1,Kinouchi2018}-- or may become not critical at all
\cite{JABO2}.

To overcome these difficulties an influential model was proposed to
explain self-organized criticality without assuming conservative
dynamics nor an infinite separation of timescales \cite{Millman}. This
model (consisting of a network of leaky integrate-and-fire neurons
with synaptic plasticity) exhibits a discontinuous phase transition
--rather than a continuous one with a critical point-- between states
of high and low activity, respectively. This is neurobiologically
sound as similar ``up'' and ``down'' states are empirically known to
emerge under deep sleep or anesthesia \cite{Steriade1993,Holcman}.
Remarkably, the model was also found to display scale-free avalanches
all across its active phase. This is puzzling from the viewpoint of
models of activity propagation, which generate scale-free avalanches
only at criticality.  

This apparent paradox has been recently solved: avalanches in the
model of Millman et al. are not the result of criticality; they appear
owing to the existence of generic scale invariance, which is a
consequence of an underlying neutral dynamics (see Appendix
A). Importantly, such neutral avalanches are detected in computational
models by employing information about causal relationships on which
neuron triggers the firing of which other \cite{neutral-neural}, and
this type of information is usually not accessible in
experiments.\footnote{ See, however, \textcite{Beggs-causal}.}
Furthermore, if avalanches in the model of Millman et al. are measured
as in experiments (employing a time binning) they are not scale-free
\cite{neutral-neural}. Thus, this model --as well as some other
similar ones \cite{Plenz2015}-- do not describe empirical
temporally-defined scale-free avalanches. More generally, these
results reveal a gap in the literature between time-binned defined
avalanches (in experiments) and causally defined avalanches (in
models).

All the above-discussed approaches have in common that they identify
neural criticality with the edge of an activity-propagation phase
transition. Recently, some other theoretical models have provided
theoretical evidence that neural dynamics should exhibit a
synchronization phase transition, at which neuronal avalanches and
incipient oscillations coexist
\cite{Gireesh,Shew2012,Linken2012,Serena-LG}. However, these models
provide no explanation --other than a possible fine tuning-- of why
the dynamics should operate precisely at the edge synchronization.

Last but not least, the amazingly detailed computational model built
within the large-scale collaborative Blue brain project \cite{BBP}
suggests that the cortical dynamics operates at the edge of a phase
transition between an asynchronous phase and a synchronous one with
emerging oscillations \cite{Cell-Markram}. The regulation of calcium
dynamics has been cited as a possible responsible mechanism for
keeping the system close to such a critical state, operating at a
point at which a whole set of empirical results can be quantitatively
explained by the model \cite{Cell-Markram}.

Finally, let us comment on two theoretical approaches --not relying on
criticality-- proposed to account for scale-free neuronal
avalanches. The first one is a mechanism called ``stochastic
amplification of fluctuations'' which is able to produce highly
variable avalanches with an (approximate but not perfect/critical)
balance between excitatory and inhibitory couplings together with
inherent stochasticity \cite{Cowan2010,Miller}. However, this
mechanism is not able to reproduce the empirically observed exponent
values \cite{balanced}.

The second is a recent work, \textcite{Touboul2}, where it is proposed
that scale-free avalanches can naturally emerge in networks of neurons
(described e.g. as a balanced network with excitation and inhibition
\cite{Brunel1}, or even as simple Poissonian point processes)
operating in synchronous irregular regimes away from criticality. In
our opinion, further work needs to be done to understand how and under
which circumstances this is true, and what are the corresponding
values of the resulting avalanche exponents.  Summing up, appealing
empirical evidences as well as sound dynamical models supporting the
idea of criticality in the brain exists; however, in many cases
empirical results are not fully convincing and alternative theoretical
interpretations are still under debate.  Fully clarifying the nature
of the overall cortical dynamical state remains an open challenge.

\subsection{Gene regulatory networks}
\label{s-gene}
Leaving aside neural networks, we move on to another type of biological
information-processing networks that also exhibit signatures of
criticality: genetic networks.

Living cells exhibit stable characteristic features which are robust
even under highly variable conditions. In parallel, they also exhibit
flexibility to adapt to environmental changes. These two aspects are
compatible owing to the fact that a given  set of genes (i.e. a
``genotype'') can give rise to different cellular states
(``phenotypes''), consisting of diverse gene-expression patterns in
which some genes are differentially expressed or silenced.
Since the pioneering work of \textcite{Origins}, cellular states have been
identified as attractors of the dynamics of gene regulatory networks,
where the genes are the network nodes and their mutual regulatory
(activation/repression) interactions are represented as directed links
between them.  Cells can be thought as ``machines'' executing complex
gene-expression programs that involve the coordinated expression of
thousands of genes\footnote{Individual genes are the basic information
  units of the genetic code and occupy a central role in biological
  inheritance and evolution.  Gene information is \emph{transcribed}
  into RNA molecules and from them \emph{translated} into proteins
  (i.e. ``expressed'') which are the final result of gene expression
  and the building blocks of functionality
  \cite{Dogma}.}\cite{Dogma,Kitano,Alon-book,Buchanan,Koonin-book1,Koonin-book2}.
Consequently, the study of information processing in cells shifted
progressively from single genes to increasingly complex
circuits/networks of genes and regulatory interactions, shedding light
on collective cellular states \cite{Stan1999,Ilya-review,Ojalvo-review}.  The
development of powerful experimental high-throughput technologies in
molecular biology paved the way
to the experimental investigation of gene-expression patterns in large
regulatory networks \cite{Filkov} and, in particular, provided
empirical evidence that sequences of cell states (apoptosis,
proliferation, differentiation, etc.) can be viewed as programs
encoded in the dynamical attractors of gene regulatory networks
\cite{Reka2,Buylla,Attractors1,Attractors2}.
\begin{figure}
  \includegraphics[scale=0.16]{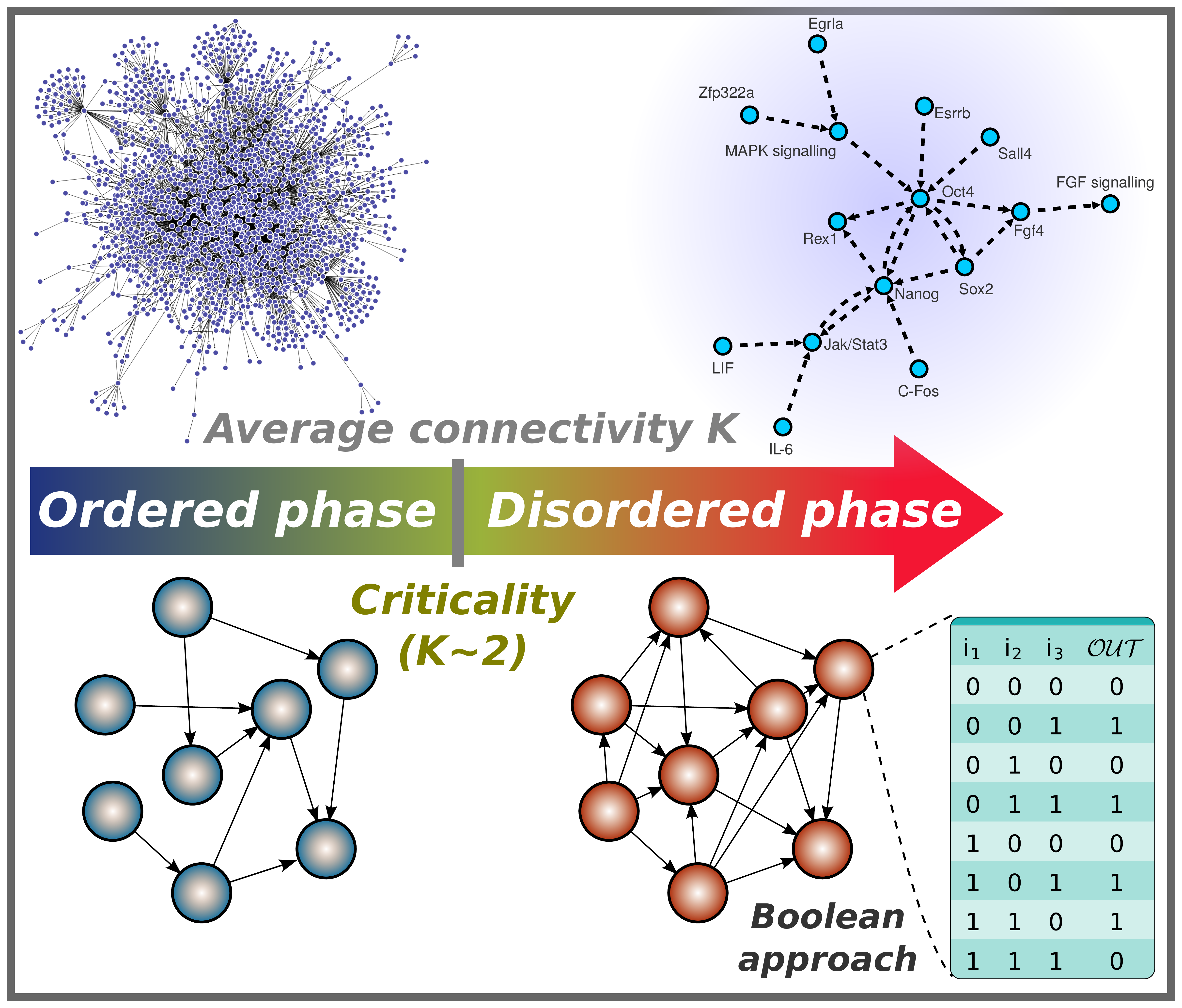}
  \caption{ The upper panels represent two gene regulatory networks:
    (Left) a large scale one (for E. Coli \cite{Red1}) and (Right) a
    small-scale one (mouse embryonic stem-cell subnetwork
    \cite{Red2}). In both cases, nodes stand for genes and links
    between them for regulatory interactions (see main text).  The
    lower panel shows a sketch of random Boolean networks as simple
    models of gene regulation. For low (high) average connectivities
    they lie in the ordered (disordered) phase, with a critical point
    occurring close to $K=2$. The table illustrates a set of logical
    operations (associating an output to a set of $3$ different
    inputs) for a given node in a Boolean setting.}
\label{GRN}
\end{figure}

\subsubsection{Models of genetic regulatory networks}
Many genes are empirically observed to exhibit bistability, i.e. their
gene-expression levels can be approximated as either ``high'' (on) or
very ``low'' (off) depending on conditions. Such binary states are
believed to be the building blocks of genetic logical circuits
\cite{Tyson}. Thus, genetic regulatory networks have been
traditionally modeled as binary information-processing systems in
which the expression level of each gene is represented by a Boolean
(on/off) variable and their interactions are modeled as Boolean
functions whose inputs are the states of other genes (see
Fig.\ref{GRN}) \cite{Origins,Ilya-review}.\footnote{Alternatively, it
  is also standard to use continuous approaches, based on
  reaction-kinetics differential equations
  \cite{Kaneko1992,Kaneko-stem}. See \textcite{DeJong} for a review.}
Boolean descriptions constitute the most basic and crudest approach to
gene regulatory networks; still, they are particularly adequate to
analyze large networks as they reduce the overwhelming complexity of
the real problem to a logical one, and they have been shown to
successfully explain e.g. cell cycles
\cite{Origins,Home,Gros-book,Bornholdt-RBN1,Bornholdt-RBN2,Drossel2008,Aldana2003,DeJong,Serra2007}.

In the simplest setup, the network architecture is described as a
random directed network\footnote{More realistic network architectures
  including, for example, node heterogeneity and modularity have also
  been considered \cite{RBN-modular,Aldana2003}.}  and regulatory
interactions are described as random Boolean functions
\cite{Kauffman69,Origins,Gros-book,DeJong,Alon-book,Reka} (see
Fig.\ref{GRN}).  So defined random Boolean networks (RBNs) can operate
in different regimes, depending on e.g. their averaged connectivity.
The ordered (low connectivity) is characterized by a small set of
stable attractors which are largely robust to perturbations, while in
the disordered phase (large connectivity) perturbations rapidly
propagate and proliferate hindering the existence of truly stable
states.  Separating these two phases there is a critical point at
which perturbations propagate marginally \cite{Derrida}. More complex
models, with e.g. stochasticity and/or continuous levels of activity,
exhibit also such two alternative phases \cite{Rohlf2002}.

Kauffman conjectured that models operating at their critical point
might provide the best possible representation of real gene regulatory
networks \cite{Origins,Home}, and that this might entail a large
variety of essential functional advantages
\cite{Ribeiro2008,Basin,Kauffman2003,Aldana2007,Aldana2012,Nykter2}.
In the ordered regime, convergence in state space implies that
distinctions between different inputs are readily erased, precluding
reliable discrimination of them. In the disordered phase, even small
perturbations lead to a very large divergence of trajectories in state
space precluding reliable action \cite{Kauffman2003}.  Hence,
criticality might confer on such networks an optimal tradeoff between
the robustness and accuracy that biological machinery demands and
responsiveness to environmental clues \cite{Kauffman2003}.  At larger
evolutionary scales, criticality might provide networks with an
optimal balance between robustness and evolvability under changing
conditions
\cite{Kaneko-evolution,Wagner,Aldana2007,Aldana2012,Gershenson}. 

It remains to be clarified how could adaptive
\cite{Gellmann,Gros-book} and/or evolutionary \cite{Nowak-book}
mechanisms, specific of living systems\footnote{This is, beyond purely
  self-organization mechanisms, such as SOC, also exhibited by
  inanimate systems \cite{2facets}.}, lead biological networks to
operate close to criticality. Theoretical approaches tackling this
question are discussed in Appendix C.

\subsubsection{Gene knock-out and damage spreading}
DNA microarrays or ``DNA chips'' are devices allowing to measure the
expression levels of large numbers of genes within a cell as well as
to quantify the differences in expression levels between two cells
\cite{DNAmicro}. Also, novel technologies made it possible to perform
gene knock-out experiments in which individual genes are silenced.
Combining these two techniques it became possible to perform ``damage
spreading'' experiments \cite{Derrida,Rohlf2007} in which the
difference in gene-expression levels between perturbed and unperturbed
cells in quantitatively monitored.  The statistics of the sizes of
``avalanches'' caused by single-gene knock-out experiments has been
analyzed using empirical gene-expression data from the yeast
(\emph{Saccharomyces cerevisae}) \cite{Hughes2000}, with the
conclusion that the best correspondence between empirical results and
(RBN) model predictions is obtained for the model operating close to
its critical point \cite{Serra2004,Serra2007,Ramo2}. However, as a
word of caution, let us stress that, given that expression levels are
noisy, it is necessary to introduce a threshold expression level to
declare when a gene is differentially expressed in the two cells.  A
caveat is that it is not clear what the influence of thresholding is
on the definition of avalanches and on their associated (size and
duration) probabilities. Thus, even if results are promising, more
precise analyses are still much needed.

\subsubsection{Networks from DNA microarray data}
In a parallel endeavor, empirical analyses of hundreds of DNA
microarray experiments allowed researchers to infer the whole network
of regulatory interactions among genes, i.e. who regulates whom in a
given cell \cite{Filkov}.  It has been consistently found that the
in-degree distribution of such regulatory networks is Poissonian,
while the out-degree distribution is scale-free (see
\textcite{Aldana2003}, \textcite{Drossel2009}, and refs. therein).
Performing damage-spreading computational analyses of dynamical RBN
models running on top of such networks --with the
empirically-determined architecture of diverse organisms such as
\textit{S. cerevisiae} and \textit{E. coli} \cite{Reka2}-- it was
concluded that they all are indeed very close to criticality, in the
sense of marginal propagation of perturbations
\cite{Balleza2008,Aldana2007,Chowdhury,Darabos}.

Alternatively to inferring the architecture of the underlying network
of interactions --which is a difficult problem \cite{Filkov}--
algorithmic information theory has also been employed to asses the
dynamical state directly from empirical measurements from
DNA-microarray data in a model-free way
\cite{Shmu2005,Kauffman2003}.\footnote{For example, one such method
  relies on computing estimators of the Kolmogorov complexity
  \cite{Kolmogorov} of sets of gene-expression time series in diverse
  microarrays, and computing how the difference between the
  information content of any two system states \cite{Loreto} changes
  over time.}  Analyses of empirical data (for, e.g.  the macrophage)
following these protocols produced results compatible with marginality
in the information flow, i.e. with critical dynamics
\cite{Nykter1}\footnote{Similar analyses for Eukaryotic cells gave
  results compatible with the dynamics being either ordered or
  critical but not disordered \cite{Shmu2005}.}.

\subsubsection{Zipf's law in gene-expression data}
Inspection of gene expression databases of diverse unicellular
organisms (such as yeast) reveals that the (continuous-valued)
abundances of expressed genes are distributed as a power-law with
exponent close to $-1$, obeying the Zipf's law
\cite{Furu2003}\footnote{Indeed, clonal populations of unicellular
  organisms such as viruses or bacteria often exhibit phenotypic
  diversity, which constitutes a sort of
  ``bet-hedging'' strategy to cope with unpredictable environmental
  changes \cite{Bet-hedging,Bet-hedging2,Leibler}.}.
\textcite{Furu2012} analyzed an abstract dynamical (not Boolean) model
describing a cellular network (the network formed by the set of
molecules which interact with others to give products within the cell)
with nutrient uptake, and showed that the Zipf's law is a universal
feature of self-regulated cells optimizing their growth rate in
nutrient-rich environments. In this setting, cells seem to adapt to
criticality to achieve the maximal capacity to assimilate and use
nutrients for recursive formation of other products
\cite{Furu2012,Kaneko-book,Erez2017,Thurner1,Thurner2}.

\subsection{Collective behavior of cells}
We have discussed the possibility of criticality within the internal
networks of individual cells. But, also ensembles of cells --both in
social unicellular organisms and in multicellular ones-- may exhibit
signatures of scale invariance and criticality in their collective
behavior \cite{Nadell}. For instance, in aggregates of the
(unicellular) amoeba Dictyostelium discoideum (the ``slime mold''),
local cell-cell coupling via secreted chemicals may be tuned to a
critical value \cite{Endres}, resulting in emergent long-range
communication and enhanced sensitivity. In the following we discuss
some other examples in multicellular organisms.

\subsubsection{Stem cell pluripotency}
Large diversity in gene-expression levels has been observed in
multipotent stem-cell populations of multicellular organisms
\cite{diversity-stemcell}.  Multipotent (hematopoietic) stem cells can
differentiate onto either erythroid or myeloid blood cells --with
rather different functionalities-- depending on the expression level
of a gene called \emph{Sca1} \cite{Pluripotency}.  The empirically
measured distribution of expression levels of \emph{Sca1} within a
population of stem cells turns out to be very broad and with various
local maxima.  This has been modeled as a stochastic process, and it
has been found that the model can exhibit two different regimes:
either a stable low-\emph{Sca1} or a stable high-\emph{Sca1}
regime. Separating these phases there is a line of discontinuous
transitions (with bistability) finishing at a critical point.
Remarkably, the best fit to gene-expression empirical data is obtained
fixing model parameters close to criticality, where maximal
variability of the two possible phenotypes is obtained.  Thus, it
seems that by adjusting near to criticality, the stem-cell population
is prompt to react and produce either erythroid or myeloid cells in
response to changing demands in an optimal way \cite{Pluripotency}.
Similar ideas have been discussed in the more general context of
collective cell decision making
\cite{decision,decision-Ojalvo,Simons,Yamaguchi}, as well as in cancer
progression \cite{Yoshikawa1,Yoshikawa2}.

\subsubsection{Morphogenesis I: Hydra regeneration} 
Morphogenesis is the biological process at the basis of the
development of multicellular organisms. It is achieved by a precise
control of cell growth, proliferation, and differentiation.  As first
suggested in the seminal work of \textcite{Turing}, morphogenesis
involves the creation of self-organized patterns and shapes in the
embryo.  A prototypical organism studied in this context is the Hydra
polyp, which has a remarkable regeneration power, as an entire new
individual can be spontaneously re-assembled even from dissociated
cells from an adult individual \cite{Hydra}.  Along such a
regeneration process, first a cell bilayer is formed with a spherical
(hollow) shape. How does the spherical symmetry break down to form a
well-defined foot-head axis in adults?  During this process, there is
a gene called \emph{ks1} that becomes progressively expressed and that
can be transferred to neighboring cells. It was empirically found that
right at the time when the spherical symmetry is broken, the size
distribution of \emph{ks1}-rich domains of cells across the sphere
becomes scale-free and that a spanning cluster emerges, much as in a
percolation phase transition \cite{Soriano2006,Soriano2012}. Thus a
critical percolation-like state with collective fluctuations of
gene-expression levels is exploited to break the symmetry, defining a
head-tail axis \cite{Soriano2006,Soriano2009}.

\subsubsection{Morphogenesis II: Gap genes in Drosophila} 
A set of so-called \emph{``gap''} genes is responsible for the
emergence of spatial patterns of gene-expression, that are at the
origin of the formation of different segments along the head-tail axis
in the development of the fruit-fly (Drosophyla) embryo. Empirical
scrutiny of the expression levels of gap genes along the head-tail
axis revealed a number of remarkable features that include: slow
dynamics, correlations of expression-level fluctuations over large
distances, non-Gaussianity in the distribution of such fluctuations,
etc.  \textcite{Bialek-morpho} proposed a simple dynamical model in
which the process is controlled by only two mutually repressing gap
genes.  Assuming that a fixed point exists, and performing a linear
stability analysis to describe the fate of fluctuations, one readily
finds that there is an instability point as the interaction strength
between the two genes is varied.  \textcite{Bialek-morpho} argued that
if the dynamics of the coupled system is tuned to operate at such an
instability point, then it constitutes an excellent qualitative
description of all the above-mentioned empirical findings, implying
that the gene dynamics operates at criticality.  This suggests that
criticality helps defining patterns without a characteristic scale, as
required for expanding/developing systems (see
\textcite{Bose1,Bose-review} for a pedagogical discussion of these
ideas in the general context of cell differentiation).

\subsection{Collective motion}
\label{collective}
Collective motion of large groups of individuals is a phenomenon
observed in a variety of social organisms such as flocks of birds,
fish schools, insect swarms, herds of mammals, human crowds
\cite{Collective1,Collective2,Collective3,Couzin-review,Couzin-science}
and also, at smaller scales, in bacterial colonies
\cite{Peruani,Goldstein-bacteria,Nadell,Ramaswamy}, and groups of
cells in general \cite{Vicsek2014}.  Flocking, schooling, swarming,
milling, and herding constitute outstanding examples of collective
phases where simple interactions between individuals give rise to
fascinating emergent behavior at larger scales, even in the absence of
central coordination. Flock of birds and fish schools behave as
plastic entities able to exhibit coherent motion, including e.g. rapid
escape manoeuvres when attacked by predators, which confers obvious
fitness advantages to the group as a whole
\cite{Couzin2007,Couzin2009}.

Such collective phenomena have attracted the attention of statistical
physicists who have tackled the problem employing: (i)
individual-based models of self-propelled particles such as the one in
\textcite{Vicsek} which models collective motion by assuming that an
individual in a group essentially follows the trajectory of its
neighbors, with some deviations treated as noise,\footnote{See
  \textcite{Chate2004,Chate2008,Chate2008b,Ginelli2016} for detailed
  statistical-mechanics analyses of Vicsek models and variants of it.}
and (ii) continuum (hydrodynamic) theories, more amenable to
theoretical analysis \cite{Yuhai1995,Yuhai2005}.  These approaches
have in common the existence of phase transitions between phases of
coherent and incoherent motion. For example, in the Vicsek model, a
phase transition from an ordered ``flocking phase'' to a disordered
``swarming phase'' occurs when the density of individuals goes below a
given threshold or, for a fixed density, when the level of
stochasticity is large.  This is consistent with experimental
findings; e.g. \textcite{Couzin2006} investigated the social behavior
of locusts and reported on a density-driven phase transition from
disordered movement of individuals to highly aligned collective motion
as density is increased \cite{Locust}.  At a conceptual level,
marginally coordinated (critical) motion can be hypothesized to
constitute an optimal tradeoff to deal with conflicting
imperatives such as e.g. (i) the need to behave cohesively as a unique
entity and (ii) being highly responsive to information from
transitorily well-informed individuals
\cite{Couzin2005,Couzin2011,Grigolini-groups,Vanni}. Similar
dichotomies exist in the empirical examples we discuss now.

\subsubsection{Flocks of birds}
On the empirical side, pioneering work by Cavagna, Giardina and
collaborators \cite{Cavagna2008,Cavagna2010} on starling flocks
allowed to record individual trajectories (with purposely devised
tracking technology). By analyzing the fluctuations in individual
velocity with respect to the average velocity of the group, these
studies provided remarkable evidence that long-range scale-invariant
correlations may be a general feature in systems exhibiting collective
motion. In particular, experimentally measured correlations --both in
orientation and speed fluctuations-- were found to grow with flock
size in large flocks, suggesting that a correlation length much larger
than the interaction range, could be a common trait of self-organized
groups needing to achieve large-scale coordination
\cite{Cavagna2010}. Let us note that the scale-free correlations in
the orientation might be attributed to the broken continuous
(rotational) symmetry, which as discussed in Appendix A leads to
generic scale-invariance. However, the presence of scale-free
correlations in the (scalar) speed fluctuations cannot be explained in
this way, suggesting that the flock might be tuned to a critical point
with maximal susceptibility.\footnote{Similar results have been
  obtained for aggregates of a social amoeba (slime mold)
  \cite{Endres}, as well as for colonies of the bacteria
  \emph{Bacillus subtilis} in the experimental setup of
  \textcite{Chen-bacteria} (but not in that of
  \textcite{Goldstein-bacteria}, which reveals only short-range
  correlations).} Furthermore, experiments on starling flocks also
allow to measure how the information of the turning of one individual
propagates across the flock, revealing that this occurs in a very fast
and efficient way, which can be taken as a direct evidence of the
existence of scale-free correlations in flocks \cite{Attanasi2}.

\textcite{Bialek-birds} applied a maximum entropy method to construct
a statistical model consistent with the empirically measured
correlations (see Appendix B). They concluded that the interaction
strength and the number of interacting neighbors do not change with
flock size in the probabilistic model; and, more importantly, the
model was able to reproduce scale-free correlations in velocity
fluctuations. It was observed (i.e. inferred from data) that this
occurs as a result of the effective model's operating close to its critical
point \cite{Bialek-social,Mora-Bialek,Mora2016}.

Furthermore, performed

 \subsubsection{Insect swarms}
 Extensive field analyses of insect (midge) swarms --which, unlike
 birds traveling in a flock, hover around a spot on the ground-- have
 also been performed \cite{Attanasi}.  By employing finite-size
 analyses of the data, Attanasi et al. showed that both the correlation
 length and the susceptibility grow with the swarm size, while the
 spacing between midges decreases. Moreover, such changes with swarm
 size occur as in the Vicsek model for finite-size systems sitting
 near the maximally correlated point of their transition region at
 each finite size\footnote{The Vicsek model exhibits, at least for
   not-too-large sizes, a wide regime where correlations peak at the
   transition and finite-size-scaling holds
   \cite{Vicsek,Chate2004,Chate2008b,Baglietto}.}. Thus, midges obey
 spatio-temporal scaling and, to achieve it, they seem to regulate
 their average distance or density (which acts as a control parameter)
 so as to function close to criticality
 \cite{Attanasi,Cavagna-NP,Us-Physics}.  On the contrary, laboratory
 experiments of small swarms do not indicate critical behavior, which
 may signal that it only arises in ``natural conditions'' or for
 larger sizes \cite{Ouellette2013,Ouellette2014}.

\subsubsection{Mammal herds}
Social herbivores (Merino sheep) have also been studied in
well-controlled environments, revealing the existence of two
conflicting needs: (i) the protection from predators offered by being
part of large cohesive group and (ii) the exploration of foraging
space by wandering individuals. Sheep resolve this conflict by
alternating a slow foraging phase, during which the group spreads out,
with fast packing events triggered by individual behavioral shifts,
leading to intermittent collective dynamics with packing events of all
accessible scales, i.e. a ``near critical'' state \cite{Sheep}.

\subsubsection{Social-insect foraging strategies}
Studies of ant foraging strategies have been recently
performed \cite{Sole-book,Ants2001,Ants2008,
  Ants-Kurths-2014,Vicsek-foraging,Ants-Chialvo}. For ant colonies to
achieve an efficient foraging strategy, a tradeoff needs to be reached
between exploratory behavior of some individuals and predominant
compliance with the rules \cite{Ants2017}.  It has been found by using
a combination of experiments and theory that some ant groups optimize
their overall performance by sitting at the edge of a phase transition
between random exploration and gregarious strategies, thus resulting
in effective criticality. This entails efficient group-level
processing of information emerging out of an optimal amplification of
transient individual information \cite{Ants2015}. Similar ideas are
being presently explored for the design of artificial systems, i.e. in
\emph{swarm robotics} \cite{swarm-robotics,swarm-Herrmann}.

\vspace{0.5cm}

To further enrich this bird's-eye view over different aspects of
criticality in biological systems, a miscellaneous collection of other
examples from the literature is presented in Appendix C.

\section{Discussion}
The hypothesis that living systems may operate in the vicinity of
critical points, with concomitant scale-invariance, has long inspired
scientists. From a theoretical viewpoint this conjecture is certainly
appealing, as it suggests an overarching mechanism exploited by
biological systems to derive important functional benefits essential
in their strive to survive and proliferate. The list of possible
critical features susceptible to be harnessed by living systems
include the unparalleled sensitivity to stimuli, the existence of huge
dynamical repertoires, maximal transmission and storage of
information, as well as optimal computational capabilities, among
others. When living systems are interpreted as information-processing
devices --needing to operate robustly but, at the same time, having to
cope with diverse environmental changes-- the virtues of critical
behavior are undeniable. Criticality represents a simple strategy to
achieve a balance between robustness (order) and flexibility
(disorder) needed to derive functionality. Similar tradeoffs, as
discussed along the paper (e.g. between stability and evolvability)
underline the potential of operating at the edge between different
types of order.

Throughout this essay we focused \emph{dynamical aspects of
  criticality}, meaning that in most of the discussed examples it is
assumed --either directly or indirectly-- that there is an underlying
dynamical process at work, and that such a process --susceptible to be
mathematically modeled-- operates in the vicinity of a continuous
phase transition, at the borderline between two alternative
regimes. Such a dynamical perspective is essentially different from
the purely statistical (or \emph{static}) one, as described e.g. in
\textcite{Mora-Bialek}. In this latter, the focus is on analyzing the
statistics of existing patterns; it has the great advantage that it
harnesses existing high-quality empirical datasets. On the
other hand, it disregards the possible dynamical generative mechanisms
behind them and focused on an effective description (which can be very
useful). Even if both approaches have deep interconnections,
here we chose to focus mostly on the dynamical one.

Synthesizing (maybe oversynthesizing), one could argue that the
ultimate reason why putative criticality appears so often in the
scrutiny of complex biological systems is that it constitutes the
simplest physical/dynamical mechanism generating complex spatio-temporal
patterns spanning through many different scales, that are all
correlated, implying system-wide coherence and large responses to
perturbations. From this perspective, critical-like behavior --and the
nested hierarchy of spatio-temporal structures it spontaneously
generates-- can be identified as a scaffold upon which (multiscale)
biological systems may build up further complexity.

Statistical physics teaches us that under some
circumstances--including e.g. systems with some form of heterogeneity
(relevant for e.g. the study of brain networks), or in systems with
continuous symmetries (relevant in collective motion) the standard
scenario of a unique critical point separating diverse phases needs to
be replaced by that of extended critical-like regions (such as
e.g. Griffiths phases discussed for the overall brain dynamics) where
some form of scale invariance emerges in a generic way. In such cases,
it might suffice for biological systems to operate in such phases
without the need to invoke precise tuning to the edge of a phase
transition to obtain functional benefits stemming from spatio-temporal
scale invariance.

From the experimental viewpoint, along the presentation we tried to
summarize existing empirical pieces of evidence for each of the
discussed examples, stressing possible drawbacks and interpretative
problems, and underlining criticisms raised in the literature. Readers
will extract their own conclusions on whether each of the examples is
sufficiently convincing or not.  Our general impression is that, in
most of the cases, larger systems, more accurate measurements, and
less ambiguous analyses would be needed to further confirm or disprove
the existence of an underlying dynamical critical process.  For most
of the leading examples (i.e. neural systems, genetic regulatory
networks, and collective motion), our opinion is that, as of today,
there is not a fully convincing proof, where experimental evidence and
mathematical theory/modeling match perfectly; i.e. we still do not
have a ``smoking gun''. Still, the existing collection of remarkable
pieces of evidence is extremely appealing and hard to neglect.

Two important aspects should be considered in future empirical
analyses to make solid progress. One is that, given that biological
systems are finite, they cannot be truly critical in the precise sense
of statistical physics; thus it is important to perform, whenever
possible, finite-size analyses to prove the existence of
scale-invariance within the experimentally accessible ranges. A second
aspect is that the two alternative phases that the alleged criticality
separates should be clearly identified in each case. From this view,
we find particularly appealing pieces of evidence (e.g. in
neuroscience) in which, by experimentally inducing alterations to
standard conditions, deviations from criticality are measured in
otherwise critical-looking systems.

A general criticism can be raised to some of the analyses discussed
along this work, specifically, to those in which the evidence relies
on the existence of a theoretical model that provides, when tuned
close to its critical point, the best possible fit to empirical
observations. The criticism is that, if feature-rich empirical data
with structures spanning over broadly diverse scales are considered,
then it seems almost a tautology to conclude that the best possible
representation of them is obtained by fitting the proposed dynamical
model to operate close its critical point, as this is typically the
only region in parameter space where complex (feature-rich) patterns,
with many scales, are generated. In contrast, from an opposite
perspective, if actual biological data are structured across many
scales, it does not seem too far fetched to assume --applying the
Occam's razor-- that a general common mechanism may underlie the
emergence of such a hierarchy of scales, and the main candidate
mechanism for this consists in operating at the edge of a continuous
phase transition, i.e. being close to criticality. Thus, we are
confronted with a (epistemological) dichotomy: Is the putative
criticality of living systems just a reflection of the limitation of
our models which can possibly resemble large levels of ``complexity''
only at criticality?  or, on the contrary, is criticality actually a
common organizing principle at the roots of the generation of many
levels of organization required for complex biological behavior to
emerge?  Providing a satisfactory answer to these questions is a
problem of outmost importance to advance in the theoretical
understanding and modeling of complex living systems.

Even if diverse biological systems were finally proved to be genuinely
critical, some researchers might still retain this conclusion as
largely uninformative or even irrelevant.  It could be asked: ``so
what?''. What practical implications could be derived from such a
knowledge?  In our opinion, the design of strategies to control
neural/genetic networks --especially those aiming at resolving
pathologies-- based on notions of criticality, the construction of
algorithms of artificial intelligence exploiting scale-invariance at
different layers, or the application of ideas of collective
motion/intelligence to the design of e.g. swarms of robots, could
constitute important avenues to provide constructive answers to the
above question.

Novel advances, both at the experimental and theoretical
sides, will help elucidating what is the actual role played by
criticality and scale invariance in biological systems; meanwhile the
mere possibility remains as inspiring as ever and, definitely, worth
pursuing.

\begin{acknowledgments}
  I am deeply indebted to the collaborators with whom I had the
  pleasure and privilege of studying some of the topics discussed
  here; among them: G. Grinstein, P.L. Garrido, J.A. Bonachela,
  A. Maritan, J. Marro, F. de los Santos, P.I. Hurtado, J.J. Torres,
  J. Cort{\'e}s. S. Johnson, R. Burioni, S. Suweis, A. Gabrielli,
  C. Castellano, S. Pigolotti, M. Cencini, A. Vespignani, R. Dickman,
  S. Zapperi, Y. Tu, D. Plenz, as well as to H. Chat{\'e} for his
  collaboration and encouragement in the early stages of this project.
  I am especially thankful to P. Moretti, J.  Hidalgo, P. Villegas,
  L. Seoane, J. Soriano, S. di Santo and V. Buend{\'\i}a, for extremely
  useful comments on early versions of the manuscript.  I acknowledge
  the Spanish-MINECO, grant FIS2017-84256-P (FEDER funds) for
  financial support.
\end{acknowledgments}
\vspace{0.5cm}

\section*{Appendix A: Generic Scale invariance}
There are situations in which spatio-temporal scaling may emerge
without the need of parameter fine tuning \cite{GG91}.  (i) A
well-known example is the breaking of a continuous symmetry in
low-dimensional systems, as it happens e.g. in some models of
magnetism in which each spin can point in any arbitrary direction in a
plane \cite{Binney}. These systems, instead of the usual ordered phase
at low temperature, exhibit a broad ``quasi-ordered'' phase
characterized by local order together with generic power-law decaying
correlations \cite{GG91}. This type of ordering is relevant for
bacterial-colony patterns \cite{Bugs} as well as in the analysis of
collective motion (see Sect. IV-C).

(ii) Generic scale invariance can also emerge in the presence of
structural disorder or heterogeneity. In statistical physics, one
refers to ``quenched disorder'' as the form of spatial-dependent
heterogeneity which is intrinsic to the microscopic components and
remains frozen in time, reflecting structural
heterogeneities. Quenched disorder can alter the nature of phase
transitions \cite{Vojta-Review,Paula1} and can also induce novel
phases absent in homogeneous systems. For instance, in the contact
process, quenched disorder can be implemented by considering a lattice
with some missing links, a more complex (disordered) network of
connections, and/or a node-dependent propagation rate $\lambda$. In
all these cases, a novel phase called a \emph{Griffiths phase}
--characterized by critical-like features appearing all across the
phase and not just at a unique point-- emerges
\cite{Vojta-Review,Adriana,Cafiero,GPCN}. 

(iii) Another mechanism that produces a type of generic
scale-invariance relevant in biological systems (see e.g. Sect. IV-A7)
is ``neutral dynamics''.  Neutral theories play a key role in
population genetics \cite{Kimura}, population ecology
\cite{Hubbell,Review-neutral}, epidemics \cite{Pinto}, etc. They have
in common the fact that differences among possible coexisting
``species'' (let them be alleles of a gene, types of trees, bacterial
strains,...) are neglected. In other words, all ``species'' are
dynamically equivalent or \emph{neutral}
\cite{Blythe,Liggett,Dornic01}.  It has been recently shown that in a
(``multispecies'') contact process that includes different
types/species of activity, if a new species --neutral to the exiting
ones-- is introduced, it experiences a stochastic process in which its
total population does not have a net tendency to either grow or
shrink. This generates generically scale-free avalanches of the focal
species unfolding in a sea of activity of the other species, without
the need to invoque criticality \cite{neutral-neural}.

\section*{Appendix B: Probabilistic models and statistical criticality}
Probabilistic models can be constructed such that they match the
statistics of observed empirical data \cite{Bialek1995}.  Without loss
of generality, an empirically observed pattern at a given time window
can be codified as a sequence of binary variables of length $N$:
$s_i=0,1$ for $i=1,2,...N$.  Denoting $P({\bf{s}})$ the (unknown) probability of
finding the system in the state ${\mathbf{s}}=(s_1,s_2,...s_N)$ it is
possible to approximate it by a distribution function with the
constraint that it reproduces the empirically-measured mean values
$\langle s_i \rangle$ for all $i$, as well as the covariances
$ \langle s_i s_j \rangle$ for all $i$ and $j$.  Imposing a maximum
entropy principle (i.e. selecting the model with the smallest number
of assumptions\footnote{In information theory, the entropy of a
  probability distribution quantifies the ignorance about the
  variable; thus, making no assumptions about the distribution is
  equivalent to maximizing the entropy \cite{Cover-Thomas,max-ent}.})
it is straightforward to derive the explicit form of the optimal model
\begin{equation} P(\mathbf{s})=\frac{1}{Z} \exp \left[
    \sum_{i<j} J_{ij} s_i s_j + \sum_i h_i s_i \right]
\label{Ising}
\end{equation}
where $Z$ ensures normalization and which coincides with the Boltzmann
equilibrium distribution of the Ising-like model, and where the free parameters $h_i$ and
$J_{ij}$ need to be fitted, so that the imposed constraints are
satisfied \cite{Boltzmann-machine}.\footnote{Obtaining the optimal
parameter set --i.e. inferring effective interactions from
correlations-- is a computationally costly task, usually referred as
``inverse Ising problem'' \cite{Cocco,Inverse,Bialek-weak}.}

Bialek and coworkers introduced an efective parameter $\beta$ --much
as an inverse temperature in equilibrium statistical mechanics--
multiplying each estimated parameter in the inferred model
Eq.(\ref{Ising}).  Clearly, varying $\beta$ a relative change of the
weights assigned to different configurations is produced. In this way
one generates a family of $\beta$-dependent probability distributions,
$P(\mathbf{s}|\beta g)$, interpolating between the low and high
temperature phases\footnote{At $\beta=0$ (infinite temperature) all
  configurations are equiprobable, while in the opposite limit all the
  weight concentrates on the most likely (fully ordered)
  configuration.}. At some intermediate value, $\beta_c$, there is a
critical point (as identified e.g. by a peak in the susceptibility or
the specific heat).  It has been found that diverse inference problems
(from retinal populations
\cite{Bialek-weak,Bialek-thermo1,Bialek-thermo2,Bialek-searching} to
flocks of birds \cite{Bialek-birds}, and the immune system
\cite{Bialek-antibody}) produce models in which $\beta_c \approx 1$,
--or converges to $1$ as the system size is enlarged-- i.e. that
inferred models appear to be close to the very critical point of the
underlying Ising-like problem (see \textcite{Mora-Bialek} for a clear
and pedagogical discussion of these issues).

Let us also mention that there is an ongoing debate on the
interpretation of these results. In particular, it has been shown that
signatures of criticality may emerge naturally in inferred models if
there is a marginalization over non-observed variables, such as
e.g. correlated external inputs, even without the need for direct
interactions among units \textcite{Mehta,Latham}.  More in general,
Marsili and collaborators pointed out that the alleged
criticality of such models can be a rather general consequence of the
inference procedure, meaning that inferred models fitting real-world
(``feature-rich'' or ``informative'' ) data do, most likely, look
critical when an effective probabilistic model is constructed
\cite{Marsili1,Marsili2,Marsili3,Marsili4}.  We shall not delve
further into the controversy about the meaning and significance of
this type of purely statistical approaches to criticality (see also
the Discussion section).

\section*{Appendix C: Adaptation and evolution towards criticality}

To shed light onto the general problem of how information-processing
(living) systems tune themselves to operate near critical points,
\textcite{Goudarzi} considered an ensemble of individuals or
``agents'', each represented as an internal RBN, including some input
nodes (able to read information from the environment) and some readout
nodes (providing outputs/responses). Such agents evolve though a
genetic algorithm \cite{GA} that allocates larger ``fitness'' values
to agents that perform better a series of computational tasks (each
one consisting in assigning a given output to each specific input),
which are alternated in time. The conclusion is that agents converge
to a state close to criticality; i.e. critical dynamics emerge as the
optimal solution under the combined selective pressures of having to
learn different tasks and being able to readily shift among them,
following changes in the tasks.  Instead, in the presence of noise,
optimal agents tend to be slightly subcritical, rather than critical,
thus compensating for extrinsic sources of variability
\cite{Villegas-Boolean}. In a similar approach, \textcite{Hidalgo2014}
showed that communities of similar adaptive agents, whose task is to
communicate with each other (inferring their respective internal
states) in an efficient way, converge to quasi-critical states.  This
result constitutes a possible parsimonious explanation for the
emergence of critical-like behavior in groups of individuals
coordinating themselves as a collective entity \cite{Hidalgo2016} (see
also \textcite{Adami2010}).

\section*{Appendix D: Other putatively critical living systems}
\label{s-mix}

Here we briefly discuss, a sample of other biological systems for
which empirical evidences of criticality exist.

{\bf Cell membranes.}  Cell membranes are not just rigid impenetrable
walls separating the interior of cells from the outside environment;
they regulate the kind, direction, and amount of substances that can
pass across them. Cell membranes are permeable only at some locations
and, for this, their local composition needs to be heterogeneous
\cite{oil,Cicuta1,Julicher-cells}. There is compelling empirical
evidence that the mixture of lipids that constitute the skeleton of
cell membranes operates very close to the (de-mixing) phase transition
at which their different components segregate
\cite{membranes1,membranes2,Cicuta1,Cicuta2,Cicuta3}.  In this way,
composition fluctuations are extremely large, enabling very diverse
structural domains to appear, thus providing the membrane with a large
spectrum of possible local structures, at which different processes
may occur, and entailing a rich repertoire of functionalities.

{\bf RNA viruses.}  RNA viruses are believed to replicate at the edge
of an ``error catastrophe''. If the error rates for copying the viral
genome were very small RNA viruses would have little variability,
hindering adaptation and evolution. Instead, if they were too large
then the fidelity of the replication machinery would be compromised
and it would not be possible to maintain important genetic elements
nor the identity of the (quasi)species itself
\cite{Eigen1979,Eigen1989}.  It was conjectured
\cite{Drake,Eigen2000,Sole-Manrubia,Sole5} and has been (partially)
verified in recent experiments \cite{Error-experiments,Error-HIV} that
RNA viruses might operate right at the edge of the catastrophe,
providing them with maximal variability compatible with genotypic
robustness\footnote{Error catastrophe has been considered for
  treatment of viral infections employing drugs that push the error
  rate beyond this threshold; see \textcite{Summers} for a critical
  review.}.

{\bf Physiological rhythms.}  The presence of temporal
scale-invariance in physiological rhythms of healthy subjects, as well
as its break-down in abnormal conditions, have been long explored
\cite{Stanley2002,West-fractal,Pathology}.  In particular, to mention
one example, a specific connection between the complex fluctuations of
human heart-rate variability and criticality has been put forward
\cite{Ivanov1999,Struzik1,Struzik2,Ivanov2007}. In the related context
of blood-pressure regulation, vaso-vagal syncopes have been identified
as large ``avalanches'' in a self-organized cardiovascular regulatory
system poised at criticality \cite{Blood}.  In general, such a
regulation to scale-free behavior seems to impart health advantages,
including system integrity and adaptability \cite{Stanley2002}.

{\bf Miscellanea.}
Criticality has also been claimed to play a relevant role in the immune system
\cite{Burgos-immune,Bialek-antibody}, cancer and carcenogenesis
\cite{Sole-cancer1,Sole-cancer2,Davies-cancer,Rosenfeld-cancer},
proteins \cite{Chialvo-proteins,Phillips2009}, mitochondria
\cite{Mito,Chialvo-Mito}, etc. Also, quantum criticality and its
relevance for the origin of life at the microscopic scale has been the
subject of a recent proposal \cite{quantum}.  Finally, let us mention
that ecosystems as a whole have been studied --from a
macroevolutionary viewpoint-- as dynamical structures lying at the
edge of instability
\cite{Sole-Manrubia,Sneppen1995,Bak1993,Adami1995,Suweis2013,Biroli,McKane},
illustrating that the ideas discussed here can be extended to larger
scales in the hierarchy of biological complexity.

\bibliography{biblio_RMP}

%merlin.mbs apsrmp4-1.bst 2010-07-25 4.21a (PWD, AO, DPC) hacked
%Control: author (75) reversed first initials jnrlst
\begin{thebibliography}{595}%
\makeatletter
\providecommand \@ifxundefined [1]{%
 \@ifx{#1\undefined}
}%
\providecommand \@ifnum [1]{%
 \ifnum #1\expandafter \@firstoftwo
 \else \expandafter \@secondoftwo
 \fi
}%
\providecommand \@ifx [1]{%
 \ifx #1\expandafter \@firstoftwo
 \else \expandafter \@secondoftwo
 \fi
}%
\providecommand \natexlab [1]{#1}%
\providecommand \enquote  [1]{``#1''}%
\providecommand \bibnamefont  [1]{#1}%
\providecommand \bibfnamefont [1]{#1}%
\providecommand \citenamefont [1]{#1}%
\providecommand \href@noop [0]{\@secondoftwo}%
\providecommand \href [0]{\begingroup \@sanitize@url \@href}%
\providecommand \@href[1]{\@@startlink{#1}\@@href}%
\providecommand \@@href[1]{\endgroup#1\@@endlink}%
\providecommand \@sanitize@url [0]{\catcode `\\12\catcode `\$12\catcode
  `\&12\catcode `\#12\catcode `\^12\catcode `\_12\catcode `\%12\relax}%
\providecommand \@@startlink[1]{}%
\providecommand \@@endlink[0]{}%
\providecommand \url  [0]{\begingroup\@sanitize@url \@url }%
\providecommand \@url [1]{\endgroup\@href {#1}{\urlprefix }}%
\providecommand \urlprefix  [0]{URL }%
\providecommand \Eprint [0]{\href }%
\providecommand \doibase [0]{http://dx.doi.org/}%
\providecommand \selectlanguage [0]{\@gobble}%
\providecommand \bibinfo  [0]{\@secondoftwo}%
\providecommand \bibfield  [0]{\@secondoftwo}%
\providecommand \translation [1]{[#1]}%
\providecommand \BibitemOpen [0]{}%
\providecommand \bibitemStop [0]{}%
\providecommand \bibitemNoStop [0]{.\EOS\space}%
\providecommand \EOS [0]{\spacefactor3000\relax}%
\providecommand \BibitemShut  [1]{\csname bibitem#1\endcsname}%
\let\auto@bib@innerbib\@empty
%</preamble>
\bibitem [{\citenamefont {Abbott}\ and\ \citenamefont {van
  Vreeswijk}(1993)}]{Abbott-asynchronous}%
  \BibitemOpen
  \bibfield  {author} {\bibinfo {author} {\bibnamefont {Abbott}, \bibfnamefont
  {L.}}, \ and\ \bibinfo {author} {\bibfnamefont {C.}~\bibnamefont {van
  Vreeswijk}}} (\bibinfo {year} {1993}),\ \href@noop {} {\bibfield  {journal}
  {\bibinfo  {journal} {Phys. Rev. E}\ }\textbf {\bibinfo {volume}
  {48}}~(\bibinfo {number} {2}),\ \bibinfo {pages} {1483}}\BibitemShut
  {NoStop}%
\bibitem [{\citenamefont {{Acebr\'on}}\ \emph {et~al.}(2005)\citenamefont
  {{Acebr\'on}}, \citenamefont {{Bonilla}}, \citenamefont {{P\'erez-Vicente}},
  \citenamefont {{Ritort}},\ and\ \citenamefont {{Spigler}}}]{Kuramoto-review}%
  \BibitemOpen
  \bibfield  {author} {\bibinfo {author} {\bibnamefont {{Acebr\'on}},
  \bibfnamefont {J.~A.}}, \bibinfo {author} {\bibfnamefont {L.~L.}\
  \bibnamefont {{Bonilla}}}, \bibinfo {author} {\bibfnamefont {C.~J.}\
  \bibnamefont {{P\'erez-Vicente}}}, \bibinfo {author} {\bibfnamefont
  {F.}~\bibnamefont {{Ritort}}}, \ and\ \bibinfo {author} {\bibfnamefont
  {R.}~\bibnamefont {{Spigler}}}} (\bibinfo {year} {2005}),\ \href {\doibase
  10.1103/RevModPhys.77.137} {\bibfield  {journal} {\bibinfo  {journal} {Rev.
  Mod. Phys.}\ }\textbf {\bibinfo {volume} {77}},\ \bibinfo {pages}
  {137}}\BibitemShut {NoStop}%
\bibitem [{\citenamefont {Ackley}\ \emph {et~al.}(1985)\citenamefont {Ackley},
  \citenamefont {Hinton},\ and\ \citenamefont {Sejnowski}}]{Boltzmann-machine}%
  \BibitemOpen
  \bibfield  {author} {\bibinfo {author} {\bibnamefont {Ackley}, \bibfnamefont
  {D.}}, \bibinfo {author} {\bibfnamefont {G.}~\bibnamefont {Hinton}}, \ and\
  \bibinfo {author} {\bibfnamefont {T.}~\bibnamefont {Sejnowski}}} (\bibinfo
  {year} {1985}),\ \href@noop {} {\bibfield  {journal} {\bibinfo  {journal}
  {Cognitive Science}\ }\textbf {\bibinfo {volume} {9}}~(\bibinfo {number}
  {1}),\ \bibinfo {pages} {147}}\BibitemShut {NoStop}%
\bibitem [{\citenamefont {Adami}(1995)}]{Adami1995}%
  \BibitemOpen
  \bibfield  {author} {\bibinfo {author} {\bibnamefont {Adami}, \bibfnamefont
  {C.}}} (\bibinfo {year} {1995}),\ \href@noop {} {\bibfield  {journal}
  {\bibinfo  {journal} {Phys. Lett. A}\ }\textbf {\bibinfo {volume} {203}},\
  \bibinfo {pages} {29}}\BibitemShut {NoStop}%
\bibitem [{\citenamefont {Aitchison}\ \emph {et~al.}(2016)\citenamefont
  {Aitchison}, \citenamefont {Corradi},\ and\ \citenamefont {Latham}}]{Latham}%
  \BibitemOpen
  \bibfield  {author} {\bibinfo {author} {\bibnamefont {Aitchison},
  \bibfnamefont {L.}}, \bibinfo {author} {\bibfnamefont {N.}~\bibnamefont
  {Corradi}}, \ and\ \bibinfo {author} {\bibfnamefont {P.}~\bibnamefont
  {Latham}}} (\bibinfo {year} {2016}),\ \href@noop {} {\bibfield  {journal}
  {\bibinfo  {journal} {PLoS Comp. Biol.}\ }\textbf {\bibinfo {volume}
  {12}}~(\bibinfo {number} {12}),\ \bibinfo {pages} {e1005110}}\BibitemShut
  {NoStop}%
\bibitem [{\citenamefont {Akaike}(1969)}]{AR}%
  \BibitemOpen
  \bibfield  {author} {\bibinfo {author} {\bibnamefont {Akaike}, \bibfnamefont
  {H.}}} (\bibinfo {year} {1969}),\ \href@noop {} {\bibfield  {journal}
  {\bibinfo  {journal} {Ann. Inst. of Stat. Math.}\ }\textbf {\bibinfo {volume}
  {21}}~(\bibinfo {number} {1}),\ \bibinfo {pages} {243}}\BibitemShut {NoStop}%
\bibitem [{\citenamefont {Albert}(2004)}]{Reka}%
  \BibitemOpen
  \bibfield  {author} {\bibinfo {author} {\bibnamefont {Albert}, \bibfnamefont
  {R.}}} (\bibinfo {year} {2004}),\ in\ \href@noop {} {\emph {\bibinfo
  {booktitle} {Complex networks}}}\ (\bibinfo  {publisher} {Springer})\ pp.\
  \bibinfo {pages} {459--481}\BibitemShut {NoStop}%
\bibitem [{\citenamefont {Albert}\ and\ \citenamefont
  {Barab\'asi}(2002)}]{Barabasi-Review}%
  \BibitemOpen
  \bibfield  {author} {\bibinfo {author} {\bibnamefont {Albert}, \bibfnamefont
  {R.}}, \ and\ \bibinfo {author} {\bibfnamefont {A.-L.}\ \bibnamefont
  {Barab\'asi}}} (\bibinfo {year} {2002}),\ \href@noop {} {\bibfield  {journal}
  {\bibinfo  {journal} {Rev. Mod. Phys.}\ }\textbf {\bibinfo {volume} {74}},\
  \bibinfo {pages} {47}}\BibitemShut {NoStop}%
\bibitem [{\citenamefont {Albert}\ and\ \citenamefont {Othmer}(2003)}]{Reka2}%
  \BibitemOpen
  \bibfield  {author} {\bibinfo {author} {\bibnamefont {Albert}, \bibfnamefont
  {R.}}, \ and\ \bibinfo {author} {\bibfnamefont {H.}~\bibnamefont {Othmer}}}
  (\bibinfo {year} {2003}),\ \href@noop {} {\bibfield  {journal} {\bibinfo
  {journal} {J.Theor.Biol.}\ }\textbf {\bibinfo {volume} {223}}~(\bibinfo
  {number} {1}),\ \bibinfo {pages} {1}}\BibitemShut {NoStop}%
\bibitem [{\citenamefont {Aldana}(2003)}]{Aldana2003}%
  \BibitemOpen
  \bibfield  {author} {\bibinfo {author} {\bibnamefont {Aldana}, \bibfnamefont
  {M.}}} (\bibinfo {year} {2003}),\ \href@noop {} {\bibfield  {journal}
  {\bibinfo  {journal} {Physica D}\ }\textbf {\bibinfo {volume}
  {185}}~(\bibinfo {number} {1}),\ \bibinfo {pages} {45}}\BibitemShut {NoStop}%
\bibitem [{\citenamefont {Aldana}\ \emph {et~al.}(2007)\citenamefont {Aldana},
  \citenamefont {Balleza}, \citenamefont {Kauffman},\ and\ \citenamefont
  {Resendiz}}]{Aldana2007}%
  \BibitemOpen
  \bibfield  {author} {\bibinfo {author} {\bibnamefont {Aldana}, \bibfnamefont
  {M.}}, \bibinfo {author} {\bibfnamefont {E.}~\bibnamefont {Balleza}},
  \bibinfo {author} {\bibfnamefont {S.}~\bibnamefont {Kauffman}}, \ and\
  \bibinfo {author} {\bibfnamefont {O.}~\bibnamefont {Resendiz}}} (\bibinfo
  {year} {2007}),\ \href@noop {} {\bibfield  {journal} {\bibinfo  {journal} {J.
  Theor. Biol.}\ }\textbf {\bibinfo {volume} {245}}~(\bibinfo {number} {3}),\
  \bibinfo {pages} {433}}\BibitemShut {NoStop}%
\bibitem [{\citenamefont {Allegrini}\ \emph {et~al.}(2010)\citenamefont
  {Allegrini}, \citenamefont {Paradisi}, \citenamefont {Menicucci},\ and\
  \citenamefont {Gemignani}}]{EEG-Allegrini}%
  \BibitemOpen
  \bibfield  {author} {\bibinfo {author} {\bibnamefont {Allegrini},
  \bibfnamefont {P.}}, \bibinfo {author} {\bibfnamefont {P.}~\bibnamefont
  {Paradisi}}, \bibinfo {author} {\bibfnamefont {D.}~\bibnamefont {Menicucci}},
  \ and\ \bibinfo {author} {\bibfnamefont {A.}~\bibnamefont {Gemignani}}}
  (\bibinfo {year} {2010}),\ \href@noop {} {\bibfield  {journal} {\bibinfo
  {journal} {Front. Physiol.}\ }\textbf {\bibinfo {volume} {1}},\ \bibinfo
  {pages} {128}}\BibitemShut {NoStop}%
\bibitem [{\citenamefont {Alon}(2003)}]{Alon-bionetworks}%
  \BibitemOpen
  \bibfield  {author} {\bibinfo {author} {\bibnamefont {Alon}, \bibfnamefont
  {U.}}} (\bibinfo {year} {2003}),\ \href@noop {} {\bibfield  {journal}
  {\bibinfo  {journal} {Science}\ }\textbf {\bibinfo {volume} {301}}~(\bibinfo
  {number} {5641}),\ \bibinfo {pages} {1866}}\BibitemShut {NoStop}%
\bibitem [{\citenamefont {Alon}(2006)}]{Alon-book}%
  \BibitemOpen
  \bibfield  {author} {\bibinfo {author} {\bibnamefont {Alon}, \bibfnamefont
  {U.}}} (\bibinfo {year} {2006}),\ \href@noop {} {\emph {\bibinfo {title} {An
  introduction to systems biology: design principles of biological circuits}}}\
  (\bibinfo  {publisher} {CRC press},\ \bibinfo {address} {London})\BibitemShut
  {NoStop}%
\bibitem [{\citenamefont {Alonso}\ \emph {et~al.}(2014)\citenamefont {Alonso},
  \citenamefont {Proekt}, \citenamefont {Schwartz}, \citenamefont {Pryor},
  \citenamefont {Cecchi},\ and\ \citenamefont {Magnasco}}]{Magnasco2}%
  \BibitemOpen
  \bibfield  {author} {\bibinfo {author} {\bibnamefont {Alonso}, \bibfnamefont
  {L.~M.}}, \bibinfo {author} {\bibfnamefont {A.}~\bibnamefont {Proekt}},
  \bibinfo {author} {\bibfnamefont {T.~H.}\ \bibnamefont {Schwartz}}, \bibinfo
  {author} {\bibfnamefont {K.~O.}\ \bibnamefont {Pryor}}, \bibinfo {author}
  {\bibfnamefont {G.~A.}\ \bibnamefont {Cecchi}}, \ and\ \bibinfo {author}
  {\bibfnamefont {M.~O.}\ \bibnamefont {Magnasco}}} (\bibinfo {year} {2014}),\
  \href@noop {} {\bibfield  {journal} {\bibinfo  {journal} {Front. Neural
  Circuits}\ }\textbf {\bibinfo {volume} {8}},\ \bibinfo {pages}
  {20}}\BibitemShut {NoStop}%
\bibitem [{\citenamefont {Alvarado}\ \emph {et~al.}(2013)\citenamefont
  {Alvarado}, \citenamefont {Sheinman}, \citenamefont {Sharma}, \citenamefont
  {MacKintosh},\ and\ \citenamefont {Koenderink}}]{McIntosh}%
  \BibitemOpen
  \bibfield  {author} {\bibinfo {author} {\bibnamefont {Alvarado},
  \bibfnamefont {J.}}, \bibinfo {author} {\bibfnamefont {M.}~\bibnamefont
  {Sheinman}}, \bibinfo {author} {\bibfnamefont {A.}~\bibnamefont {Sharma}},
  \bibinfo {author} {\bibfnamefont {F.~C.}\ \bibnamefont {MacKintosh}}, \ and\
  \bibinfo {author} {\bibfnamefont {G.~H.}\ \bibnamefont {Koenderink}}}
  (\bibinfo {year} {2013}),\ \href@noop {} {\bibfield  {journal} {\bibinfo
  {journal} {Nat. Phys.}\ }\textbf {\bibinfo {volume} {9}}~(\bibinfo {number}
  {9}),\ \bibinfo {pages} {591}}\BibitemShut {NoStop}%
\bibitem [{\citenamefont {Amari}(1972)}]{Amari}%
  \BibitemOpen
  \bibfield  {author} {\bibinfo {author} {\bibnamefont {Amari}, \bibfnamefont
  {S.}}} (\bibinfo {year} {1972}),\ \href@noop {} {\bibfield  {journal}
  {\bibinfo  {journal} {IEEE Trans. Syst. Man. Cybern.}\ }\textbf {\bibinfo
  {volume} {2}},\ \bibinfo {pages} {643}}\BibitemShut {NoStop}%
\bibitem [{\citenamefont {Amit}(1992)}]{Amit-book2}%
  \BibitemOpen
  \bibfield  {author} {\bibinfo {author} {\bibnamefont {Amit}, \bibfnamefont
  {D.}}} (\bibinfo {year} {1992}),\ \href@noop {} {\emph {\bibinfo {title}
  {Modeling brain function: The world of attractor neural networks}}}\
  (\bibinfo  {publisher} {Cambridge University Press})\BibitemShut {NoStop}%
\bibitem [{\citenamefont {Amit}\ and\ \citenamefont
  {Brunel}(1997)}]{Amit-Brunel}%
  \BibitemOpen
  \bibfield  {author} {\bibinfo {author} {\bibnamefont {Amit}, \bibfnamefont
  {D.}}, \ and\ \bibinfo {author} {\bibfnamefont {N.}~\bibnamefont {Brunel}}}
  (\bibinfo {year} {1997}),\ \href@noop {} {\bibfield  {journal} {\bibinfo
  {journal} {Cerebral cortex}\ }\textbf {\bibinfo {volume} {7}}~(\bibinfo
  {number} {3}),\ \bibinfo {pages} {237}}\BibitemShut {NoStop}%
\bibitem [{\citenamefont {Anderson}\ \emph {et~al.}(1972)\citenamefont
  {Anderson} \emph {et~al.}}]{Anderson}%
  \BibitemOpen
  \bibfield  {author} {\bibinfo {author} {\bibnamefont {Anderson},
  \bibfnamefont {P.~W.}},  \emph {et~al.}} (\bibinfo {year} {1972}),\
  \href@noop {} {\bibfield  {journal} {\bibinfo  {journal} {Science}\ }\textbf
  {\bibinfo {volume} {177}}~(\bibinfo {number} {4047}),\ \bibinfo {pages}
  {393}}\BibitemShut {NoStop}%
\bibitem [{\citenamefont {de~Andrade~Costa}\ \emph {et~al.}(2015)\citenamefont
  {de~Andrade~Costa}, \citenamefont {Copelli},\ and\ \citenamefont
  {Kinouchi}}]{Copelli2015}%
  \BibitemOpen
  \bibfield  {author} {\bibinfo {author} {\bibnamefont {de~Andrade~Costa},
  \bibfnamefont {A.}}, \bibinfo {author} {\bibfnamefont {M.}~\bibnamefont
  {Copelli}}, \ and\ \bibinfo {author} {\bibfnamefont {O.}~\bibnamefont
  {Kinouchi}}} (\bibinfo {year} {2015}),\ \href@noop {} {\bibfield  {journal}
  {\bibinfo  {journal} {J. Stat. Mech.}\ }\textbf {\bibinfo {volume}
  {2015}}~(\bibinfo {number} {6}),\ \bibinfo {pages} {P06004}}\BibitemShut
  {NoStop}%
\bibitem [{\citenamefont {Aon}\ \emph {et~al.}(2004)\citenamefont {Aon},
  \citenamefont {Cortassa},\ and\ \citenamefont {O'Rourke}}]{Mito}%
  \BibitemOpen
  \bibfield  {author} {\bibinfo {author} {\bibnamefont {Aon}, \bibfnamefont
  {M.~A.}}, \bibinfo {author} {\bibfnamefont {S.}~\bibnamefont {Cortassa}}, \
  and\ \bibinfo {author} {\bibfnamefont {B.}~\bibnamefont {O'Rourke}}}
  (\bibinfo {year} {2004}),\ \href@noop {} {\bibfield  {journal} {\bibinfo
  {journal} {Proc. Nat. Acad. of Sci. USA}\ }\textbf {\bibinfo {volume}
  {101}}~(\bibinfo {number} {13}),\ \bibinfo {pages} {4447}}\BibitemShut
  {NoStop}%
\bibitem [{\citenamefont {de~Arcangelis}(2011)}]{Lucilla2011}%
  \BibitemOpen
  \bibfield  {author} {\bibinfo {author} {\bibnamefont {de~Arcangelis},
  \bibfnamefont {L.}}} (\bibinfo {year} {2011}),\ \href@noop {} {\bibfield
  {journal} {\bibinfo  {journal} {J. of Physics: Conf. Series}\ }\textbf
  {\bibinfo {volume} {297}}~(\bibinfo {number} {1}),\ \bibinfo {pages}
  {012001}}\BibitemShut {NoStop}%
\bibitem [{\citenamefont {de~Arcangelis}(2012)}]{Lucilla-dragon}%
  \BibitemOpen
  \bibfield  {author} {\bibinfo {author} {\bibnamefont {de~Arcangelis},
  \bibfnamefont {L.}}} (\bibinfo {year} {2012}),\ \href@noop {} {\bibfield
  {journal} {\bibinfo  {journal} {Eur. Phys. J. Spec.Top.}\ }\textbf {\bibinfo
  {volume} {205}}~(\bibinfo {number} {1}),\ \bibinfo {pages} {243}}\BibitemShut
  {NoStop}%
\bibitem [{\citenamefont {de~Arcangelis}\ and\ \citenamefont
  {Herrmann}(2010)}]{Lucilla2010}%
  \BibitemOpen
  \bibfield  {author} {\bibinfo {author} {\bibnamefont {de~Arcangelis},
  \bibfnamefont {L.}}, \ and\ \bibinfo {author} {\bibfnamefont {H.~J.}\
  \bibnamefont {Herrmann}}} (\bibinfo {year} {2010}),\ \href@noop {} {\bibfield
   {journal} {\bibinfo  {journal} {Proc. Natl. Acad. Sci. USA.}\ }\textbf
  {\bibinfo {volume} {107}}~(\bibinfo {number} {9}),\ \bibinfo {pages}
  {3977}}\BibitemShut {NoStop}%
\bibitem [{\citenamefont {de~Arcangelis}\ and\ \citenamefont
  {Herrmann}(2012)}]{Lucilla2012}%
  \BibitemOpen
  \bibfield  {author} {\bibinfo {author} {\bibnamefont {de~Arcangelis},
  \bibfnamefont {L.}}, \ and\ \bibinfo {author} {\bibfnamefont {H.~J.}\
  \bibnamefont {Herrmann}}} (\bibinfo {year} {2012}),\ \href@noop {} {\bibfield
   {journal} {\bibinfo  {journal} {Front. Physiol.}\ }\textbf {\bibinfo
  {volume} {3}},\ \bibinfo {pages} {0062}}\BibitemShut {NoStop}%
\bibitem [{\citenamefont {de~Arcangelis}\ \emph {et~al.}(2014)\citenamefont
  {de~Arcangelis}, \citenamefont {Lombardi},\ and\ \citenamefont
  {Herrmann}}]{Lucilla2014}%
  \BibitemOpen
  \bibfield  {author} {\bibinfo {author} {\bibnamefont {de~Arcangelis},
  \bibfnamefont {L.}}, \bibinfo {author} {\bibfnamefont {F.}~\bibnamefont
  {Lombardi}}, \ and\ \bibinfo {author} {\bibfnamefont {H.}~\bibnamefont
  {Herrmann}}} (\bibinfo {year} {2014}),\ \href@noop {} {\bibfield  {journal}
  {\bibinfo  {journal} {J. Stat. Mech.}\ }\textbf {\bibinfo {volume}
  {2014}}~(\bibinfo {number} {3}),\ \bibinfo {pages} {P03026}}\BibitemShut
  {NoStop}%
\bibitem [{\citenamefont {de~Arcangelis}\ \emph {et~al.}(2006)\citenamefont
  {de~Arcangelis}, \citenamefont {Perrone-Capano},\ and\ \citenamefont
  {Herrmann}}]{Lucilla2006}%
  \BibitemOpen
  \bibfield  {author} {\bibinfo {author} {\bibnamefont {de~Arcangelis},
  \bibfnamefont {L.}}, \bibinfo {author} {\bibfnamefont {C.}~\bibnamefont
  {Perrone-Capano}}, \ and\ \bibinfo {author} {\bibfnamefont {H.~J.}\
  \bibnamefont {Herrmann}}} (\bibinfo {year} {2006}),\ \href {\doibase
  10.1103/PhysRevLett.96.028107} {\bibfield  {journal} {\bibinfo  {journal}
  {Phys. Rev. Lett.}\ }\textbf {\bibinfo {volume} {96}},\ \bibinfo {pages}
  {028107}}\BibitemShut {NoStop}%
\bibitem [{\citenamefont {Arenas}\ \emph {et~al.}(2008)\citenamefont {Arenas},
  \citenamefont {D{\'\i}az-Guilera}, \citenamefont {Kurths}, \citenamefont
  {Moreno},\ and\ \citenamefont {Zhou}}]{Arenas-synchro}%
  \BibitemOpen
  \bibfield  {author} {\bibinfo {author} {\bibnamefont {Arenas}, \bibfnamefont
  {A.}}, \bibinfo {author} {\bibfnamefont {A.}~\bibnamefont
  {D{\'\i}az-Guilera}}, \bibinfo {author} {\bibfnamefont {J.}~\bibnamefont
  {Kurths}}, \bibinfo {author} {\bibfnamefont {Y.}~\bibnamefont {Moreno}}, \
  and\ \bibinfo {author} {\bibfnamefont {C.}~\bibnamefont {Zhou}}} (\bibinfo
  {year} {2008}),\ \href@noop {} {\bibfield  {journal} {\bibinfo  {journal}
  {Phys. Rep.}\ }\textbf {\bibinfo {volume} {469}}~(\bibinfo {number} {3}),\
  \bibinfo {pages} {93}}\BibitemShut {NoStop}%
\bibitem [{\citenamefont {Arieli}\ \emph {et~al.}(1996)\citenamefont {Arieli},
  \citenamefont {Sterkin}, \citenamefont {Grinvald},\ and\ \citenamefont
  {Aertsen}}]{Arieli}%
  \BibitemOpen
  \bibfield  {author} {\bibinfo {author} {\bibnamefont {Arieli}, \bibfnamefont
  {A.}}, \bibinfo {author} {\bibfnamefont {A.}~\bibnamefont {Sterkin}},
  \bibinfo {author} {\bibfnamefont {A.}~\bibnamefont {Grinvald}}, \ and\
  \bibinfo {author} {\bibfnamefont {A.}~\bibnamefont {Aertsen}}} (\bibinfo
  {year} {1996}),\ \href@noop {} {\bibfield  {journal} {\bibinfo  {journal}
  {Science}\ }\textbf {\bibinfo {volume} {273}}~(\bibinfo {number} {5283}),\
  \bibinfo {pages} {1868}}\BibitemShut {NoStop}%
\bibitem [{\citenamefont {Ashby}(1960)}]{Ashby}%
  \BibitemOpen
  \bibfield  {author} {\bibinfo {author} {\bibnamefont {Ashby}, \bibfnamefont
  {W.}}} (\bibinfo {year} {1960}),\ \href@noop {} {\emph {\bibinfo {title}
  {Design for a Brain. The origin of adaptive behaviour}}}\ (\bibinfo
  {publisher} {New York, Willey})\BibitemShut {NoStop}%
\bibitem [{\citenamefont {Attanasi}\ \emph
  {et~al.}(2014{\natexlab{a}})\citenamefont {Attanasi}, \citenamefont
  {Cavagna}, \citenamefont {Del~Castello}, \citenamefont {Giardina},
  \citenamefont {Grigera}, \citenamefont {Jeli{\'c}}, \citenamefont {Melillo},
  \citenamefont {Parisi}, \citenamefont {Pohl}, \citenamefont {Shen} \emph
  {et~al.}}]{Attanasi2}%
  \BibitemOpen
  \bibfield  {author} {\bibinfo {author} {\bibnamefont {Attanasi},
  \bibfnamefont {A.}}, \bibinfo {author} {\bibfnamefont {A.}~\bibnamefont
  {Cavagna}}, \bibinfo {author} {\bibfnamefont {L.}~\bibnamefont
  {Del~Castello}}, \bibinfo {author} {\bibfnamefont {I.}~\bibnamefont
  {Giardina}}, \bibinfo {author} {\bibfnamefont {T.~S.}\ \bibnamefont
  {Grigera}}, \bibinfo {author} {\bibfnamefont {A.}~\bibnamefont {Jeli{\'c}}},
  \bibinfo {author} {\bibfnamefont {S.}~\bibnamefont {Melillo}}, \bibinfo
  {author} {\bibfnamefont {L.}~\bibnamefont {Parisi}}, \bibinfo {author}
  {\bibfnamefont {O.}~\bibnamefont {Pohl}}, \bibinfo {author} {\bibfnamefont
  {E.}~\bibnamefont {Shen}},  \emph {et~al.}} (\bibinfo {year}
  {2014}{\natexlab{a}}),\ \href@noop {} {\bibfield  {journal} {\bibinfo
  {journal} {Nat. Phys.}\ }\textbf {\bibinfo {volume} {10}}~(\bibinfo {number}
  {9}),\ \bibinfo {pages} {691}}\BibitemShut {NoStop}%
\bibitem [{\citenamefont {Attanasi}\ \emph
  {et~al.}(2014{\natexlab{b}})\citenamefont {Attanasi}, \citenamefont
  {Cavagna}, \citenamefont {Del~Castello}, \citenamefont {Giardina},
  \citenamefont {Melillo}, \citenamefont {Parisi}, \citenamefont {Pohl},
  \citenamefont {Rossaro}, \citenamefont {Shen}, \citenamefont {Silvestri}
  \emph {et~al.}}]{Attanasi}%
  \BibitemOpen
  \bibfield  {author} {\bibinfo {author} {\bibnamefont {Attanasi},
  \bibfnamefont {A.}}, \bibinfo {author} {\bibfnamefont {A.}~\bibnamefont
  {Cavagna}}, \bibinfo {author} {\bibfnamefont {L.}~\bibnamefont
  {Del~Castello}}, \bibinfo {author} {\bibfnamefont {I.}~\bibnamefont
  {Giardina}}, \bibinfo {author} {\bibfnamefont {S.}~\bibnamefont {Melillo}},
  \bibinfo {author} {\bibfnamefont {L.}~\bibnamefont {Parisi}}, \bibinfo
  {author} {\bibfnamefont {O.}~\bibnamefont {Pohl}}, \bibinfo {author}
  {\bibfnamefont {B.}~\bibnamefont {Rossaro}}, \bibinfo {author} {\bibfnamefont
  {E.}~\bibnamefont {Shen}}, \bibinfo {author} {\bibfnamefont {E.}~\bibnamefont
  {Silvestri}},  \emph {et~al.}} (\bibinfo {year} {2014}{\natexlab{b}}),\
  \href@noop {} {\bibfield  {journal} {\bibinfo  {journal} {Phys. Rev. Lett.}\
  }\textbf {\bibinfo {volume} {113}}~(\bibinfo {number} {23}),\ \bibinfo
  {pages} {238102}}\BibitemShut {NoStop}%
\bibitem [{\citenamefont {Aurell}\ and\ \citenamefont
  {Ekeberg}(2012)}]{Inverse}%
  \BibitemOpen
  \bibfield  {author} {\bibinfo {author} {\bibnamefont {Aurell}, \bibfnamefont
  {E.}}, \ and\ \bibinfo {author} {\bibfnamefont {M.}~\bibnamefont {Ekeberg}}}
  (\bibinfo {year} {2012}),\ \href@noop {} {\bibfield  {journal} {\bibinfo
  {journal} {Phys. Rev. Lett.}\ }\textbf {\bibinfo {volume} {108}}~(\bibinfo
  {number} {9}),\ \bibinfo {pages} {090201}}\BibitemShut {NoStop}%
\bibitem [{\citenamefont {Azaele}\ \emph {et~al.}(2016)\citenamefont {Azaele},
  \citenamefont {Suweis}, \citenamefont {Grilli}, \citenamefont {Volkov},
  \citenamefont {Banavar},\ and\ \citenamefont {Maritan}}]{Review-neutral}%
  \BibitemOpen
  \bibfield  {author} {\bibinfo {author} {\bibnamefont {Azaele}, \bibfnamefont
  {S.}}, \bibinfo {author} {\bibfnamefont {S.}~\bibnamefont {Suweis}}, \bibinfo
  {author} {\bibfnamefont {J.}~\bibnamefont {Grilli}}, \bibinfo {author}
  {\bibfnamefont {I.}~\bibnamefont {Volkov}}, \bibinfo {author} {\bibfnamefont
  {J.~R.}\ \bibnamefont {Banavar}}, \ and\ \bibinfo {author} {\bibfnamefont
  {A.}~\bibnamefont {Maritan}}} (\bibinfo {year} {2016}),\ \href {\doibase
  10.1103/RevModPhys.88.035003} {\bibfield  {journal} {\bibinfo  {journal}
  {Rev. Mod. Phys.}\ }\textbf {\bibinfo {volume} {88}},\ \bibinfo {pages}
  {035003}}\BibitemShut {NoStop}%
\bibitem [{\citenamefont {Baek}\ \emph {et~al.}(2011)\citenamefont {Baek},
  \citenamefont {Bernhardsson},\ and\ \citenamefont {Minnhagen}}]{Minnhagen}%
  \BibitemOpen
  \bibfield  {author} {\bibinfo {author} {\bibnamefont {Baek}, \bibfnamefont
  {S.~K.}}, \bibinfo {author} {\bibfnamefont {S.}~\bibnamefont {Bernhardsson}},
  \ and\ \bibinfo {author} {\bibfnamefont {P.}~\bibnamefont {Minnhagen}}}
  (\bibinfo {year} {2011}),\ \href@noop {} {\bibfield  {journal} {\bibinfo
  {journal} {New J. Phys.}\ }\textbf {\bibinfo {volume} {13}}~(\bibinfo
  {number} {4}),\ \bibinfo {pages} {043004}}\BibitemShut {NoStop}%
\bibitem [{\citenamefont {Baglietto}\ \emph {et~al.}(2012)\citenamefont
  {Baglietto}, \citenamefont {Albano},\ and\ \citenamefont
  {Candia}}]{Baglietto}%
  \BibitemOpen
  \bibfield  {author} {\bibinfo {author} {\bibnamefont {Baglietto},
  \bibfnamefont {G.}}, \bibinfo {author} {\bibfnamefont {E.~V.}\ \bibnamefont
  {Albano}}, \ and\ \bibinfo {author} {\bibfnamefont {J.}~\bibnamefont
  {Candia}}} (\bibinfo {year} {2012}),\ \href@noop {} {\bibfield  {journal}
  {\bibinfo  {journal} {Interface Focus}\ }\textbf {\bibinfo {volume}
  {2}}~(\bibinfo {number} {6}),\ \bibinfo {pages} {708}}\BibitemShut {NoStop}%
\bibitem [{\citenamefont {Bagnoli}\ \emph {et~al.}(2003)\citenamefont
  {Bagnoli}, \citenamefont {Cecconi}, \citenamefont {Flammini},\ and\
  \citenamefont {Vespignani}}]{anomalies3}%
  \BibitemOpen
  \bibfield  {author} {\bibinfo {author} {\bibnamefont {Bagnoli}, \bibfnamefont
  {F.}}, \bibinfo {author} {\bibfnamefont {F.}~\bibnamefont {Cecconi}},
  \bibinfo {author} {\bibfnamefont {A.}~\bibnamefont {Flammini}}, \ and\
  \bibinfo {author} {\bibfnamefont {A.}~\bibnamefont {Vespignani}}} (\bibinfo
  {year} {2003}),\ \href@noop {} {\bibfield  {journal} {\bibinfo  {journal}
  {EPL (Europhys. Lett.)}\ }\textbf {\bibinfo {volume} {63}}~(\bibinfo {number}
  {4}),\ \bibinfo {pages} {512}}\BibitemShut {NoStop}%
\bibitem [{\citenamefont {Bak}(1996)}]{Bak}%
  \BibitemOpen
  \bibfield  {author} {\bibinfo {author} {\bibnamefont {Bak}, \bibfnamefont
  {P.}}} (\bibinfo {year} {1996}),\ \href@noop {} {\emph {\bibinfo {title} {How
  nature works: the science of self-organized criticality}}}\ (\bibinfo
  {publisher} {Copernicus, New York})\BibitemShut {NoStop}%
\bibitem [{\citenamefont {Bak}\ \emph {et~al.}(1990)\citenamefont {Bak},
  \citenamefont {Chen},\ and\ \citenamefont {Tang}}]{FF1}%
  \BibitemOpen
  \bibfield  {author} {\bibinfo {author} {\bibnamefont {Bak}, \bibfnamefont
  {P.}}, \bibinfo {author} {\bibfnamefont {K.}~\bibnamefont {Chen}}, \ and\
  \bibinfo {author} {\bibfnamefont {C.}~\bibnamefont {Tang}}} (\bibinfo {year}
  {1990}),\ \href@noop {} {\bibfield  {journal} {\bibinfo  {journal} {Phys.
  Lett. A}\ }\textbf {\bibinfo {volume} {147}}~(\bibinfo {number} {5}),\
  \bibinfo {pages} {297}}\BibitemShut {NoStop}%
\bibitem [{\citenamefont {Bak}\ and\ \citenamefont
  {Chialvo}(2001)}]{Chialvo-Bak2}%
  \BibitemOpen
  \bibfield  {author} {\bibinfo {author} {\bibnamefont {Bak}, \bibfnamefont
  {P.}}, \ and\ \bibinfo {author} {\bibfnamefont {D.~R.}\ \bibnamefont
  {Chialvo}}} (\bibinfo {year} {2001}),\ \href {\doibase
  10.1103/PhysRevE.63.031912} {\bibfield  {journal} {\bibinfo  {journal} {Phys.
  Rev. E}\ }\textbf {\bibinfo {volume} {63}},\ \bibinfo {pages}
  {031912}}\BibitemShut {NoStop}%
\bibitem [{\citenamefont {Bak}\ and\ \citenamefont {Sneppen}(1993)}]{Bak1993}%
  \BibitemOpen
  \bibfield  {author} {\bibinfo {author} {\bibnamefont {Bak}, \bibfnamefont
  {P.}}, \ and\ \bibinfo {author} {\bibfnamefont {K.}~\bibnamefont {Sneppen}}}
  (\bibinfo {year} {1993}),\ \href@noop {} {\bibfield  {journal} {\bibinfo
  {journal} {Phys. Rev. Lett.}\ }\textbf {\bibinfo {volume} {71}}~(\bibinfo
  {number} {24}),\ \bibinfo {pages} {4083}}\BibitemShut {NoStop}%
\bibitem [{\citenamefont {Bak}\ and\ \citenamefont {Tang}(1989)}]{BT}%
  \BibitemOpen
  \bibfield  {author} {\bibinfo {author} {\bibnamefont {Bak}, \bibfnamefont
  {P.}}, \ and\ \bibinfo {author} {\bibfnamefont {C.}~\bibnamefont {Tang}}}
  (\bibinfo {year} {1989}),\ \href@noop {} {\bibfield  {journal} {\bibinfo
  {journal} {J. Geophys. Res}\ }\textbf {\bibinfo {volume} {94}}~(\bibinfo
  {number} {15}),\ \bibinfo {pages} {635}}\BibitemShut {NoStop}%
\bibitem [{\citenamefont {Bak}\ \emph {et~al.}(1987)\citenamefont {Bak},
  \citenamefont {Tang},\ and\ \citenamefont {Wiesenfeld}}]{BTW}%
  \BibitemOpen
  \bibfield  {author} {\bibinfo {author} {\bibnamefont {Bak}, \bibfnamefont
  {P.}}, \bibinfo {author} {\bibfnamefont {C.}~\bibnamefont {Tang}}, \ and\
  \bibinfo {author} {\bibfnamefont {K.}~\bibnamefont {Wiesenfeld}}} (\bibinfo
  {year} {1987}),\ \href@noop {} {\bibfield  {journal} {\bibinfo  {journal}
  {Phys. Rev. Lett.}\ }\textbf {\bibinfo {volume} {59}}~(\bibinfo {number}
  {4}),\ \bibinfo {pages} {381}}\BibitemShut {NoStop}%
\bibitem [{\citenamefont {Ballerini}\ \emph {et~al.}(2008)\citenamefont
  {Ballerini}, \citenamefont {Cabibbo}, \citenamefont {Candelier},
  \citenamefont {Cavagna}, \citenamefont {Cisbani}, \citenamefont {Giardina},
  \citenamefont {Lecomte}, \citenamefont {Orlandi}, \citenamefont {Parisi},
  \citenamefont {Procaccini} \emph {et~al.}}]{Cavagna2008}%
  \BibitemOpen
  \bibfield  {author} {\bibinfo {author} {\bibnamefont {Ballerini},
  \bibfnamefont {M.}}, \bibinfo {author} {\bibfnamefont {N.}~\bibnamefont
  {Cabibbo}}, \bibinfo {author} {\bibfnamefont {R.}~\bibnamefont {Candelier}},
  \bibinfo {author} {\bibfnamefont {A.}~\bibnamefont {Cavagna}}, \bibinfo
  {author} {\bibfnamefont {E.}~\bibnamefont {Cisbani}}, \bibinfo {author}
  {\bibfnamefont {I.}~\bibnamefont {Giardina}}, \bibinfo {author}
  {\bibfnamefont {V.}~\bibnamefont {Lecomte}}, \bibinfo {author} {\bibfnamefont
  {A.}~\bibnamefont {Orlandi}}, \bibinfo {author} {\bibfnamefont
  {G.}~\bibnamefont {Parisi}}, \bibinfo {author} {\bibfnamefont
  {A.}~\bibnamefont {Procaccini}},  \emph {et~al.}} (\bibinfo {year} {2008}),\
  \href@noop {} {\bibfield  {journal} {\bibinfo  {journal} {Proc. Natl. Acad.
  Sci. USA.}\ }\textbf {\bibinfo {volume} {105}}~(\bibinfo {number} {4}),\
  \bibinfo {pages} {1232}}\BibitemShut {NoStop}%
\bibitem [{\citenamefont {Balleza}\ \emph {et~al.}(2008)\citenamefont
  {Balleza}, \citenamefont {Alvarez-Buylla}, \citenamefont {Chaos},
  \citenamefont {Kauffman}, \citenamefont {Shmulevich},\ and\ \citenamefont
  {Aldana}}]{Balleza2008}%
  \BibitemOpen
  \bibfield  {author} {\bibinfo {author} {\bibnamefont {Balleza}, \bibfnamefont
  {E.}}, \bibinfo {author} {\bibfnamefont {E.~R.}\ \bibnamefont
  {Alvarez-Buylla}}, \bibinfo {author} {\bibfnamefont {A.}~\bibnamefont
  {Chaos}}, \bibinfo {author} {\bibfnamefont {S.}~\bibnamefont {Kauffman}},
  \bibinfo {author} {\bibfnamefont {I.}~\bibnamefont {Shmulevich}}, \ and\
  \bibinfo {author} {\bibfnamefont {M.}~\bibnamefont {Aldana}}} (\bibinfo
  {year} {2008}),\ \href@noop {} {\bibfield  {journal} {\bibinfo  {journal}
  {PLoS One}\ }\textbf {\bibinfo {volume} {3}}~(\bibinfo {number} {6}),\
  \bibinfo {pages} {e2456}}\BibitemShut {NoStop}%
\bibitem [{\citenamefont {Banavar}\ \emph {et~al.}(2014)\citenamefont
  {Banavar}, \citenamefont {Cooke}, \citenamefont {Rinaldo},\ and\
  \citenamefont {Maritan}}]{Amos-form}%
  \BibitemOpen
  \bibfield  {author} {\bibinfo {author} {\bibnamefont {Banavar}, \bibfnamefont
  {J.~R.}}, \bibinfo {author} {\bibfnamefont {T.~J.}\ \bibnamefont {Cooke}},
  \bibinfo {author} {\bibfnamefont {A.}~\bibnamefont {Rinaldo}}, \ and\
  \bibinfo {author} {\bibfnamefont {A.}~\bibnamefont {Maritan}}} (\bibinfo
  {year} {2014}),\ \href@noop {} {\bibfield  {journal} {\bibinfo  {journal}
  {Proc. Natl. Acad. Sci. USA.}\ }\textbf {\bibinfo {volume} {111}}~(\bibinfo
  {number} {9}),\ \bibinfo {pages} {3332}}\BibitemShut {NoStop}%
\bibitem [{\citenamefont {Banavar}\ \emph {et~al.}(1999)\citenamefont
  {Banavar}, \citenamefont {Maritan},\ and\ \citenamefont
  {Rinaldo}}]{Amos-networks}%
  \BibitemOpen
  \bibfield  {author} {\bibinfo {author} {\bibnamefont {Banavar}, \bibfnamefont
  {J.~R.}}, \bibinfo {author} {\bibfnamefont {A.}~\bibnamefont {Maritan}}, \
  and\ \bibinfo {author} {\bibfnamefont {A.}~\bibnamefont {Rinaldo}}} (\bibinfo
  {year} {1999}),\ \href@noop {} {\bibfield  {journal} {\bibinfo  {journal}
  {Nature}\ }\textbf {\bibinfo {volume} {399}}~(\bibinfo {number} {6732}),\
  \bibinfo {pages} {130}}\BibitemShut {NoStop}%
\bibitem [{\citenamefont {Banavar}\ \emph
  {et~al.}(2010{\natexlab{a}})\citenamefont {Banavar}, \citenamefont
  {Maritan},\ and\ \citenamefont {Volkov}}]{max-ent}%
  \BibitemOpen
  \bibfield  {author} {\bibinfo {author} {\bibnamefont {Banavar}, \bibfnamefont
  {J.~R.}}, \bibinfo {author} {\bibfnamefont {A.}~\bibnamefont {Maritan}}, \
  and\ \bibinfo {author} {\bibfnamefont {I.}~\bibnamefont {Volkov}}} (\bibinfo
  {year} {2010}{\natexlab{a}}),\ \href@noop {} {\bibfield  {journal} {\bibinfo
  {journal} {J. of Phys.: Cond. Matt.}\ }\textbf {\bibinfo {volume}
  {22}}~(\bibinfo {number} {6}),\ \bibinfo {pages} {063101}}\BibitemShut
  {NoStop}%
\bibitem [{\citenamefont {Banavar}\ \emph
  {et~al.}(2010{\natexlab{b}})\citenamefont {Banavar}, \citenamefont {Moses},
  \citenamefont {Brown}, \citenamefont {Damuth}, \citenamefont {Rinaldo},
  \citenamefont {Sibly},\ and\ \citenamefont {Maritan}}]{Amos-form2}%
  \BibitemOpen
  \bibfield  {author} {\bibinfo {author} {\bibnamefont {Banavar}, \bibfnamefont
  {J.~R.}}, \bibinfo {author} {\bibfnamefont {M.~E.}\ \bibnamefont {Moses}},
  \bibinfo {author} {\bibfnamefont {J.~H.}\ \bibnamefont {Brown}}, \bibinfo
  {author} {\bibfnamefont {J.}~\bibnamefont {Damuth}}, \bibinfo {author}
  {\bibfnamefont {A.}~\bibnamefont {Rinaldo}}, \bibinfo {author} {\bibfnamefont
  {R.~M.}\ \bibnamefont {Sibly}}, \ and\ \bibinfo {author} {\bibfnamefont
  {A.}~\bibnamefont {Maritan}}} (\bibinfo {year} {2010}{\natexlab{b}}),\
  \href@noop {} {\bibfield  {journal} {\bibinfo  {journal} {Proc. Natl. Acad.
  Sci. USA.}\ }\textbf {\bibinfo {volume} {107}}~(\bibinfo {number} {36}),\
  \bibinfo {pages} {15816}}\BibitemShut {NoStop}%
\bibitem [{\citenamefont {Barabasi}(2005)}]{Bara-human}%
  \BibitemOpen
  \bibfield  {author} {\bibinfo {author} {\bibnamefont {Barabasi},
  \bibfnamefont {A.-L.}}} (\bibinfo {year} {2005}),\ \href@noop {} {\bibfield
  {journal} {\bibinfo  {journal} {Nature}\ }\textbf {\bibinfo {volume} {435}},\
  \bibinfo {pages} {207}}\BibitemShut {NoStop}%
\bibitem [{\citenamefont {Barab{\'a}si}\ and\ \citenamefont
  {Albert}(1999)}]{BA}%
  \BibitemOpen
  \bibfield  {author} {\bibinfo {author} {\bibnamefont {Barab{\'a}si},
  \bibfnamefont {A.-L.}}, \ and\ \bibinfo {author} {\bibfnamefont
  {R.}~\bibnamefont {Albert}}} (\bibinfo {year} {1999}),\ \href@noop {}
  {\bibfield  {journal} {\bibinfo  {journal} {Science}\ }\textbf {\bibinfo
  {volume} {286}}~(\bibinfo {number} {5439}),\ \bibinfo {pages}
  {509}}\BibitemShut {NoStop}%
\bibitem [{\citenamefont {Barnett}\ \emph {et~al.}(2013)\citenamefont
  {Barnett}, \citenamefont {Lizier}, \citenamefont {Harr{\'e}}, \citenamefont
  {Seth},\ and\ \citenamefont {Bossomaier}}]{Barnett}%
  \BibitemOpen
  \bibfield  {author} {\bibinfo {author} {\bibnamefont {Barnett}, \bibfnamefont
  {L.}}, \bibinfo {author} {\bibfnamefont {J.~T.}\ \bibnamefont {Lizier}},
  \bibinfo {author} {\bibfnamefont {M.}~\bibnamefont {Harr{\'e}}}, \bibinfo
  {author} {\bibfnamefont {A.~K.}\ \bibnamefont {Seth}}, \ and\ \bibinfo
  {author} {\bibfnamefont {T.}~\bibnamefont {Bossomaier}}} (\bibinfo {year}
  {2013}),\ \href@noop {} {\bibfield  {journal} {\bibinfo  {journal} {Phys.
  Rev. Lett.}\ }\textbf {\bibinfo {volume} {111}}~(\bibinfo {number} {17}),\
  \bibinfo {pages} {177203}}\BibitemShut {NoStop}%
\bibitem [{\citenamefont {Barral}\ and\ \citenamefont {Reyes}(2016)}]{Reyes}%
  \BibitemOpen
  \bibfield  {author} {\bibinfo {author} {\bibnamefont {Barral}, \bibfnamefont
  {J.}}, \ and\ \bibinfo {author} {\bibfnamefont {A.~D.}\ \bibnamefont
  {Reyes}}} (\bibinfo {year} {2016}),\ \href@noop {} {\bibfield  {journal}
  {\bibinfo  {journal} {Nature Neurosci.}\ }\textbf {\bibinfo {volume}
  {19}}~(\bibinfo {number} {12}),\ \bibinfo {pages} {1690}}\BibitemShut
  {NoStop}%
\bibitem [{\citenamefont {Barrat}\ \emph {et~al.}(2008)\citenamefont {Barrat},
  \citenamefont {Barthelemy},\ and\ \citenamefont {Vespignani}}]{Vespi-book}%
  \BibitemOpen
  \bibfield  {author} {\bibinfo {author} {\bibnamefont {Barrat}, \bibfnamefont
  {A.}}, \bibinfo {author} {\bibfnamefont {M.}~\bibnamefont {Barthelemy}}, \
  and\ \bibinfo {author} {\bibfnamefont {A.}~\bibnamefont {Vespignani}}}
  (\bibinfo {year} {2008}),\ \href@noop {} {\emph {\bibinfo {title} {Dynamical
  processes on complex networks}}}\ (\bibinfo  {publisher} {Cambridge Univ.
  press})\BibitemShut {NoStop}%
\bibitem [{\citenamefont {Bassingthwaighte}\ \emph {et~al.}(1994)\citenamefont
  {Bassingthwaighte}, \citenamefont {Liebovitch},\ and\ \citenamefont
  {West}}]{West-fractal}%
  \BibitemOpen
  \bibfield  {author} {\bibinfo {author} {\bibnamefont {Bassingthwaighte},
  \bibfnamefont {J.~B.}}, \bibinfo {author} {\bibfnamefont {L.~S.}\
  \bibnamefont {Liebovitch}}, \ and\ \bibinfo {author} {\bibfnamefont {B.~J.}\
  \bibnamefont {West}}} (\bibinfo {year} {1994}),\ in\ \href@noop {} {\emph
  {\bibinfo {booktitle} {Fractal physiology}}}\ (\bibinfo  {publisher}
  {Springer})\ pp.\ \bibinfo {pages} {11--44}\BibitemShut {NoStop}%
\bibitem [{\citenamefont {Beckmann}\ \emph {et~al.}(2005)\citenamefont
  {Beckmann}, \citenamefont {DeLuca}, \citenamefont {Devlin},\ and\
  \citenamefont {Smith}}]{RSN4}%
  \BibitemOpen
  \bibfield  {author} {\bibinfo {author} {\bibnamefont {Beckmann},
  \bibfnamefont {C.~F.}}, \bibinfo {author} {\bibfnamefont {M.}~\bibnamefont
  {DeLuca}}, \bibinfo {author} {\bibfnamefont {J.~T.}\ \bibnamefont {Devlin}},
  \ and\ \bibinfo {author} {\bibfnamefont {S.~M.}\ \bibnamefont {Smith}}}
  (\bibinfo {year} {2005}),\ \href@noop {} {\bibfield  {journal} {\bibinfo
  {journal} {Philos. Trans. R. Soc. London, Ser. B}\ }\textbf {\bibinfo
  {volume} {360}}~(\bibinfo {number} {1457}),\ \bibinfo {pages}
  {1001}}\BibitemShut {NoStop}%
\bibitem [{\citenamefont {B\'{e}dard}\ \emph {et~al.}(2006)\citenamefont
  {B\'{e}dard}, \citenamefont {Kr\"{o}ger},\ and\ \citenamefont
  {Destexhe}}]{Destexhe2006}%
  \BibitemOpen
  \bibfield  {author} {\bibinfo {author} {\bibnamefont {B\'{e}dard},
  \bibfnamefont {C.}}, \bibinfo {author} {\bibfnamefont {H.}~\bibnamefont
  {Kr\"{o}ger}}, \ and\ \bibinfo {author} {\bibfnamefont {A.}~\bibnamefont
  {Destexhe}}} (\bibinfo {year} {2006}),\ \href {\doibase
  10.1103/PhysRevLett.97.118102} {\bibfield  {journal} {\bibinfo  {journal}
  {Phys. Rev. Lett.}\ }\textbf {\bibinfo {volume} {97}}~(\bibinfo {number}
  {11}),\ \bibinfo {pages} {118102}}\BibitemShut {NoStop}%
\bibitem [{\citenamefont {Beekman}\ \emph {et~al.}(2001)\citenamefont
  {Beekman}, \citenamefont {Sumpter},\ and\ \citenamefont
  {Ratnieks}}]{Ants2001}%
  \BibitemOpen
  \bibfield  {author} {\bibinfo {author} {\bibnamefont {Beekman}, \bibfnamefont
  {M.}}, \bibinfo {author} {\bibfnamefont {D.~J.}\ \bibnamefont {Sumpter}}, \
  and\ \bibinfo {author} {\bibfnamefont {F.~L.}\ \bibnamefont {Ratnieks}}}
  (\bibinfo {year} {2001}),\ \href@noop {} {\bibfield  {journal} {\bibinfo
  {journal} {Proc. Natl. Acad. Sci. USA.}\ }\textbf {\bibinfo {volume}
  {98}}~(\bibinfo {number} {17}),\ \bibinfo {pages} {9703}}\BibitemShut
  {NoStop}%
\bibitem [{\citenamefont {Beggs}(2008)}]{Beggs-criticality}%
  \BibitemOpen
  \bibfield  {author} {\bibinfo {author} {\bibnamefont {Beggs}, \bibfnamefont
  {J.~M.}}} (\bibinfo {year} {2008}),\ \href@noop {} {\bibfield  {journal}
  {\bibinfo  {journal} {Phil Trans R Soc A}\ }\textbf {\bibinfo {volume}
  {366}}~(\bibinfo {number} {1864}),\ \bibinfo {pages} {329}}\BibitemShut
  {NoStop}%
\bibitem [{\citenamefont {Beggs}\ and\ \citenamefont {Plenz}(2003)}]{BP2003}%
  \BibitemOpen
  \bibfield  {author} {\bibinfo {author} {\bibnamefont {Beggs}, \bibfnamefont
  {J.~M.}}, \ and\ \bibinfo {author} {\bibfnamefont {D.}~\bibnamefont {Plenz}}}
  (\bibinfo {year} {2003}),\ \href@noop {} {\bibfield  {journal} {\bibinfo
  {journal} {J. Neurosci.}\ }\textbf {\bibinfo {volume} {23}}~(\bibinfo
  {number} {35}),\ \bibinfo {pages} {11167}}\BibitemShut {NoStop}%
\bibitem [{\citenamefont {Beggs}\ and\ \citenamefont {Plenz}(2004)}]{BP2004}%
  \BibitemOpen
  \bibfield  {author} {\bibinfo {author} {\bibnamefont {Beggs}, \bibfnamefont
  {J.~M.}}, \ and\ \bibinfo {author} {\bibfnamefont {D.}~\bibnamefont {Plenz}}}
  (\bibinfo {year} {2004}),\ \href@noop {} {\bibfield  {journal} {\bibinfo
  {journal} {J. Neurosci.}\ }\textbf {\bibinfo {volume} {24}}~(\bibinfo
  {number} {22}),\ \bibinfo {pages} {5216}}\BibitemShut {NoStop}%
\bibitem [{\citenamefont {Bellay}\ \emph {et~al.}(2015)\citenamefont {Bellay},
  \citenamefont {Klaus}, \citenamefont {Seshadri},\ and\ \citenamefont
  {Plenz}}]{2-photon}%
  \BibitemOpen
  \bibfield  {author} {\bibinfo {author} {\bibnamefont {Bellay}, \bibfnamefont
  {T.}}, \bibinfo {author} {\bibfnamefont {A.}~\bibnamefont {Klaus}}, \bibinfo
  {author} {\bibfnamefont {S.}~\bibnamefont {Seshadri}}, \ and\ \bibinfo
  {author} {\bibfnamefont {D.}~\bibnamefont {Plenz}}} (\bibinfo {year}
  {2015}),\ \href@noop {} {\bibfield  {journal} {\bibinfo  {journal} {Elife}\
  }\textbf {\bibinfo {volume} {4}},\ \bibinfo {pages} {e07224}}\BibitemShut
  {NoStop}%
\bibitem [{\citenamefont {Benayoun}\ \emph {et~al.}(2010)\citenamefont
  {Benayoun}, \citenamefont {Cowan}, \citenamefont {Van~Drongelen},\ and\
  \citenamefont {Wallace}}]{Cowan2010}%
  \BibitemOpen
  \bibfield  {author} {\bibinfo {author} {\bibnamefont {Benayoun},
  \bibfnamefont {M.}}, \bibinfo {author} {\bibfnamefont {J.~D.}\ \bibnamefont
  {Cowan}}, \bibinfo {author} {\bibfnamefont {W.}~\bibnamefont
  {Van~Drongelen}}, \ and\ \bibinfo {author} {\bibfnamefont {E.}~\bibnamefont
  {Wallace}}} (\bibinfo {year} {2010}),\ \href
  {http://dx.plos.org/10.1371/journal.pcbi.1000846} {\bibfield  {journal}
  {\bibinfo  {journal} {PLoS Comput. Biol.}\ }\textbf {\bibinfo {volume}
  {6}}~(\bibinfo {number} {7}),\ \bibinfo {pages} {13}}\BibitemShut {NoStop}%
\bibitem [{\citenamefont {Benedetto}\ \emph {et~al.}(2002)\citenamefont
  {Benedetto}, \citenamefont {Caglioti},\ and\ \citenamefont
  {Loreto}}]{Loreto}%
  \BibitemOpen
  \bibfield  {author} {\bibinfo {author} {\bibnamefont {Benedetto},
  \bibfnamefont {D.}}, \bibinfo {author} {\bibfnamefont {E.}~\bibnamefont
  {Caglioti}}, \ and\ \bibinfo {author} {\bibfnamefont {V.}~\bibnamefont
  {Loreto}}} (\bibinfo {year} {2002}),\ \href@noop {} {\bibfield  {journal}
  {\bibinfo  {journal} {Phys. Rev. Lett.}\ }\textbf {\bibinfo {volume}
  {88}}~(\bibinfo {number} {4}),\ \bibinfo {pages} {048702}}\BibitemShut
  {NoStop}%
\bibitem [{\citenamefont {Benford}(1938)}]{Benford}%
  \BibitemOpen
  \bibfield  {author} {\bibinfo {author} {\bibnamefont {Benford}, \bibfnamefont
  {F.}}} (\bibinfo {year} {1938}),\ \href@noop {} {\bibinfo  {journal} {Proc.
  Am. Philos. Soc.}\ ,\ \bibinfo {pages} {551}}\BibitemShut {NoStop}%
\bibitem [{\citenamefont {Beni}(2004)}]{swarm-robotics}%
  \BibitemOpen
\bibfield  {journal} {  }\bibfield  {author} {\bibinfo {author} {\bibnamefont
  {Beni}, \bibfnamefont {G.}}} (\bibinfo {year} {2004}),\ in\ \href@noop {}
  {\emph {\bibinfo {booktitle} {International Workshop on Swarm Robotics}}}\
  (\bibinfo {organization} {Springer})\ pp.\ \bibinfo {pages}
  {1--9}\BibitemShut {NoStop}%
\bibitem [{\citenamefont {Berdahl}\ \emph {et~al.}(2013)\citenamefont
  {Berdahl}, \citenamefont {Torney}, \citenamefont {Ioannou}, \citenamefont
  {Faria},\ and\ \citenamefont {Couzin}}]{Couzin-science}%
  \BibitemOpen
  \bibfield  {author} {\bibinfo {author} {\bibnamefont {Berdahl}, \bibfnamefont
  {A.}}, \bibinfo {author} {\bibfnamefont {C.~J.}\ \bibnamefont {Torney}},
  \bibinfo {author} {\bibfnamefont {C.~C.}\ \bibnamefont {Ioannou}}, \bibinfo
  {author} {\bibfnamefont {J.~J.}\ \bibnamefont {Faria}}, \ and\ \bibinfo
  {author} {\bibfnamefont {I.~D.}\ \bibnamefont {Couzin}}} (\bibinfo {year}
  {2013}),\ \href@noop {} {\bibfield  {journal} {\bibinfo  {journal} {Science}\
  }\textbf {\bibinfo {volume} {339}}~(\bibinfo {number} {6119}),\ \bibinfo
  {pages} {574}}\BibitemShut {NoStop}%
\bibitem [{\citenamefont {Berg}(1993)}]{Berg}%
  \BibitemOpen
  \bibfield  {author} {\bibinfo {author} {\bibnamefont {Berg}, \bibfnamefont
  {H.~C.}}} (\bibinfo {year} {1993}),\ \href@noop {} {\emph {\bibinfo {title}
  {Random walks in biology}}}\ (\bibinfo  {publisher} {Princeton University
  Press})\BibitemShut {NoStop}%
\bibitem [{\citenamefont {Bertschinger}\ and\ \citenamefont
  {Natschlager}(2004)}]{Ber-Nat}%
  \BibitemOpen
  \bibfield  {author} {\bibinfo {author} {\bibnamefont {Bertschinger},
  \bibfnamefont {N.}}, \ and\ \bibinfo {author} {\bibfnamefont
  {T.}~\bibnamefont {Natschlager}}} (\bibinfo {year} {2004}),\ \href@noop {}
  {\bibfield  {journal} {\bibinfo  {journal} {Neural Comput.}\ }\textbf
  {\bibinfo {volume} {16}}~(\bibinfo {number} {7}),\ \bibinfo {pages}
  {1413}}\BibitemShut {NoStop}%
\bibitem [{\citenamefont {Betzel}\ \emph {et~al.}(2013)\citenamefont {Betzel},
  \citenamefont {Griffa}, \citenamefont {Avena-Koenigsberger}, \citenamefont
  {Go\~ni}, \citenamefont {Thiran}, \citenamefont {Hagmann},\ and\
  \citenamefont {Sporns}}]{Sporns2014}%
  \BibitemOpen
  \bibfield  {author} {\bibinfo {author} {\bibnamefont {Betzel}, \bibfnamefont
  {R.~F.}}, \bibinfo {author} {\bibfnamefont {A.}~\bibnamefont {Griffa}},
  \bibinfo {author} {\bibfnamefont {A.}~\bibnamefont {Avena-Koenigsberger}},
  \bibinfo {author} {\bibfnamefont {J.}~\bibnamefont {Go\~ni}}, \bibinfo
  {author} {\bibfnamefont {J.-P.}\ \bibnamefont {Thiran}}, \bibinfo {author}
  {\bibfnamefont {P.}~\bibnamefont {Hagmann}}, \ and\ \bibinfo {author}
  {\bibfnamefont {O.}~\bibnamefont {Sporns}}} (\bibinfo {year} {2013}),\
  \href@noop {} {\bibfield  {journal} {\bibinfo  {journal} {Network Science}\
  }\textbf {\bibinfo {volume} {1}},\ \bibinfo {pages} {353}}\BibitemShut
  {NoStop}%
\bibitem [{\citenamefont {Bhattacharya}\ and\ \citenamefont
  {Vicsek}(2014)}]{Vicsek-foraging}%
  \BibitemOpen
  \bibfield  {author} {\bibinfo {author} {\bibnamefont {Bhattacharya},
  \bibfnamefont {K.}}, \ and\ \bibinfo {author} {\bibfnamefont
  {T.}~\bibnamefont {Vicsek}}} (\bibinfo {year} {2014}),\ \href@noop {}
  {\bibfield  {journal} {\bibinfo  {journal} {J. Royal Soc. Interface}\
  }\textbf {\bibinfo {volume} {11}}~(\bibinfo {number} {100})}\BibitemShut
  {NoStop}%
\bibitem [{\citenamefont {Bialek}(2012)}]{Bialek-book}%
  \BibitemOpen
  \bibfield  {author} {\bibinfo {author} {\bibnamefont {Bialek}, \bibfnamefont
  {W.}}} (\bibinfo {year} {2012}),\ \href@noop {} {\emph {\bibinfo {title}
  {Biophysics: searching for principles}}}\ (\bibinfo  {publisher} {Princeton
  University Press})\BibitemShut {NoStop}%
\bibitem [{\citenamefont {Bialek}(2018)}]{Bialek-perspectives}%
  \BibitemOpen
  \bibfield  {author} {\bibinfo {author} {\bibnamefont {Bialek}, \bibfnamefont
  {W.}}} (\bibinfo {year} {2018}),\ \href
  {http://stacks.iop.org/0034-4885/81/i=1/a=012601} {\bibfield  {journal}
  {\bibinfo  {journal} {Rep. Prog. Phys.}\ }\textbf {\bibinfo {volume}
  {81}}~(\bibinfo {number} {1}),\ \bibinfo {pages} {012601}}\BibitemShut
  {NoStop}%
\bibitem [{\citenamefont {Bialek}\ \emph {et~al.}(2014)\citenamefont {Bialek},
  \citenamefont {Cavagna}, \citenamefont {Giardina}, \citenamefont {Mora},
  \citenamefont {Pohl}, \citenamefont {Silvestri}, \citenamefont {Viale},\ and\
  \citenamefont {Walczak}}]{Bialek-social}%
  \BibitemOpen
  \bibfield  {author} {\bibinfo {author} {\bibnamefont {Bialek}, \bibfnamefont
  {W.}}, \bibinfo {author} {\bibfnamefont {A.}~\bibnamefont {Cavagna}},
  \bibinfo {author} {\bibfnamefont {I.}~\bibnamefont {Giardina}}, \bibinfo
  {author} {\bibfnamefont {T.}~\bibnamefont {Mora}}, \bibinfo {author}
  {\bibfnamefont {O.}~\bibnamefont {Pohl}}, \bibinfo {author} {\bibfnamefont
  {E.}~\bibnamefont {Silvestri}}, \bibinfo {author} {\bibfnamefont
  {M.}~\bibnamefont {Viale}}, \ and\ \bibinfo {author} {\bibfnamefont {A.~M.}\
  \bibnamefont {Walczak}}} (\bibinfo {year} {2014}),\ \href@noop {} {\bibfield
  {journal} {\bibinfo  {journal} {Proc. Natl. Acad. Sci. USA.}\ }\textbf
  {\bibinfo {volume} {111}}~(\bibinfo {number} {20}),\ \bibinfo {pages}
  {7212}}\BibitemShut {NoStop}%
\bibitem [{\citenamefont {Bialek}\ \emph {et~al.}(2012)\citenamefont {Bialek},
  \citenamefont {Cavagna}, \citenamefont {Giardina}, \citenamefont {Mora},
  \citenamefont {Silvestri}, \citenamefont {Viale},\ and\ \citenamefont
  {Walczak}}]{Bialek-birds}%
  \BibitemOpen
  \bibfield  {author} {\bibinfo {author} {\bibnamefont {Bialek}, \bibfnamefont
  {W.}}, \bibinfo {author} {\bibfnamefont {A.}~\bibnamefont {Cavagna}},
  \bibinfo {author} {\bibfnamefont {I.}~\bibnamefont {Giardina}}, \bibinfo
  {author} {\bibfnamefont {T.}~\bibnamefont {Mora}}, \bibinfo {author}
  {\bibfnamefont {E.}~\bibnamefont {Silvestri}}, \bibinfo {author}
  {\bibfnamefont {M.}~\bibnamefont {Viale}}, \ and\ \bibinfo {author}
  {\bibfnamefont {A.~M.}\ \bibnamefont {Walczak}}} (\bibinfo {year} {2012}),\
  \href@noop {} {\bibfield  {journal} {\bibinfo  {journal} {Proc. Natl. Acad.
  Sci. USA.}\ }\textbf {\bibinfo {volume} {109}}~(\bibinfo {number} {13}),\
  \bibinfo {pages} {4786}}\BibitemShut {NoStop}%
\bibitem [{\citenamefont {Bianconi}\ and\ \citenamefont
  {Marsili}(2004)}]{Bianconi2004}%
  \BibitemOpen
  \bibfield  {author} {\bibinfo {author} {\bibnamefont {Bianconi},
  \bibfnamefont {G.}}, \ and\ \bibinfo {author} {\bibfnamefont
  {M.}~\bibnamefont {Marsili}}} (\bibinfo {year} {2004}),\ \href@noop {}
  {\bibfield  {journal} {\bibinfo  {journal} {Phys. Rev. E}\ }\textbf {\bibinfo
  {volume} {70}}~(\bibinfo {number} {3}),\ \bibinfo {pages}
  {035105}}\BibitemShut {NoStop}%
\bibitem [{\citenamefont {Bienenstock}\ and\ \citenamefont
  {Lehmann}(1998)}]{Bienenstock}%
  \BibitemOpen
  \bibfield  {author} {\bibinfo {author} {\bibnamefont {Bienenstock},
  \bibfnamefont {E.}}, \ and\ \bibinfo {author} {\bibfnamefont
  {D.}~\bibnamefont {Lehmann}}} (\bibinfo {year} {1998}),\ \href@noop {}
  {\bibfield  {journal} {\bibinfo  {journal} {Adv. Complex Systems}\ }\textbf
  {\bibinfo {volume} {1}}~(\bibinfo {number} {04}),\ \bibinfo {pages}
  {361}}\BibitemShut {NoStop}%
\bibitem [{\citenamefont {Binder}(1981)}]{Binder}%
  \BibitemOpen
  \bibfield  {author} {\bibinfo {author} {\bibnamefont {Binder}, \bibfnamefont
  {K.}}} (\bibinfo {year} {1981}),\ \href@noop {} {\bibfield  {journal}
  {\bibinfo  {journal} {Z. Phys. B}\ }\textbf {\bibinfo {volume}
  {43}}~(\bibinfo {number} {2}),\ \bibinfo {pages} {119}}\BibitemShut {NoStop}%
\bibitem [{\citenamefont {Binney}\ \emph {et~al.}(1993)\citenamefont {Binney},
  \citenamefont {Dowrick}, \citenamefont {Fisher},\ and\ \citenamefont
  {Newman}}]{Binney}%
  \BibitemOpen
  \bibfield  {author} {\bibinfo {author} {\bibnamefont {Binney}, \bibfnamefont
  {J.}}, \bibinfo {author} {\bibfnamefont {N.}~\bibnamefont {Dowrick}},
  \bibinfo {author} {\bibfnamefont {A.}~\bibnamefont {Fisher}}, \ and\ \bibinfo
  {author} {\bibfnamefont {M.}~\bibnamefont {Newman}}} (\bibinfo {year}
  {1993}),\ \href@noop {} {\emph {\bibinfo {title} {The Theory of Critical
  Phenomena}}}\ (\bibinfo  {publisher} {Oxford University Press},\ \bibinfo
  {address} {Oxford})\BibitemShut {NoStop}%
\bibitem [{\citenamefont {Biroli}\ \emph {et~al.}(2017)\citenamefont {Biroli},
  \citenamefont {Bunin},\ and\ \citenamefont {Cammarota}}]{Biroli}%
  \BibitemOpen
  \bibfield  {author} {\bibinfo {author} {\bibnamefont {Biroli}, \bibfnamefont
  {G.}}, \bibinfo {author} {\bibfnamefont {G.}~\bibnamefont {Bunin}}, \ and\
  \bibinfo {author} {\bibfnamefont {C.}~\bibnamefont {Cammarota}}} (\bibinfo
  {year} {2017}),\ \href@noop {} {\bibinfo  {journal} {arXiv:1710.03606}\
  }\BibitemShut {NoStop}%
\bibitem [{\citenamefont {Biswal}\ \emph {et~al.}(1995)\citenamefont {Biswal},
  \citenamefont {Zerrin~Yetkin}, \citenamefont {Haughton},\ and\ \citenamefont
  {Hyde}}]{RSN1}%
  \BibitemOpen
\bibfield  {journal} {  }\bibfield  {author} {\bibinfo {author} {\bibnamefont
  {Biswal}, \bibfnamefont {B.}}, \bibinfo {author} {\bibfnamefont
  {F.}~\bibnamefont {Zerrin~Yetkin}}, \bibinfo {author} {\bibfnamefont {V.~M.}\
  \bibnamefont {Haughton}}, \ and\ \bibinfo {author} {\bibfnamefont {J.~S.}\
  \bibnamefont {Hyde}}} (\bibinfo {year} {1995}),\ \href@noop {} {\bibfield
  {journal} {\bibinfo  {journal} {Magnetic resonance in medicine}\ }\textbf
  {\bibinfo {volume} {34}}~(\bibinfo {number} {4}),\ \bibinfo {pages}
  {537}}\BibitemShut {NoStop}%
\bibitem [{\citenamefont {Blythe}\ and\ \citenamefont {McKane}(2007)}]{Blythe}%
  \BibitemOpen
  \bibfield  {author} {\bibinfo {author} {\bibnamefont {Blythe}, \bibfnamefont
  {R.~A.}}, \ and\ \bibinfo {author} {\bibfnamefont {A.~J.}\ \bibnamefont
  {McKane}}} (\bibinfo {year} {2007}),\ \href@noop {} {\bibfield  {journal}
  {\bibinfo  {journal} {J. Stat. Mech.}\ }\textbf {\bibinfo {volume}
  {2007}}~(\bibinfo {number} {07}),\ \bibinfo {pages} {P07018}}\BibitemShut
  {NoStop}%
\bibitem [{\citenamefont {Boccaletti}\ \emph {et~al.}(2006)\citenamefont
  {Boccaletti}, \citenamefont {Latora}, \citenamefont {Moreno}, \citenamefont
  {Chavez},\ and\ \citenamefont {Hwang}}]{Yamir-review}%
  \BibitemOpen
  \bibfield  {author} {\bibinfo {author} {\bibnamefont {Boccaletti},
  \bibfnamefont {S.}}, \bibinfo {author} {\bibfnamefont {V.}~\bibnamefont
  {Latora}}, \bibinfo {author} {\bibfnamefont {Y.}~\bibnamefont {Moreno}},
  \bibinfo {author} {\bibfnamefont {M.}~\bibnamefont {Chavez}}, \ and\ \bibinfo
  {author} {\bibfnamefont {D.-U.}\ \bibnamefont {Hwang}}} (\bibinfo {year}
  {2006}),\ \href@noop {} {\bibfield  {journal} {\bibinfo  {journal} {Phys.
  Rep.}\ }\textbf {\bibinfo {volume} {424}}~(\bibinfo {number} {4}),\ \bibinfo
  {pages} {175}}\BibitemShut {NoStop}%
\bibitem [{\citenamefont {Boedecker}\ \emph {et~al.}(2012)\citenamefont
  {Boedecker}, \citenamefont {Obst}, \citenamefont {Lizier}, \citenamefont
  {Mayer},\ and\ \citenamefont {Asada}}]{Asada}%
  \BibitemOpen
  \bibfield  {author} {\bibinfo {author} {\bibnamefont {Boedecker},
  \bibfnamefont {J.}}, \bibinfo {author} {\bibfnamefont {O.}~\bibnamefont
  {Obst}}, \bibinfo {author} {\bibfnamefont {J.~T.}\ \bibnamefont {Lizier}},
  \bibinfo {author} {\bibfnamefont {N.~M.}\ \bibnamefont {Mayer}}, \ and\
  \bibinfo {author} {\bibfnamefont {M.}~\bibnamefont {Asada}}} (\bibinfo {year}
  {2012}),\ \href@noop {} {\bibfield  {journal} {\bibinfo  {journal} {Theory in
  Biosci.}\ }\textbf {\bibinfo {volume} {131}}~(\bibinfo {number} {3}),\
  \bibinfo {pages} {205}}\BibitemShut {NoStop}%
\bibitem [{\citenamefont {Bonabeau}\ \emph {et~al.}(1999)\citenamefont
  {Bonabeau}, \citenamefont {Dorigo},\ and\ \citenamefont
  {Theraulaz}}]{Collective3}%
  \BibitemOpen
  \bibfield  {author} {\bibinfo {author} {\bibnamefont {Bonabeau},
  \bibfnamefont {E.}}, \bibinfo {author} {\bibfnamefont {M.}~\bibnamefont
  {Dorigo}}, \ and\ \bibinfo {author} {\bibfnamefont {G.}~\bibnamefont
  {Theraulaz}}} (\bibinfo {year} {1999}),\ \href@noop {} {\emph {\bibinfo
  {title} {Swarm intelligence: from natural to artificial systems}}},\ \bibinfo
  {number} {1}\ (\bibinfo  {publisher} {Oxford University Press})\BibitemShut
  {NoStop}%
\bibitem [{\citenamefont {Bonachela}\ \emph {et~al.}(2010)\citenamefont
  {Bonachela}, \citenamefont {de~Franciscis}, \citenamefont {Torres},\ and\
  \citenamefont {Mu{\~n}oz}}]{JABO2}%
  \BibitemOpen
  \bibfield  {author} {\bibinfo {author} {\bibnamefont {Bonachela},
  \bibfnamefont {J.}}, \bibinfo {author} {\bibfnamefont {S.}~\bibnamefont
  {de~Franciscis}}, \bibinfo {author} {\bibfnamefont {J.}~\bibnamefont
  {Torres}}, \ and\ \bibinfo {author} {\bibfnamefont {M.~A.}\ \bibnamefont
  {Mu{\~n}oz}}} (\bibinfo {year} {2010}),\ \href
  {http://stacks.iop.org/1742-5468/2010/i=02/a=P02015} {\bibfield  {journal}
  {\bibinfo  {journal} {J. Stat. Mech.}\ }\textbf {\bibinfo {volume}
  {2010}}~(\bibinfo {number} {02}),\ \bibinfo {pages} {P02015}}\BibitemShut
  {NoStop}%
\bibitem [{\citenamefont {Bonachela}\ and\ \citenamefont
  {Mu{\~n}oz}(2009)}]{JABO1}%
  \BibitemOpen
  \bibfield  {author} {\bibinfo {author} {\bibnamefont {Bonachela},
  \bibfnamefont {J.~A.}}, \ and\ \bibinfo {author} {\bibfnamefont {M.~A.}\
  \bibnamefont {Mu{\~n}oz}}} (\bibinfo {year} {2009}),\ \href@noop {} {\bibinfo
   {journal} {J. Stat. Mech.}\ ,\ \bibinfo {pages} {P09009}}\BibitemShut
  {NoStop}%
\bibitem [{\citenamefont {Bornholdt}(2005)}]{Bornholdt-RBN1}%
  \BibitemOpen
\bibfield  {journal} {  }\bibfield  {author} {\bibinfo {author} {\bibnamefont
  {Bornholdt}, \bibfnamefont {S.}}} (\bibinfo {year} {2005}),\ \href@noop {}
  {\bibfield  {journal} {\bibinfo  {journal} {Science}\ }\textbf {\bibinfo
  {volume} {310}}~(\bibinfo {number} {5747}),\ \bibinfo {pages}
  {449}}\BibitemShut {NoStop}%
\bibitem [{\citenamefont {Bornholdt}(2008)}]{Bornholdt-RBN2}%
  \BibitemOpen
  \bibfield  {author} {\bibinfo {author} {\bibnamefont {Bornholdt},
  \bibfnamefont {S.}}} (\bibinfo {year} {2008}),\ \href@noop {} {\bibfield
  {journal} {\bibinfo  {journal} {J. R. Soc. Interface}\ }\textbf {\bibinfo
  {volume} {5}}~(\bibinfo {number} {Suppl 1}),\ \bibinfo {pages}
  {S85}}\BibitemShut {NoStop}%
\bibitem [{\citenamefont {Bornholdt}\ and\ \citenamefont
  {Rohlf}(2000)}]{BR2000}%
  \BibitemOpen
  \bibfield  {author} {\bibinfo {author} {\bibnamefont {Bornholdt},
  \bibfnamefont {S.}}, \ and\ \bibinfo {author} {\bibfnamefont
  {T.}~\bibnamefont {Rohlf}}} (\bibinfo {year} {2000}),\ \href@noop {}
  {\bibfield  {journal} {\bibinfo  {journal} {Phys. Rev. Lett.}\ }\textbf
  {\bibinfo {volume} {84}}~(\bibinfo {number} {26}),\ \bibinfo {pages}
  {6114}}\BibitemShut {NoStop}%
\bibitem [{\citenamefont {Bosch}(2007)}]{Hydra}%
  \BibitemOpen
  \bibfield  {author} {\bibinfo {author} {\bibnamefont {Bosch}, \bibfnamefont
  {T.}}} (\bibinfo {year} {2007}),\ \href@noop {} {\bibfield  {journal}
  {\bibinfo  {journal} {Devel. Biol.}\ }\textbf {\bibinfo {volume}
  {303}}~(\bibinfo {number} {2}),\ \bibinfo {pages} {421}}\BibitemShut
  {NoStop}%
\bibitem [{\citenamefont {Bose}\ and\ \citenamefont {Pal}(2017)}]{Bose-review}%
  \BibitemOpen
  \bibfield  {author} {\bibinfo {author} {\bibnamefont {Bose}, \bibfnamefont
  {I.}}, \ and\ \bibinfo {author} {\bibfnamefont {M.}~\bibnamefont {Pal}}}
  (\bibinfo {year} {2017}),\ \href@noop {} {\bibfield  {journal} {\bibinfo
  {journal} {J. of Biosci.}\ }\textbf {\bibinfo {volume} {42}}~(\bibinfo
  {number} {4}),\ \bibinfo {pages} {683}}\BibitemShut {NoStop}%
\bibitem [{\citenamefont {Breakspear}(2017)}]{Breakspear2017}%
  \BibitemOpen
  \bibfield  {author} {\bibinfo {author} {\bibnamefont {Breakspear},
  \bibfnamefont {M.}}} (\bibinfo {year} {2017}),\ \href@noop {} {\bibfield
  {journal} {\bibinfo  {journal} {Nature neuroscience}\ }\textbf {\bibinfo
  {volume} {20}}~(\bibinfo {number} {3}),\ \bibinfo {pages} {340}}\BibitemShut
  {NoStop}%
\bibitem [{\citenamefont {Brockmann}\ \emph {et~al.}(2006)\citenamefont
  {Brockmann}, \citenamefont {Hufnagel},\ and\ \citenamefont
  {Geisel}}]{Geisel-human}%
  \BibitemOpen
  \bibfield  {author} {\bibinfo {author} {\bibnamefont {Brockmann},
  \bibfnamefont {D.}}, \bibinfo {author} {\bibfnamefont {L.}~\bibnamefont
  {Hufnagel}}, \ and\ \bibinfo {author} {\bibfnamefont {T.}~\bibnamefont
  {Geisel}}} (\bibinfo {year} {2006}),\ \href {\doibase 10.1038/nature04292}
  {\bibfield  {journal} {\bibinfo  {journal} {Nature}\ }\textbf {\bibinfo
  {volume} {439}}~(\bibinfo {number} {7075}),\ \bibinfo {pages}
  {462}}\BibitemShut {NoStop}%
\bibitem [{\citenamefont {Brown}\ and\ \citenamefont
  {Botstein}(1999)}]{DNAmicro}%
  \BibitemOpen
  \bibfield  {author} {\bibinfo {author} {\bibnamefont {Brown}, \bibfnamefont
  {P.~O.}}, \ and\ \bibinfo {author} {\bibfnamefont {D.}~\bibnamefont
  {Botstein}}} (\bibinfo {year} {1999}),\ \href@noop {} {\bibfield  {journal}
  {\bibinfo  {journal} {Nature Genet.}\ }\textbf {\bibinfo {volume} {21}},\
  \bibinfo {pages} {33}}\BibitemShut {NoStop}%
\bibitem [{\citenamefont {Brunel}(2000)}]{Brunel1}%
  \BibitemOpen
  \bibfield  {author} {\bibinfo {author} {\bibnamefont {Brunel}, \bibfnamefont
  {N.}}} (\bibinfo {year} {2000}),\ \href@noop {} {\bibfield  {journal}
  {\bibinfo  {journal} {J. Comput. Neurosci.}\ }\textbf {\bibinfo {volume}
  {8}},\ \bibinfo {pages} {183}}\BibitemShut {NoStop}%
\bibitem [{\citenamefont {Brunel}\ and\ \citenamefont {Hakim}(2008)}]{Brunel2}%
  \BibitemOpen
  \bibfield  {author} {\bibinfo {author} {\bibnamefont {Brunel}, \bibfnamefont
  {N.}}, \ and\ \bibinfo {author} {\bibfnamefont {V.}~\bibnamefont {Hakim}}}
  (\bibinfo {year} {2008}),\ \href@noop {} {\bibfield  {journal} {\bibinfo
  {journal} {Chaos}\ }\textbf {\bibinfo {volume} {18}}~(\bibinfo {number}
  {1}),\ \bibinfo {pages} {015113}}\BibitemShut {NoStop}%
\bibitem [{\citenamefont {Buchanan}(2010)}]{Buchanan}%
  \BibitemOpen
  \bibfield  {author} {\bibinfo {author} {\bibnamefont {Buchanan},
  \bibfnamefont {M.}}} (\bibinfo {year} {2010}),\ \href@noop {} {\emph
  {\bibinfo {title} {Networks in cell biology}}}\ (\bibinfo  {publisher}
  {Cambridge University Press})\BibitemShut {NoStop}%
\bibitem [{\citenamefont {Buhl}\ \emph {et~al.}(2006)\citenamefont {Buhl},
  \citenamefont {Sumpter}, \citenamefont {Couzin}, \citenamefont {Hale},
  \citenamefont {Despland}, \citenamefont {Miller},\ and\ \citenamefont
  {Simpson}}]{Couzin2006}%
  \BibitemOpen
  \bibfield  {author} {\bibinfo {author} {\bibnamefont {Buhl}, \bibfnamefont
  {J.}}, \bibinfo {author} {\bibfnamefont {D.~J.}\ \bibnamefont {Sumpter}},
  \bibinfo {author} {\bibfnamefont {I.~D.}\ \bibnamefont {Couzin}}, \bibinfo
  {author} {\bibfnamefont {J.~J.}\ \bibnamefont {Hale}}, \bibinfo {author}
  {\bibfnamefont {E.}~\bibnamefont {Despland}}, \bibinfo {author}
  {\bibfnamefont {E.}~\bibnamefont {Miller}}, \ and\ \bibinfo {author}
  {\bibfnamefont {S.~J.}\ \bibnamefont {Simpson}}} (\bibinfo {year} {2006}),\
  \href@noop {} {\bibfield  {journal} {\bibinfo  {journal} {Science}\ }\textbf
  {\bibinfo {volume} {312}}~(\bibinfo {number} {5778}),\ \bibinfo {pages}
  {1402}}\BibitemShut {NoStop}%
\bibitem [{\citenamefont {Burgos}\ and\ \citenamefont
  {Moreno-Tovar}(1996)}]{Burgos-immune}%
  \BibitemOpen
  \bibfield  {author} {\bibinfo {author} {\bibnamefont {Burgos}, \bibfnamefont
  {J.}}, \ and\ \bibinfo {author} {\bibfnamefont {P.}~\bibnamefont
  {Moreno-Tovar}}} (\bibinfo {year} {1996}),\ \href@noop {} {\bibfield
  {journal} {\bibinfo  {journal} {Biosystems}\ }\textbf {\bibinfo {volume}
  {39}}~(\bibinfo {number} {3}),\ \bibinfo {pages} {227}}\BibitemShut {NoStop}%
\bibitem [{\citenamefont {Bushdid}\ \emph {et~al.}(2014)\citenamefont
  {Bushdid}, \citenamefont {Magnasco}, \citenamefont {Vosshall},\ and\
  \citenamefont {Keller}}]{Olfactory}%
  \BibitemOpen
  \bibfield  {author} {\bibinfo {author} {\bibnamefont {Bushdid}, \bibfnamefont
  {C.}}, \bibinfo {author} {\bibfnamefont {M.~O.}\ \bibnamefont {Magnasco}},
  \bibinfo {author} {\bibfnamefont {L.~B.}\ \bibnamefont {Vosshall}}, \ and\
  \bibinfo {author} {\bibfnamefont {A.}~\bibnamefont {Keller}}} (\bibinfo
  {year} {2014}),\ \href@noop {} {\bibfield  {journal} {\bibinfo  {journal}
  {Science}\ }\textbf {\bibinfo {volume} {343}}~(\bibinfo {number} {6177}),\
  \bibinfo {pages} {1370}}\BibitemShut {NoStop}%
\bibitem [{\citenamefont {Buzsaki}(2009)}]{Buzsaki}%
  \BibitemOpen
  \bibfield  {author} {\bibinfo {author} {\bibnamefont {Buzsaki}, \bibfnamefont
  {G.}}} (\bibinfo {year} {2009}),\ \href@noop {} {\emph {\bibinfo {title}
  {Rhythms of the Brain}}}\ (\bibinfo  {publisher} {Oxford University Press,
  USA})\BibitemShut {NoStop}%
\bibitem [{\citenamefont {Cabral}\ \emph {et~al.}(2011)\citenamefont {Cabral},
  \citenamefont {Hugues}, \citenamefont {Sporns},\ and\ \citenamefont
  {Deco}}]{Cabral}%
  \BibitemOpen
  \bibfield  {author} {\bibinfo {author} {\bibnamefont {Cabral}, \bibfnamefont
  {J.}}, \bibinfo {author} {\bibfnamefont {E.}~\bibnamefont {Hugues}}, \bibinfo
  {author} {\bibfnamefont {O.}~\bibnamefont {Sporns}}, \ and\ \bibinfo {author}
  {\bibfnamefont {G.}~\bibnamefont {Deco}}} (\bibinfo {year} {2011}),\
  \href@noop {} {\bibfield  {journal} {\bibinfo  {journal} {Neuroimage}\
  }\textbf {\bibinfo {volume} {57}}~(\bibinfo {number} {1}),\ \bibinfo {pages}
  {130}}\BibitemShut {NoStop}%
\bibitem [{\citenamefont {Cafiero}\ \emph {et~al.}(1998)\citenamefont
  {Cafiero}, \citenamefont {Gabrielli},\ and\ \citenamefont
  {Mu{\~n}oz}}]{Cafiero}%
  \BibitemOpen
  \bibfield  {author} {\bibinfo {author} {\bibnamefont {Cafiero}, \bibfnamefont
  {R.}}, \bibinfo {author} {\bibfnamefont {A.}~\bibnamefont {Gabrielli}}, \
  and\ \bibinfo {author} {\bibfnamefont {M.~A.}\ \bibnamefont {Mu{\~n}oz}}}
  (\bibinfo {year} {1998}),\ \href@noop {} {\bibfield  {journal} {\bibinfo
  {journal} {Phys. Rev. E}\ }\textbf {\bibinfo {volume} {57}}~(\bibinfo
  {number} {5}),\ \bibinfo {pages} {5060}}\BibitemShut {NoStop}%
\bibitem [{\citenamefont {Caldarelli}(2007)}]{Caldarelli-book}%
  \BibitemOpen
  \bibfield  {author} {\bibinfo {author} {\bibnamefont {Caldarelli},
  \bibfnamefont {G.}}} (\bibinfo {year} {2007}),\ \href@noop {} {\emph
  {\bibinfo {title} {Scale-free networks: complex webs in nature and
  technology}}}\ (\bibinfo  {publisher} {Oxford University Press})\BibitemShut
  {NoStop}%
\bibitem [{\citenamefont {Camalet}\ \emph {et~al.}(2000)\citenamefont
  {Camalet}, \citenamefont {Duke}, \citenamefont {J{\"u}licher},\ and\
  \citenamefont {Prost}}]{Camalet}%
  \BibitemOpen
  \bibfield  {author} {\bibinfo {author} {\bibnamefont {Camalet}, \bibfnamefont
  {S.}}, \bibinfo {author} {\bibfnamefont {T.}~\bibnamefont {Duke}}, \bibinfo
  {author} {\bibfnamefont {F.}~\bibnamefont {J{\"u}licher}}, \ and\ \bibinfo
  {author} {\bibfnamefont {J.}~\bibnamefont {Prost}}} (\bibinfo {year}
  {2000}),\ \href@noop {} {\bibfield  {journal} {\bibinfo  {journal} {Proc.
  Natl. Acad. Sci. USA.}\ }\textbf {\bibinfo {volume} {97}}~(\bibinfo {number}
  {7}),\ \bibinfo {pages} {3183}}\BibitemShut {NoStop}%
\bibitem [{\citenamefont {Carlson}\ and\ \citenamefont {Doyle}(2000)}]{HOT}%
  \BibitemOpen
  \bibfield  {author} {\bibinfo {author} {\bibnamefont {Carlson}, \bibfnamefont
  {J.~M.}}, \ and\ \bibinfo {author} {\bibfnamefont {J.}~\bibnamefont {Doyle}}}
  (\bibinfo {year} {2000}),\ \href@noop {} {\bibfield  {journal} {\bibinfo
  {journal} {Phys. Rev. Lett.}\ }\textbf {\bibinfo {volume} {84}}~(\bibinfo
  {number} {11}),\ \bibinfo {pages} {2529}}\BibitemShut {NoStop}%
\bibitem [{\citenamefont {Carpenter}\ and\ \citenamefont
  {Grossberg}(2016)}]{Grossberg2}%
  \BibitemOpen
  \bibfield  {author} {\bibinfo {author} {\bibnamefont {Carpenter},
  \bibfnamefont {G.~A.}}, \ and\ \bibinfo {author} {\bibfnamefont
  {S.}~\bibnamefont {Grossberg}}} (\bibinfo {year} {2016}),\ in\ \href@noop {}
  {\emph {\bibinfo {booktitle} {Encyclopedia of Machine Learning and Data
  Mining}}}\ (\bibinfo  {publisher} {Springer})\ pp.\ \bibinfo {pages}
  {1--17}\BibitemShut {NoStop}%
\bibitem [{\citenamefont {Cavagna}\ \emph {et~al.}(2010)\citenamefont
  {Cavagna}, \citenamefont {Cimarelli}, \citenamefont {Giardina}, \citenamefont
  {Parisi}, \citenamefont {Santagati}, \citenamefont {Stefanini},\ and\
  \citenamefont {Viale}}]{Cavagna2010}%
  \BibitemOpen
  \bibfield  {author} {\bibinfo {author} {\bibnamefont {Cavagna}, \bibfnamefont
  {A.}}, \bibinfo {author} {\bibfnamefont {A.}~\bibnamefont {Cimarelli}},
  \bibinfo {author} {\bibfnamefont {I.}~\bibnamefont {Giardina}}, \bibinfo
  {author} {\bibfnamefont {G.}~\bibnamefont {Parisi}}, \bibinfo {author}
  {\bibfnamefont {R.}~\bibnamefont {Santagati}}, \bibinfo {author}
  {\bibfnamefont {F.}~\bibnamefont {Stefanini}}, \ and\ \bibinfo {author}
  {\bibfnamefont {M.}~\bibnamefont {Viale}}} (\bibinfo {year} {2010}),\
  \href@noop {} {\bibfield  {journal} {\bibinfo  {journal} {Proc. Natl. Acad.
  Sci. USA.}\ }\textbf {\bibinfo {volume} {107}}~(\bibinfo {number} {26}),\
  \bibinfo {pages} {11865}}\BibitemShut {NoStop}%
\bibitem [{\citenamefont {Cavagna}\ \emph {et~al.}(2017)\citenamefont
  {Cavagna}, \citenamefont {Conti}, \citenamefont {Creato}, \citenamefont
  {Del~Castello}, \citenamefont {Giardina}, \citenamefont {Grigera},
  \citenamefont {Melillo}, \citenamefont {Parisi},\ and\ \citenamefont
  {Viale}}]{Cavagna-NP}%
  \BibitemOpen
  \bibfield  {author} {\bibinfo {author} {\bibnamefont {Cavagna}, \bibfnamefont
  {A.}}, \bibinfo {author} {\bibfnamefont {D.}~\bibnamefont {Conti}}, \bibinfo
  {author} {\bibfnamefont {C.}~\bibnamefont {Creato}}, \bibinfo {author}
  {\bibfnamefont {L.}~\bibnamefont {Del~Castello}}, \bibinfo {author}
  {\bibfnamefont {I.}~\bibnamefont {Giardina}}, \bibinfo {author}
  {\bibfnamefont {T.~S.}\ \bibnamefont {Grigera}}, \bibinfo {author}
  {\bibfnamefont {S.}~\bibnamefont {Melillo}}, \bibinfo {author} {\bibfnamefont
  {L.}~\bibnamefont {Parisi}}, \ and\ \bibinfo {author} {\bibfnamefont
  {M.}~\bibnamefont {Viale}}} (\bibinfo {year} {2017}),\ \href@noop {}
  {\bibinfo  {journal} {Nat. Phys.}\ }\BibitemShut {NoStop}%
\bibitem [{\citenamefont {Cavagna}\ \emph {et~al.}(2008)\citenamefont
  {Cavagna}, \citenamefont {Giardina}, \citenamefont {Orlandi}, \citenamefont
  {Parisi}, \citenamefont {Procaccini}, \citenamefont {Viale},\ and\
  \citenamefont {Zdravkovic}}]{Starflag}%
  \BibitemOpen
\bibfield  {journal} {  }\bibfield  {author} {\bibinfo {author} {\bibnamefont
  {Cavagna}, \bibfnamefont {A.}}, \bibinfo {author} {\bibfnamefont
  {I.}~\bibnamefont {Giardina}}, \bibinfo {author} {\bibfnamefont
  {A.}~\bibnamefont {Orlandi}}, \bibinfo {author} {\bibfnamefont
  {G.}~\bibnamefont {Parisi}}, \bibinfo {author} {\bibfnamefont
  {A.}~\bibnamefont {Procaccini}}, \bibinfo {author} {\bibfnamefont
  {M.}~\bibnamefont {Viale}}, \ and\ \bibinfo {author} {\bibfnamefont
  {V.}~\bibnamefont {Zdravkovic}}} (\bibinfo {year} {2008}),\ \href@noop {}
  {\bibinfo  {journal} {arXiv:0802.1668}\ }\BibitemShut {NoStop}%
\bibitem [{\citenamefont {Chaikin}\ and\ \citenamefont
  {Lubensky}(2000)}]{Chaikin}%
  \BibitemOpen
\bibfield  {journal} {  }\bibfield  {author} {\bibinfo {author} {\bibnamefont
  {Chaikin}, \bibfnamefont {P.~M.}}, \ and\ \bibinfo {author} {\bibfnamefont
  {T.~C.}\ \bibnamefont {Lubensky}}} (\bibinfo {year} {2000}),\ \href@noop {}
  {\emph {\bibinfo {title} {Principles of condensed matter physics}}},\
  Vol.~\bibinfo {volume} {1}\ (\bibinfo  {publisher} {Cambridge Univ
  Press})\BibitemShut {NoStop}%
\bibitem [{\citenamefont {Chat{\'e}}\ \emph {et~al.}(2008)\citenamefont
  {Chat{\'e}}, \citenamefont {Ginelli}, \citenamefont {Gr{\'e}goire},
  \citenamefont {Peruani},\ and\ \citenamefont {Raynaud}}]{Chate2008}%
  \BibitemOpen
  \bibfield  {author} {\bibinfo {author} {\bibnamefont {Chat{\'e}},
  \bibfnamefont {H.}}, \bibinfo {author} {\bibfnamefont {F.}~\bibnamefont
  {Ginelli}}, \bibinfo {author} {\bibfnamefont {G.}~\bibnamefont
  {Gr{\'e}goire}}, \bibinfo {author} {\bibfnamefont {F.}~\bibnamefont
  {Peruani}}, \ and\ \bibinfo {author} {\bibfnamefont {F.}~\bibnamefont
  {Raynaud}}} (\bibinfo {year} {2008}),\ \href@noop {} {\bibfield  {journal}
  {\bibinfo  {journal} {Europ. Phys. Jour. B}\ }\textbf {\bibinfo {volume}
  {64}}~(\bibinfo {number} {3-4}),\ \bibinfo {pages} {451}}\BibitemShut
  {NoStop}%
\bibitem [{\citenamefont {Chat\'e}\ \emph {et~al.}(2008)\citenamefont
  {Chat\'e}, \citenamefont {Ginelli}, \citenamefont {Gr\'egoire},\ and\
  \citenamefont {Raynaud}}]{Chate2008b}%
  \BibitemOpen
  \bibfield  {author} {\bibinfo {author} {\bibnamefont {Chat\'e}, \bibfnamefont
  {H.}}, \bibinfo {author} {\bibfnamefont {F.}~\bibnamefont {Ginelli}},
  \bibinfo {author} {\bibfnamefont {G.}~\bibnamefont {Gr\'egoire}}, \ and\
  \bibinfo {author} {\bibfnamefont {F.}~\bibnamefont {Raynaud}}} (\bibinfo
  {year} {2008}),\ \href {\doibase 10.1103/PhysRevE.77.046113} {\bibfield
  {journal} {\bibinfo  {journal} {Phys. Rev. E}\ }\textbf {\bibinfo {volume}
  {77}},\ \bibinfo {pages} {046113}}\BibitemShut {NoStop}%
\bibitem [{\citenamefont {Chat{\'e}}\ and\ \citenamefont
  {Mu{\~n}oz}(2014)}]{Us-Physics}%
  \BibitemOpen
  \bibfield  {author} {\bibinfo {author} {\bibnamefont {Chat{\'e}},
  \bibfnamefont {H.}}, \ and\ \bibinfo {author} {\bibfnamefont
  {M.}~\bibnamefont {Mu{\~n}oz}}} (\bibinfo {year} {2014}),\ \href@noop {}
  {\bibfield  {journal} {\bibinfo  {journal} {Physics}\ }\textbf {\bibinfo
  {volume} {7}},\ \bibinfo {pages} {120}}\BibitemShut {NoStop}%
\bibitem [{\citenamefont {Chen}\ \emph {et~al.}(2012)\citenamefont {Chen},
  \citenamefont {Dong}, \citenamefont {Be’er}, \citenamefont {Swinney},\ and\
  \citenamefont {Zhang}}]{Chen-bacteria}%
  \BibitemOpen
  \bibfield  {author} {\bibinfo {author} {\bibnamefont {Chen}, \bibfnamefont
  {X.}}, \bibinfo {author} {\bibfnamefont {X.}~\bibnamefont {Dong}}, \bibinfo
  {author} {\bibfnamefont {A.}~\bibnamefont {Be’er}}, \bibinfo {author}
  {\bibfnamefont {H.}~\bibnamefont {Swinney}}, \ and\ \bibinfo {author}
  {\bibfnamefont {H.}~\bibnamefont {Zhang}}} (\bibinfo {year} {2012}),\
  \href@noop {} {\bibfield  {journal} {\bibinfo  {journal} {Phys. Rev. Lett.}\
  }\textbf {\bibinfo {volume} {108}}~(\bibinfo {number} {14}),\ \bibinfo
  {pages} {148101}}\BibitemShut {NoStop}%
\bibitem [{\citenamefont {Chialvo}(2004)}]{Chialvo2004}%
  \BibitemOpen
  \bibfield  {author} {\bibinfo {author} {\bibnamefont {Chialvo}, \bibfnamefont
  {D.~R.}}} (\bibinfo {year} {2004}),\ \href@noop {} {\bibfield  {journal}
  {\bibinfo  {journal} {Physica A}\ }\textbf {\bibinfo {volume}
  {340}}~(\bibinfo {number} {4}),\ \bibinfo {pages} {756}}\BibitemShut
  {NoStop}%
\bibitem [{\citenamefont {Chialvo}(2006)}]{Chialvo-senses}%
  \BibitemOpen
  \bibfield  {author} {\bibinfo {author} {\bibnamefont {Chialvo}, \bibfnamefont
  {D.~R.}}} (\bibinfo {year} {2006}),\ \href@noop {} {\bibfield  {journal}
  {\bibinfo  {journal} {Nat. Phys.}\ }\textbf {\bibinfo {volume} {2}},\
  \bibinfo {pages} {301}}\BibitemShut {NoStop}%
\bibitem [{\citenamefont {Chialvo}(2010)}]{Chialvo2010}%
  \BibitemOpen
  \bibfield  {author} {\bibinfo {author} {\bibnamefont {Chialvo}, \bibfnamefont
  {D.~R.}}} (\bibinfo {year} {2010}),\ \href@noop {} {\bibfield  {journal}
  {\bibinfo  {journal} {Nat. Phys.}\ }\textbf {\bibinfo {volume} {6}},\
  \bibinfo {pages} {744}}\BibitemShut {NoStop}%
\bibitem [{\citenamefont {Chialvo}\ and\ \citenamefont
  {Bak}(1999)}]{Chialvo-Bak1}%
  \BibitemOpen
  \bibfield  {author} {\bibinfo {author} {\bibnamefont {Chialvo}, \bibfnamefont
  {D.~R.}}, \ and\ \bibinfo {author} {\bibfnamefont {P.}~\bibnamefont {Bak}}}
  (\bibinfo {year} {1999}),\ \href@noop {} {\bibfield  {journal} {\bibinfo
  {journal} {Neuroscience}\ }\textbf {\bibinfo {volume} {90}}~(\bibinfo
  {number} {4}),\ \bibinfo {pages} {1137}}\BibitemShut {NoStop}%
\bibitem [{\citenamefont {Chialvo}\ \emph {et~al.}(2008)\citenamefont
  {Chialvo}, \citenamefont {Balenzuela},\ and\ \citenamefont
  {Fraiman}}]{Chialvo2008}%
  \BibitemOpen
  \bibfield  {author} {\bibinfo {author} {\bibnamefont {Chialvo}, \bibfnamefont
  {D.~R.}}, \bibinfo {author} {\bibfnamefont {P.}~\bibnamefont {Balenzuela}}, \
  and\ \bibinfo {author} {\bibfnamefont {D.}~\bibnamefont {Fraiman}}} (\bibinfo
  {year} {2008}),\ \href@noop {} {\bibfield  {journal} {\bibinfo  {journal}
  {AIP Conference Proceedings}\ }\textbf {\bibinfo {volume} {1028}}~(\bibinfo
  {number} {1}),\ \bibinfo {pages} {28}}\BibitemShut {NoStop}%
\bibitem [{\citenamefont {Choe}\ \emph {et~al.}(1998)\citenamefont {Choe},
  \citenamefont {Magnasco},\ and\ \citenamefont {Hudspeth}}]{Hudspeth1}%
  \BibitemOpen
  \bibfield  {author} {\bibinfo {author} {\bibnamefont {Choe}, \bibfnamefont
  {Y.}}, \bibinfo {author} {\bibfnamefont {M.~O.}\ \bibnamefont {Magnasco}}, \
  and\ \bibinfo {author} {\bibfnamefont {A.}~\bibnamefont {Hudspeth}}}
  (\bibinfo {year} {1998}),\ \href@noop {} {\bibfield  {journal} {\bibinfo
  {journal} {Proc. Natl. Acad. Sci. USA.}\ }\textbf {\bibinfo {volume}
  {95}}~(\bibinfo {number} {26}),\ \bibinfo {pages} {15321}}\BibitemShut
  {NoStop}%
\bibitem [{\citenamefont {Chowdhury}\ \emph {et~al.}(2010)\citenamefont
  {Chowdhury}, \citenamefont {Lloyd-Price}, \citenamefont {Smolander},
  \citenamefont {Baici}, \citenamefont {Hughes}, \citenamefont {Yli-Harja},
  \citenamefont {Chua},\ and\ \citenamefont {Ribeiro}}]{Chowdhury}%
  \BibitemOpen
  \bibfield  {author} {\bibinfo {author} {\bibnamefont {Chowdhury},
  \bibfnamefont {S.}}, \bibinfo {author} {\bibfnamefont {J.}~\bibnamefont
  {Lloyd-Price}}, \bibinfo {author} {\bibfnamefont {O.-P.}\ \bibnamefont
  {Smolander}}, \bibinfo {author} {\bibfnamefont {W.~C.}\ \bibnamefont
  {Baici}}, \bibinfo {author} {\bibfnamefont {T.~R.}\ \bibnamefont {Hughes}},
  \bibinfo {author} {\bibfnamefont {O.}~\bibnamefont {Yli-Harja}}, \bibinfo
  {author} {\bibfnamefont {G.}~\bibnamefont {Chua}}, \ and\ \bibinfo {author}
  {\bibfnamefont {A.~S.}\ \bibnamefont {Ribeiro}}} (\bibinfo {year} {2010}),\
  \href@noop {} {\bibfield  {journal} {\bibinfo  {journal} {BMC Systems
  Biology}\ }\textbf {\bibinfo {volume} {4}}~(\bibinfo {number} {1}),\ \bibinfo
  {pages} {1}}\BibitemShut {NoStop}%
\bibitem [{\citenamefont {Christensen}\ \emph {et~al.}(1996)\citenamefont
  {Christensen}, \citenamefont {Corral}, \citenamefont {Frette}, \citenamefont
  {Feder},\ and\ \citenamefont {J{\o}ssang}}]{Oslo}%
  \BibitemOpen
  \bibfield  {author} {\bibinfo {author} {\bibnamefont {Christensen},
  \bibfnamefont {K.}}, \bibinfo {author} {\bibfnamefont {{\'A}.}~\bibnamefont
  {Corral}}, \bibinfo {author} {\bibfnamefont {V.}~\bibnamefont {Frette}},
  \bibinfo {author} {\bibfnamefont {J.}~\bibnamefont {Feder}}, \ and\ \bibinfo
  {author} {\bibfnamefont {T.}~\bibnamefont {J{\o}ssang}}} (\bibinfo {year}
  {1996}),\ \href@noop {} {\bibfield  {journal} {\bibinfo  {journal} {Phys.
  Rev. Lett.}\ }\textbf {\bibinfo {volume} {77}}~(\bibinfo {number} {1}),\
  \bibinfo {pages} {107}}\BibitemShut {NoStop}%
\bibitem [{\citenamefont {Christensen}\ and\ \citenamefont
  {Moloney}(2005)}]{Moloney-book}%
  \BibitemOpen
  \bibfield  {author} {\bibinfo {author} {\bibnamefont {Christensen},
  \bibfnamefont {K.}}, \ and\ \bibinfo {author} {\bibfnamefont
  {N.}~\bibnamefont {Moloney}}} (\bibinfo {year} {2005}),\ \href@noop {} {\emph
  {\bibinfo {title} {Complexity and criticality}}}\ (\bibinfo  {publisher}
  {Imperial College Press},\ \bibinfo {address} {London})\BibitemShut {NoStop}%
\bibitem [{\citenamefont {Cicuta}(2013)}]{Cicuta1}%
  \BibitemOpen
  \bibfield  {author} {\bibinfo {author} {\bibnamefont {Cicuta}, \bibfnamefont
  {P.}}} (\bibinfo {year} {2013}),\ in\ \href@noop {} {\emph {\bibinfo
  {booktitle} {Encyclopedia of Biophysics}}}\ (\bibinfo  {publisher}
  {Springer})\ pp.\ \bibinfo {pages} {387--390}\BibitemShut {NoStop}%
\bibitem [{\citenamefont {Clauset}\ \emph {et~al.}(2009)\citenamefont
  {Clauset}, \citenamefont {Shalizi},\ and\ \citenamefont {Newman}}]{Clauset}%
  \BibitemOpen
  \bibfield  {author} {\bibinfo {author} {\bibnamefont {Clauset}, \bibfnamefont
  {A.}}, \bibinfo {author} {\bibfnamefont {C.~R.}\ \bibnamefont {Shalizi}}, \
  and\ \bibinfo {author} {\bibfnamefont {M.~E.~J.}\ \bibnamefont {Newman}}}
  (\bibinfo {year} {2009}),\ \href@noop {} {\bibfield  {journal} {\bibinfo
  {journal} {SIAM Rev.}\ }\textbf {\bibinfo {volume} {51}}~(\bibinfo {number}
  {4}),\ \bibinfo {pages} {661}}\BibitemShut {NoStop}%
\bibitem [{\citenamefont {Cocchi}\ \emph {et~al.}(2017)\citenamefont {Cocchi},
  \citenamefont {Gollo}, \citenamefont {Zalesky},\ and\ \citenamefont
  {Breakspear}}]{Breakspear-review}%
  \BibitemOpen
  \bibfield  {author} {\bibinfo {author} {\bibnamefont {Cocchi}, \bibfnamefont
  {L.}}, \bibinfo {author} {\bibfnamefont {L.~L.}\ \bibnamefont {Gollo}},
  \bibinfo {author} {\bibfnamefont {A.}~\bibnamefont {Zalesky}}, \ and\
  \bibinfo {author} {\bibfnamefont {M.}~\bibnamefont {Breakspear}}} (\bibinfo
  {year} {2017}),\ \href@noop {} {\bibinfo  {journal} {Progress in
  Neurobiology}\ }\BibitemShut {NoStop}%
\bibitem [{\citenamefont {Cocco}\ \emph {et~al.}(2009)\citenamefont {Cocco},
  \citenamefont {Leibler},\ and\ \citenamefont {Monasson}}]{Cocco}%
  \BibitemOpen
\bibfield  {journal} {  }\bibfield  {author} {\bibinfo {author} {\bibnamefont
  {Cocco}, \bibfnamefont {S.}}, \bibinfo {author} {\bibfnamefont
  {S.}~\bibnamefont {Leibler}}, \ and\ \bibinfo {author} {\bibfnamefont
  {R.}~\bibnamefont {Monasson}}} (\bibinfo {year} {2009}),\ \href@noop {}
  {\bibfield  {journal} {\bibinfo  {journal} {Proc. Natl. Acad. Sci. USA.}\
  }\textbf {\bibinfo {volume} {106}}~(\bibinfo {number} {33}),\ \bibinfo
  {pages} {14058}}\BibitemShut {NoStop}%
\bibitem [{\citenamefont {Cohen}\ \emph {et~al.}(2012)\citenamefont {Cohen},
  \citenamefont {Xu},\ and\ \citenamefont {Schuster}}]{Cohen}%
  \BibitemOpen
  \bibfield  {author} {\bibinfo {author} {\bibnamefont {Cohen}, \bibfnamefont
  {J.~E.}}, \bibinfo {author} {\bibfnamefont {M.}~\bibnamefont {Xu}}, \ and\
  \bibinfo {author} {\bibfnamefont {W.~S.}\ \bibnamefont {Schuster}}} (\bibinfo
  {year} {2012}),\ \href@noop {} {\bibfield  {journal} {\bibinfo  {journal}
  {Proc. Natl. Acad. Sci. USA}\ }\textbf {\bibinfo {volume} {109}}~(\bibinfo
  {number} {39}),\ \bibinfo {pages} {15829}}\BibitemShut {NoStop}%
\bibitem [{\citenamefont {Coniglio}\ \emph {et~al.}(1989)\citenamefont
  {Coniglio}, \citenamefont {de~Arcangelis}, \citenamefont {Herrmann},\ and\
  \citenamefont {Jan}}]{damage1}%
  \BibitemOpen
  \bibfield  {author} {\bibinfo {author} {\bibnamefont {Coniglio},
  \bibfnamefont {A.}}, \bibinfo {author} {\bibfnamefont {L.}~\bibnamefont
  {de~Arcangelis}}, \bibinfo {author} {\bibfnamefont {H.~J.}\ \bibnamefont
  {Herrmann}}, \ and\ \bibinfo {author} {\bibfnamefont {N.}~\bibnamefont
  {Jan}}} (\bibinfo {year} {1989}),\ \href
  {http://stacks.iop.org/0295-5075/8/i=4/a=003} {\bibfield  {journal} {\bibinfo
   {journal} {EPL (Europhys. Lett.)}\ }\textbf {\bibinfo {volume}
  {8}}~(\bibinfo {number} {4}),\ \bibinfo {pages} {315}}\BibitemShut {NoStop}%
\bibitem [{\citenamefont {Corominas-Murtra}\ \emph {et~al.}(2013)\citenamefont
  {Corominas-Murtra}, \citenamefont {Go{\~n}i}, \citenamefont {Sol{\'e}},\ and\
  \citenamefont {Rodr{\'\i}guez-Caso}}]{Corominas1}%
  \BibitemOpen
  \bibfield  {author} {\bibinfo {author} {\bibnamefont {Corominas-Murtra},
  \bibfnamefont {B.}}, \bibinfo {author} {\bibfnamefont {J.}~\bibnamefont
  {Go{\~n}i}}, \bibinfo {author} {\bibfnamefont {R.~V.}\ \bibnamefont
  {Sol{\'e}}}, \ and\ \bibinfo {author} {\bibfnamefont {C.}~\bibnamefont
  {Rodr{\'\i}guez-Caso}}} (\bibinfo {year} {2013}),\ \href@noop {} {\bibfield
  {journal} {\bibinfo  {journal} {Proc. Natl. Acad. Sci. USA.}\ }\textbf
  {\bibinfo {volume} {110}}~(\bibinfo {number} {33}),\ \bibinfo {pages}
  {13316}}\BibitemShut {NoStop}%
\bibitem [{\citenamefont {Corral}(2004)}]{Corral-quakes}%
  \BibitemOpen
  \bibfield  {author} {\bibinfo {author} {\bibnamefont {Corral}, \bibfnamefont
  {A.}}} (\bibinfo {year} {2004}),\ \href@noop {} {\bibfield  {journal}
  {\bibinfo  {journal} {Phys. Rev. Lett.}\ }\textbf {\bibinfo {volume}
  {92}}~(\bibinfo {number} {10}),\ \bibinfo {pages} {108501}}\BibitemShut
  {NoStop}%
\bibitem [{\citenamefont {Corral}\ \emph {et~al.}(1995)\citenamefont {Corral},
  \citenamefont {P{\'e}rez}, \citenamefont {D{\'\i}az-Guilera},\ and\
  \citenamefont {Arenas}}]{Corral}%
  \BibitemOpen
  \bibfield  {author} {\bibinfo {author} {\bibnamefont {Corral}, \bibfnamefont
  {A.}}, \bibinfo {author} {\bibfnamefont {C.~J.}\ \bibnamefont {P{\'e}rez}},
  \bibinfo {author} {\bibfnamefont {A.}~\bibnamefont {D{\'\i}az-Guilera}}, \
  and\ \bibinfo {author} {\bibfnamefont {A.}~\bibnamefont {Arenas}}} (\bibinfo
  {year} {1995}),\ \href@noop {} {\bibfield  {journal} {\bibinfo  {journal}
  {Phys. Rev. Lett.}\ }\textbf {\bibinfo {volume} {74}}~(\bibinfo {number}
  {1}),\ \bibinfo {pages} {118}}\BibitemShut {NoStop}%
\bibitem [{\citenamefont {Couzin}(2007)}]{Couzin2007}%
  \BibitemOpen
  \bibfield  {author} {\bibinfo {author} {\bibnamefont {Couzin}, \bibfnamefont
  {I.}}} (\bibinfo {year} {2007}),\ \href@noop {} {\bibfield  {journal}
  {\bibinfo  {journal} {Nature}\ }\textbf {\bibinfo {volume} {445}}~(\bibinfo
  {number} {7129}),\ \bibinfo {pages} {715}}\BibitemShut {NoStop}%
\bibitem [{\citenamefont {Couzin}(2009)}]{Couzin2009}%
  \BibitemOpen
  \bibfield  {author} {\bibinfo {author} {\bibnamefont {Couzin}, \bibfnamefont
  {I.~D.}}} (\bibinfo {year} {2009}),\ \href@noop {} {\bibfield  {journal}
  {\bibinfo  {journal} {Trends in cognitive sciences}\ }\textbf {\bibinfo
  {volume} {13}}~(\bibinfo {number} {1}),\ \bibinfo {pages} {36}}\BibitemShut
  {NoStop}%
\bibitem [{\citenamefont {Couzin}\ \emph {et~al.}(2011)\citenamefont {Couzin},
  \citenamefont {Ioannou}, \citenamefont {Demirel}, \citenamefont {Gross},
  \citenamefont {Torney}, \citenamefont {Hartnett}, \citenamefont {Conradt},
  \citenamefont {Levin},\ and\ \citenamefont {Leonard}}]{Couzin2011}%
  \BibitemOpen
  \bibfield  {author} {\bibinfo {author} {\bibnamefont {Couzin}, \bibfnamefont
  {I.~D.}}, \bibinfo {author} {\bibfnamefont {C.~C.}\ \bibnamefont {Ioannou}},
  \bibinfo {author} {\bibfnamefont {G.}~\bibnamefont {Demirel}}, \bibinfo
  {author} {\bibfnamefont {T.}~\bibnamefont {Gross}}, \bibinfo {author}
  {\bibfnamefont {C.~J.}\ \bibnamefont {Torney}}, \bibinfo {author}
  {\bibfnamefont {A.}~\bibnamefont {Hartnett}}, \bibinfo {author}
  {\bibfnamefont {L.}~\bibnamefont {Conradt}}, \bibinfo {author} {\bibfnamefont
  {S.~A.}\ \bibnamefont {Levin}}, \ and\ \bibinfo {author} {\bibfnamefont
  {N.~E.}\ \bibnamefont {Leonard}}} (\bibinfo {year} {2011}),\ \href@noop {}
  {\bibfield  {journal} {\bibinfo  {journal} {Science}\ }\textbf {\bibinfo
  {volume} {334}}~(\bibinfo {number} {6062}),\ \bibinfo {pages}
  {1578}}\BibitemShut {NoStop}%
\bibitem [{\citenamefont {Couzin}\ and\ \citenamefont
  {Krause}(2003)}]{Couzin-review}%
  \BibitemOpen
  \bibfield  {author} {\bibinfo {author} {\bibnamefont {Couzin}, \bibfnamefont
  {I.~D.}}, \ and\ \bibinfo {author} {\bibfnamefont {J.}~\bibnamefont
  {Krause}}} (\bibinfo {year} {2003}),\ \href@noop {} {\bibfield  {journal}
  {\bibinfo  {journal} {Advances in the Study of Behavior}\ }\textbf {\bibinfo
  {volume} {32}},\ \bibinfo {pages} {1}}\BibitemShut {NoStop}%
\bibitem [{\citenamefont {Couzin}\ \emph {et~al.}(2005)\citenamefont {Couzin},
  \citenamefont {Krause}, \citenamefont {Franks},\ and\ \citenamefont
  {Levin}}]{Couzin2005}%
  \BibitemOpen
  \bibfield  {author} {\bibinfo {author} {\bibnamefont {Couzin}, \bibfnamefont
  {I.~D.}}, \bibinfo {author} {\bibfnamefont {J.}~\bibnamefont {Krause}},
  \bibinfo {author} {\bibfnamefont {N.~R.}\ \bibnamefont {Franks}}, \ and\
  \bibinfo {author} {\bibfnamefont {S.~A.}\ \bibnamefont {Levin}}} (\bibinfo
  {year} {2005}),\ \href@noop {} {\bibfield  {journal} {\bibinfo  {journal}
  {Nature}\ }\textbf {\bibinfo {volume} {433}}~(\bibinfo {number} {7025}),\
  \bibinfo {pages} {513}}\BibitemShut {NoStop}%
\bibitem [{\citenamefont {Cover}\ and\ \citenamefont
  {Thomas}(1991)}]{Cover-Thomas}%
  \BibitemOpen
  \bibfield  {author} {\bibinfo {author} {\bibnamefont {Cover}, \bibfnamefont
  {T.~M.}}, \ and\ \bibinfo {author} {\bibfnamefont {J.}~\bibnamefont
  {Thomas}}} (\bibinfo {year} {1991}),\ \href@noop {} {\emph {\bibinfo {title}
  {Elements of Information Theory}}}\ (\bibinfo  {publisher}
  {Wiley})\BibitemShut {NoStop}%
\bibitem [{\citenamefont {Crick}(1970)}]{Dogma}%
  \BibitemOpen
  \bibfield  {author} {\bibinfo {author} {\bibnamefont {Crick}, \bibfnamefont
  {F.}}} (\bibinfo {year} {1970}),\ \href@noop {} {\bibfield  {journal}
  {\bibinfo  {journal} {Nature}\ }\textbf {\bibinfo {volume} {227}}~(\bibinfo
  {number} {5258}),\ \bibinfo {pages} {561}}\BibitemShut {NoStop}%
\bibitem [{\citenamefont {Crotty}\ \emph {et~al.}(2001)\citenamefont {Crotty},
  \citenamefont {Cameron},\ and\ \citenamefont {Andino}}]{Error-experiments}%
  \BibitemOpen
  \bibfield  {author} {\bibinfo {author} {\bibnamefont {Crotty}, \bibfnamefont
  {S.}}, \bibinfo {author} {\bibfnamefont {C.~E.}\ \bibnamefont {Cameron}}, \
  and\ \bibinfo {author} {\bibfnamefont {R.}~\bibnamefont {Andino}}} (\bibinfo
  {year} {2001}),\ \href@noop {} {\bibfield  {journal} {\bibinfo  {journal}
  {Proc. Natl. Acad. Sci. USA.}\ }\textbf {\bibinfo {volume} {98}}~(\bibinfo
  {number} {12}),\ \bibinfo {pages} {6895}}\BibitemShut {NoStop}%
\bibitem [{\citenamefont {Crutchfield}(2012)}]{Crutchfield2012}%
  \BibitemOpen
  \bibfield  {author} {\bibinfo {author} {\bibnamefont {Crutchfield},
  \bibfnamefont {J.~P.}}} (\bibinfo {year} {2012}),\ \href@noop {} {\bibfield
  {journal} {\bibinfo  {journal} {Nat. Phys.}\ }\textbf {\bibinfo {volume}
  {8}}~(\bibinfo {number} {1}),\ \bibinfo {pages} {17}}\BibitemShut {NoStop}%
\bibitem [{\citenamefont {Crutchfield}\ and\ \citenamefont
  {Young}(1988)}]{Crutchfield1988}%
  \BibitemOpen
  \bibfield  {author} {\bibinfo {author} {\bibnamefont {Crutchfield},
  \bibfnamefont {J.~P.}}, \ and\ \bibinfo {author} {\bibfnamefont
  {K.}~\bibnamefont {Young}}} (\bibinfo {year} {1988}),\ in\ \href@noop {}
  {\emph {\bibinfo {booktitle} {Entropy, Complexity, and the Physics of
  Information}}},\ \bibinfo {editor} {edited by\ \bibinfo {editor}
  {\bibfnamefont {W.}~\bibnamefont {Zurek}}}\ (\bibinfo {organization}
  {Addison-Wesley, Reading})\BibitemShut {NoStop}%
\bibitem [{\citenamefont {Darabos}\ \emph {et~al.}(2009)\citenamefont
  {Darabos}, \citenamefont {Giacobini}, \citenamefont {Tomassini},
  \citenamefont {Provero},\ and\ \citenamefont {Di~Cunto}}]{Darabos}%
  \BibitemOpen
  \bibfield  {author} {\bibinfo {author} {\bibnamefont {Darabos}, \bibfnamefont
  {C.}}, \bibinfo {author} {\bibfnamefont {M.}~\bibnamefont {Giacobini}},
  \bibinfo {author} {\bibfnamefont {M.}~\bibnamefont {Tomassini}}, \bibinfo
  {author} {\bibfnamefont {P.}~\bibnamefont {Provero}}, \ and\ \bibinfo
  {author} {\bibfnamefont {F.}~\bibnamefont {Di~Cunto}}} (\bibinfo {year}
  {2009}),\ in\ \href@noop {} {\emph {\bibinfo {booktitle} {European Conf. on
  Artificial Life}}}\ (\bibinfo {organization} {Springer})\ pp.\ \bibinfo
  {pages} {281--288}\BibitemShut {NoStop}%
\bibitem [{\citenamefont {Davies}\ \emph {et~al.}(2011)\citenamefont {Davies},
  \citenamefont {Demetrius},\ and\ \citenamefont {Tuszynski}}]{Davies-cancer}%
  \BibitemOpen
  \bibfield  {author} {\bibinfo {author} {\bibnamefont {Davies}, \bibfnamefont
  {P.~C.}}, \bibinfo {author} {\bibfnamefont {L.}~\bibnamefont {Demetrius}}, \
  and\ \bibinfo {author} {\bibfnamefont {J.~A.}\ \bibnamefont {Tuszynski}}}
  (\bibinfo {year} {2011}),\ \href@noop {} {\bibfield  {journal} {\bibinfo
  {journal} {Theoretical Biology and Medical Modelling}\ }\textbf {\bibinfo
  {volume} {8}}~(\bibinfo {number} {1}),\ \bibinfo {pages} {1}}\BibitemShut
  {NoStop}%
\bibitem [{\citenamefont {Dayan}\ and\ \citenamefont {Abbott}(2006)}]{Dayan}%
  \BibitemOpen
  \bibfield  {author} {\bibinfo {author} {\bibnamefont {Dayan}, \bibfnamefont
  {P.}}, \ and\ \bibinfo {author} {\bibfnamefont {L.~F.}\ \bibnamefont
  {Abbott}}} (\bibinfo {year} {2006}),\ \href@noop {} {\emph {\bibinfo {title}
  {{ Theoretical Neuroscience: Computational and Mathematical Modeling of
  Neural Systems}}}}\ (\bibinfo  {publisher} {Cambridge, MIT Press,
  USA})\BibitemShut {NoStop}%
\bibitem [{\citenamefont {De~Gennes}(1979)}]{DeGennes}%
  \BibitemOpen
  \bibfield  {author} {\bibinfo {author} {\bibnamefont {De~Gennes},
  \bibfnamefont {P.-G.}}} (\bibinfo {year} {1979}),\ \href@noop {} {\emph
  {\bibinfo {title} {Scaling concepts in polymer physics}}}\ (\bibinfo
  {publisher} {Cornell university press})\BibitemShut {NoStop}%
\bibitem [{\citenamefont {De~Jong}(2002)}]{DeJong}%
  \BibitemOpen
  \bibfield  {author} {\bibinfo {author} {\bibnamefont {De~Jong}, \bibfnamefont
  {H.}}} (\bibinfo {year} {2002}),\ \href@noop {} {\bibfield  {journal}
  {\bibinfo  {journal} {J. Comput. Biol.}\ }\textbf {\bibinfo {volume}
  {9}}~(\bibinfo {number} {1}),\ \bibinfo {pages} {67}}\BibitemShut {NoStop}%
\bibitem [{\citenamefont {De~Palo}\ \emph {et~al.}(2017)\citenamefont
  {De~Palo}, \citenamefont {Yi},\ and\ \citenamefont {Endres}}]{Endres}%
  \BibitemOpen
  \bibfield  {author} {\bibinfo {author} {\bibnamefont {De~Palo}, \bibfnamefont
  {G.}}, \bibinfo {author} {\bibfnamefont {D.}~\bibnamefont {Yi}}, \ and\
  \bibinfo {author} {\bibfnamefont {R.~G.}\ \bibnamefont {Endres}}} (\bibinfo
  {year} {2017}),\ \href@noop {} {\bibfield  {journal} {\bibinfo  {journal}
  {PLoS biology}\ }\textbf {\bibinfo {volume} {15}}~(\bibinfo {number} {4}),\
  \bibinfo {pages} {e1002602}}\BibitemShut {NoStop}%
\bibitem [{\citenamefont {De~Vincenzo}\ \emph {et~al.}(2017)\citenamefont
  {De~Vincenzo}, \citenamefont {Giannoccaro}, \citenamefont {Carbone},\ and\
  \citenamefont {Grigolini}}]{Grigolini-groups}%
  \BibitemOpen
  \bibfield  {author} {\bibinfo {author} {\bibnamefont {De~Vincenzo},
  \bibfnamefont {I.}}, \bibinfo {author} {\bibfnamefont {I.}~\bibnamefont
  {Giannoccaro}}, \bibinfo {author} {\bibfnamefont {G.}~\bibnamefont
  {Carbone}}, \ and\ \bibinfo {author} {\bibfnamefont {P.}~\bibnamefont
  {Grigolini}}} (\bibinfo {year} {2017}),\ \href {\doibase
  10.1103/PhysRevE.96.022309} {\bibfield  {journal} {\bibinfo  {journal} {Phys.
  Rev. E}\ }\textbf {\bibinfo {volume} {96}},\ \bibinfo {pages}
  {022309}}\BibitemShut {NoStop}%
\bibitem [{\citenamefont {Deco}\ and\ \citenamefont
  {Jirsa}(2012)}]{Deco-Jirsa}%
  \BibitemOpen
  \bibfield  {author} {\bibinfo {author} {\bibnamefont {Deco}, \bibfnamefont
  {G.}}, \ and\ \bibinfo {author} {\bibfnamefont {V.~K.}\ \bibnamefont
  {Jirsa}}} (\bibinfo {year} {2012}),\ \href@noop {} {\bibfield  {journal}
  {\bibinfo  {journal} {J. Neurosci.}\ }\textbf {\bibinfo {volume}
  {32}}~(\bibinfo {number} {10}),\ \bibinfo {pages} {3366}}\BibitemShut
  {NoStop}%
\bibitem [{\citenamefont {Deco}\ \emph {et~al.}(2011)\citenamefont {Deco},
  \citenamefont {Jirsa},\ and\ \citenamefont {McIntosh}}]{Deco-RSN1}%
  \BibitemOpen
  \bibfield  {author} {\bibinfo {author} {\bibnamefont {Deco}, \bibfnamefont
  {G.}}, \bibinfo {author} {\bibfnamefont {V.~K.}\ \bibnamefont {Jirsa}}, \
  and\ \bibinfo {author} {\bibfnamefont {A.~R.}\ \bibnamefont {McIntosh}}}
  (\bibinfo {year} {2011}),\ \href@noop {} {\bibfield  {journal} {\bibinfo
  {journal} {Nature Rev. Neurosci.}\ }\textbf {\bibinfo {volume}
  {12}}~(\bibinfo {number} {1}),\ \bibinfo {pages} {43}}\BibitemShut {NoStop}%
\bibitem [{\citenamefont {Deco}\ \emph
  {et~al.}(2013{\natexlab{a}})\citenamefont {Deco}, \citenamefont {Jirsa},\
  and\ \citenamefont {McIntosh}}]{Deco-RSN2}%
  \BibitemOpen
  \bibfield  {author} {\bibinfo {author} {\bibnamefont {Deco}, \bibfnamefont
  {G.}}, \bibinfo {author} {\bibfnamefont {V.~K.}\ \bibnamefont {Jirsa}}, \
  and\ \bibinfo {author} {\bibfnamefont {A.~R.}\ \bibnamefont {McIntosh}}}
  (\bibinfo {year} {2013}{\natexlab{a}}),\ \href@noop {} {\bibfield  {journal}
  {\bibinfo  {journal} {Trends in Neurosci.}\ }\textbf {\bibinfo {volume}
  {36}}~(\bibinfo {number} {5}),\ \bibinfo {pages} {268}}\BibitemShut {NoStop}%
\bibitem [{\citenamefont {Deco}\ \emph {et~al.}(2014)\citenamefont {Deco},
  \citenamefont {Ponce-Alvarez}, \citenamefont {Hagmann}, \citenamefont
  {Romani}, \citenamefont {Mantini},\ and\ \citenamefont
  {Corbetta}}]{Deco-balanced}%
  \BibitemOpen
  \bibfield  {author} {\bibinfo {author} {\bibnamefont {Deco}, \bibfnamefont
  {G.}}, \bibinfo {author} {\bibfnamefont {A.}~\bibnamefont {Ponce-Alvarez}},
  \bibinfo {author} {\bibfnamefont {P.}~\bibnamefont {Hagmann}}, \bibinfo
  {author} {\bibfnamefont {G.~L.}\ \bibnamefont {Romani}}, \bibinfo {author}
  {\bibfnamefont {D.}~\bibnamefont {Mantini}}, \ and\ \bibinfo {author}
  {\bibfnamefont {M.}~\bibnamefont {Corbetta}}} (\bibinfo {year} {2014}),\
  \href@noop {} {\bibfield  {journal} {\bibinfo  {journal} {J. Neurosci.}\
  }\textbf {\bibinfo {volume} {34}}~(\bibinfo {number} {23}),\ \bibinfo {pages}
  {7886}}\BibitemShut {NoStop}%
\bibitem [{\citenamefont {Deco}\ \emph
  {et~al.}(2013{\natexlab{b}})\citenamefont {Deco}, \citenamefont
  {Ponce-Alvarez}, \citenamefont {Mantini}, \citenamefont {Romani},
  \citenamefont {Hagmann},\ and\ \citenamefont {Corbetta}}]{Deco2013}%
  \BibitemOpen
  \bibfield  {author} {\bibinfo {author} {\bibnamefont {Deco}, \bibfnamefont
  {G.}}, \bibinfo {author} {\bibfnamefont {A.}~\bibnamefont {Ponce-Alvarez}},
  \bibinfo {author} {\bibfnamefont {D.}~\bibnamefont {Mantini}}, \bibinfo
  {author} {\bibfnamefont {G.~L.}\ \bibnamefont {Romani}}, \bibinfo {author}
  {\bibfnamefont {P.}~\bibnamefont {Hagmann}}, \ and\ \bibinfo {author}
  {\bibfnamefont {M.}~\bibnamefont {Corbetta}}} (\bibinfo {year}
  {2013}{\natexlab{b}}),\ \href@noop {} {\bibfield  {journal} {\bibinfo
  {journal} {J. Neurosci.}\ }\textbf {\bibinfo {volume} {33}}~(\bibinfo
  {number} {27}),\ \bibinfo {pages} {11239}}\BibitemShut {NoStop}%
\bibitem [{\citenamefont {Delamotte}(2012)}]{Delamotte}%
  \BibitemOpen
  \bibfield  {author} {\bibinfo {author} {\bibnamefont {Delamotte},
  \bibfnamefont {B.}}} (\bibinfo {year} {2012}),\ in\ \href@noop {} {\emph
  {\bibinfo {booktitle} {Renormalization Group and Effective Field Theory
  Approaches to Many-Body Systems}}}\ (\bibinfo  {publisher} {Springer})\ pp.\
  \bibinfo {pages} {49--132}\BibitemShut {NoStop}%
\bibitem [{\citenamefont {Derrida}\ and\ \citenamefont
  {Pomeau}(1986)}]{Derrida}%
  \BibitemOpen
  \bibfield  {author} {\bibinfo {author} {\bibnamefont {Derrida}, \bibfnamefont
  {B.}}, \ and\ \bibinfo {author} {\bibfnamefont {Y.}~\bibnamefont {Pomeau}}}
  (\bibinfo {year} {1986}),\ \href@noop {} {\bibfield  {journal} {\bibinfo
  {journal} {EPL (Europhys. Lett.)}\ }\textbf {\bibinfo {volume} {1}}~(\bibinfo
  {number} {2}),\ \bibinfo {pages} {45}}\BibitemShut {NoStop}%
\bibitem [{\citenamefont {Destexhe}(2009)}]{Destexhe2009}%
  \BibitemOpen
  \bibfield  {author} {\bibinfo {author} {\bibnamefont {Destexhe},
  \bibfnamefont {A.}}} (\bibinfo {year} {2009}),\ \href@noop {} {\bibfield
  {journal} {\bibinfo  {journal} {J. Comput. Neurosci.}\ }\textbf {\bibinfo
  {volume} {27}}~(\bibinfo {number} {3}),\ \bibinfo {pages} {493}}\BibitemShut
  {NoStop}%
\bibitem [{\citenamefont {Dhar}(1999)}]{Dhar}%
  \BibitemOpen
  \bibfield  {author} {\bibinfo {author} {\bibnamefont {Dhar}, \bibfnamefont
  {D.}}} (\bibinfo {year} {1999}),\ \href@noop {} {\bibfield  {journal}
  {\bibinfo  {journal} {Physica A}\ }\textbf {\bibinfo {volume}
  {263}}~(\bibinfo {number} {1}),\ \bibinfo {pages} {4}}\BibitemShut {NoStop}%
\bibitem [{\citenamefont {Dickman}\ \emph {et~al.}(2000)\citenamefont
  {Dickman}, \citenamefont {Mu{\~n}oz}, \citenamefont {Vespignani},\ and\
  \citenamefont {Zapperi}}]{BJP}%
  \BibitemOpen
  \bibfield  {author} {\bibinfo {author} {\bibnamefont {Dickman}, \bibfnamefont
  {R.}}, \bibinfo {author} {\bibfnamefont {M.~A.}\ \bibnamefont {Mu{\~n}oz}},
  \bibinfo {author} {\bibfnamefont {A.}~\bibnamefont {Vespignani}}, \ and\
  \bibinfo {author} {\bibfnamefont {S.}~\bibnamefont {Zapperi}}} (\bibinfo
  {year} {2000}),\ \href@noop {} {\bibfield  {journal} {\bibinfo  {journal}
  {Braz. J. Phys.}\ }\textbf {\bibinfo {volume} {30}}~(\bibinfo {number} {1}),\
  \bibinfo {pages} {27}}\BibitemShut {NoStop}%
\bibitem [{\citenamefont {Diez}\ \emph {et~al.}(2015)\citenamefont {Diez},
  \citenamefont {Bonifazi}, \citenamefont {Escudero}, \citenamefont {Mateos},
  \citenamefont {Mu{\~n}oz}, \citenamefont {Stramaglia},\ and\ \citenamefont
  {Cortes}}]{Ibai}%
  \BibitemOpen
  \bibfield  {author} {\bibinfo {author} {\bibnamefont {Diez}, \bibfnamefont
  {I.}}, \bibinfo {author} {\bibfnamefont {P.}~\bibnamefont {Bonifazi}},
  \bibinfo {author} {\bibfnamefont {I.}~\bibnamefont {Escudero}}, \bibinfo
  {author} {\bibfnamefont {B.}~\bibnamefont {Mateos}}, \bibinfo {author}
  {\bibfnamefont {M.~A.}\ \bibnamefont {Mu{\~n}oz}}, \bibinfo {author}
  {\bibfnamefont {S.}~\bibnamefont {Stramaglia}}, \ and\ \bibinfo {author}
  {\bibfnamefont {J.~M.}\ \bibnamefont {Cortes}}} (\bibinfo {year} {2015}),\
  \href@noop {} {\bibfield  {journal} {\bibinfo  {journal} {Sci. Rep.}\
  }\textbf {\bibinfo {volume} {5}}}\BibitemShut {NoStop}%
\bibitem [{\citenamefont {Dornic}\ \emph {et~al.}(2001)\citenamefont {Dornic},
  \citenamefont {Chat\'e}, \citenamefont {Chave},\ and\ \citenamefont
  {Hinrichsen}}]{Dornic01}%
  \BibitemOpen
  \bibfield  {author} {\bibinfo {author} {\bibnamefont {Dornic}, \bibfnamefont
  {I.}}, \bibinfo {author} {\bibfnamefont {H.}~\bibnamefont {Chat\'e}},
  \bibinfo {author} {\bibfnamefont {J.}~\bibnamefont {Chave}}, \ and\ \bibinfo
  {author} {\bibfnamefont {H.}~\bibnamefont {Hinrichsen}}} (\bibinfo {year}
  {2001}),\ \href {\doibase 10.1103/PhysRevLett.87.045701} {\bibfield
  {journal} {\bibinfo  {journal} {Phys. Rev. Lett.}\ }\textbf {\bibinfo
  {volume} {87}},\ \bibinfo {pages} {045701}}\BibitemShut {NoStop}%
\bibitem [{\citenamefont {Dorogovtsev}\ \emph {et~al.}(2008)\citenamefont
  {Dorogovtsev}, \citenamefont {Goltsev},\ and\ \citenamefont
  {Mendes}}]{Doro-critical}%
  \BibitemOpen
  \bibfield  {author} {\bibinfo {author} {\bibnamefont {Dorogovtsev},
  \bibfnamefont {S.~N.}}, \bibinfo {author} {\bibfnamefont {A.~V.}\
  \bibnamefont {Goltsev}}, \ and\ \bibinfo {author} {\bibfnamefont {J.~F.}\
  \bibnamefont {Mendes}}} (\bibinfo {year} {2008}),\ \href@noop {} {\bibfield
  {journal} {\bibinfo  {journal} {Rev. Mod. Phys.}\ }\textbf {\bibinfo {volume}
  {80}}~(\bibinfo {number} {4}),\ \bibinfo {pages} {1275}}\BibitemShut
  {NoStop}%
\bibitem [{\citenamefont {Dorogovtsev}\ and\ \citenamefont
  {Mendes}(2002)}]{Doro-evol}%
  \BibitemOpen
  \bibfield  {author} {\bibinfo {author} {\bibnamefont {Dorogovtsev},
  \bibfnamefont {S.~N.}}, \ and\ \bibinfo {author} {\bibfnamefont {J.~F.}\
  \bibnamefont {Mendes}}} (\bibinfo {year} {2002}),\ \href@noop {} {\bibfield
  {journal} {\bibinfo  {journal} {Adv. in Phys.}\ }\textbf {\bibinfo {volume}
  {51}}~(\bibinfo {number} {4}),\ \bibinfo {pages} {1079}}\BibitemShut
  {NoStop}%
\bibitem [{\citenamefont {Drake}\ and\ \citenamefont {Holland}(1999)}]{Drake}%
  \BibitemOpen
  \bibfield  {author} {\bibinfo {author} {\bibnamefont {Drake}, \bibfnamefont
  {J.~W.}}, \ and\ \bibinfo {author} {\bibfnamefont {J.~J.}\ \bibnamefont
  {Holland}}} (\bibinfo {year} {1999}),\ \href@noop {} {\bibfield  {journal}
  {\bibinfo  {journal} {Proc. Natl. Acad. Sci. USA.}\ }\textbf {\bibinfo
  {volume} {96}}~(\bibinfo {number} {24}),\ \bibinfo {pages}
  {13910}}\BibitemShut {NoStop}%
\bibitem [{\citenamefont {Drossel}(2008)}]{Drossel2008}%
  \BibitemOpen
  \bibfield  {author} {\bibinfo {author} {\bibnamefont {Drossel}, \bibfnamefont
  {B.}}} (\bibinfo {year} {2008}),\ in\ \href@noop {} {\emph {\bibinfo
  {booktitle} {Reviews of nonlinear dynamics and complexity}}},\ Vol.~\bibinfo
  {volume} {1},\ \bibinfo {editor} {edited by\ \bibinfo {editor} {\bibfnamefont
  {H.~G.}\ \bibnamefont {Schuster}}},\ Chap.~\bibinfo {chapter} {3}\ (\bibinfo
  {publisher} {John Wiley \& Sons})\ pp.\ \bibinfo {pages}
  {69--110}\BibitemShut {NoStop}%
\bibitem [{\citenamefont {Drossel}\ and\ \citenamefont
  {Greil}(2009)}]{Drossel2009}%
  \BibitemOpen
  \bibfield  {author} {\bibinfo {author} {\bibnamefont {Drossel}, \bibfnamefont
  {B.}}, \ and\ \bibinfo {author} {\bibfnamefont {F.}~\bibnamefont {Greil}}}
  (\bibinfo {year} {2009}),\ \href@noop {} {\bibfield  {journal} {\bibinfo
  {journal} {Phys. Rev. E}\ }\textbf {\bibinfo {volume} {80}},\ \bibinfo
  {pages} {026102}}\BibitemShut {NoStop}%
\bibitem [{\citenamefont {Drossel}\ and\ \citenamefont {Schwabl}(1992)}]{FF2}%
  \BibitemOpen
  \bibfield  {author} {\bibinfo {author} {\bibnamefont {Drossel}, \bibfnamefont
  {B.}}, \ and\ \bibinfo {author} {\bibfnamefont {F.}~\bibnamefont {Schwabl}}}
  (\bibinfo {year} {1992}),\ \href@noop {} {\bibfield  {journal} {\bibinfo
  {journal} {Phys. Rev. Lett.}\ }\textbf {\bibinfo {volume} {69}}~(\bibinfo
  {number} {11}),\ \bibinfo {pages} {1629}}\BibitemShut {NoStop}%
\bibitem [{\citenamefont {Droste}\ \emph {et~al.}(2013)\citenamefont {Droste},
  \citenamefont {Do},\ and\ \citenamefont {Gross}}]{Anne-Ly}%
  \BibitemOpen
  \bibfield  {author} {\bibinfo {author} {\bibnamefont {Droste}, \bibfnamefont
  {F.}}, \bibinfo {author} {\bibfnamefont {A.-L.}\ \bibnamefont {Do}}, \ and\
  \bibinfo {author} {\bibfnamefont {T.}~\bibnamefont {Gross}}} (\bibinfo {year}
  {2013}),\ \href@noop {} {\bibinfo  {journal} {J. R. Soc. Interface}\ ,\
  \bibinfo {pages} {20120558}}\BibitemShut {NoStop}%
\bibitem [{\citenamefont {Duke}\ and\ \citenamefont
  {J{\"u}licher}(2003)}]{Duke2003}%
  \BibitemOpen
\bibfield  {journal} {  }\bibfield  {author} {\bibinfo {author} {\bibnamefont
  {Duke}, \bibfnamefont {T.}}, \ and\ \bibinfo {author} {\bibfnamefont
  {F.}~\bibnamefont {J{\"u}licher}}} (\bibinfo {year} {2003}),\ \href@noop {}
  {\bibfield  {journal} {\bibinfo  {journal} {Phys. Rev. Lett.}\ }\textbf
  {\bibinfo {volume} {90}}~(\bibinfo {number} {15}),\ \bibinfo {pages}
  {158101}}\BibitemShut {NoStop}%
\bibitem [{\citenamefont {Dyson}\ \emph {et~al.}(2015)\citenamefont {Dyson},
  \citenamefont {Yates}, \citenamefont {Buhl},\ and\ \citenamefont
  {McKane}}]{Locust}%
  \BibitemOpen
  \bibfield  {author} {\bibinfo {author} {\bibnamefont {Dyson}, \bibfnamefont
  {L.}}, \bibinfo {author} {\bibfnamefont {C.~A.}\ \bibnamefont {Yates}},
  \bibinfo {author} {\bibfnamefont {J.}~\bibnamefont {Buhl}}, \ and\ \bibinfo
  {author} {\bibfnamefont {A.~J.}\ \bibnamefont {McKane}}} (\bibinfo {year}
  {2015}),\ \href@noop {} {\bibfield  {journal} {\bibinfo  {journal} {Phys.
  Rev. E}\ }\textbf {\bibinfo {volume} {92}}~(\bibinfo {number} {5}),\ \bibinfo
  {pages} {052708}}\BibitemShut {NoStop}%
\bibitem [{\citenamefont {Effenberger}\ \emph {et~al.}(2015)\citenamefont
  {Effenberger}, \citenamefont {Jost},\ and\ \citenamefont
  {Levina}}]{Levina2015}%
  \BibitemOpen
  \bibfield  {author} {\bibinfo {author} {\bibnamefont {Effenberger},
  \bibfnamefont {F.}}, \bibinfo {author} {\bibfnamefont {J.}~\bibnamefont
  {Jost}}, \ and\ \bibinfo {author} {\bibfnamefont {A.}~\bibnamefont {Levina}}}
  (\bibinfo {year} {2015}),\ \href@noop {} {\bibfield  {journal} {\bibinfo
  {journal} {PLoS Comput. Biol.}\ }\textbf {\bibinfo {volume} {11}}~(\bibinfo
  {number} {9}),\ \bibinfo {pages} {e1004420}}\BibitemShut {NoStop}%
\bibitem [{\citenamefont {Egu{\'\i}luz}\ \emph {et~al.}(2000)\citenamefont
  {Egu{\'\i}luz}, \citenamefont {Ospeck}, \citenamefont {Choe}, \citenamefont
  {Hudspeth},\ and\ \citenamefont {Magnasco}}]{Eguiluz1}%
  \BibitemOpen
  \bibfield  {author} {\bibinfo {author} {\bibnamefont {Egu{\'\i}luz},
  \bibfnamefont {V.~M.}}, \bibinfo {author} {\bibfnamefont {M.}~\bibnamefont
  {Ospeck}}, \bibinfo {author} {\bibfnamefont {Y.}~\bibnamefont {Choe}},
  \bibinfo {author} {\bibfnamefont {A.}~\bibnamefont {Hudspeth}}, \ and\
  \bibinfo {author} {\bibfnamefont {M.~O.}\ \bibnamefont {Magnasco}}} (\bibinfo
  {year} {2000}),\ \href@noop {} {\bibfield  {journal} {\bibinfo  {journal}
  {Phys. Rev. Lett.}\ }\textbf {\bibinfo {volume} {84}}~(\bibinfo {number}
  {22}),\ \bibinfo {pages} {5232}}\BibitemShut {NoStop}%
\bibitem [{\citenamefont {Ehrig}\ \emph {et~al.}(2011)\citenamefont {Ehrig},
  \citenamefont {Petrov},\ and\ \citenamefont {Schwille}}]{membranes2}%
  \BibitemOpen
  \bibfield  {author} {\bibinfo {author} {\bibnamefont {Ehrig}, \bibfnamefont
  {J.}}, \bibinfo {author} {\bibfnamefont {E.~P.}\ \bibnamefont {Petrov}}, \
  and\ \bibinfo {author} {\bibfnamefont {P.}~\bibnamefont {Schwille}}}
  (\bibinfo {year} {2011}),\ \href@noop {} {\bibfield  {journal} {\bibinfo
  {journal} {Biophys. Jour.}\ }\textbf {\bibinfo {volume} {100}}~(\bibinfo
  {number} {1}),\ \bibinfo {pages} {80}}\BibitemShut {NoStop}%
\bibitem [{\citenamefont {Eigen}(2002)}]{Eigen2000}%
  \BibitemOpen
  \bibfield  {author} {\bibinfo {author} {\bibnamefont {Eigen}, \bibfnamefont
  {M.}}} (\bibinfo {year} {2002}),\ \href@noop {} {\bibfield  {journal}
  {\bibinfo  {journal} {Proc. Natl. Acad. Sci. USA.}\ }\textbf {\bibinfo
  {volume} {99}}~(\bibinfo {number} {21}),\ \bibinfo {pages}
  {13374}}\BibitemShut {NoStop}%
\bibitem [{\citenamefont {Eigen}\ \emph {et~al.}(1989)\citenamefont {Eigen},
  \citenamefont {McCaskill},\ and\ \citenamefont {Schuster}}]{Eigen1989}%
  \BibitemOpen
  \bibfield  {author} {\bibinfo {author} {\bibnamefont {Eigen}, \bibfnamefont
  {M.}}, \bibinfo {author} {\bibfnamefont {J.}~\bibnamefont {McCaskill}}, \
  and\ \bibinfo {author} {\bibfnamefont {P.}~\bibnamefont {Schuster}}}
  (\bibinfo {year} {1989}),\ \href@noop {} {\bibfield  {journal} {\bibinfo
  {journal} {Adv. Chem. Phys}\ }\textbf {\bibinfo {volume} {75}},\ \bibinfo
  {pages} {149}}\BibitemShut {NoStop}%
\bibitem [{\citenamefont {Eigen}\ and\ \citenamefont
  {Schuster}(1979)}]{Eigen1979}%
  \BibitemOpen
  \bibfield  {author} {\bibinfo {author} {\bibnamefont {Eigen}, \bibfnamefont
  {M.}}, \ and\ \bibinfo {author} {\bibfnamefont {P.}~\bibnamefont {Schuster}}}
  (\bibinfo {year} {1979}),\ \href@noop {} {\bibinfo  {journal} {New York}\
  }\BibitemShut {NoStop}%
\bibitem [{\citenamefont {Erez}\ \emph {et~al.}(2017)\citenamefont {Erez},
  \citenamefont {Byrd}, \citenamefont {Vogel}, \citenamefont {Altan-Bonnet},\
  and\ \citenamefont {Mugler}}]{Erez2017}%
  \BibitemOpen
\bibfield  {journal} {  }\bibfield  {author} {\bibinfo {author} {\bibnamefont
  {Erez}, \bibfnamefont {A.}}, \bibinfo {author} {\bibfnamefont {T.~A.}\
  \bibnamefont {Byrd}}, \bibinfo {author} {\bibfnamefont {R.~M.}\ \bibnamefont
  {Vogel}}, \bibinfo {author} {\bibfnamefont {G.}~\bibnamefont {Altan-Bonnet}},
  \ and\ \bibinfo {author} {\bibfnamefont {A.}~\bibnamefont {Mugler}}}
  (\bibinfo {year} {2017}),\ \href@noop {} {\bibinfo  {journal}
  {arXiv:1703.04194}\ }\BibitemShut {NoStop}%
\bibitem [{\citenamefont {Erskine}\ and\ \citenamefont
  {Herrmann}(2014)}]{swarm-Herrmann}%
  \BibitemOpen
\bibfield  {journal} {  }\bibfield  {author} {\bibinfo {author} {\bibnamefont
  {Erskine}, \bibfnamefont {A.}}, \ and\ \bibinfo {author} {\bibfnamefont
  {J.~M.}\ \bibnamefont {Herrmann}}} (\bibinfo {year} {2014}),\ \href@noop {}
  {\bibinfo  {journal} {arXiv:1402.6888}\ }\BibitemShut {NoStop}%
\bibitem [{\citenamefont {Espinosa-Soto}\ \emph {et~al.}(2004)\citenamefont
  {Espinosa-Soto}, \citenamefont {Padilla-Longoria},\ and\ \citenamefont
  {Alvarez-Buylla}}]{Buylla}%
  \BibitemOpen
\bibfield  {journal} {  }\bibfield  {author} {\bibinfo {author} {\bibnamefont
  {Espinosa-Soto}, \bibfnamefont {C.}}, \bibinfo {author} {\bibfnamefont
  {P.}~\bibnamefont {Padilla-Longoria}}, \ and\ \bibinfo {author}
  {\bibfnamefont {E.~R.}\ \bibnamefont {Alvarez-Buylla}}} (\bibinfo {year}
  {2004}),\ \href@noop {} {\bibfield  {journal} {\bibinfo  {journal} {The Plant
  Cell}\ }\textbf {\bibinfo {volume} {16}}~(\bibinfo {number} {11}),\ \bibinfo
  {pages} {2923}}\BibitemShut {NoStop}%
\bibitem [{\citenamefont {Expert}\ \emph {et~al.}(2011)\citenamefont {Expert},
  \citenamefont {Lambiotte}, \citenamefont {Chialvo}, \citenamefont
  {Christensen}, \citenamefont {Jensen}, \citenamefont {Sharp},\ and\
  \citenamefont {Turkheimer}}]{Expert2011}%
  \BibitemOpen
  \bibfield  {author} {\bibinfo {author} {\bibnamefont {Expert}, \bibfnamefont
  {P.}}, \bibinfo {author} {\bibfnamefont {R.}~\bibnamefont {Lambiotte}},
  \bibinfo {author} {\bibfnamefont {D.~R.}\ \bibnamefont {Chialvo}}, \bibinfo
  {author} {\bibfnamefont {K.}~\bibnamefont {Christensen}}, \bibinfo {author}
  {\bibfnamefont {H.~J.}\ \bibnamefont {Jensen}}, \bibinfo {author}
  {\bibfnamefont {D.~J.}\ \bibnamefont {Sharp}}, \ and\ \bibinfo {author}
  {\bibfnamefont {F.}~\bibnamefont {Turkheimer}}} (\bibinfo {year} {2011}),\
  \href@noop {} {\bibfield  {journal} {\bibinfo  {journal} {J. R. Soc.
  Interface}\ }\textbf {\bibinfo {volume} {8}}~(\bibinfo {number} {57}),\
  \bibinfo {pages} {472}}\BibitemShut {NoStop}%
\bibitem [{\citenamefont {Eytan}\ and\ \citenamefont {Marom}(2006)}]{Eytan}%
  \BibitemOpen
  \bibfield  {author} {\bibinfo {author} {\bibnamefont {Eytan}, \bibfnamefont
  {D.}}, \ and\ \bibinfo {author} {\bibfnamefont {S.}~\bibnamefont {Marom}}}
  (\bibinfo {year} {2006}),\ \href@noop {} {\bibfield  {journal} {\bibinfo
  {journal} {J. Neurosci.}\ }\textbf {\bibinfo {volume} {26}}~(\bibinfo
  {number} {33}),\ \bibinfo {pages} {8465}}\BibitemShut {NoStop}%
\bibitem [{\citenamefont {Feinerman}\ and\ \citenamefont
  {Korman}(2017)}]{Ants2017}%
  \BibitemOpen
  \bibfield  {author} {\bibinfo {author} {\bibnamefont {Feinerman},
  \bibfnamefont {O.}}, \ and\ \bibinfo {author} {\bibfnamefont
  {A.}~\bibnamefont {Korman}}} (\bibinfo {year} {2017}),\ \href@noop {}
  {\bibfield  {journal} {\bibinfo  {journal} {J. Exp.Biol.}\ }\textbf {\bibinfo
  {volume} {220}}~(\bibinfo {number} {1}),\ \bibinfo {pages} {73}}\BibitemShut
  {NoStop}%
\bibitem [{\citenamefont {Filkov}(2005)}]{Filkov}%
  \BibitemOpen
  \bibfield  {author} {\bibinfo {author} {\bibnamefont {Filkov}, \bibfnamefont
  {V.}}} (\bibinfo {year} {2005}),\ \href@noop {} {\bibinfo  {journal}
  {Handbook of Computational Molecular Biology}\ ,\ \bibinfo {pages}
  {27}}\BibitemShut {NoStop}%
\bibitem [{\citenamefont {Fisher}(1974)}]{Fisher}%
  \BibitemOpen
\bibfield  {journal} {  }\bibfield  {author} {\bibinfo {author} {\bibnamefont
  {Fisher}, \bibfnamefont {M.~E.}}} (\bibinfo {year} {1974}),\ \href@noop {}
  {\bibfield  {journal} {\bibinfo  {journal} {Rev. Mod. Phys.}\ }\textbf
  {\bibinfo {volume} {46}}~(\bibinfo {number} {4}),\ \bibinfo {pages}
  {597}}\BibitemShut {NoStop}%
\bibitem [{\citenamefont {Font-Clos}\ \emph {et~al.}(2015)\citenamefont
  {Font-Clos}, \citenamefont {Pruessner}, \citenamefont {Moloney},\ and\
  \citenamefont {Deluca}}]{Perils2}%
  \BibitemOpen
  \bibfield  {author} {\bibinfo {author} {\bibnamefont {Font-Clos},
  \bibfnamefont {F.}}, \bibinfo {author} {\bibfnamefont {G.}~\bibnamefont
  {Pruessner}}, \bibinfo {author} {\bibfnamefont {N.~R.}\ \bibnamefont
  {Moloney}}, \ and\ \bibinfo {author} {\bibfnamefont {A.}~\bibnamefont
  {Deluca}}} (\bibinfo {year} {2015}),\ \href@noop {} {\bibfield  {journal}
  {\bibinfo  {journal} {New J. Phys}\ }\textbf {\bibinfo {volume}
  {17}}~(\bibinfo {number} {4}),\ \bibinfo {pages} {043066}}\BibitemShut
  {NoStop}%
\bibitem [{\citenamefont {Forgacs}\ \emph {et~al.}(1991)\citenamefont
  {Forgacs}, \citenamefont {Newman}, \citenamefont {Obukhov},\ and\
  \citenamefont {Birk}}]{Forgacs1}%
  \BibitemOpen
  \bibfield  {author} {\bibinfo {author} {\bibnamefont {Forgacs}, \bibfnamefont
  {G.}}, \bibinfo {author} {\bibfnamefont {S.~A.}\ \bibnamefont {Newman}},
  \bibinfo {author} {\bibfnamefont {S.~P.}\ \bibnamefont {Obukhov}}, \ and\
  \bibinfo {author} {\bibfnamefont {D.~E.}\ \bibnamefont {Birk}}} (\bibinfo
  {year} {1991}),\ \href {\doibase 10.1103/PhysRevLett.67.2399} {\bibfield
  {journal} {\bibinfo  {journal} {Phys. Rev. Lett.}\ }\textbf {\bibinfo
  {volume} {67}},\ \bibinfo {pages} {2399}}\BibitemShut {NoStop}%
\bibitem [{\citenamefont {Fortrat}\ and\ \citenamefont {Gharib}(2016)}]{Blood}%
  \BibitemOpen
  \bibfield  {author} {\bibinfo {author} {\bibnamefont {Fortrat}, \bibfnamefont
  {J.-O.}}, \ and\ \bibinfo {author} {\bibfnamefont {C.}~\bibnamefont
  {Gharib}}} (\bibinfo {year} {2016}),\ \href@noop {} {\bibfield  {journal}
  {\bibinfo  {journal} {Front. Physiol.}\ }\textbf {\bibinfo {volume}
  {7}}}\BibitemShut {NoStop}%
\bibitem [{\citenamefont {Fox}\ and\ \citenamefont {Raichle}(2007)}]{Fox}%
  \BibitemOpen
  \bibfield  {author} {\bibinfo {author} {\bibnamefont {Fox}, \bibfnamefont
  {M.~D.}}, \ and\ \bibinfo {author} {\bibfnamefont {M.~E.}\ \bibnamefont
  {Raichle}}} (\bibinfo {year} {2007}),\ \href@noop {} {\bibfield  {journal}
  {\bibinfo  {journal} {Nature Reviews Neuroscience}\ }\textbf {\bibinfo
  {volume} {8}}~(\bibinfo {number} {9}),\ \bibinfo {pages} {700}}\BibitemShut
  {NoStop}%
\bibitem [{\citenamefont {Fraiman}\ \emph {et~al.}(2009)\citenamefont
  {Fraiman}, \citenamefont {Balenzuela}, \citenamefont {Jennifer},\ and\
  \citenamefont {Chialvo}}]{Fraiman}%
  \BibitemOpen
  \bibfield  {author} {\bibinfo {author} {\bibnamefont {Fraiman}, \bibfnamefont
  {D.}}, \bibinfo {author} {\bibfnamefont {P.}~\bibnamefont {Balenzuela}},
  \bibinfo {author} {\bibfnamefont {J.}~\bibnamefont {Jennifer}}, \ and\
  \bibinfo {author} {\bibfnamefont {D.~R.}\ \bibnamefont {Chialvo}}} (\bibinfo
  {year} {2009}),\ \href@noop {} {\bibfield  {journal} {\bibinfo  {journal}
  {Phys. Rev. E}\ }\textbf {\bibinfo {volume} {79}}~(\bibinfo {number} {6}),\
  \bibinfo {pages} {061922}}\BibitemShut {NoStop}%
\bibitem [{\citenamefont {Frauenfelder}(2014)}]{Ask}%
  \BibitemOpen
  \bibfield  {author} {\bibinfo {author} {\bibnamefont {Frauenfelder},
  \bibfnamefont {H.}}} (\bibinfo {year} {2014}),\ \href
  {http://stacks.iop.org/1478-3975/11/i=5/a=053004} {\bibfield  {journal}
  {\bibinfo  {journal} {Phys. Biol.}\ }\textbf {\bibinfo {volume}
  {11}}~(\bibinfo {number} {5}),\ \bibinfo {pages} {053004}}\BibitemShut
  {NoStop}%
\bibitem [{\citenamefont {Freeman}(2013)}]{Freeman2013}%
  \BibitemOpen
  \bibfield  {author} {\bibinfo {author} {\bibnamefont {Freeman}, \bibfnamefont
  {W.~J.}}} (\bibinfo {year} {2013}),\ in\ \href@noop {} {\emph {\bibinfo
  {booktitle} {Chaos, CNN, Memristors and Beyond: A Festschrift for
  L.Chua.}}},\ Vol.~\bibinfo {volume} {1},\ \bibinfo {editor} {edited by\
  \bibinfo {editor} {\bibfnamefont {A.~A.}\ \bibnamefont {et~al.}}}\ (\bibinfo
  {publisher} {World Scientific Publishing Co.})\ pp.\ \bibinfo {pages}
  {271--284}\BibitemShut {NoStop}%
\bibitem [{\citenamefont {Freeman}\ and\ \citenamefont
  {Holmes}(2005)}]{Freeman2005}%
  \BibitemOpen
  \bibfield  {author} {\bibinfo {author} {\bibnamefont {Freeman}, \bibfnamefont
  {W.~J.}}, \ and\ \bibinfo {author} {\bibfnamefont {M.~D.}\ \bibnamefont
  {Holmes}}} (\bibinfo {year} {2005}),\ \href@noop {} {\bibfield  {journal}
  {\bibinfo  {journal} {Neural Networks}\ }\textbf {\bibinfo {volume}
  {18}}~(\bibinfo {number} {5}),\ \bibinfo {pages} {497}}\BibitemShut {NoStop}%
\bibitem [{\citenamefont {Freeman}\ \emph {et~al.}(2003)\citenamefont
  {Freeman}, \citenamefont {Holmes}, \citenamefont {Burke},\ and\ \citenamefont
  {Vanhatalo}}]{EEG-Freeman}%
  \BibitemOpen
  \bibfield  {author} {\bibinfo {author} {\bibnamefont {Freeman}, \bibfnamefont
  {W.~J.}}, \bibinfo {author} {\bibfnamefont {M.~D.}\ \bibnamefont {Holmes}},
  \bibinfo {author} {\bibfnamefont {B.~C.}\ \bibnamefont {Burke}}, \ and\
  \bibinfo {author} {\bibfnamefont {S.}~\bibnamefont {Vanhatalo}}} (\bibinfo
  {year} {2003}),\ \href@noop {} {\bibfield  {journal} {\bibinfo  {journal}
  {Clinical Neurophys.}\ }\textbf {\bibinfo {volume} {114}}~(\bibinfo {number}
  {6}),\ \bibinfo {pages} {1053}}\BibitemShut {NoStop}%
\bibitem [{\citenamefont {Frette}\ \emph {et~al.}(1996)\citenamefont {Frette},
  \citenamefont {Christensen}, \citenamefont {Malthe-S{\o}renssen},
  \citenamefont {Feder}, \citenamefont {J{\o}ssang},\ and\ \citenamefont
  {Meakin}}]{Frette}%
  \BibitemOpen
  \bibfield  {author} {\bibinfo {author} {\bibnamefont {Frette}, \bibfnamefont
  {V.}}, \bibinfo {author} {\bibfnamefont {K.}~\bibnamefont {Christensen}},
  \bibinfo {author} {\bibfnamefont {A.}~\bibnamefont {Malthe-S{\o}renssen}},
  \bibinfo {author} {\bibfnamefont {J.}~\bibnamefont {Feder}}, \bibinfo
  {author} {\bibfnamefont {T.}~\bibnamefont {J{\o}ssang}}, \ and\ \bibinfo
  {author} {\bibfnamefont {P.}~\bibnamefont {Meakin}}} (\bibinfo {year}
  {1996}),\ \href@noop {} {\bibfield  {journal} {\bibinfo  {journal} {Nature}\
  }\textbf {\bibinfo {volume} {379}}~(\bibinfo {number} {6560}),\ \bibinfo
  {pages} {49}}\BibitemShut {NoStop}%
\bibitem [{\citenamefont {Friedman}\ \emph {et~al.}(2012)\citenamefont
  {Friedman}, \citenamefont {Ito}, \citenamefont {Brinkman}, \citenamefont
  {Shimono}, \citenamefont {DeVille}, \citenamefont {Dahmen}, \citenamefont
  {Beggs},\ and\ \citenamefont {Butler}}]{Dahmen}%
  \BibitemOpen
  \bibfield  {author} {\bibinfo {author} {\bibnamefont {Friedman},
  \bibfnamefont {N.}}, \bibinfo {author} {\bibfnamefont {S.}~\bibnamefont
  {Ito}}, \bibinfo {author} {\bibfnamefont {B.~A.~W.}\ \bibnamefont
  {Brinkman}}, \bibinfo {author} {\bibfnamefont {M.}~\bibnamefont {Shimono}},
  \bibinfo {author} {\bibfnamefont {R.~E.~L.}\ \bibnamefont {DeVille}},
  \bibinfo {author} {\bibfnamefont {K.~A.}\ \bibnamefont {Dahmen}}, \bibinfo
  {author} {\bibfnamefont {J.~M.}\ \bibnamefont {Beggs}}, \ and\ \bibinfo
  {author} {\bibfnamefont {T.~C.}\ \bibnamefont {Butler}}} (\bibinfo {year}
  {2012}),\ \href {\doibase 10.1103/PhysRevLett.108.208102} {\bibfield
  {journal} {\bibinfo  {journal} {Phys. Rev. Lett.}\ }\textbf {\bibinfo
  {volume} {108}},\ \bibinfo {pages} {208102}}\BibitemShut {NoStop}%
\bibitem [{\citenamefont {Friston}\ \emph {et~al.}(2012)\citenamefont
  {Friston}, \citenamefont {Breakspear},\ and\ \citenamefont
  {Deco}}]{Deco2012}%
  \BibitemOpen
  \bibfield  {author} {\bibinfo {author} {\bibnamefont {Friston}, \bibfnamefont
  {K.}}, \bibinfo {author} {\bibfnamefont {M.}~\bibnamefont {Breakspear}}, \
  and\ \bibinfo {author} {\bibfnamefont {G.}~\bibnamefont {Deco}}} (\bibinfo
  {year} {2012}),\ \href@noop {} {\bibfield  {journal} {\bibinfo  {journal}
  {Front. Comput. Neurosci.}\ }\textbf {\bibinfo {volume} {6}},\ \bibinfo
  {pages} {44}}\BibitemShut {NoStop}%
\bibitem [{\citenamefont {Furusawa}\ and\ \citenamefont
  {Kaneko}(2003)}]{Furu2003}%
  \BibitemOpen
  \bibfield  {author} {\bibinfo {author} {\bibnamefont {Furusawa},
  \bibfnamefont {C.}}, \ and\ \bibinfo {author} {\bibfnamefont
  {K.}~\bibnamefont {Kaneko}}} (\bibinfo {year} {2003}),\ \href {\doibase
  10.1103/PhysRevLett.90.088102} {\bibfield  {journal} {\bibinfo  {journal}
  {Phys. Rev. Lett.}\ }\textbf {\bibinfo {volume} {90}},\ \bibinfo {pages}
  {088102}}\BibitemShut {NoStop}%
\bibitem [{\citenamefont {Furusawa}\ and\ \citenamefont
  {Kaneko}(2012{\natexlab{a}})}]{Furu2012}%
  \BibitemOpen
  \bibfield  {author} {\bibinfo {author} {\bibnamefont {Furusawa},
  \bibfnamefont {C.}}, \ and\ \bibinfo {author} {\bibfnamefont
  {K.}~\bibnamefont {Kaneko}}} (\bibinfo {year} {2012}{\natexlab{a}}),\
  \href@noop {} {\bibfield  {journal} {\bibinfo  {journal} {Phys. Rev. Lett.}\
  }\textbf {\bibinfo {volume} {108}}~(\bibinfo {number} {20}),\ \bibinfo
  {pages} {208103}}\BibitemShut {NoStop}%
\bibitem [{\citenamefont {Furusawa}\ and\ \citenamefont
  {Kaneko}(2012{\natexlab{b}})}]{Kaneko-stem}%
  \BibitemOpen
  \bibfield  {author} {\bibinfo {author} {\bibnamefont {Furusawa},
  \bibfnamefont {C.}}, \ and\ \bibinfo {author} {\bibfnamefont
  {K.}~\bibnamefont {Kaneko}}} (\bibinfo {year} {2012}{\natexlab{b}}),\
  \href@noop {} {\bibfield  {journal} {\bibinfo  {journal} {Science}\ }\textbf
  {\bibinfo {volume} {338}}~(\bibinfo {number} {6104}),\ \bibinfo {pages}
  {215}}\BibitemShut {NoStop}%
\bibitem [{\citenamefont {Gal}\ and\ \citenamefont
  {Marom}(2013)}]{single-Marom}%
  \BibitemOpen
  \bibfield  {author} {\bibinfo {author} {\bibnamefont {Gal}, \bibfnamefont
  {A.}}, \ and\ \bibinfo {author} {\bibfnamefont {S.}~\bibnamefont {Marom}}}
  (\bibinfo {year} {2013}),\ \href@noop {} {\bibfield  {journal} {\bibinfo
  {journal} {Phys. Rev. E}\ }\textbf {\bibinfo {volume} {88}}~(\bibinfo
  {number} {6}),\ \bibinfo {pages} {062717}}\BibitemShut {NoStop}%
\bibitem [{\citenamefont {Gallotti}\ and\ \citenamefont
  {Chialvo}(2017)}]{Ants-Chialvo}%
  \BibitemOpen
  \bibfield  {author} {\bibinfo {author} {\bibnamefont {Gallotti},
  \bibfnamefont {R.}}, \ and\ \bibinfo {author} {\bibfnamefont {D.~R.}\
  \bibnamefont {Chialvo}}} (\bibinfo {year} {2017}),\ \href@noop {} {\bibinfo
  {journal} {arXiv:1707.07135}\ }\BibitemShut {NoStop}%
\bibitem [{\citenamefont {Gama-Castro}\ \emph {et~al.}(2015)\citenamefont
  {Gama-Castro}, \citenamefont {Salgado}, \citenamefont {Santos-Zavaleta},
  \citenamefont {Ledezma-Tejeida}, \citenamefont {Mu{\~n}iz-Rascado},
  \citenamefont {Garc{\'\i}a-Sotelo}, \citenamefont {Alquicira-Hern{\'a}ndez},
  \citenamefont {Mart{\'\i}nez-Flores}, \citenamefont {Pannier}, \citenamefont
  {Castro-Mondrag{\'o}n} \emph {et~al.}}]{Red1}%
  \BibitemOpen
\bibfield  {journal} {  }\bibfield  {author} {\bibinfo {author} {\bibnamefont
  {Gama-Castro}, \bibfnamefont {S.}}, \bibinfo {author} {\bibfnamefont
  {H.}~\bibnamefont {Salgado}}, \bibinfo {author} {\bibfnamefont
  {A.}~\bibnamefont {Santos-Zavaleta}}, \bibinfo {author} {\bibfnamefont
  {D.}~\bibnamefont {Ledezma-Tejeida}}, \bibinfo {author} {\bibfnamefont
  {L.}~\bibnamefont {Mu{\~n}iz-Rascado}}, \bibinfo {author} {\bibfnamefont
  {J.~S.}\ \bibnamefont {Garc{\'\i}a-Sotelo}}, \bibinfo {author} {\bibfnamefont
  {K.}~\bibnamefont {Alquicira-Hern{\'a}ndez}}, \bibinfo {author}
  {\bibfnamefont {I.}~\bibnamefont {Mart{\'\i}nez-Flores}}, \bibinfo {author}
  {\bibfnamefont {L.}~\bibnamefont {Pannier}}, \bibinfo {author} {\bibfnamefont
  {J.~A.}\ \bibnamefont {Castro-Mondrag{\'o}n}},  \emph {et~al.}} (\bibinfo
  {year} {2015}),\ \href@noop {} {\bibfield  {journal} {\bibinfo  {journal}
  {Nucleic Acids Res.}\ }\textbf {\bibinfo {volume} {44}}~(\bibinfo {number}
  {D1}),\ \bibinfo {pages} {D133}}\BibitemShut {NoStop}%
\bibitem [{\citenamefont {Gamba}\ \emph {et~al.}(2012)\citenamefont {Gamba},
  \citenamefont {Nicodemi}, \citenamefont {Soriano},\ and\ \citenamefont
  {Ott}}]{Soriano2012}%
  \BibitemOpen
  \bibfield  {author} {\bibinfo {author} {\bibnamefont {Gamba}, \bibfnamefont
  {A.}}, \bibinfo {author} {\bibfnamefont {M.}~\bibnamefont {Nicodemi}},
  \bibinfo {author} {\bibfnamefont {J.}~\bibnamefont {Soriano}}, \ and\
  \bibinfo {author} {\bibfnamefont {A.}~\bibnamefont {Ott}}} (\bibinfo {year}
  {2012}),\ \href@noop {} {\bibfield  {journal} {\bibinfo  {journal} {Phys.
  Rev. Lett.}\ }\textbf {\bibinfo {volume} {108}}~(\bibinfo {number} {15}),\
  \bibinfo {pages} {158103}}\BibitemShut {NoStop}%
\bibitem [{\citenamefont {Garcia-Ojalvo}(2011)}]{Ojalvo-review}%
  \BibitemOpen
  \bibfield  {author} {\bibinfo {author} {\bibnamefont {Garcia-Ojalvo},
  \bibfnamefont {J.}}} (\bibinfo {year} {2011}),\ \href@noop {} {\bibfield
  {journal} {\bibinfo  {journal} {Contemp. Phys.}\ }\textbf {\bibinfo {volume}
  {52}}~(\bibinfo {number} {5}),\ \bibinfo {pages} {439}}\BibitemShut {NoStop}%
\bibitem [{\citenamefont {Garcia-Ojalvo}\ and\ \citenamefont
  {Arias}(2012)}]{decision-Ojalvo}%
  \BibitemOpen
  \bibfield  {author} {\bibinfo {author} {\bibnamefont {Garcia-Ojalvo},
  \bibfnamefont {J.}}, \ and\ \bibinfo {author} {\bibfnamefont {A.~M.}\
  \bibnamefont {Arias}}} (\bibinfo {year} {2012}),\ \href@noop {} {\bibfield
  {journal} {\bibinfo  {journal} {Current opinion in genetics \& development}\
  }\textbf {\bibinfo {volume} {22}}~(\bibinfo {number} {6}),\ \bibinfo {pages}
  {619}}\BibitemShut {NoStop}%
\bibitem [{\citenamefont {Garcia-Ojalvo}\ \emph {et~al.}(2004)\citenamefont
  {Garcia-Ojalvo}, \citenamefont {Elowitz},\ and\ \citenamefont
  {Strogatz}}]{PNAS-Ojalvo}%
  \BibitemOpen
  \bibfield  {author} {\bibinfo {author} {\bibnamefont {Garcia-Ojalvo},
  \bibfnamefont {J.}}, \bibinfo {author} {\bibfnamefont {M.~B.}\ \bibnamefont
  {Elowitz}}, \ and\ \bibinfo {author} {\bibfnamefont {S.~H.}\ \bibnamefont
  {Strogatz}}} (\bibinfo {year} {2004}),\ \href@noop {} {\bibfield  {journal}
  {\bibinfo  {journal} {Proc. Natl. Acad. Sci. USA.}\ }\textbf {\bibinfo
  {volume} {101}}~(\bibinfo {number} {30}),\ \bibinfo {pages}
  {10955}}\BibitemShut {NoStop}%
\bibitem [{\citenamefont {Gautam}\ \emph {et~al.}(2015)\citenamefont {Gautam},
  \citenamefont {Hoang}, \citenamefont {McClanahan}, \citenamefont {Grady},\
  and\ \citenamefont {Shew}}]{Shew2015a}%
  \BibitemOpen
  \bibfield  {author} {\bibinfo {author} {\bibnamefont {Gautam}, \bibfnamefont
  {S.~H.}}, \bibinfo {author} {\bibfnamefont {T.~T.}\ \bibnamefont {Hoang}},
  \bibinfo {author} {\bibfnamefont {K.}~\bibnamefont {McClanahan}}, \bibinfo
  {author} {\bibfnamefont {S.~K.}\ \bibnamefont {Grady}}, \ and\ \bibinfo
  {author} {\bibfnamefont {W.~L.}\ \bibnamefont {Shew}}} (\bibinfo {year}
  {2015}),\ \href@noop {} {\bibfield  {journal} {\bibinfo  {journal} {PLoS
  Comput. Biol.}\ }\textbf {\bibinfo {volume} {11}}~(\bibinfo {number} {12}),\
  \bibinfo {pages} {e1004576}}\BibitemShut {NoStop}%
\bibitem [{\citenamefont {Gelblum}\ \emph {et~al.}(2015)\citenamefont
  {Gelblum}, \citenamefont {Pinkoviezky}, \citenamefont {Fonio}, \citenamefont
  {Ghosh}, \citenamefont {Gov},\ and\ \citenamefont {Feinerman}}]{Ants2015}%
  \BibitemOpen
  \bibfield  {author} {\bibinfo {author} {\bibnamefont {Gelblum}, \bibfnamefont
  {A.}}, \bibinfo {author} {\bibfnamefont {I.}~\bibnamefont {Pinkoviezky}},
  \bibinfo {author} {\bibfnamefont {E.}~\bibnamefont {Fonio}}, \bibinfo
  {author} {\bibfnamefont {A.}~\bibnamefont {Ghosh}}, \bibinfo {author}
  {\bibfnamefont {N.}~\bibnamefont {Gov}}, \ and\ \bibinfo {author}
  {\bibfnamefont {O.}~\bibnamefont {Feinerman}}} (\bibinfo {year} {2015}),\
  \href@noop {} {\bibfield  {journal} {\bibinfo  {journal} {Nat. Comm.}\
  }\textbf {\bibinfo {volume} {6}}}\BibitemShut {NoStop}%
\bibitem [{\citenamefont {GellMann}(1994)}]{Gellmann}%
  \BibitemOpen
  \bibfield  {author} {\bibinfo {author} {\bibnamefont {GellMann},
  \bibfnamefont {M.}}} (\bibinfo {year} {1994}),\ \href@noop {} {\bibinfo
  {journal} {Complexity: Metaphors, models and reality}\ ,\ \bibinfo {pages}
  {17}}\BibitemShut {NoStop}%
\bibitem [{\citenamefont {Gershenson}(2012)}]{Gershenson}%
  \BibitemOpen
\bibfield  {journal} {  }\bibfield  {author} {\bibinfo {author} {\bibnamefont
  {Gershenson}, \bibfnamefont {C.}}} (\bibinfo {year} {2012}),\ \href@noop {}
  {\bibfield  {journal} {\bibinfo  {journal} {Theory in Biosci.}\ }\textbf
  {\bibinfo {volume} {131}}~(\bibinfo {number} {3}),\ \bibinfo {pages}
  {181}}\BibitemShut {NoStop}%
\bibitem [{\citenamefont {Gerstein}\ and\ \citenamefont
  {Mandelbrot}(1964)}]{Mandelbrot-RW}%
  \BibitemOpen
  \bibfield  {author} {\bibinfo {author} {\bibnamefont {Gerstein},
  \bibfnamefont {G.~L.}}, \ and\ \bibinfo {author} {\bibfnamefont
  {B.}~\bibnamefont {Mandelbrot}}} (\bibinfo {year} {1964}),\ \href@noop {}
  {\bibfield  {journal} {\bibinfo  {journal} {Biophys. Jour.}\ }\textbf
  {\bibinfo {volume} {4}}~(\bibinfo {number} {1}),\ \bibinfo {pages}
  {41}}\BibitemShut {NoStop}%
\bibitem [{\citenamefont {Ghosh}\ \emph {et~al.}(2008)\citenamefont {Ghosh},
  \citenamefont {Rho}, \citenamefont {McIntosh}, \citenamefont {K{\"o}tter},\
  and\ \citenamefont {Jirsa}}]{Ghosh}%
  \BibitemOpen
  \bibfield  {author} {\bibinfo {author} {\bibnamefont {Ghosh}, \bibfnamefont
  {A.}}, \bibinfo {author} {\bibfnamefont {Y.}~\bibnamefont {Rho}}, \bibinfo
  {author} {\bibfnamefont {A.~R.}\ \bibnamefont {McIntosh}}, \bibinfo {author}
  {\bibfnamefont {R.}~\bibnamefont {K{\"o}tter}}, \ and\ \bibinfo {author}
  {\bibfnamefont {V.~K.}\ \bibnamefont {Jirsa}}} (\bibinfo {year} {2008}),\
  \href@noop {} {\bibfield  {journal} {\bibinfo  {journal} {PLoS Comput.
  Biol.}\ }\textbf {\bibinfo {volume} {4}}~(\bibinfo {number} {10}),\ \bibinfo
  {pages} {e1000196}}\BibitemShut {NoStop}%
\bibitem [{\citenamefont {Ginelli}(2016)}]{Ginelli2016}%
  \BibitemOpen
  \bibfield  {author} {\bibinfo {author} {\bibnamefont {Ginelli}, \bibfnamefont
  {F.}}} (\bibinfo {year} {2016}),\ \href@noop {} {\bibfield  {journal}
  {\bibinfo  {journal} {Eur. Phys. Jour. Special Topics}\ }\textbf {\bibinfo
  {volume} {225}}~(\bibinfo {number} {11-12}),\ \bibinfo {pages}
  {2099}}\BibitemShut {NoStop}%
\bibitem [{\citenamefont {Ginelli}\ \emph {et~al.}(2015)\citenamefont
  {Ginelli}, \citenamefont {Peruani}, \citenamefont {Pillot}, \citenamefont
  {Chat{\'e}}, \citenamefont {Theraulaz},\ and\ \citenamefont {Bon}}]{Sheep}%
  \BibitemOpen
  \bibfield  {author} {\bibinfo {author} {\bibnamefont {Ginelli}, \bibfnamefont
  {F.}}, \bibinfo {author} {\bibfnamefont {F.}~\bibnamefont {Peruani}},
  \bibinfo {author} {\bibfnamefont {M.-H.}\ \bibnamefont {Pillot}}, \bibinfo
  {author} {\bibfnamefont {H.}~\bibnamefont {Chat{\'e}}}, \bibinfo {author}
  {\bibfnamefont {G.}~\bibnamefont {Theraulaz}}, \ and\ \bibinfo {author}
  {\bibfnamefont {R.}~\bibnamefont {Bon}}} (\bibinfo {year} {2015}),\
  \href@noop {} {\bibfield  {journal} {\bibinfo  {journal} {Proc. Natl. Acad.
  Sci. USA.}\ }\textbf {\bibinfo {volume} {112}}~(\bibinfo {number} {41}),\
  \bibinfo {pages} {12729}}\BibitemShut {NoStop}%
\bibitem [{\citenamefont {Giometto}\ \emph {et~al.}(2015)\citenamefont
  {Giometto}, \citenamefont {Formentin}, \citenamefont {Rinaldo}, \citenamefont
  {Cohen},\ and\ \citenamefont {Maritan}}]{Cohen+Amos}%
  \BibitemOpen
  \bibfield  {author} {\bibinfo {author} {\bibnamefont {Giometto},
  \bibfnamefont {A.}}, \bibinfo {author} {\bibfnamefont {M.}~\bibnamefont
  {Formentin}}, \bibinfo {author} {\bibfnamefont {A.}~\bibnamefont {Rinaldo}},
  \bibinfo {author} {\bibfnamefont {J.~E.}\ \bibnamefont {Cohen}}, \ and\
  \bibinfo {author} {\bibfnamefont {A.}~\bibnamefont {Maritan}}} (\bibinfo
  {year} {2015}),\ \href {\doibase 10.1073/pnas.1505882112} {\bibfield
  {journal} {\bibinfo  {journal} {Proc. Natl. Acad. Sci. USA}\ }\textbf
  {\bibinfo {volume} {112}}~(\bibinfo {number} {25}),\ \bibinfo {pages}
  {7755}}\BibitemShut {NoStop}%
\bibitem [{\citenamefont {Gireesh}\ and\ \citenamefont
  {Plenz}(2008)}]{Gireesh}%
  \BibitemOpen
  \bibfield  {author} {\bibinfo {author} {\bibnamefont {Gireesh}, \bibfnamefont
  {E.~D.}}, \ and\ \bibinfo {author} {\bibfnamefont {D.}~\bibnamefont {Plenz}}}
  (\bibinfo {year} {2008}),\ \href {\doibase 10.1073/pnas.0800537105}
  {\bibfield  {journal} {\bibinfo  {journal} {Proc. Natl. Acad. Sci. USA}\
  }\textbf {\bibinfo {volume} {105}}~(\bibinfo {number} {21}),\ \bibinfo
  {pages} {7576}}\BibitemShut {NoStop}%
\bibitem [{\citenamefont {Gisiger}(2001)}]{Gisiger}%
  \BibitemOpen
  \bibfield  {author} {\bibinfo {author} {\bibnamefont {Gisiger}, \bibfnamefont
  {T.}}} (\bibinfo {year} {2001}),\ \href@noop {} {\bibfield  {journal}
  {\bibinfo  {journal} {Biol. Rev Cambridge Philosophical Soc.}\ }\textbf
  {\bibinfo {volume} {76}}~(\bibinfo {number} {02}),\ \bibinfo {pages}
  {161}}\BibitemShut {NoStop}%
\bibitem [{\citenamefont {Gold}(1948)}]{Gold}%
  \BibitemOpen
  \bibfield  {author} {\bibinfo {author} {\bibnamefont {Gold}, \bibfnamefont
  {T.}}} (\bibinfo {year} {1948}),\ \href@noop {} {\bibfield  {journal}
  {\bibinfo  {journal} {Proc. R. Soc. Ser. B}\ }\textbf {\bibinfo {volume}
  {135}}~(\bibinfo {number} {881}),\ \bibinfo {pages} {492}}\BibitemShut
  {NoStop}%
\bibitem [{\citenamefont {Goldberg}\ and\ \citenamefont {Holland}(1988)}]{GA}%
  \BibitemOpen
  \bibfield  {author} {\bibinfo {author} {\bibnamefont {Goldberg},
  \bibfnamefont {D.~E.}}, \ and\ \bibinfo {author} {\bibfnamefont {J.~H.}\
  \bibnamefont {Holland}}} (\bibinfo {year} {1988}),\ \href@noop {} {\bibfield
  {journal} {\bibinfo  {journal} {Machine learning}\ }\textbf {\bibinfo
  {volume} {3}}~(\bibinfo {number} {2}),\ \bibinfo {pages} {95}}\BibitemShut
  {NoStop}%
\bibitem [{\citenamefont {Goldberger}(1992)}]{heart}%
  \BibitemOpen
  \bibfield  {author} {\bibinfo {author} {\bibnamefont {Goldberger},
  \bibfnamefont {A.~L.}}} (\bibinfo {year} {1992}),\ \href@noop {} {\bibfield
  {journal} {\bibinfo  {journal} {IEEE Engineering in Medicine and Biology
  Magazine}\ }\textbf {\bibinfo {volume} {11}}~(\bibinfo {number} {2}),\
  \bibinfo {pages} {47}}\BibitemShut {NoStop}%
\bibitem [{\citenamefont {Goldberger}\ \emph {et~al.}(2002)\citenamefont
  {Goldberger}, \citenamefont {Amaral}, \citenamefont {Hausdorff},
  \citenamefont {Ivanov}, \citenamefont {Peng},\ and\ \citenamefont
  {Stanley}}]{Stanley2002}%
  \BibitemOpen
  \bibfield  {author} {\bibinfo {author} {\bibnamefont {Goldberger},
  \bibfnamefont {A.~L.}}, \bibinfo {author} {\bibfnamefont {L.~A.}\
  \bibnamefont {Amaral}}, \bibinfo {author} {\bibfnamefont {J.~M.}\
  \bibnamefont {Hausdorff}}, \bibinfo {author} {\bibfnamefont {P.~C.}\
  \bibnamefont {Ivanov}}, \bibinfo {author} {\bibfnamefont {C.-K.}\
  \bibnamefont {Peng}}, \ and\ \bibinfo {author} {\bibfnamefont {H.~E.}\
  \bibnamefont {Stanley}}} (\bibinfo {year} {2002}),\ \href@noop {} {\bibfield
  {journal} {\bibinfo  {journal} {Proc. Natl. Acad. Sci. USA.}\ }\textbf
  {\bibinfo {volume} {99}}~(\bibinfo {number} {suppl 1}),\ \bibinfo {pages}
  {2466}}\BibitemShut {NoStop}%
\bibitem [{\citenamefont {Goldenfeld}\ and\ \citenamefont
  {Woese}(2011)}]{Goldenfeld}%
  \BibitemOpen
  \bibfield  {author} {\bibinfo {author} {\bibnamefont {Goldenfeld},
  \bibfnamefont {N.}}, \ and\ \bibinfo {author} {\bibfnamefont
  {C.}~\bibnamefont {Woese}}} (\bibinfo {year} {2011}),\ \href@noop {}
  {\bibfield  {journal} {\bibinfo  {journal} {Annu. Rev. Condens. Matter
  Phys.}\ }\textbf {\bibinfo {volume} {2}}~(\bibinfo {number} {1}),\ \bibinfo
  {pages} {375}}\BibitemShut {NoStop}%
\bibitem [{\citenamefont {Gollo}\ \emph {et~al.}(2013)\citenamefont {Gollo},
  \citenamefont {Kinouchi},\ and\ \citenamefont {Copelli}}]{single-Gollo}%
  \BibitemOpen
  \bibfield  {author} {\bibinfo {author} {\bibnamefont {Gollo}, \bibfnamefont
  {L.}}, \bibinfo {author} {\bibfnamefont {O.}~\bibnamefont {Kinouchi}}, \ and\
  \bibinfo {author} {\bibfnamefont {M.}~\bibnamefont {Copelli}}} (\bibinfo
  {year} {2013}),\ \href@noop {} {\bibfield  {journal} {\bibinfo  {journal}
  {Sci.Rep.}\ }\textbf {\bibinfo {volume} {3}}}\BibitemShut {NoStop}%
\bibitem [{\citenamefont {Gomez}\ \emph {et~al.}(2015)\citenamefont {Gomez},
  \citenamefont {Lorimer},\ and\ \citenamefont {Stoop}}]{Stoop2015}%
  \BibitemOpen
  \bibfield  {author} {\bibinfo {author} {\bibnamefont {Gomez}, \bibfnamefont
  {F.}}, \bibinfo {author} {\bibfnamefont {T.}~\bibnamefont {Lorimer}}, \ and\
  \bibinfo {author} {\bibfnamefont {R.}~\bibnamefont {Stoop}}} (\bibinfo {year}
  {2015}),\ \href@noop {} {\bibinfo  {journal} {arXiv:1510.03241}\
  }\BibitemShut {NoStop}%
\bibitem [{\citenamefont {G{\'o}mez}\ \emph {et~al.}(2008)\citenamefont
  {G{\'o}mez}, \citenamefont {Kaltenbrunner}, \citenamefont {L{\'o}pez},\ and\
  \citenamefont {Kappen}}]{Kappen}%
  \BibitemOpen
\bibfield  {journal} {  }\bibfield  {author} {\bibinfo {author} {\bibnamefont
  {G{\'o}mez}, \bibfnamefont {V.}}, \bibinfo {author} {\bibfnamefont
  {A.}~\bibnamefont {Kaltenbrunner}}, \bibinfo {author} {\bibfnamefont
  {V.}~\bibnamefont {L{\'o}pez}}, \ and\ \bibinfo {author} {\bibfnamefont
  {H.~J.}\ \bibnamefont {Kappen}}} (\bibinfo {year} {2008}),\ in\ \href@noop {}
  {\emph {\bibinfo {booktitle} {NIPS conference}}},\ pp.\ \bibinfo {pages}
  {513--520}\BibitemShut {NoStop}%
\bibitem [{\citenamefont {Goodell}\ \emph {et~al.}(2015)\citenamefont
  {Goodell}, \citenamefont {Nguyen},\ and\ \citenamefont
  {Shroyer}}]{diversity-stemcell}%
  \BibitemOpen
  \bibfield  {author} {\bibinfo {author} {\bibnamefont {Goodell}, \bibfnamefont
  {M.~A.}}, \bibinfo {author} {\bibfnamefont {H.}~\bibnamefont {Nguyen}}, \
  and\ \bibinfo {author} {\bibfnamefont {N.}~\bibnamefont {Shroyer}}} (\bibinfo
  {year} {2015}),\ \href@noop {} {\bibfield  {journal} {\bibinfo  {journal}
  {Nature reviews. Molecular cell biology}\ }\textbf {\bibinfo {volume}
  {16}}~(\bibinfo {number} {5}),\ \bibinfo {pages} {299}}\BibitemShut {NoStop}%
\bibitem [{\citenamefont {Goudarzi}\ \emph {et~al.}(2012)\citenamefont
  {Goudarzi}, \citenamefont {Teuscher}, \citenamefont {Gulbahce},\ and\
  \citenamefont {Rohlf}}]{Goudarzi}%
  \BibitemOpen
  \bibfield  {author} {\bibinfo {author} {\bibnamefont {Goudarzi},
  \bibfnamefont {A.}}, \bibinfo {author} {\bibfnamefont {C.}~\bibnamefont
  {Teuscher}}, \bibinfo {author} {\bibfnamefont {N.}~\bibnamefont {Gulbahce}},
  \ and\ \bibinfo {author} {\bibfnamefont {T.}~\bibnamefont {Rohlf}}} (\bibinfo
  {year} {2012}),\ \href@noop {} {\bibfield  {journal} {\bibinfo  {journal}
  {Phys. Rev. Lett.}\ }\textbf {\bibinfo {volume} {108}}~(\bibinfo {number}
  {12}),\ \bibinfo {pages} {128702}}\BibitemShut {NoStop}%
\bibitem [{\citenamefont {Grassberger}(1995)}]{damage3}%
  \BibitemOpen
  \bibfield  {author} {\bibinfo {author} {\bibnamefont {Grassberger},
  \bibfnamefont {P.}}} (\bibinfo {year} {1995}),\ \href@noop {} {\bibfield
  {journal} {\bibinfo  {journal} {J. Stat. Phys.}\ }\textbf {\bibinfo {volume}
  {79}}~(\bibinfo {number} {1-2}),\ \bibinfo {pages} {13}}\BibitemShut
  {NoStop}%
\bibitem [{\citenamefont {Gr{\'e}goire}\ and\ \citenamefont
  {Chat{\'e}}(2004)}]{Chate2004}%
  \BibitemOpen
  \bibfield  {author} {\bibinfo {author} {\bibnamefont {Gr{\'e}goire},
  \bibfnamefont {G.}}, \ and\ \bibinfo {author} {\bibfnamefont
  {H.}~\bibnamefont {Chat{\'e}}}} (\bibinfo {year} {2004}),\ \href@noop {}
  {\bibfield  {journal} {\bibinfo  {journal} {Phys. Rev. Lett.}\ }\textbf
  {\bibinfo {volume} {92}}~(\bibinfo {number} {2}),\ \bibinfo {pages}
  {025702}}\BibitemShut {NoStop}%
\bibitem [{\citenamefont {Greicius}\ \emph {et~al.}(2003)\citenamefont
  {Greicius}, \citenamefont {Krasnow}, \citenamefont {Reiss},\ and\
  \citenamefont {Menon}}]{RSN3}%
  \BibitemOpen
  \bibfield  {author} {\bibinfo {author} {\bibnamefont {Greicius},
  \bibfnamefont {M.~D.}}, \bibinfo {author} {\bibfnamefont {B.}~\bibnamefont
  {Krasnow}}, \bibinfo {author} {\bibfnamefont {A.~L.}\ \bibnamefont {Reiss}},
  \ and\ \bibinfo {author} {\bibfnamefont {V.}~\bibnamefont {Menon}}} (\bibinfo
  {year} {2003}),\ \href@noop {} {\bibfield  {journal} {\bibinfo  {journal}
  {Proc. Natl. Acad. Sci. USA.}\ }\textbf {\bibinfo {volume} {100}}~(\bibinfo
  {number} {1}),\ \bibinfo {pages} {253}}\BibitemShut {NoStop}%
\bibitem [{\citenamefont {Grinstein}(1991)}]{GG91}%
  \BibitemOpen
  \bibfield  {author} {\bibinfo {author} {\bibnamefont {Grinstein},
  \bibfnamefont {G.}}} (\bibinfo {year} {1991}),\ \href@noop {} {\bibfield
  {journal} {\bibinfo  {journal} {J. Appl. Phys.}\ }\textbf {\bibinfo {volume}
  {69}}~(\bibinfo {number} {8}),\ \bibinfo {pages} {5441}}\BibitemShut
  {NoStop}%
\bibitem [{\citenamefont {Gros}(2008)}]{Gros-book}%
  \BibitemOpen
  \bibfield  {author} {\bibinfo {author} {\bibnamefont {Gros}, \bibfnamefont
  {C.}}} (\bibinfo {year} {2008}),\ \href@noop {} {\emph {\bibinfo {title}
  {Complex and adaptive dynamical systems: A primer}}}\ (\bibinfo  {publisher}
  {Springer})\BibitemShut {NoStop}%
\bibitem [{\citenamefont {Gross}\ and\ \citenamefont
  {Blasius}(2008)}]{Blasius}%
  \BibitemOpen
  \bibfield  {author} {\bibinfo {author} {\bibnamefont {Gross}, \bibfnamefont
  {T.}}, \ and\ \bibinfo {author} {\bibfnamefont {B.}~\bibnamefont {Blasius}}}
  (\bibinfo {year} {2008}),\ \href@noop {} {\bibfield  {journal} {\bibinfo
  {journal} {J. R. Soc. Interface}\ }\textbf {\bibinfo {volume} {5}}~(\bibinfo
  {number} {20}),\ \bibinfo {pages} {259}}\BibitemShut {NoStop}%
\bibitem [{\citenamefont {Grossberg}(1982)}]{Grossberg1}%
  \BibitemOpen
  \bibfield  {author} {\bibinfo {author} {\bibnamefont {Grossberg},
  \bibfnamefont {S.}}} (\bibinfo {year} {1982}),\ in\ \href@noop {} {\emph
  {\bibinfo {booktitle} {A Portrait of Twenty-five Years}}}\ (\bibinfo
  {publisher} {Springer})\ pp.\ \bibinfo {pages} {257--276}\BibitemShut
  {NoStop}%
\bibitem [{\citenamefont {Hagmann}\ \emph {et~al.}(2008)\citenamefont
  {Hagmann}, \citenamefont {Cammoun}, \citenamefont {Gigandet}, \citenamefont
  {Meuli}, \citenamefont {Honey}, \citenamefont {Wedeen},\ and\ \citenamefont
  {Sporns}}]{Hagmann}%
  \BibitemOpen
  \bibfield  {author} {\bibinfo {author} {\bibnamefont {Hagmann}, \bibfnamefont
  {P.}}, \bibinfo {author} {\bibfnamefont {L.}~\bibnamefont {Cammoun}},
  \bibinfo {author} {\bibfnamefont {X.}~\bibnamefont {Gigandet}}, \bibinfo
  {author} {\bibfnamefont {R.}~\bibnamefont {Meuli}}, \bibinfo {author}
  {\bibfnamefont {C.~J.}\ \bibnamefont {Honey}}, \bibinfo {author}
  {\bibfnamefont {V.~J.}\ \bibnamefont {Wedeen}}, \ and\ \bibinfo {author}
  {\bibfnamefont {O.}~\bibnamefont {Sporns}}} (\bibinfo {year} {2008}),\ \href
  {\doibase 10.1371/journal.pbio.0060159} {\bibfield  {journal} {\bibinfo
  {journal} {PLoS Biol}\ }\textbf {\bibinfo {volume} {6}}~(\bibinfo {number}
  {7}),\ \bibinfo {pages} {e159}}\BibitemShut {NoStop}%
\bibitem [{\citenamefont {Hahn}\ \emph {et~al.}(2010)\citenamefont {Hahn},
  \citenamefont {Petermann}, \citenamefont {Havenith}, \citenamefont {Yu},
  \citenamefont {Singer}, \citenamefont {Plenz},\ and\ \citenamefont
  {Nikoli{\'c}}}]{Hahn2010}%
  \BibitemOpen
  \bibfield  {author} {\bibinfo {author} {\bibnamefont {Hahn}, \bibfnamefont
  {G.}}, \bibinfo {author} {\bibfnamefont {T.}~\bibnamefont {Petermann}},
  \bibinfo {author} {\bibfnamefont {M.~N.}\ \bibnamefont {Havenith}}, \bibinfo
  {author} {\bibfnamefont {S.}~\bibnamefont {Yu}}, \bibinfo {author}
  {\bibfnamefont {W.}~\bibnamefont {Singer}}, \bibinfo {author} {\bibfnamefont
  {D.}~\bibnamefont {Plenz}}, \ and\ \bibinfo {author} {\bibfnamefont
  {D.}~\bibnamefont {Nikoli{\'c}}}} (\bibinfo {year} {2010}),\ \href@noop {}
  {\bibfield  {journal} {\bibinfo  {journal} {J. Neurophysiol.}\ }\textbf
  {\bibinfo {volume} {104}}~(\bibinfo {number} {6}),\ \bibinfo {pages}
  {3312}}\BibitemShut {NoStop}%
\bibitem [{\citenamefont {Hahn}\ \emph {et~al.}(2017)\citenamefont {Hahn},
  \citenamefont {Ponce-Alvarez}, \citenamefont {Monier}, \citenamefont
  {Benvenuti}, \citenamefont {Kumar}, \citenamefont {Chavane}, \citenamefont
  {Deco},\ and\ \citenamefont {Fr{\'e}gnac}}]{Deco2017}%
  \BibitemOpen
  \bibfield  {author} {\bibinfo {author} {\bibnamefont {Hahn}, \bibfnamefont
  {G.}}, \bibinfo {author} {\bibfnamefont {A.}~\bibnamefont {Ponce-Alvarez}},
  \bibinfo {author} {\bibfnamefont {C.}~\bibnamefont {Monier}}, \bibinfo
  {author} {\bibfnamefont {G.}~\bibnamefont {Benvenuti}}, \bibinfo {author}
  {\bibfnamefont {A.}~\bibnamefont {Kumar}}, \bibinfo {author} {\bibfnamefont
  {F.}~\bibnamefont {Chavane}}, \bibinfo {author} {\bibfnamefont
  {G.}~\bibnamefont {Deco}}, \ and\ \bibinfo {author} {\bibfnamefont
  {Y.}~\bibnamefont {Fr{\'e}gnac}}} (\bibinfo {year} {2017}),\ \href@noop {}
  {\bibfield  {journal} {\bibinfo  {journal} {PLoS Comput. Biol.}\ }\textbf
  {\bibinfo {volume} {13}}~(\bibinfo {number} {5}),\ \bibinfo {pages}
  {e1005543}}\BibitemShut {NoStop}%
\bibitem [{\citenamefont {Haimovici}\ and\ \citenamefont
  {Marsili}(2015)}]{Marsili3}%
  \BibitemOpen
  \bibfield  {author} {\bibinfo {author} {\bibnamefont {Haimovici},
  \bibfnamefont {A.}}, \ and\ \bibinfo {author} {\bibfnamefont
  {M.}~\bibnamefont {Marsili}}} (\bibinfo {year} {2015}),\ \href@noop {}
  {\bibfield  {journal} {\bibinfo  {journal} {J. Stat. Mech.}\ }\textbf
  {\bibinfo {volume} {2015}}~(\bibinfo {number} {10}),\ \bibinfo {pages}
  {P10013}}\BibitemShut {NoStop}%
\bibitem [{\citenamefont {Haimovici}\ \emph {et~al.}(2013)\citenamefont
  {Haimovici}, \citenamefont {Tagliazucchi}, \citenamefont {Balenzuela},\ and\
  \citenamefont {Chialvo}}]{Haimovici}%
  \BibitemOpen
  \bibfield  {author} {\bibinfo {author} {\bibnamefont {Haimovici},
  \bibfnamefont {A.}}, \bibinfo {author} {\bibfnamefont {E.}~\bibnamefont
  {Tagliazucchi}}, \bibinfo {author} {\bibfnamefont {P.}~\bibnamefont
  {Balenzuela}}, \ and\ \bibinfo {author} {\bibfnamefont {D.~R.}\ \bibnamefont
  {Chialvo}}} (\bibinfo {year} {2013}),\ \href@noop {} {\bibfield  {journal}
  {\bibinfo  {journal} {Phys. Rev. Lett}\ }\textbf {\bibinfo {volume} {110}},\
  \bibinfo {pages} {178101}}\BibitemShut {NoStop}%
\bibitem [{\citenamefont {Haken}(1977)}]{Haken1977}%
  \BibitemOpen
  \bibfield  {author} {\bibinfo {author} {\bibnamefont {Haken}, \bibfnamefont
  {H.}}} (\bibinfo {year} {1977}),\ \href@noop {} {\bibfield  {journal}
  {\bibinfo  {journal} {Physics Bulletin}\ }\textbf {\bibinfo {volume}
  {28}}~(\bibinfo {number} {9}),\ \bibinfo {pages} {412}}\BibitemShut {NoStop}%
\bibitem [{\citenamefont {Haken}(2013)}]{Haken2013}%
  \BibitemOpen
  \bibfield  {author} {\bibinfo {author} {\bibnamefont {Haken}, \bibfnamefont
  {H.}}} (\bibinfo {year} {2013}),\ \href@noop {} {\emph {\bibinfo {title}
  {Principles of brain functioning: a synergetic approach to brain activity,
  behavior and cognition}}},\ Vol.~\bibinfo {volume} {67}\ (\bibinfo
  {publisher} {Springer Science \& Business Media})\BibitemShut {NoStop}%
\bibitem [{\citenamefont {Haken}\ \emph {et~al.}(1985)\citenamefont {Haken},
  \citenamefont {Kelso},\ and\ \citenamefont {Bunz}}]{Kelso1985}%
  \BibitemOpen
  \bibfield  {author} {\bibinfo {author} {\bibnamefont {Haken}, \bibfnamefont
  {H.}}, \bibinfo {author} {\bibfnamefont {J.~S.}\ \bibnamefont {Kelso}}, \
  and\ \bibinfo {author} {\bibfnamefont {H.}~\bibnamefont {Bunz}}} (\bibinfo
  {year} {1985}),\ \href@noop {} {\bibfield  {journal} {\bibinfo  {journal}
  {Biological cybernetics}\ }\textbf {\bibinfo {volume} {51}}~(\bibinfo
  {number} {5}),\ \bibinfo {pages} {347}}\BibitemShut {NoStop}%
\bibitem [{\citenamefont {Haldeman}\ and\ \citenamefont
  {Beggs}(2005)}]{Haldeman}%
  \BibitemOpen
  \bibfield  {author} {\bibinfo {author} {\bibnamefont {Haldeman},
  \bibfnamefont {C.}}, \ and\ \bibinfo {author} {\bibfnamefont {J.~M.}\
  \bibnamefont {Beggs}}} (\bibinfo {year} {2005}),\ \href@noop {} {\bibfield
  {journal} {\bibinfo  {journal} {Phys. Rev. Lett.}\ }\textbf {\bibinfo
  {volume} {94}}~(\bibinfo {number} {5}),\ \bibinfo {pages}
  {058101}}\BibitemShut {NoStop}%
\bibitem [{\citenamefont {Halley}\ \emph {et~al.}(2009)\citenamefont {Halley},
  \citenamefont {Burden},\ and\ \citenamefont {Winkler}}]{decision}%
  \BibitemOpen
  \bibfield  {author} {\bibinfo {author} {\bibnamefont {Halley}, \bibfnamefont
  {J.~D.}}, \bibinfo {author} {\bibfnamefont {F.~R.}\ \bibnamefont {Burden}}, \
  and\ \bibinfo {author} {\bibfnamefont {D.~A.}\ \bibnamefont {Winkler}}}
  (\bibinfo {year} {2009}),\ \href@noop {} {\bibfield  {journal} {\bibinfo
  {journal} {Stem Cell Research}\ }\textbf {\bibinfo {volume} {2}}~(\bibinfo
  {number} {3}),\ \bibinfo {pages} {165}}\BibitemShut {NoStop}%
\bibitem [{\citenamefont {Halley}\ and\ \citenamefont
  {Winkler}(2008)}]{2facets}%
  \BibitemOpen
  \bibfield  {author} {\bibinfo {author} {\bibnamefont {Halley}, \bibfnamefont
  {J.~D.}}, \ and\ \bibinfo {author} {\bibfnamefont {D.~A.}\ \bibnamefont
  {Winkler}}} (\bibinfo {year} {2008}),\ \href@noop {} {\bibfield  {journal}
  {\bibinfo  {journal} {Biosystems}\ }\textbf {\bibinfo {volume}
  {92}}~(\bibinfo {number} {2}),\ \bibinfo {pages} {148}}\BibitemShut {NoStop}%
\bibitem [{\citenamefont {Hanel}\ \emph {et~al.}(2010)\citenamefont {Hanel},
  \citenamefont {P{\"o}chacker},\ and\ \citenamefont {Thurner}}]{Thurner2}%
  \BibitemOpen
  \bibfield  {author} {\bibinfo {author} {\bibnamefont {Hanel}, \bibfnamefont
  {R.}}, \bibinfo {author} {\bibfnamefont {M.}~\bibnamefont {P{\"o}chacker}}, \
  and\ \bibinfo {author} {\bibfnamefont {S.}~\bibnamefont {Thurner}}} (\bibinfo
  {year} {2010}),\ \href@noop {} {\bibfield  {journal} {\bibinfo  {journal}
  {Philos. Trans. R. Soc. London, Ser. A}\ }\textbf {\bibinfo {volume}
  {368}}~(\bibinfo {number} {1933}),\ \bibinfo {pages} {5583}}\BibitemShut
  {NoStop}%
\bibitem [{\citenamefont {Harris}(2002)}]{Harris}%
  \BibitemOpen
  \bibfield  {author} {\bibinfo {author} {\bibnamefont {Harris}, \bibfnamefont
  {T.~E.}}} (\bibinfo {year} {2002}),\ \href@noop {} {\emph {\bibinfo {title}
  {The theory of branching processes}}}\ (\bibinfo  {publisher} {Courier
  Corporation})\BibitemShut {NoStop}%
\bibitem [{\citenamefont {Hart}\ and\ \citenamefont
  {Ferguson}(2015)}]{Error-HIV}%
  \BibitemOpen
  \bibfield  {author} {\bibinfo {author} {\bibnamefont {Hart}, \bibfnamefont
  {G.~R.}}, \ and\ \bibinfo {author} {\bibfnamefont {A.~L.}\ \bibnamefont
  {Ferguson}}} (\bibinfo {year} {2015}),\ \href@noop {} {\bibfield  {journal}
  {\bibinfo  {journal} {Phys. Rev. E}\ }\textbf {\bibinfo {volume}
  {91}}~(\bibinfo {number} {3}),\ \bibinfo {pages} {032705}}\BibitemShut
  {NoStop}%
\bibitem [{\citenamefont {Hartwell}\ \emph {et~al.}(1999)\citenamefont
  {Hartwell}, \citenamefont {Hopfield}, \citenamefont {Leibler},\ and\
  \citenamefont {Murray}}]{Stan1999}%
  \BibitemOpen
  \bibfield  {author} {\bibinfo {author} {\bibnamefont {Hartwell},
  \bibfnamefont {L.~H.}}, \bibinfo {author} {\bibfnamefont {J.~J.}\
  \bibnamefont {Hopfield}}, \bibinfo {author} {\bibfnamefont {S.}~\bibnamefont
  {Leibler}}, \ and\ \bibinfo {author} {\bibfnamefont {A.~W.}\ \bibnamefont
  {Murray}}} (\bibinfo {year} {1999}),\ \href@noop {} {\bibfield  {journal}
  {\bibinfo  {journal} {Nature}\ }\textbf {\bibinfo {volume} {402}},\ \bibinfo
  {pages} {C47}}\BibitemShut {NoStop}%
\bibitem [{\citenamefont {He}(2011)}]{fMRI-He}%
  \BibitemOpen
  \bibfield  {author} {\bibinfo {author} {\bibnamefont {He}, \bibfnamefont
  {B.~J.}}} (\bibinfo {year} {2011}),\ \href@noop {} {\bibfield  {journal}
  {\bibinfo  {journal} {J. Neurosci.}\ }\textbf {\bibinfo {volume}
  {31}}~(\bibinfo {number} {39}),\ \bibinfo {pages} {13786}}\BibitemShut
  {NoStop}%
\bibitem [{\citenamefont {He}(2014)}]{He}%
  \BibitemOpen
  \bibfield  {author} {\bibinfo {author} {\bibnamefont {He}, \bibfnamefont
  {B.~J.}}} (\bibinfo {year} {2014}),\ \href@noop {} {\bibfield  {journal}
  {\bibinfo  {journal} {Trends in cognitive sciences}\ }\textbf {\bibinfo
  {volume} {18}}~(\bibinfo {number} {9}),\ \bibinfo {pages} {480}}\BibitemShut
  {NoStop}%
\bibitem [{\citenamefont {Henkel}\ \emph {et~al.}(2008)\citenamefont {Henkel},
  \citenamefont {Hinrichsen},\ and\ \citenamefont {L{\"u}beck}}]{Henkel}%
  \BibitemOpen
  \bibfield  {author} {\bibinfo {author} {\bibnamefont {Henkel}, \bibfnamefont
  {M.}}, \bibinfo {author} {\bibfnamefont {H.}~\bibnamefont {Hinrichsen}}, \
  and\ \bibinfo {author} {\bibfnamefont {S.}~\bibnamefont {L{\"u}beck}}}
  (\bibinfo {year} {2008}),\ \href@noop {} {\emph {\bibinfo {title}
  {Non-equilibrium Phase Transitions: Absorbing phase transitions}}},\ Theor.
  and Math. Phys.\ (\bibinfo  {publisher} {Springer London},\ \bibinfo
  {address} {Berlin})\BibitemShut {NoStop}%
\bibitem [{\citenamefont {Hernandez-Urbina}\ and\ \citenamefont
  {Herrmann}(2017)}]{Herrmann2017}%
  \BibitemOpen
  \bibfield  {author} {\bibinfo {author} {\bibnamefont {Hernandez-Urbina},
  \bibfnamefont {V.}}, \ and\ \bibinfo {author} {\bibfnamefont {J.~M.}\
  \bibnamefont {Herrmann}}} (\bibinfo {year} {2017}),\ \href {\doibase
  10.3389/fphy.2016.00054} {\bibfield  {journal} {\bibinfo  {journal} {Front.
  Phys.}\ }\textbf {\bibinfo {volume} {4}},\ \bibinfo {pages} {54}}\BibitemShut
  {NoStop}%
\bibitem [{\citenamefont {Hertz}\ \emph {et~al.}(1991)\citenamefont {Hertz},
  \citenamefont {Krogh},\ and\ \citenamefont {Palmer}}]{Hertz-book}%
  \BibitemOpen
  \bibfield  {author} {\bibinfo {author} {\bibnamefont {Hertz}, \bibfnamefont
  {J.}}, \bibinfo {author} {\bibfnamefont {A.}~\bibnamefont {Krogh}}, \ and\
  \bibinfo {author} {\bibfnamefont {R.~G.}\ \bibnamefont {Palmer}}} (\bibinfo
  {year} {1991}),\ \href@noop {} {\emph {\bibinfo {title} {Introduction to the
  theory of neural computation}}},\ Vol.~\bibinfo {volume} {1}\ (\bibinfo
  {publisher} {Basic Books})\BibitemShut {NoStop}%
\bibitem [{\citenamefont {Herz}\ and\ \citenamefont {Hopfield}(1995)}]{Herz}%
  \BibitemOpen
  \bibfield  {author} {\bibinfo {author} {\bibnamefont {Herz}, \bibfnamefont
  {A.~V.}}, \ and\ \bibinfo {author} {\bibfnamefont {J.~J.}\ \bibnamefont
  {Hopfield}}} (\bibinfo {year} {1995}),\ \href@noop {} {\bibfield  {journal}
  {\bibinfo  {journal} {Phys. Rev. Lett.}\ }\textbf {\bibinfo {volume}
  {75}}~(\bibinfo {number} {6}),\ \bibinfo {pages} {1222}}\BibitemShut
  {NoStop}%
\bibitem [{\citenamefont {Hesse}\ and\ \citenamefont
  {Gross}(2014)}]{Gross2014}%
  \BibitemOpen
  \bibfield  {author} {\bibinfo {author} {\bibnamefont {Hesse}, \bibfnamefont
  {J.}}, \ and\ \bibinfo {author} {\bibfnamefont {T.}~\bibnamefont {Gross}}}
  (\bibinfo {year} {2014}),\ \href@noop {} {\bibfield  {journal} {\bibinfo
  {journal} {Front. Comput. Neurosci.}\ }\textbf {\bibinfo {volume}
  {8}}}\BibitemShut {NoStop}%
\bibitem [{\citenamefont {Hidalgo}\ \emph {et~al.}(2016)\citenamefont
  {Hidalgo}, \citenamefont {Grilli}, \citenamefont {Suweis}, \citenamefont
  {Maritan},\ and\ \citenamefont {Mu{\~n}oz}}]{Hidalgo2016}%
  \BibitemOpen
  \bibfield  {author} {\bibinfo {author} {\bibnamefont {Hidalgo}, \bibfnamefont
  {J.}}, \bibinfo {author} {\bibfnamefont {J.}~\bibnamefont {Grilli}}, \bibinfo
  {author} {\bibfnamefont {S.}~\bibnamefont {Suweis}}, \bibinfo {author}
  {\bibfnamefont {A.}~\bibnamefont {Maritan}}, \ and\ \bibinfo {author}
  {\bibfnamefont {M.~A.}\ \bibnamefont {Mu{\~n}oz}}} (\bibinfo {year} {2016}),\
  \href@noop {} {\bibfield  {journal} {\bibinfo  {journal} {J. of Stat. Mech.}\
  }\textbf {\bibinfo {volume} {2016}}~(\bibinfo {number} {3}),\ \bibinfo
  {pages} {033203}}\BibitemShut {NoStop}%
\bibitem [{\citenamefont {Hidalgo}\ \emph {et~al.}(2014)\citenamefont
  {Hidalgo}, \citenamefont {Grilli}, \citenamefont {Suweis}, \citenamefont
  {Mu{\~n}oz}, \citenamefont {Ba~navar},\ and\ \citenamefont
  {Maritan}}]{Hidalgo2014}%
  \BibitemOpen
  \bibfield  {author} {\bibinfo {author} {\bibnamefont {Hidalgo}, \bibfnamefont
  {J.}}, \bibinfo {author} {\bibfnamefont {J.}~\bibnamefont {Grilli}}, \bibinfo
  {author} {\bibfnamefont {S.}~\bibnamefont {Suweis}}, \bibinfo {author}
  {\bibfnamefont {M.~A.}\ \bibnamefont {Mu{\~n}oz}}, \bibinfo {author}
  {\bibfnamefont {J.~R.}\ \bibnamefont {Ba~navar}}, \ and\ \bibinfo {author}
  {\bibfnamefont {A.}~\bibnamefont {Maritan}}} (\bibinfo {year} {2014}),\
  \href@noop {} {\bibfield  {journal} {\bibinfo  {journal} {Proc. Natl. Acad.
  Sci. USA.}\ }\textbf {\bibinfo {volume} {111}}~(\bibinfo {number} {28}),\
  \bibinfo {pages} {10095}}\BibitemShut {NoStop}%
\bibitem [{\citenamefont {Hidalgo}\ \emph {et~al.}(2012)\citenamefont
  {Hidalgo}, \citenamefont {Seoane}, \citenamefont {Cort\'es},\ and\
  \citenamefont {Mu{\~n}oz}}]{Hidalgo2012}%
  \BibitemOpen
  \bibfield  {author} {\bibinfo {author} {\bibnamefont {Hidalgo}, \bibfnamefont
  {J.}}, \bibinfo {author} {\bibfnamefont {L.}~\bibnamefont {Seoane}}, \bibinfo
  {author} {\bibfnamefont {J.}~\bibnamefont {Cort\'es}}, \ and\ \bibinfo
  {author} {\bibfnamefont {M.}~\bibnamefont {Mu{\~n}oz}}} (\bibinfo {year}
  {2012}),\ \href@noop {} {\bibfield  {journal} {\bibinfo  {journal} {PLoS
  One}\ }\textbf {\bibinfo {volume} {7(8)}},\ \bibinfo {pages}
  {e40710}}\BibitemShut {NoStop}%
\bibitem [{\citenamefont {Hinrichsen}(2000)}]{Hinrichsen}%
  \BibitemOpen
  \bibfield  {author} {\bibinfo {author} {\bibnamefont {Hinrichsen},
  \bibfnamefont {H.}}} (\bibinfo {year} {2000}),\ \href@noop {} {\bibfield
  {journal} {\bibinfo  {journal} {Adv. in Phys.}\ }\textbf {\bibinfo {volume}
  {49}},\ \bibinfo {pages} {815}}\BibitemShut {NoStop}%
\bibitem [{\citenamefont {Hinrichsen}\ and\ \citenamefont
  {Domany}(1997)}]{damage4}%
  \BibitemOpen
  \bibfield  {author} {\bibinfo {author} {\bibnamefont {Hinrichsen},
  \bibfnamefont {H.}}, \ and\ \bibinfo {author} {\bibfnamefont
  {E.}~\bibnamefont {Domany}}} (\bibinfo {year} {1997}),\ \href@noop {}
  {\bibfield  {journal} {\bibinfo  {journal} {Phys. Rev. E}\ }\textbf {\bibinfo
  {volume} {56}}~(\bibinfo {number} {1}),\ \bibinfo {pages} {94}}\BibitemShut
  {NoStop}%
\bibitem [{\citenamefont {Hobbs}\ \emph {et~al.}(2010)\citenamefont {Hobbs},
  \citenamefont {Smith},\ and\ \citenamefont {Beggs}}]{epilepsy}%
  \BibitemOpen
  \bibfield  {author} {\bibinfo {author} {\bibnamefont {Hobbs}, \bibfnamefont
  {J.~P.}}, \bibinfo {author} {\bibfnamefont {J.~L.}\ \bibnamefont {Smith}}, \
  and\ \bibinfo {author} {\bibfnamefont {J.~M.}\ \bibnamefont {Beggs}}}
  (\bibinfo {year} {2010}),\ \href@noop {} {\bibfield  {journal} {\bibinfo
  {journal} {J. of Clin. Neurophys.}\ }\textbf {\bibinfo {volume}
  {27}}~(\bibinfo {number} {6}),\ \bibinfo {pages} {380}}\BibitemShut {NoStop}%
\bibitem [{\citenamefont {Hohenberg}\ and\ \citenamefont
  {Halperin}(1977)}]{HH}%
  \BibitemOpen
  \bibfield  {author} {\bibinfo {author} {\bibnamefont {Hohenberg},
  \bibfnamefont {P.~C.}}, \ and\ \bibinfo {author} {\bibfnamefont {B.~I.}\
  \bibnamefont {Halperin}}} (\bibinfo {year} {1977}),\ \href@noop {} {\bibfield
   {journal} {\bibinfo  {journal} {Rev. Mod. Phys.}\ }\textbf {\bibinfo
  {volume} {49}}~(\bibinfo {number} {3}),\ \bibinfo {pages} {435}}\BibitemShut
  {NoStop}%
\bibitem [{\citenamefont {Holcman}\ and\ \citenamefont
  {Tsodyks}(2006)}]{Holcman}%
  \BibitemOpen
  \bibfield  {author} {\bibinfo {author} {\bibnamefont {Holcman}, \bibfnamefont
  {D.}}, \ and\ \bibinfo {author} {\bibfnamefont {M.}~\bibnamefont {Tsodyks}}}
  (\bibinfo {year} {2006}),\ \href@noop {} {\bibfield  {journal} {\bibinfo
  {journal} {PLoS Comput. Biol.}\ }\textbf {\bibinfo {volume} {2}}~(\bibinfo
  {number} {3}),\ \bibinfo {pages} {e23}}\BibitemShut {NoStop}%
\bibitem [{\citenamefont {Honerkamp-Smith}\ \emph {et~al.}(2008)\citenamefont
  {Honerkamp-Smith}, \citenamefont {Cicuta}, \citenamefont {Collins},
  \citenamefont {Veatch}, \citenamefont {den Nijs}, \citenamefont {Schick},\
  and\ \citenamefont {Keller}}]{Cicuta3}%
  \BibitemOpen
  \bibfield  {author} {\bibinfo {author} {\bibnamefont {Honerkamp-Smith},
  \bibfnamefont {A.~R.}}, \bibinfo {author} {\bibfnamefont {P.}~\bibnamefont
  {Cicuta}}, \bibinfo {author} {\bibfnamefont {M.~D.}\ \bibnamefont {Collins}},
  \bibinfo {author} {\bibfnamefont {S.~L.}\ \bibnamefont {Veatch}}, \bibinfo
  {author} {\bibfnamefont {M.}~\bibnamefont {den Nijs}}, \bibinfo {author}
  {\bibfnamefont {M.}~\bibnamefont {Schick}}, \ and\ \bibinfo {author}
  {\bibfnamefont {S.~L.}\ \bibnamefont {Keller}}} (\bibinfo {year} {2008}),\
  \href@noop {} {\bibfield  {journal} {\bibinfo  {journal} {Biophysical
  journal}\ }\textbf {\bibinfo {volume} {95}}~(\bibinfo {number} {1}),\
  \bibinfo {pages} {236}}\BibitemShut {NoStop}%
\bibitem [{\citenamefont {Hopfield}(1994)}]{Hopfield1994}%
  \BibitemOpen
  \bibfield  {author} {\bibinfo {author} {\bibnamefont {Hopfield},
  \bibfnamefont {J.}}} (\bibinfo {year} {1994}),\ \href@noop {} {\bibfield
  {journal} {\bibinfo  {journal} {J. Theor. Biol.}\ }\textbf {\bibinfo {volume}
  {171}}~(\bibinfo {number} {1}),\ \bibinfo {pages} {53}}\BibitemShut {NoStop}%
\bibitem [{\citenamefont {Hopfield}(1982)}]{Hopfield1982}%
  \BibitemOpen
  \bibfield  {author} {\bibinfo {author} {\bibnamefont {Hopfield},
  \bibfnamefont {J.~J.}}} (\bibinfo {year} {1982}),\ \href@noop {} {\bibfield
  {journal} {\bibinfo  {journal} {Proc. Natl. Acad. Sci. USA.}\ }\textbf
  {\bibinfo {volume} {79}}~(\bibinfo {number} {8}),\ \bibinfo {pages}
  {2554}}\BibitemShut {NoStop}%
\bibitem [{\citenamefont {Hu}\ \emph {et~al.}(2012)\citenamefont {Hu},
  \citenamefont {Meijer}, \citenamefont {Shea}, \citenamefont
  {Tjebbe~vanderLeest}, \citenamefont {Pittman-Polletta}, \citenamefont
  {Houben}, \citenamefont {van Oosterhout}, \citenamefont {Deboer},\ and\
  \citenamefont {Scheer}}]{Meijer}%
  \BibitemOpen
  \bibfield  {author} {\bibinfo {author} {\bibnamefont {Hu}, \bibfnamefont
  {K.}}, \bibinfo {author} {\bibfnamefont {J.~H.}\ \bibnamefont {Meijer}},
  \bibinfo {author} {\bibfnamefont {S.~A.}\ \bibnamefont {Shea}}, \bibinfo
  {author} {\bibfnamefont {H.}~\bibnamefont {Tjebbe~vanderLeest}}, \bibinfo
  {author} {\bibfnamefont {B.}~\bibnamefont {Pittman-Polletta}}, \bibinfo
  {author} {\bibfnamefont {T.}~\bibnamefont {Houben}}, \bibinfo {author}
  {\bibfnamefont {F.}~\bibnamefont {van Oosterhout}}, \bibinfo {author}
  {\bibfnamefont {T.}~\bibnamefont {Deboer}}, \ and\ \bibinfo {author}
  {\bibfnamefont {F.~A.}\ \bibnamefont {Scheer}}} (\bibinfo {year} {2012}),\
  \href@noop {} {\bibfield  {journal} {\bibinfo  {journal} {PLoS One}\ }\textbf
  {\bibinfo {volume} {7}}~(\bibinfo {number} {11}),\ \bibinfo {pages}
  {e48927}}\BibitemShut {NoStop}%
\bibitem [{\citenamefont {Huang}\ \emph {et~al.}(2005)\citenamefont {Huang},
  \citenamefont {Eichler}, \citenamefont {Bar-Yam},\ and\ \citenamefont
  {Ingber}}]{Attractors1}%
  \BibitemOpen
  \bibfield  {author} {\bibinfo {author} {\bibnamefont {Huang}, \bibfnamefont
  {S.}}, \bibinfo {author} {\bibfnamefont {G.}~\bibnamefont {Eichler}},
  \bibinfo {author} {\bibfnamefont {Y.}~\bibnamefont {Bar-Yam}}, \ and\
  \bibinfo {author} {\bibfnamefont {D.~E.}\ \bibnamefont {Ingber}}} (\bibinfo
  {year} {2005}),\ \href@noop {} {\bibfield  {journal} {\bibinfo  {journal}
  {Phys. Rev. Lett.}\ }\textbf {\bibinfo {volume} {94}}~(\bibinfo {number}
  {12}),\ \bibinfo {pages} {128701}}\BibitemShut {NoStop}%
\bibitem [{\citenamefont {Hubbell}(2001)}]{Hubbell}%
  \BibitemOpen
  \bibfield  {author} {\bibinfo {author} {\bibnamefont {Hubbell}, \bibfnamefont
  {S.~P.}}} (\bibinfo {year} {2001}),\ \href@noop {} {\emph {\bibinfo {title}
  {The unified neutral theory of biodiversity and biogeography (MPB-32)}}},\
  Vol.~\bibinfo {volume} {32}\ (\bibinfo  {publisher} {Princeton University
  Press})\BibitemShut {NoStop}%
\bibitem [{\citenamefont {Hudspeth}(2014)}]{Hud-Review}%
  \BibitemOpen
  \bibfield  {author} {\bibinfo {author} {\bibnamefont {Hudspeth},
  \bibfnamefont {A.}}} (\bibinfo {year} {2014}),\ \href@noop {} {\bibfield
  {journal} {\bibinfo  {journal} {Nature Rev. Neurosci.}\ }\textbf {\bibinfo
  {volume} {15}}~(\bibinfo {number} {9}),\ \bibinfo {pages} {600}}\BibitemShut
  {NoStop}%
\bibitem [{\citenamefont {Hudspeth}\ \emph {et~al.}(2010)\citenamefont
  {Hudspeth}, \citenamefont {J{\"u}licher},\ and\ \citenamefont
  {Martin}}]{Julicher2010}%
  \BibitemOpen
  \bibfield  {author} {\bibinfo {author} {\bibnamefont {Hudspeth},
  \bibfnamefont {A.}}, \bibinfo {author} {\bibfnamefont {F.}~\bibnamefont
  {J{\"u}licher}}, \ and\ \bibinfo {author} {\bibfnamefont {P.}~\bibnamefont
  {Martin}}} (\bibinfo {year} {2010}),\ \href@noop {} {\bibfield  {journal}
  {\bibinfo  {journal} {Journal of neurophysiology}\ }\textbf {\bibinfo
  {volume} {104}}~(\bibinfo {number} {3}),\ \bibinfo {pages}
  {1219}}\BibitemShut {NoStop}%
\bibitem [{\citenamefont {Hughes}\ \emph {et~al.}(2000)\citenamefont {Hughes},
  \citenamefont {Marton}, \citenamefont {Jones}, \citenamefont {Roberts},
  \citenamefont {Stoughton}, \citenamefont {Armour}, \citenamefont {Bennett},
  \citenamefont {Coffey}, \citenamefont {Dai}, \citenamefont {He} \emph
  {et~al.}}]{Hughes2000}%
  \BibitemOpen
  \bibfield  {author} {\bibinfo {author} {\bibnamefont {Hughes}, \bibfnamefont
  {T.~R.}}, \bibinfo {author} {\bibfnamefont {M.~J.}\ \bibnamefont {Marton}},
  \bibinfo {author} {\bibfnamefont {A.~R.}\ \bibnamefont {Jones}}, \bibinfo
  {author} {\bibfnamefont {C.~J.}\ \bibnamefont {Roberts}}, \bibinfo {author}
  {\bibfnamefont {R.}~\bibnamefont {Stoughton}}, \bibinfo {author}
  {\bibfnamefont {C.~D.}\ \bibnamefont {Armour}}, \bibinfo {author}
  {\bibfnamefont {H.~A.}\ \bibnamefont {Bennett}}, \bibinfo {author}
  {\bibfnamefont {E.}~\bibnamefont {Coffey}}, \bibinfo {author} {\bibfnamefont
  {H.}~\bibnamefont {Dai}}, \bibinfo {author} {\bibfnamefont {Y.~D.}\
  \bibnamefont {He}},  \emph {et~al.}} (\bibinfo {year} {2000}),\ \href@noop {}
  {\bibfield  {journal} {\bibinfo  {journal} {Cell}\ }\textbf {\bibinfo
  {volume} {102}}~(\bibinfo {number} {1}),\ \bibinfo {pages} {109}}\BibitemShut
  {NoStop}%
\bibitem [{\citenamefont {Hyman}\ and\ \citenamefont {Simons}(2012)}]{oil}%
  \BibitemOpen
  \bibfield  {author} {\bibinfo {author} {\bibnamefont {Hyman}, \bibfnamefont
  {A.~A.}}, \ and\ \bibinfo {author} {\bibfnamefont {K.}~\bibnamefont
  {Simons}}} (\bibinfo {year} {2012}),\ \href@noop {} {\bibfield  {journal}
  {\bibinfo  {journal} {Science}\ }\textbf {\bibinfo {volume} {337}}~(\bibinfo
  {number} {6098}),\ \bibinfo {pages} {1047}}\BibitemShut {NoStop}%
\bibitem [{\citenamefont {Iliopoulos}\ \emph {et~al.}(2010)\citenamefont
  {Iliopoulos}, \citenamefont {Hintze},\ and\ \citenamefont
  {Adami}}]{Adami2010}%
  \BibitemOpen
  \bibfield  {author} {\bibinfo {author} {\bibnamefont {Iliopoulos},
  \bibfnamefont {D.}}, \bibinfo {author} {\bibfnamefont {A.}~\bibnamefont
  {Hintze}}, \ and\ \bibinfo {author} {\bibfnamefont {C.}~\bibnamefont
  {Adami}}} (\bibinfo {year} {2010}),\ \href@noop {} {\bibfield  {journal}
  {\bibinfo  {journal} {PLoS Comput. Biol.}\ }\textbf {\bibinfo {volume}
  {6}}~(\bibinfo {number} {10})}\BibitemShut {NoStop}%
\bibitem [{\citenamefont {Ivanov}(2007)}]{Ivanov2007}%
  \BibitemOpen
  \bibfield  {author} {\bibinfo {author} {\bibnamefont {Ivanov}, \bibfnamefont
  {P.}}} (\bibinfo {year} {2007}),\ \href@noop {} {\bibfield  {journal}
  {\bibinfo  {journal} {IEEE Engineering in Med. and Biol. Magazine}\ }\textbf
  {\bibinfo {volume} {26}}~(\bibinfo {number} {6}),\ \bibinfo {pages}
  {33}}\BibitemShut {NoStop}%
\bibitem [{\citenamefont {Ivanov}\ \emph {et~al.}(1999)\citenamefont {Ivanov},
  \citenamefont {Amaral}, \citenamefont {Goldberger}, \citenamefont {Havlin},
  \citenamefont {Rosenblum}, \citenamefont {Struzik},\ and\ \citenamefont
  {Stanley}}]{Ivanov1999}%
  \BibitemOpen
  \bibfield  {author} {\bibinfo {author} {\bibnamefont {Ivanov}, \bibfnamefont
  {P.~C.}}, \bibinfo {author} {\bibfnamefont {L.~A.~N.}\ \bibnamefont
  {Amaral}}, \bibinfo {author} {\bibfnamefont {A.~L.}\ \bibnamefont
  {Goldberger}}, \bibinfo {author} {\bibfnamefont {S.}~\bibnamefont {Havlin}},
  \bibinfo {author} {\bibfnamefont {M.~G.}\ \bibnamefont {Rosenblum}}, \bibinfo
  {author} {\bibfnamefont {Z.~R.}\ \bibnamefont {Struzik}}, \ and\ \bibinfo
  {author} {\bibfnamefont {H.~E.}\ \bibnamefont {Stanley}}} (\bibinfo {year}
  {1999}),\ \href@noop {} {\bibfield  {journal} {\bibinfo  {journal} {Nature}\
  }\textbf {\bibinfo {volume} {399}}~(\bibinfo {number} {6735}),\ \bibinfo
  {pages} {461}}\BibitemShut {NoStop}%
\bibitem [{\citenamefont {Izhikevich}(2004)}]{Izhi2}%
  \BibitemOpen
  \bibfield  {author} {\bibinfo {author} {\bibnamefont {Izhikevich},
  \bibfnamefont {E.~M.}}} (\bibinfo {year} {2004}),\ \href@noop {} {\bibfield
  {journal} {\bibinfo  {journal} {IEEE transactions on Neural Networks}\
  }\textbf {\bibinfo {volume} {15}}~(\bibinfo {number} {5}),\ \bibinfo {pages}
  {1063}}\BibitemShut {NoStop}%
\bibitem [{\citenamefont {Izhikevich}(2007)}]{Izhi1}%
  \BibitemOpen
  \bibfield  {author} {\bibinfo {author} {\bibnamefont {Izhikevich},
  \bibfnamefont {E.~M.}}} (\bibinfo {year} {2007}),\ \href@noop {} {\emph
  {\bibinfo {title} {Dynamical Systems in Neuroscience: The Geometry of
  Excitability and Bursting}}}\ (\bibinfo  {publisher} {MIT Press},\ \bibinfo
  {address} {Cambridge MA})\BibitemShut {NoStop}%
\bibitem [{\citenamefont {Jaeger}(2007)}]{Jaeger}%
  \BibitemOpen
  \bibfield  {author} {\bibinfo {author} {\bibnamefont {Jaeger}, \bibfnamefont
  {H.}}} (\bibinfo {year} {2007}),\ \href {\doibase 10.4249/scholarpedia.2330}
  {\bibfield  {journal} {\bibinfo  {journal} {Scholarpedia}\ }\textbf {\bibinfo
  {volume} {2}}~(\bibinfo {number} {9}),\ \bibinfo {pages} {2330}}\BibitemShut
  {NoStop}%
\bibitem [{\citenamefont {Jani{\'c}evi{\'c}}\ \emph {et~al.}(2016)\citenamefont
  {Jani{\'c}evi{\'c}}, \citenamefont {Laurson}, \citenamefont {M{\aa}l{\o}y},
  \citenamefont {Santucci},\ and\ \citenamefont {Alava}}]{Perils3}%
  \BibitemOpen
  \bibfield  {author} {\bibinfo {author} {\bibnamefont {Jani{\'c}evi{\'c}},
  \bibfnamefont {S.}}, \bibinfo {author} {\bibfnamefont {L.}~\bibnamefont
  {Laurson}}, \bibinfo {author} {\bibfnamefont {K.~J.}\ \bibnamefont
  {M{\aa}l{\o}y}}, \bibinfo {author} {\bibfnamefont {S.}~\bibnamefont
  {Santucci}}, \ and\ \bibinfo {author} {\bibfnamefont {M.~J.}\ \bibnamefont
  {Alava}}} (\bibinfo {year} {2016}),\ \href@noop {} {\bibfield  {journal}
  {\bibinfo  {journal} {Phys. Rev. Lett.}\ }\textbf {\bibinfo {volume}
  {117}}~(\bibinfo {number} {23}),\ \bibinfo {pages} {230601}}\BibitemShut
  {NoStop}%
\bibitem [{\citenamefont {Jensen}(1998)}]{Jensen}%
  \BibitemOpen
  \bibfield  {author} {\bibinfo {author} {\bibnamefont {Jensen}, \bibfnamefont
  {H.~J.}}} (\bibinfo {year} {1998}),\ \href@noop {} {\emph {\bibinfo {title}
  {Self-organized criticality: emergent complex behavior in physical and
  biological systems}}}\ (\bibinfo  {publisher} {Cambridge Univ.
  press})\BibitemShut {NoStop}%
\bibitem [{\citenamefont {Kaiser}(2011)}]{Review-Kaiser}%
  \BibitemOpen
  \bibfield  {author} {\bibinfo {author} {\bibnamefont {Kaiser}, \bibfnamefont
  {M.}}} (\bibinfo {year} {2011}),\ \href@noop {} {\bibfield  {journal}
  {\bibinfo  {journal} {NeuroImage}\ }\textbf {\bibinfo {volume}
  {57}}~(\bibinfo {number} {3}),\ \bibinfo {pages} {892}}\BibitemShut {NoStop}%
\bibitem [{\citenamefont {Kamenev}(2011)}]{Kamenev}%
  \BibitemOpen
  \bibfield  {author} {\bibinfo {author} {\bibnamefont {Kamenev}, \bibfnamefont
  {A.}}} (\bibinfo {year} {2011}),\ \href@noop {} {\emph {\bibinfo {title}
  {Field theory of non-equilibrium systems}}}\ (\bibinfo  {publisher}
  {Cambridge University Press})\BibitemShut {NoStop}%
\bibitem [{\citenamefont {Kandel}\ \emph {et~al.}(2000)\citenamefont {Kandel},
  \citenamefont {Schwartz},\ and\ \citenamefont {Jessel}}]{Kandel}%
  \BibitemOpen
  \bibfield  {author} {\bibinfo {author} {\bibnamefont {Kandel}, \bibfnamefont
  {E.}}, \bibinfo {author} {\bibfnamefont {J.}~\bibnamefont {Schwartz}}, \ and\
  \bibinfo {author} {\bibfnamefont {T.}~\bibnamefont {Jessel}}} (\bibinfo
  {year} {2000}),\ \href@noop {} {\emph {\bibinfo {title} {{Principles of
  Neural Science}}}}\ (\bibinfo  {publisher} {McGraw-Hill, New
  York})\BibitemShut {NoStop}%
\bibitem [{\citenamefont {Kaneko}(2006)}]{Kaneko-book}%
  \BibitemOpen
  \bibfield  {author} {\bibinfo {author} {\bibnamefont {Kaneko}, \bibfnamefont
  {K.}}} (\bibinfo {year} {2006}),\ \href@noop {} {\emph {\bibinfo {title}
  {Life: an introduction to complex systems biology}}}\ (\bibinfo  {publisher}
  {Springer})\BibitemShut {NoStop}%
\bibitem [{\citenamefont {Kaneko}(2012)}]{Kaneko-evolution}%
  \BibitemOpen
  \bibfield  {author} {\bibinfo {author} {\bibnamefont {Kaneko}, \bibfnamefont
  {K.}}} (\bibinfo {year} {2012}),\ \href@noop {} {\bibfield  {journal}
  {\bibinfo  {journal} {J. Stat. Phys.}\ }\textbf {\bibinfo {volume}
  {148}}~(\bibinfo {number} {4}),\ \bibinfo {pages} {687}}\BibitemShut
  {NoStop}%
\bibitem [{\citenamefont {Kaneko}\ and\ \citenamefont
  {Ikegami}(1992)}]{Kaneko1992}%
  \BibitemOpen
  \bibfield  {author} {\bibinfo {author} {\bibnamefont {Kaneko}, \bibfnamefont
  {K.}}, \ and\ \bibinfo {author} {\bibfnamefont {T.}~\bibnamefont {Ikegami}}}
  (\bibinfo {year} {1992}),\ \href@noop {} {\bibfield  {journal} {\bibinfo
  {journal} {Physica D: Nonlinear Phenomena}\ }\textbf {\bibinfo {volume}
  {56}}~(\bibinfo {number} {4}),\ \bibinfo {pages} {406}}\BibitemShut {NoStop}%
\bibitem [{\citenamefont {Kauffman}(1969)}]{Kauffman69}%
  \BibitemOpen
  \bibfield  {author} {\bibinfo {author} {\bibnamefont {Kauffman},
  \bibfnamefont {S.}}} (\bibinfo {year} {1969}),\ \href@noop {} {\bibfield
  {journal} {\bibinfo  {journal} {J. Theor. Biol.}\ }\textbf {\bibinfo {volume}
  {22}}~(\bibinfo {number} {3}),\ \bibinfo {pages} {437}}\BibitemShut {NoStop}%
\bibitem [{\citenamefont {Kauffman}(1996)}]{Home}%
  \BibitemOpen
  \bibfield  {author} {\bibinfo {author} {\bibnamefont {Kauffman},
  \bibfnamefont {S.}}} (\bibinfo {year} {1996}),\ \href@noop {} {\emph
  {\bibinfo {title} {At home in the universe: The search for the laws of
  self-organization and complexity}}}\ (\bibinfo  {publisher} {Oxford
  university press})\BibitemShut {NoStop}%
\bibitem [{\citenamefont {Kauffman}\ \emph {et~al.}(2003)\citenamefont
  {Kauffman}, \citenamefont {Peterson}, \citenamefont {Samuelsson},\ and\
  \citenamefont {Troein}}]{Kauffman2003}%
  \BibitemOpen
  \bibfield  {author} {\bibinfo {author} {\bibnamefont {Kauffman},
  \bibfnamefont {S.}}, \bibinfo {author} {\bibfnamefont {C.}~\bibnamefont
  {Peterson}}, \bibinfo {author} {\bibfnamefont {B.}~\bibnamefont
  {Samuelsson}}, \ and\ \bibinfo {author} {\bibfnamefont {C.}~\bibnamefont
  {Troein}}} (\bibinfo {year} {2003}),\ \href@noop {} {\bibfield  {journal}
  {\bibinfo  {journal} {Proc. Natl. Acad. Sci. USA}\ }\textbf {\bibinfo
  {volume} {100}}~(\bibinfo {number} {25}),\ \bibinfo {pages}
  {14796}}\BibitemShut {NoStop}%
\bibitem [{\citenamefont {Kauffman}(1993)}]{Origins}%
  \BibitemOpen
  \bibfield  {author} {\bibinfo {author} {\bibnamefont {Kauffman},
  \bibfnamefont {S.~A.}}} (\bibinfo {year} {1993}),\ \href@noop {} {\emph
  {\bibinfo {title} {The origins of order: Self organization and selection in
  evolution}}}\ (\bibinfo  {publisher} {Oxford university press})\BibitemShut
  {NoStop}%
\bibitem [{\citenamefont {Keenan}\ \emph {et~al.}(2007)\citenamefont {Keenan},
  \citenamefont {Krill}, \citenamefont {Platek},\ and\ \citenamefont
  {Shackelford}}]{Numbers}%
  \BibitemOpen
  \bibfield  {author} {\bibinfo {author} {\bibnamefont {Keenan}, \bibfnamefont
  {J.~P.}}, \bibinfo {author} {\bibfnamefont {A.~L.}\ \bibnamefont {Krill}},
  \bibinfo {author} {\bibfnamefont {S.~M.}\ \bibnamefont {Platek}}, \ and\
  \bibinfo {author} {\bibfnamefont {T.~K.}\ \bibnamefont {Shackelford}}}
  (\bibinfo {year} {2007}),\ \href@noop {} {\bibinfo  {journal} {Evolutionary
  Cognitive Neuroscience}\ ,\ \bibinfo {pages} {579}}\BibitemShut {NoStop}%
\bibitem [{\citenamefont {Kelley}\ and\ \citenamefont
  {Ouellette}(2013)}]{Ouellette2013}%
  \BibitemOpen
\bibfield  {journal} {  }\bibfield  {author} {\bibinfo {author} {\bibnamefont
  {Kelley}, \bibfnamefont {D.~H.}}, \ and\ \bibinfo {author} {\bibfnamefont
  {N.~T.}\ \bibnamefont {Ouellette}}} (\bibinfo {year} {2013}),\ \href@noop {}
  {\bibfield  {journal} {\bibinfo  {journal} {Sci. Rep.}\ }\textbf {\bibinfo
  {volume} {3}}}\BibitemShut {NoStop}%
\bibitem [{\citenamefont {Kello}\ \emph {et~al.}(2010)\citenamefont {Kello},
  \citenamefont {Brown}, \citenamefont {Ferrer-i Cancho}, \citenamefont
  {Holden}, \citenamefont {Linkenkaer-Hansen}, \citenamefont {Rhodes},\ and\
  \citenamefont {Van~Orden}}]{Linken-cognitive}%
  \BibitemOpen
  \bibfield  {author} {\bibinfo {author} {\bibnamefont {Kello}, \bibfnamefont
  {C.~T.}}, \bibinfo {author} {\bibfnamefont {G.~D.}\ \bibnamefont {Brown}},
  \bibinfo {author} {\bibfnamefont {R.}~\bibnamefont {Ferrer-i Cancho}},
  \bibinfo {author} {\bibfnamefont {J.~G.}\ \bibnamefont {Holden}}, \bibinfo
  {author} {\bibfnamefont {K.}~\bibnamefont {Linkenkaer-Hansen}}, \bibinfo
  {author} {\bibfnamefont {T.}~\bibnamefont {Rhodes}}, \ and\ \bibinfo {author}
  {\bibfnamefont {G.~C.}\ \bibnamefont {Van~Orden}}} (\bibinfo {year} {2010}),\
  \href@noop {} {\bibfield  {journal} {\bibinfo  {journal} {Trends in cognitive
  sciences}\ }\textbf {\bibinfo {volume} {14}}~(\bibinfo {number} {5}),\
  \bibinfo {pages} {223}}\BibitemShut {NoStop}%
\bibitem [{\citenamefont {Kelso}\ \emph {et~al.}(1986)\citenamefont {Kelso},
  \citenamefont {Scholz},\ and\ \citenamefont {Sch{\"o}ner}}]{Kelso1986}%
  \BibitemOpen
  \bibfield  {author} {\bibinfo {author} {\bibnamefont {Kelso}, \bibfnamefont
  {J.}}, \bibinfo {author} {\bibfnamefont {J.}~\bibnamefont {Scholz}}, \ and\
  \bibinfo {author} {\bibfnamefont {G.}~\bibnamefont {Sch{\"o}ner}}} (\bibinfo
  {year} {1986}),\ \href@noop {} {\bibfield  {journal} {\bibinfo  {journal}
  {Phys. Lett. A}\ }\textbf {\bibinfo {volume} {118}}~(\bibinfo {number} {6}),\
  \bibinfo {pages} {279}}\BibitemShut {NoStop}%
\bibitem [{\citenamefont {Kelso}(1984)}]{Kelso}%
  \BibitemOpen
  \bibfield  {author} {\bibinfo {author} {\bibnamefont {Kelso}, \bibfnamefont
  {J.~A.}}} (\bibinfo {year} {1984}),\ \href@noop {} {\bibfield  {journal}
  {\bibinfo  {journal} {Am J Physiol Regul Integr Comp Physiol}\ }\textbf
  {\bibinfo {volume} {246}}~(\bibinfo {number} {6}),\ \bibinfo {pages}
  {R1000}}\BibitemShut {NoStop}%
\bibitem [{\citenamefont {Kern}\ and\ \citenamefont {Stoop}(2003)}]{Stoop2003}%
  \BibitemOpen
  \bibfield  {author} {\bibinfo {author} {\bibnamefont {Kern}, \bibfnamefont
  {A.}}, \ and\ \bibinfo {author} {\bibfnamefont {R.}~\bibnamefont {Stoop}}}
  (\bibinfo {year} {2003}),\ \href@noop {} {\bibfield  {journal} {\bibinfo
  {journal} {Phys. Rev. Lett.}\ }\textbf {\bibinfo {volume} {91}}~(\bibinfo
  {number} {12}),\ \bibinfo {pages} {128101}}\BibitemShut {NoStop}%
\bibitem [{\citenamefont {Kimura}(1984)}]{Kimura}%
  \BibitemOpen
  \bibfield  {author} {\bibinfo {author} {\bibnamefont {Kimura}, \bibfnamefont
  {M.}}} (\bibinfo {year} {1984}),\ \href@noop {} {\emph {\bibinfo {title} {The
  neutral theory of molecular evolution}}}\ (\bibinfo  {publisher} {Cambridge
  University Press})\BibitemShut {NoStop}%
\bibitem [{\citenamefont {Kinouchi}\ \emph {et~al.}(2018)\citenamefont
  {Kinouchi}, \citenamefont {Brochini}, \citenamefont {Costa}, \citenamefont
  {Campos},\ and\ \citenamefont {Copelli}}]{Kinouchi2018}%
  \BibitemOpen
  \bibfield  {author} {\bibinfo {author} {\bibnamefont {Kinouchi},
  \bibfnamefont {O.}}, \bibinfo {author} {\bibfnamefont {L.}~\bibnamefont
  {Brochini}}, \bibinfo {author} {\bibfnamefont {A.~A.}\ \bibnamefont {Costa}},
  \bibinfo {author} {\bibfnamefont {J.~G.~F.}\ \bibnamefont {Campos}}, \ and\
  \bibinfo {author} {\bibfnamefont {M.}~\bibnamefont {Copelli}}} (\bibinfo
  {year} {2018}),\ \href@noop {} {\bibinfo  {journal} {arXiv:1803.05537}\
  }\BibitemShut {NoStop}%
\bibitem [{\citenamefont {Kinouchi}\ and\ \citenamefont
  {Copelli}(2006)}]{Kinouchi-Copelli}%
  \BibitemOpen
\bibfield  {journal} {  }\bibfield  {author} {\bibinfo {author} {\bibnamefont
  {Kinouchi}, \bibfnamefont {O.}}, \ and\ \bibinfo {author} {\bibfnamefont
  {M.}~\bibnamefont {Copelli}}} (\bibinfo {year} {2006}),\ \href@noop {}
  {\bibfield  {journal} {\bibinfo  {journal} {Nat. Phys.}\ }\textbf {\bibinfo
  {volume} {2}}~(\bibinfo {number} {5}),\ \bibinfo {pages} {348}}\BibitemShut
  {NoStop}%
\bibitem [{\citenamefont {Kitano}\ \emph {et~al.}(2001)\citenamefont {Kitano}
  \emph {et~al.}}]{Kitano}%
  \BibitemOpen
  \bibfield  {author} {\bibinfo {author} {\bibnamefont {Kitano}, \bibfnamefont
  {H.}},  \emph {et~al.}} (\bibinfo {year} {2001}),\ \href@noop {} {\emph
  {\bibinfo {title} {Foundations of systems biology}}}\ (\bibinfo  {publisher}
  {MIT press Cambridge})\BibitemShut {NoStop}%
\bibitem [{\citenamefont {Kiyono}\ \emph {et~al.}(2004)\citenamefont {Kiyono},
  \citenamefont {Struzik}, \citenamefont {Aoyagi}, \citenamefont {Sakata},
  \citenamefont {Hayano},\ and\ \citenamefont {Yamamoto}}]{Struzik1}%
  \BibitemOpen
  \bibfield  {author} {\bibinfo {author} {\bibnamefont {Kiyono}, \bibfnamefont
  {K.}}, \bibinfo {author} {\bibfnamefont {Z.}~\bibnamefont {Struzik}},
  \bibinfo {author} {\bibfnamefont {N.}~\bibnamefont {Aoyagi}}, \bibinfo
  {author} {\bibfnamefont {S.}~\bibnamefont {Sakata}}, \bibinfo {author}
  {\bibfnamefont {J.}~\bibnamefont {Hayano}}, \ and\ \bibinfo {author}
  {\bibfnamefont {Y.}~\bibnamefont {Yamamoto}}} (\bibinfo {year} {2004}),\
  \href@noop {} {\bibfield  {journal} {\bibinfo  {journal} {Phys. Rev. Lett.}\
  }\textbf {\bibinfo {volume} {93}}~(\bibinfo {number} {17}),\ \bibinfo {pages}
  {178103}}\BibitemShut {NoStop}%
\bibitem [{\citenamefont {Kiyono}\ \emph {et~al.}(2005)\citenamefont {Kiyono},
  \citenamefont {Struzik}, \citenamefont {Aoyagi}, \citenamefont {Togo},\ and\
  \citenamefont {Yamamoto}}]{Struzik2}%
  \BibitemOpen
  \bibfield  {author} {\bibinfo {author} {\bibnamefont {Kiyono}, \bibfnamefont
  {K.}}, \bibinfo {author} {\bibfnamefont {Z.~R.}\ \bibnamefont {Struzik}},
  \bibinfo {author} {\bibfnamefont {N.}~\bibnamefont {Aoyagi}}, \bibinfo
  {author} {\bibfnamefont {F.}~\bibnamefont {Togo}}, \ and\ \bibinfo {author}
  {\bibfnamefont {Y.}~\bibnamefont {Yamamoto}}} (\bibinfo {year} {2005}),\
  \href@noop {} {\bibfield  {journal} {\bibinfo  {journal} {Phys. Rev. Lett.}\
  }\textbf {\bibinfo {volume} {95}}~(\bibinfo {number} {5}),\ \bibinfo {pages}
  {058101}}\BibitemShut {NoStop}%
\bibitem [{\citenamefont {Kleiber}(1932)}]{Kleiber1932}%
  \BibitemOpen
  \bibfield  {author} {\bibinfo {author} {\bibnamefont {Kleiber}, \bibfnamefont
  {M.}}} (\bibinfo {year} {1932}),\ \href@noop {} {\bibfield  {journal}
  {\bibinfo  {journal} {Hilgardia}\ }\textbf {\bibinfo {volume} {6}},\ \bibinfo
  {pages} {315}}\BibitemShut {NoStop}%
\bibitem [{\citenamefont {Koonin}(2011)}]{Koonin-book2}%
  \BibitemOpen
  \bibfield  {author} {\bibinfo {author} {\bibnamefont {Koonin}, \bibfnamefont
  {E.~V.}}} (\bibinfo {year} {2011}),\ \href@noop {} {\emph {\bibinfo {title}
  {The logic of chance: the nature and origin of biological evolution}}}\
  (\bibinfo  {publisher} {FT press})\BibitemShut {NoStop}%
\bibitem [{\citenamefont {Koonin}\ \emph {et~al.}(2006)\citenamefont {Koonin},
  \citenamefont {Wolf},\ and\ \citenamefont {Karev}}]{Koonin-book1}%
  \BibitemOpen
  \bibfield  {author} {\bibinfo {author} {\bibnamefont {Koonin}, \bibfnamefont
  {E.~V.}}, \bibinfo {author} {\bibfnamefont {Y.~I.}\ \bibnamefont {Wolf}}, \
  and\ \bibinfo {author} {\bibfnamefont {G.~P.}\ \bibnamefont {Karev}}}
  (\bibinfo {year} {2006}),\ \href@noop {} {\emph {\bibinfo {title} {Power
  laws, scale-free networks and genome biology}}}\ (\bibinfo  {publisher}
  {Springer})\BibitemShut {NoStop}%
\bibitem [{\citenamefont {Krause}\ and\ \citenamefont
  {Ruxton}(2002)}]{Collective2}%
  \BibitemOpen
  \bibfield  {author} {\bibinfo {author} {\bibnamefont {Krause}, \bibfnamefont
  {J.}}, \ and\ \bibinfo {author} {\bibfnamefont {G.~D.}\ \bibnamefont
  {Ruxton}}} (\bibinfo {year} {2002}),\ \href@noop {} {\emph {\bibinfo {title}
  {Living in groups}}}\ (\bibinfo  {publisher} {Oxford University
  Press})\BibitemShut {NoStop}%
\bibitem [{\citenamefont {Krawitz}\ and\ \citenamefont
  {Shmulevich}(2007)}]{Basin}%
  \BibitemOpen
  \bibfield  {author} {\bibinfo {author} {\bibnamefont {Krawitz}, \bibfnamefont
  {P.}}, \ and\ \bibinfo {author} {\bibfnamefont {I.}~\bibnamefont
  {Shmulevich}}} (\bibinfo {year} {2007}),\ \href@noop {} {\bibfield  {journal}
  {\bibinfo  {journal} {Phys. Rev. Lett.}\ }\textbf {\bibinfo {volume}
  {98}}~(\bibinfo {number} {15}),\ \bibinfo {pages} {158701}}\BibitemShut
  {NoStop}%
\bibitem [{\citenamefont {Krotov}\ \emph {et~al.}(2014)\citenamefont {Krotov},
  \citenamefont {Dubuis}, \citenamefont {Gregor},\ and\ \citenamefont
  {Bialek}}]{Bialek-morpho}%
  \BibitemOpen
  \bibfield  {author} {\bibinfo {author} {\bibnamefont {Krotov}, \bibfnamefont
  {D.}}, \bibinfo {author} {\bibfnamefont {J.~O.}\ \bibnamefont {Dubuis}},
  \bibinfo {author} {\bibfnamefont {T.}~\bibnamefont {Gregor}}, \ and\ \bibinfo
  {author} {\bibfnamefont {W.}~\bibnamefont {Bialek}}} (\bibinfo {year}
  {2014}),\ \href {\doibase 10.1073/pnas.1324186111} {\bibfield  {journal}
  {\bibinfo  {journal} {Proc. Natl. Acad. Sci. USA.}\ }\textbf {\bibinfo
  {volume} {111}}~(\bibinfo {number} {10}),\ \bibinfo {pages}
  {3683}}\BibitemShut {NoStop}%
\bibitem [{\citenamefont {Ktitarev}\ \emph {et~al.}(2000)\citenamefont
  {Ktitarev}, \citenamefont {L{\"u}beck}, \citenamefont {Grassberger},\ and\
  \citenamefont {Priezzhev}}]{anomalies1}%
  \BibitemOpen
  \bibfield  {author} {\bibinfo {author} {\bibnamefont {Ktitarev},
  \bibfnamefont {D.}}, \bibinfo {author} {\bibfnamefont {S.}~\bibnamefont
  {L{\"u}beck}}, \bibinfo {author} {\bibfnamefont {P.}~\bibnamefont
  {Grassberger}}, \ and\ \bibinfo {author} {\bibfnamefont {V.}~\bibnamefont
  {Priezzhev}}} (\bibinfo {year} {2000}),\ \href@noop {} {\bibfield  {journal}
  {\bibinfo  {journal} {Phys. Rev. E}\ }\textbf {\bibinfo {volume}
  {61}}~(\bibinfo {number} {1}),\ \bibinfo {pages} {81}}\BibitemShut {NoStop}%
\bibitem [{\citenamefont {Kuehn}(2012)}]{Kuehn}%
  \BibitemOpen
  \bibfield  {author} {\bibinfo {author} {\bibnamefont {Kuehn}, \bibfnamefont
  {C.}}} (\bibinfo {year} {2012}),\ \href@noop {} {\bibfield  {journal}
  {\bibinfo  {journal} {Phys. Rev. E}\ }\textbf {\bibinfo {volume}
  {85}}~(\bibinfo {number} {2}),\ \bibinfo {pages} {026103}}\BibitemShut
  {NoStop}%
\bibitem [{\citenamefont {Kuramoto}(1975)}]{Kuramoto}%
  \BibitemOpen
  \bibfield  {author} {\bibinfo {author} {\bibnamefont {Kuramoto},
  \bibfnamefont {Y.}}} (\bibinfo {year} {1975}),\ \href@noop {} {\bibfield
  {journal} {\bibinfo  {journal} {Lecture Notes in Physics}\ }\textbf {\bibinfo
  {volume} {39}},\ \bibinfo {pages} {420}}\BibitemShut {NoStop}%
\bibitem [{\citenamefont {Kussell}\ and\ \citenamefont
  {Leibler}(2005)}]{Leibler}%
  \BibitemOpen
  \bibfield  {author} {\bibinfo {author} {\bibnamefont {Kussell}, \bibfnamefont
  {E.}}, \ and\ \bibinfo {author} {\bibfnamefont {S.}~\bibnamefont {Leibler}}}
  (\bibinfo {year} {2005}),\ \href@noop {} {\bibfield  {journal} {\bibinfo
  {journal} {Science}\ }\textbf {\bibinfo {volume} {309}}~(\bibinfo {number}
  {5743}),\ \bibinfo {pages} {2075}}\BibitemShut {NoStop}%
\bibitem [{\citenamefont {Langton}(1990)}]{Langton1990}%
  \BibitemOpen
  \bibfield  {author} {\bibinfo {author} {\bibnamefont {Langton}, \bibfnamefont
  {C.}}} (\bibinfo {year} {1990}),\ \href@noop {} {\bibfield  {journal}
  {\bibinfo  {journal} {Physica D}\ }\textbf {\bibinfo {volume} {42}}~(\bibinfo
  {number} {1}),\ \bibinfo {pages} {12}}\BibitemShut {NoStop}%
\bibitem [{\citenamefont {Laurson}\ \emph {et~al.}(2009)\citenamefont
  {Laurson}, \citenamefont {Illa},\ and\ \citenamefont {Alava}}]{Perils1}%
  \BibitemOpen
  \bibfield  {author} {\bibinfo {author} {\bibnamefont {Laurson}, \bibfnamefont
  {L.}}, \bibinfo {author} {\bibfnamefont {X.}~\bibnamefont {Illa}}, \ and\
  \bibinfo {author} {\bibfnamefont {M.~J.}\ \bibnamefont {Alava}}} (\bibinfo
  {year} {2009}),\ \href@noop {} {\bibfield  {journal} {\bibinfo  {journal} {J.
  Stat. Mech.}\ }\textbf {\bibinfo {volume} {2009}}~(\bibinfo {number} {01}),\
  \bibinfo {pages} {P01019}}\BibitemShut {NoStop}%
\bibitem [{\citenamefont {LeCun}\ \emph {et~al.}(2015)\citenamefont {LeCun},
  \citenamefont {Bengio},\ and\ \citenamefont {Hinton}}]{Deep}%
  \BibitemOpen
  \bibfield  {author} {\bibinfo {author} {\bibnamefont {LeCun}, \bibfnamefont
  {Y.}}, \bibinfo {author} {\bibfnamefont {Y.}~\bibnamefont {Bengio}}, \ and\
  \bibinfo {author} {\bibfnamefont {G.}~\bibnamefont {Hinton}}} (\bibinfo
  {year} {2015}),\ \href@noop {} {\bibfield  {journal} {\bibinfo  {journal}
  {Nature}\ }\textbf {\bibinfo {volume} {521}}~(\bibinfo {number} {7553}),\
  \bibinfo {pages} {436}}\BibitemShut {NoStop}%
\bibitem [{\citenamefont {Lee}\ \emph {et~al.}(2013)\citenamefont {Lee},
  \citenamefont {Brangwynne}, \citenamefont {Gharakhani}, \citenamefont
  {Hyman},\ and\ \citenamefont {J{\"u}licher}}]{Julicher-cells}%
  \BibitemOpen
  \bibfield  {author} {\bibinfo {author} {\bibnamefont {Lee}, \bibfnamefont
  {C.~F.}}, \bibinfo {author} {\bibfnamefont {C.~P.}\ \bibnamefont
  {Brangwynne}}, \bibinfo {author} {\bibfnamefont {J.}~\bibnamefont
  {Gharakhani}}, \bibinfo {author} {\bibfnamefont {A.~A.}\ \bibnamefont
  {Hyman}}, \ and\ \bibinfo {author} {\bibfnamefont {F.}~\bibnamefont
  {J{\"u}licher}}} (\bibinfo {year} {2013}),\ \href@noop {} {\bibfield
  {journal} {\bibinfo  {journal} {Phys. Rev. Lett.}\ }\textbf {\bibinfo
  {volume} {111}}~(\bibinfo {number} {8}),\ \bibinfo {pages}
  {088101}}\BibitemShut {NoStop}%
\bibitem [{\citenamefont {Legenstein}\ and\ \citenamefont
  {Maass}(2007)}]{Legenstein2007}%
  \BibitemOpen
  \bibfield  {author} {\bibinfo {author} {\bibnamefont {Legenstein},
  \bibfnamefont {R.}}, \ and\ \bibinfo {author} {\bibfnamefont
  {W.}~\bibnamefont {Maass}}} (\bibinfo {year} {2007}),\ \href@noop {}
  {\bibfield  {journal} {\bibinfo  {journal} {Neural Networks}\ }\textbf
  {\bibinfo {volume} {20}}~(\bibinfo {number} {3}),\ \bibinfo {pages}
  {323}}\BibitemShut {NoStop}%
\bibitem [{\citenamefont {Legenstein}(2005)}]{Ber-Nat2}%
  \BibitemOpen
  \bibfield  {author} {\bibinfo {author} {\bibnamefont {Legenstein},
  \bibfnamefont {T.}}} (\bibinfo {year} {2005}),\ in\ \href@noop {} {\emph
  {\bibinfo {booktitle} {Advances in Neural Information Processing Systems 17:
  Proceedings of the 2004 Conference}}},\ Vol.~\bibinfo {volume} {17}\
  (\bibinfo {organization} {MIT Press})\ p.\ \bibinfo {pages} {145}\BibitemShut
  {NoStop}%
\bibitem [{\citenamefont {Lesk}(2017)}]{Lesk-book}%
  \BibitemOpen
  \bibfield  {author} {\bibinfo {author} {\bibnamefont {Lesk}, \bibfnamefont
  {A.}}} (\bibinfo {year} {2017}),\ \href@noop {} {\emph {\bibinfo {title}
  {Introduction to genomics}}}\ (\bibinfo  {publisher} {Oxford University
  Press})\BibitemShut {NoStop}%
\bibitem [{\citenamefont {Levina}\ \emph {et~al.}(2007)\citenamefont {Levina},
  \citenamefont {Herrmann},\ and\ \citenamefont {Geisel}}]{Levina2007}%
  \BibitemOpen
  \bibfield  {author} {\bibinfo {author} {\bibnamefont {Levina}, \bibfnamefont
  {A.}}, \bibinfo {author} {\bibfnamefont {J.~M.}\ \bibnamefont {Herrmann}}, \
  and\ \bibinfo {author} {\bibfnamefont {T.}~\bibnamefont {Geisel}}} (\bibinfo
  {year} {2007}),\ \href@noop {} {\bibfield  {journal} {\bibinfo  {journal}
  {Nat. Phys.}\ }\textbf {\bibinfo {volume} {3}}~(\bibinfo {number} {12}),\
  \bibinfo {pages} {857}}\BibitemShut {NoStop}%
\bibitem [{\citenamefont {Levina}\ \emph {et~al.}(2009)\citenamefont {Levina},
  \citenamefont {Herrmann},\ and\ \citenamefont {Geisel}}]{Levina2009}%
  \BibitemOpen
  \bibfield  {author} {\bibinfo {author} {\bibnamefont {Levina}, \bibfnamefont
  {A.}}, \bibinfo {author} {\bibfnamefont {J.~M.}\ \bibnamefont {Herrmann}}, \
  and\ \bibinfo {author} {\bibfnamefont {T.}~\bibnamefont {Geisel}}} (\bibinfo
  {year} {2009}),\ \href@noop {} {\bibfield  {journal} {\bibinfo  {journal}
  {Phys. Rev. Lett.}\ }\textbf {\bibinfo {volume} {102}}~(\bibinfo {number}
  {11}),\ \bibinfo {pages} {118110}}\BibitemShut {NoStop}%
\bibitem [{\citenamefont {Li}\ \emph {et~al.}(2004)\citenamefont {Li},
  \citenamefont {Long}, \citenamefont {Lu}, \citenamefont {Ouyang},\ and\
  \citenamefont {Tang}}]{Attractors2}%
  \BibitemOpen
  \bibfield  {author} {\bibinfo {author} {\bibnamefont {Li}, \bibfnamefont
  {F.}}, \bibinfo {author} {\bibfnamefont {T.}~\bibnamefont {Long}}, \bibinfo
  {author} {\bibfnamefont {Y.}~\bibnamefont {Lu}}, \bibinfo {author}
  {\bibfnamefont {Q.}~\bibnamefont {Ouyang}}, \ and\ \bibinfo {author}
  {\bibfnamefont {C.}~\bibnamefont {Tang}}} (\bibinfo {year} {2004}),\
  \href@noop {} {\bibfield  {journal} {\bibinfo  {journal} {Proc. Natl. Acad.
  Sci. USA.}\ }\textbf {\bibinfo {volume} {101}}~(\bibinfo {number} {14}),\
  \bibinfo {pages} {4781}}\BibitemShut {NoStop}%
\bibitem [{\citenamefont {Li}\ \emph {et~al.}(2014)\citenamefont {Li},
  \citenamefont {Peng}, \citenamefont {Kurths}, \citenamefont {Yang},\ and\
  \citenamefont {Schellnhuber}}]{Ants-Kurths-2014}%
  \BibitemOpen
  \bibfield  {author} {\bibinfo {author} {\bibnamefont {Li}, \bibfnamefont
  {L.}}, \bibinfo {author} {\bibfnamefont {H.}~\bibnamefont {Peng}}, \bibinfo
  {author} {\bibfnamefont {J.}~\bibnamefont {Kurths}}, \bibinfo {author}
  {\bibfnamefont {Y.}~\bibnamefont {Yang}}, \ and\ \bibinfo {author}
  {\bibfnamefont {H.~J.}\ \bibnamefont {Schellnhuber}}} (\bibinfo {year}
  {2014}),\ \href@noop {} {\bibfield  {journal} {\bibinfo  {journal} {Proc.
  Natl. Acad. Sci. USA.}\ }\textbf {\bibinfo {volume} {111}}~(\bibinfo {number}
  {23}),\ \bibinfo {pages} {8392}}\BibitemShut {NoStop}%
\bibitem [{\citenamefont {Li}\ \emph {et~al.}(1990)\citenamefont {Li},
  \citenamefont {Packard},\ and\ \citenamefont {Langton}}]{Li1990}%
  \BibitemOpen
  \bibfield  {author} {\bibinfo {author} {\bibnamefont {Li}, \bibfnamefont
  {W.}}, \bibinfo {author} {\bibfnamefont {N.~H.}\ \bibnamefont {Packard}}, \
  and\ \bibinfo {author} {\bibfnamefont {C.~G.}\ \bibnamefont {Langton}}}
  (\bibinfo {year} {1990}),\ \href@noop {} {\bibfield  {journal} {\bibinfo
  {journal} {Physica D}\ }\textbf {\bibinfo {volume} {45}}~(\bibinfo {number}
  {1}),\ \bibinfo {pages} {77}}\BibitemShut {NoStop}%
\bibitem [{\citenamefont {Li}\ and\ \citenamefont {Retzloff}(2006)}]{DNA}%
  \BibitemOpen
  \bibfield  {author} {\bibinfo {author} {\bibnamefont {Li}, \bibfnamefont
  {Y.~C.}}, \ and\ \bibinfo {author} {\bibfnamefont {D.}~\bibnamefont
  {Retzloff}}} (\bibinfo {year} {2006}),\ \href@noop {} {\bibfield  {journal}
  {\bibinfo  {journal} {Math. Biosci.}\ }\textbf {\bibinfo {volume}
  {203}}~(\bibinfo {number} {1}),\ \bibinfo {pages} {137}}\BibitemShut
  {NoStop}%
\bibitem [{\citenamefont {Liggett}(2004)}]{Liggett}%
  \BibitemOpen
  \bibfield  {author} {\bibinfo {author} {\bibnamefont {Liggett}, \bibfnamefont
  {T.}}} (\bibinfo {year} {2004}),\ \href
  {http://books.google.it/books?id=I3aNPR1FursC} {\emph {\bibinfo {title}
  {Interacting Particle Systems}}},\ Classics in Mathematics\ (\bibinfo
  {publisher} {Springer})\BibitemShut {NoStop}%
\bibitem [{\citenamefont {Lin}\ and\ \citenamefont {Tegmark}(2017)}]{Tegmark}%
  \BibitemOpen
  \bibfield  {author} {\bibinfo {author} {\bibnamefont {Lin}, \bibfnamefont
  {H.~W.}}, \ and\ \bibinfo {author} {\bibfnamefont {M.}~\bibnamefont
  {Tegmark}}} (\bibinfo {year} {2017}),\ \href@noop {} {\bibfield  {journal}
  {\bibinfo  {journal} {Entropy}\ }\textbf {\bibinfo {volume} {19}}~(\bibinfo
  {number} {7}),\ \bibinfo {pages} {299}}\BibitemShut {NoStop}%
\bibitem [{\citenamefont {Liu}\ and\ \citenamefont {Bassler}(2006)}]{Bassler}%
  \BibitemOpen
  \bibfield  {author} {\bibinfo {author} {\bibnamefont {Liu}, \bibfnamefont
  {M.}}, \ and\ \bibinfo {author} {\bibfnamefont {K.~E.}\ \bibnamefont
  {Bassler}}} (\bibinfo {year} {2006}),\ \href@noop {} {\bibfield  {journal}
  {\bibinfo  {journal} {Phys. Rev. E}\ }\textbf {\bibinfo {volume}
  {74}}~(\bibinfo {number} {4}),\ \bibinfo {pages} {041910}}\BibitemShut
  {NoStop}%
\bibitem [{\citenamefont {Lizier}\ \emph
  {et~al.}(2008{\natexlab{a}})\citenamefont {Lizier}, \citenamefont
  {Prokopenko},\ and\ \citenamefont {Zomaya}}]{Lizier}%
  \BibitemOpen
  \bibfield  {author} {\bibinfo {author} {\bibnamefont {Lizier}, \bibfnamefont
  {J.~T.}}, \bibinfo {author} {\bibfnamefont {M.}~\bibnamefont {Prokopenko}}, \
  and\ \bibinfo {author} {\bibfnamefont {A.~Y.}\ \bibnamefont {Zomaya}}}
  (\bibinfo {year} {2008}{\natexlab{a}}),\ in\ \href@noop {} {\emph {\bibinfo
  {booktitle} {ALIFE}}},\ pp.\ \bibinfo {pages} {374--381}\BibitemShut
  {NoStop}%
\bibitem [{\citenamefont {Lizier}\ \emph
  {et~al.}(2008{\natexlab{b}})\citenamefont {Lizier}, \citenamefont
  {Prokopenko},\ and\ \citenamefont {Zomaya}}]{Lizier-transfer}%
  \BibitemOpen
  \bibfield  {author} {\bibinfo {author} {\bibnamefont {Lizier}, \bibfnamefont
  {J.~T.}}, \bibinfo {author} {\bibfnamefont {M.}~\bibnamefont {Prokopenko}}, \
  and\ \bibinfo {author} {\bibfnamefont {A.~Y.}\ \bibnamefont {Zomaya}}}
  (\bibinfo {year} {2008}{\natexlab{b}}),\ \href@noop {} {\bibfield  {journal}
  {\bibinfo  {journal} {Phys. Rev. E}\ }\textbf {\bibinfo {volume}
  {77}}~(\bibinfo {number} {2}),\ \bibinfo {pages} {026110}}\BibitemShut
  {NoStop}%
\bibitem [{\citenamefont {Loengarov}\ and\ \citenamefont
  {Tereshko}(2008)}]{Ants2008}%
  \BibitemOpen
  \bibfield  {author} {\bibinfo {author} {\bibnamefont {Loengarov},
  \bibfnamefont {A.}}, \ and\ \bibinfo {author} {\bibfnamefont
  {V.}~\bibnamefont {Tereshko}}} (\bibinfo {year} {2008}),\ \href {\doibase
  10.1162/artl.2008.14.1.111} {\bibfield  {journal} {\bibinfo  {journal} {Artif
  Life}\ }\textbf {\bibinfo {volume} {14}}~(\bibinfo {number} {1}),\ \bibinfo
  {pages} {111}}\BibitemShut {NoStop}%
\bibitem [{\citenamefont {Lopez-Garcia}\ \emph {et~al.}(2010)\citenamefont
  {Lopez-Garcia}, \citenamefont {Klein}, \citenamefont {Simons},\ and\
  \citenamefont {Winton}}]{Simons}%
  \BibitemOpen
  \bibfield  {author} {\bibinfo {author} {\bibnamefont {Lopez-Garcia},
  \bibfnamefont {C.}}, \bibinfo {author} {\bibfnamefont {A.~M.}\ \bibnamefont
  {Klein}}, \bibinfo {author} {\bibfnamefont {B.~D.}\ \bibnamefont {Simons}}, \
  and\ \bibinfo {author} {\bibfnamefont {D.~J.}\ \bibnamefont {Winton}}}
  (\bibinfo {year} {2010}),\ \href@noop {} {\bibfield  {journal} {\bibinfo
  {journal} {Science}\ }\textbf {\bibinfo {volume} {330}}~(\bibinfo {number}
  {6005}),\ \bibinfo {pages} {822}}\BibitemShut {NoStop}%
\bibitem [{\citenamefont {Losa}(1995)}]{Pathology}%
  \BibitemOpen
  \bibfield  {author} {\bibinfo {author} {\bibnamefont {Losa}, \bibfnamefont
  {G.~A.}}} (\bibinfo {year} {1995}),\ \href@noop {} {\bibfield  {journal}
  {\bibinfo  {journal} {Pathologica}\ }\textbf {\bibinfo {volume} {87}},\
  \bibinfo {pages} {310}}\BibitemShut {NoStop}%
\bibitem [{\citenamefont {Luko{\v{s}}evi{\v{c}}ius}\ \emph
  {et~al.}(2012)\citenamefont {Luko{\v{s}}evi{\v{c}}ius}, \citenamefont
  {Jaeger},\ and\ \citenamefont {Schrauwen}}]{Reservoir-review}%
  \BibitemOpen
  \bibfield  {author} {\bibinfo {author} {\bibnamefont
  {Luko{\v{s}}evi{\v{c}}ius}, \bibfnamefont {M.}}, \bibinfo {author}
  {\bibfnamefont {H.}~\bibnamefont {Jaeger}}, \ and\ \bibinfo {author}
  {\bibfnamefont {B.}~\bibnamefont {Schrauwen}}} (\bibinfo {year} {2012}),\
  \href@noop {} {\bibfield  {journal} {\bibinfo  {journal} {KI-K{\"u}nstliche
  Intelligenz}\ }\textbf {\bibinfo {volume} {26}}~(\bibinfo {number} {4}),\
  \bibinfo {pages} {365}}\BibitemShut {NoStop}%
\bibitem [{\citenamefont {Luque}\ and\ \citenamefont {Ferrera}(2000)}]{Luque}%
  \BibitemOpen
  \bibfield  {author} {\bibinfo {author} {\bibnamefont {Luque}, \bibfnamefont
  {B.}}, \ and\ \bibinfo {author} {\bibfnamefont {A.}~\bibnamefont {Ferrera}}}
  (\bibinfo {year} {2000}),\ \href@noop {} {\bibfield  {journal} {\bibinfo
  {journal} {Complex Systems}\ }\textbf {\bibinfo {volume} {12}},\ \bibinfo
  {pages} {241}}\BibitemShut {NoStop}%
\bibitem [{\citenamefont {Maass}\ \emph {et~al.}(2002)\citenamefont {Maass},
  \citenamefont {Natschlager},\ and\ \citenamefont {Markram}}]{Maass2002}%
  \BibitemOpen
  \bibfield  {author} {\bibinfo {author} {\bibnamefont {Maass}, \bibfnamefont
  {W.}}, \bibinfo {author} {\bibfnamefont {T.}~\bibnamefont {Natschlager}}, \
  and\ \bibinfo {author} {\bibfnamefont {H.}~\bibnamefont {Markram}}} (\bibinfo
  {year} {2002}),\ \href@noop {} {\bibfield  {journal} {\bibinfo  {journal}
  {Neural Comput.}\ }\textbf {\bibinfo {volume} {14}},\ \bibinfo {pages}
  {2531}}\BibitemShut {NoStop}%
\bibitem [{\citenamefont {MacArthur}\ \emph {et~al.}(2010)\citenamefont
  {MacArthur}, \citenamefont {S{\'a}nchez-Garc{\'\i}a},\ and\ \citenamefont
  {Ma’ayan}}]{Macarthur}%
  \BibitemOpen
  \bibfield  {author} {\bibinfo {author} {\bibnamefont {MacArthur},
  \bibfnamefont {B.~D.}}, \bibinfo {author} {\bibfnamefont {R.~J.}\
  \bibnamefont {S{\'a}nchez-Garc{\'\i}a}}, \ and\ \bibinfo {author}
  {\bibfnamefont {A.}~\bibnamefont {Ma’ayan}}} (\bibinfo {year} {2010}),\
  \href@noop {} {\bibfield  {journal} {\bibinfo  {journal} {Phys. Rev. Lett.}\
  }\textbf {\bibinfo {volume} {104}}~(\bibinfo {number} {16}),\ \bibinfo
  {pages} {168701}}\BibitemShut {NoStop}%
\bibitem [{\citenamefont {Magnasco}(2003)}]{Magnasco2003}%
  \BibitemOpen
  \bibfield  {author} {\bibinfo {author} {\bibnamefont {Magnasco},
  \bibfnamefont {M.~O.}}} (\bibinfo {year} {2003}),\ \href@noop {} {\bibfield
  {journal} {\bibinfo  {journal} {Phys. Rev. Lett.}\ }\textbf {\bibinfo
  {volume} {90}}~(\bibinfo {number} {5}),\ \bibinfo {pages}
  {058101}}\BibitemShut {NoStop}%
\bibitem [{\citenamefont {Magnasco}\ \emph {et~al.}(2009)\citenamefont
  {Magnasco}, \citenamefont {Piro},\ and\ \citenamefont
  {Cecchi}}]{Magnasco-anti}%
  \BibitemOpen
  \bibfield  {author} {\bibinfo {author} {\bibnamefont {Magnasco},
  \bibfnamefont {M.~O.}}, \bibinfo {author} {\bibfnamefont {O.}~\bibnamefont
  {Piro}}, \ and\ \bibinfo {author} {\bibfnamefont {G.~A.}\ \bibnamefont
  {Cecchi}}} (\bibinfo {year} {2009}),\ \href@noop {} {\bibfield  {journal}
  {\bibinfo  {journal} {Phys. Rev. Lett.}\ }\textbf {\bibinfo {volume}
  {102}}~(\bibinfo {number} {25}),\ \bibinfo {pages} {258102}}\BibitemShut
  {NoStop}%
\bibitem [{\citenamefont {Mandelbrot}(2002)}]{Mandelbrot-1/f}%
  \BibitemOpen
  \bibfield  {author} {\bibinfo {author} {\bibnamefont {Mandelbrot},
  \bibfnamefont {B.}}} (\bibinfo {year} {2002}),\ \href@noop {} {\emph
  {\bibinfo {title} {Gaussian self-affinity and fractals: Globality, the earth,
  1/f noise, and R/S}}},\ Vol.~\bibinfo {volume} {8}\ (\bibinfo  {publisher}
  {Springer Science \& Business Media})\BibitemShut {NoStop}%
\bibitem [{\citenamefont {Mandelbrot}(1983)}]{Mandelbrot}%
  \BibitemOpen
  \bibfield  {author} {\bibinfo {author} {\bibnamefont {Mandelbrot},
  \bibfnamefont {B.~B.}}} (\bibinfo {year} {1983}),\ \href@noop {} {\emph
  {\bibinfo {title} {The fractal geometry of nature}}},\ Vol.\ \bibinfo
  {volume} {173}\ (\bibinfo  {publisher} {Macmillan})\BibitemShut {NoStop}%
\bibitem [{\citenamefont {Manna}(1991)}]{Manna}%
  \BibitemOpen
  \bibfield  {author} {\bibinfo {author} {\bibnamefont {Manna}, \bibfnamefont
  {S.}}} (\bibinfo {year} {1991}),\ \href@noop {} {\bibfield  {journal}
  {\bibinfo  {journal} {J. of Phys. A: Mathematical and General}\ }\textbf
  {\bibinfo {volume} {24}}~(\bibinfo {number} {7}),\ \bibinfo {pages}
  {L363}}\BibitemShut {NoStop}%
\bibitem [{\citenamefont {Markovi{\'c}}\ and\ \citenamefont
  {Gros}(2014)}]{Gros-powerlaws}%
  \BibitemOpen
  \bibfield  {author} {\bibinfo {author} {\bibnamefont {Markovi{\'c}},
  \bibfnamefont {D.}}, \ and\ \bibinfo {author} {\bibfnamefont
  {C.}~\bibnamefont {Gros}}} (\bibinfo {year} {2014}),\ \href@noop {}
  {\bibfield  {journal} {\bibinfo  {journal} {Phys. Rep.}\ }\textbf {\bibinfo
  {volume} {536}}~(\bibinfo {number} {2}),\ \bibinfo {pages} {41}}\BibitemShut
  {NoStop}%
\bibitem [{\citenamefont {Markram}(2006)}]{BBP}%
  \BibitemOpen
  \bibfield  {author} {\bibinfo {author} {\bibnamefont {Markram}, \bibfnamefont
  {H.}}} (\bibinfo {year} {2006}),\ \href@noop {} {\bibfield  {journal}
  {\bibinfo  {journal} {Nature Rev. Neurosci.}\ }\textbf {\bibinfo {volume}
  {7}}~(\bibinfo {number} {2}),\ \bibinfo {pages} {153}}\BibitemShut {NoStop}%
\bibitem [{\citenamefont {Markram}\ and\ \citenamefont {Tsodyks}(1996)}]{TM96}%
  \BibitemOpen
  \bibfield  {author} {\bibinfo {author} {\bibnamefont {Markram}, \bibfnamefont
  {H.}}, \ and\ \bibinfo {author} {\bibfnamefont {M.}~\bibnamefont {Tsodyks}}}
  (\bibinfo {year} {1996}),\ \href@noop {} {\bibfield  {journal} {\bibinfo
  {journal} {Nature}\ }\textbf {\bibinfo {volume} {382}},\ \bibinfo {pages}
  {807}}\BibitemShut {NoStop}%
\bibitem [{\citenamefont {Markram}\ \emph {et~al.}(2015)\citenamefont {Markram}
  \emph {et~al.}}]{Cell-Markram}%
  \BibitemOpen
  \bibfield  {author} {\bibinfo {author} {\bibnamefont {Markram}, \bibfnamefont
  {H.}},  \emph {et~al.}} (\bibinfo {year} {2015}),\ \href@noop {} {\bibfield
  {journal} {\bibinfo  {journal} {Cell}\ }\textbf {\bibinfo {volume}
  {163}}~(\bibinfo {number} {2}),\ \bibinfo {pages} {456}}\BibitemShut
  {NoStop}%
\bibitem [{\citenamefont {Marre}\ \emph {et~al.}(2012)\citenamefont {Marre},
  \citenamefont {Amodei}, \citenamefont {Deshmukh}, \citenamefont {Sadeghi},
  \citenamefont {Soo}, \citenamefont {Holy},\ and\ \citenamefont
  {Berry}}]{Marre2012}%
  \BibitemOpen
  \bibfield  {author} {\bibinfo {author} {\bibnamefont {Marre}, \bibfnamefont
  {O.}}, \bibinfo {author} {\bibfnamefont {D.}~\bibnamefont {Amodei}}, \bibinfo
  {author} {\bibfnamefont {N.}~\bibnamefont {Deshmukh}}, \bibinfo {author}
  {\bibfnamefont {K.}~\bibnamefont {Sadeghi}}, \bibinfo {author} {\bibfnamefont
  {F.}~\bibnamefont {Soo}}, \bibinfo {author} {\bibfnamefont {T.~E.}\
  \bibnamefont {Holy}}, \ and\ \bibinfo {author} {\bibfnamefont {M.~J.}\
  \bibnamefont {Berry}}} (\bibinfo {year} {2012}),\ \href@noop {} {\bibfield
  {journal} {\bibinfo  {journal} {J. Neurosci.}\ }\textbf {\bibinfo {volume}
  {32}}~(\bibinfo {number} {43}),\ \bibinfo {pages} {14859}}\BibitemShut
  {NoStop}%
\bibitem [{\citenamefont {Marro}\ and\ \citenamefont {Dickman}(1999)}]{Marro}%
  \BibitemOpen
  \bibfield  {author} {\bibinfo {author} {\bibnamefont {Marro}, \bibfnamefont
  {J.}}, \ and\ \bibinfo {author} {\bibfnamefont {R.}~\bibnamefont {Dickman}}}
  (\bibinfo {year} {1999}),\ \href@noop {} {\emph {\bibinfo {title}
  {Nonequilibrium Phase Transition in Lattice Models}}}\ (\bibinfo  {publisher}
  {Cambridge University Press})\BibitemShut {NoStop}%
\bibitem [{\citenamefont {Marsili}\ \emph {et~al.}(2013)\citenamefont
  {Marsili}, \citenamefont {Mastromatteo},\ and\ \citenamefont
  {Roudi}}]{Marsili2}%
  \BibitemOpen
  \bibfield  {author} {\bibinfo {author} {\bibnamefont {Marsili}, \bibfnamefont
  {M.}}, \bibinfo {author} {\bibfnamefont {I.}~\bibnamefont {Mastromatteo}}, \
  and\ \bibinfo {author} {\bibfnamefont {Y.}~\bibnamefont {Roudi}}} (\bibinfo
  {year} {2013}),\ \href@noop {} {\bibfield  {journal} {\bibinfo  {journal} {J.
  Stat. Mech.}\ }\textbf {\bibinfo {volume} {2013}}~(\bibinfo {number} {09}),\
  \bibinfo {pages} {P09003}}\BibitemShut {NoStop}%
\bibitem [{\citenamefont {Marsili}\ and\ \citenamefont
  {Zhang}(1998)}]{Marsili-Zipf}%
  \BibitemOpen
  \bibfield  {author} {\bibinfo {author} {\bibnamefont {Marsili}, \bibfnamefont
  {M.}}, \ and\ \bibinfo {author} {\bibfnamefont {Y.-C.}\ \bibnamefont
  {Zhang}}} (\bibinfo {year} {1998}),\ \href@noop {} {\bibfield  {journal}
  {\bibinfo  {journal} {Phys. Rev. Lett.}\ }\textbf {\bibinfo {volume}
  {80}}~(\bibinfo {number} {12}),\ \bibinfo {pages} {2741}}\BibitemShut
  {NoStop}%
\bibitem [{\citenamefont {Martin}\ \emph {et~al.}(2001)\citenamefont {Martin},
  \citenamefont {Hudspeth},\ and\ \citenamefont {J{\"u}licher}}]{Julicher2001}%
  \BibitemOpen
  \bibfield  {author} {\bibinfo {author} {\bibnamefont {Martin}, \bibfnamefont
  {P.}}, \bibinfo {author} {\bibfnamefont {A.}~\bibnamefont {Hudspeth}}, \ and\
  \bibinfo {author} {\bibfnamefont {F.}~\bibnamefont {J{\"u}licher}}} (\bibinfo
  {year} {2001}),\ \href@noop {} {\bibfield  {journal} {\bibinfo  {journal}
  {Proc. Natl. Acad. Sci. USA.}\ }\textbf {\bibinfo {volume} {98}}~(\bibinfo
  {number} {25}),\ \bibinfo {pages} {14380}}\BibitemShut {NoStop}%
\bibitem [{\citenamefont {Martinello}\ \emph {et~al.}(2017)\citenamefont
  {Martinello}, \citenamefont {Hidalgo}, \citenamefont {Maritan}, \citenamefont
  {di~Santo}, \citenamefont {Plenz},\ and\ \citenamefont
  {Mu{\~n}oz}}]{neutral-neural}%
  \BibitemOpen
  \bibfield  {author} {\bibinfo {author} {\bibnamefont {Martinello},
  \bibfnamefont {M.}}, \bibinfo {author} {\bibfnamefont {J.}~\bibnamefont
  {Hidalgo}}, \bibinfo {author} {\bibfnamefont {A.}~\bibnamefont {Maritan}},
  \bibinfo {author} {\bibfnamefont {S.}~\bibnamefont {di~Santo}}, \bibinfo
  {author} {\bibfnamefont {D.}~\bibnamefont {Plenz}}, \ and\ \bibinfo {author}
  {\bibfnamefont {M.~A.}\ \bibnamefont {Mu{\~n}oz}}} (\bibinfo {year} {2017}),\
  \href@noop {} {\bibfield  {journal} {\bibinfo  {journal} {Physical Review X}\
  }\textbf {\bibinfo {volume} {7}}~(\bibinfo {number} {4}),\ \bibinfo {pages}
  {041071}}\BibitemShut {NoStop}%
\bibitem [{\citenamefont {Massobrio}\ \emph {et~al.}(2015)\citenamefont
  {Massobrio}, \citenamefont {de~Arcangelis}, \citenamefont {Pasquale},
  \citenamefont {Jensen},\ and\ \citenamefont {Plenz}}]{Healthy}%
  \BibitemOpen
  \bibfield  {author} {\bibinfo {author} {\bibnamefont {Massobrio},
  \bibfnamefont {P.}}, \bibinfo {author} {\bibfnamefont {L.}~\bibnamefont
  {de~Arcangelis}}, \bibinfo {author} {\bibfnamefont {V.}~\bibnamefont
  {Pasquale}}, \bibinfo {author} {\bibfnamefont {H.~J.}\ \bibnamefont
  {Jensen}}, \ and\ \bibinfo {author} {\bibfnamefont {D.}~\bibnamefont
  {Plenz}}} (\bibinfo {year} {2015}),\ \href@noop {} {\bibinfo  {journal}
  {Criticality as a signature of healthy neural systems: multi-scale
  experimental and computational studies}\ ,\ \bibinfo {pages} {4}}\BibitemShut
  {NoStop}%
\bibitem [{\citenamefont {Mastromatteo}\ and\ \citenamefont
  {Marsili}(2011)}]{Marsili1}%
  \BibitemOpen
\bibfield  {journal} {  }\bibfield  {author} {\bibinfo {author} {\bibnamefont
  {Mastromatteo}, \bibfnamefont {I.}}, \ and\ \bibinfo {author} {\bibfnamefont
  {M.}~\bibnamefont {Marsili}}} (\bibinfo {year} {2011}),\ \href@noop {}
  {\bibfield  {journal} {\bibinfo  {journal} {J. Stat. Mech.}\ }\textbf
  {\bibinfo {volume} {2011}}~(\bibinfo {number} {10}),\ \bibinfo {pages}
  {P10012}}\BibitemShut {NoStop}%
\bibitem [{\citenamefont {Mattia}\ and\ \citenamefont
  {Sanchez-Vives}(2012)}]{Vives-exploring}%
  \BibitemOpen
  \bibfield  {author} {\bibinfo {author} {\bibnamefont {Mattia}, \bibfnamefont
  {M.}}, \ and\ \bibinfo {author} {\bibfnamefont {M.~V.}\ \bibnamefont
  {Sanchez-Vives}}} (\bibinfo {year} {2012}),\ \href@noop {} {\bibfield
  {journal} {\bibinfo  {journal} {Cognitive neurodynamics}\ }\textbf {\bibinfo
  {volume} {6}}~(\bibinfo {number} {3}),\ \bibinfo {pages} {239}}\BibitemShut
  {NoStop}%
\bibitem [{\citenamefont {Mazzoni}\ \emph {et~al.}(2007)\citenamefont
  {Mazzoni}, \citenamefont {Broccard}, \citenamefont {Garcia-Perez},
  \citenamefont {Bonifazi}, \citenamefont {Ruaro},\ and\ \citenamefont
  {Torre}}]{Torre2007}%
  \BibitemOpen
  \bibfield  {author} {\bibinfo {author} {\bibnamefont {Mazzoni}, \bibfnamefont
  {A.}}, \bibinfo {author} {\bibfnamefont {F.~D.}\ \bibnamefont {Broccard}},
  \bibinfo {author} {\bibfnamefont {E.}~\bibnamefont {Garcia-Perez}}, \bibinfo
  {author} {\bibfnamefont {P.}~\bibnamefont {Bonifazi}}, \bibinfo {author}
  {\bibfnamefont {M.~E.}\ \bibnamefont {Ruaro}}, \ and\ \bibinfo {author}
  {\bibfnamefont {V.}~\bibnamefont {Torre}}} (\bibinfo {year} {2007}),\
  \href@noop {} {\bibfield  {journal} {\bibinfo  {journal} {PloS One}\ }\textbf
  {\bibinfo {volume} {2}}~(\bibinfo {number} {5}),\ \bibinfo {pages}
  {e439}}\BibitemShut {NoStop}%
\bibitem [{\citenamefont {M{\'e}hes}\ and\ \citenamefont
  {Vicsek}(2014)}]{Vicsek2014}%
  \BibitemOpen
  \bibfield  {author} {\bibinfo {author} {\bibnamefont {M{\'e}hes},
  \bibfnamefont {E.}}, \ and\ \bibinfo {author} {\bibfnamefont
  {T.}~\bibnamefont {Vicsek}}} (\bibinfo {year} {2014}),\ \href@noop {}
  {\bibfield  {journal} {\bibinfo  {journal} {Integrative Biol.}\ }\textbf
  {\bibinfo {volume} {6}}~(\bibinfo {number} {9}),\ \bibinfo {pages}
  {831}}\BibitemShut {NoStop}%
\bibitem [{\citenamefont {Mehta}\ and\ \citenamefont
  {Schwab}(2014)}]{Mehta-RG}%
  \BibitemOpen
  \bibfield  {author} {\bibinfo {author} {\bibnamefont {Mehta}, \bibfnamefont
  {P.}}, \ and\ \bibinfo {author} {\bibfnamefont {D.~J.}\ \bibnamefont
  {Schwab}}} (\bibinfo {year} {2014}),\ \href@noop {} {\bibinfo  {journal}
  {arXiv:1410.3831}\ }\BibitemShut {NoStop}%
\bibitem [{\citenamefont {Meisel}\ and\ \citenamefont
  {Gross}(2009)}]{Meisel-Gross}%
  \BibitemOpen
\bibfield  {journal} {  }\bibfield  {author} {\bibinfo {author} {\bibnamefont
  {Meisel}, \bibfnamefont {C.}}, \ and\ \bibinfo {author} {\bibfnamefont
  {T.}~\bibnamefont {Gross}}} (\bibinfo {year} {2009}),\ \href@noop {}
  {\bibfield  {journal} {\bibinfo  {journal} {Phys. Rev. E}\ }\textbf {\bibinfo
  {volume} {80}}~(\bibinfo {number} {6}),\ \bibinfo {pages}
  {061917}}\BibitemShut {NoStop}%
\bibitem [{\citenamefont {Meisel}\ \emph {et~al.}(2013)\citenamefont {Meisel},
  \citenamefont {Olbrich}, \citenamefont {Shriki},\ and\ \citenamefont
  {Achermann}}]{Meisel2013}%
  \BibitemOpen
  \bibfield  {author} {\bibinfo {author} {\bibnamefont {Meisel}, \bibfnamefont
  {C.}}, \bibinfo {author} {\bibfnamefont {E.}~\bibnamefont {Olbrich}},
  \bibinfo {author} {\bibfnamefont {O.}~\bibnamefont {Shriki}}, \ and\ \bibinfo
  {author} {\bibfnamefont {P.}~\bibnamefont {Achermann}}} (\bibinfo {year}
  {2013}),\ \href@noop {} {\bibfield  {journal} {\bibinfo  {journal} {J.
  Neurosci.}\ }\textbf {\bibinfo {volume} {33}}~(\bibinfo {number} {44}),\
  \bibinfo {pages} {17363}}\BibitemShut {NoStop}%
\bibitem [{\citenamefont {Meisel}\ \emph {et~al.}(2012)\citenamefont {Meisel},
  \citenamefont {Storch}, \citenamefont {Hallmeyer-Elgner}, \citenamefont
  {Bullmore},\ and\ \citenamefont {Gross}}]{Meisel2012}%
  \BibitemOpen
  \bibfield  {author} {\bibinfo {author} {\bibnamefont {Meisel}, \bibfnamefont
  {C.}}, \bibinfo {author} {\bibfnamefont {A.}~\bibnamefont {Storch}}, \bibinfo
  {author} {\bibfnamefont {S.}~\bibnamefont {Hallmeyer-Elgner}}, \bibinfo
  {author} {\bibfnamefont {E.}~\bibnamefont {Bullmore}}, \ and\ \bibinfo
  {author} {\bibfnamefont {T.}~\bibnamefont {Gross}}} (\bibinfo {year}
  {2012}),\ \href@noop {} {\bibfield  {journal} {\bibinfo  {journal} {PLoS
  Comput. Biol.}\ }\textbf {\bibinfo {volume} {8}}~(\bibinfo {number} {1}),\
  \bibinfo {pages} {e1002312}}\BibitemShut {NoStop}%
\bibitem [{\citenamefont {Meister}\ \emph {et~al.}(1991)\citenamefont
  {Meister}, \citenamefont {Wong}, \citenamefont {Baylor},\ and\ \citenamefont
  {Shatz}}]{Meister1991}%
  \BibitemOpen
  \bibfield  {author} {\bibinfo {author} {\bibnamefont {Meister}, \bibfnamefont
  {M.}}, \bibinfo {author} {\bibfnamefont {R.}~\bibnamefont {Wong}}, \bibinfo
  {author} {\bibfnamefont {D.~A.}\ \bibnamefont {Baylor}}, \ and\ \bibinfo
  {author} {\bibfnamefont {C.~J.}\ \bibnamefont {Shatz}}} (\bibinfo {year}
  {1991}),\ \href@noop {} {\bibfield  {journal} {\bibinfo  {journal} {Science}\
  }\textbf {\bibinfo {volume} {252}}~(\bibinfo {number} {5008}),\ \bibinfo
  {pages} {939}}\BibitemShut {NoStop}%
\bibitem [{\citenamefont {Mejias}\ \emph {et~al.}(2010)\citenamefont {Mejias},
  \citenamefont {Kappen},\ and\ \citenamefont {Torres}}]{Mejias0}%
  \BibitemOpen
  \bibfield  {author} {\bibinfo {author} {\bibnamefont {Mejias}, \bibfnamefont
  {J.~F.}}, \bibinfo {author} {\bibfnamefont {H.~J.}\ \bibnamefont {Kappen}}, \
  and\ \bibinfo {author} {\bibfnamefont {J.~J.}\ \bibnamefont {Torres}}}
  (\bibinfo {year} {2010}),\ \href@noop {} {\bibfield  {journal} {\bibinfo
  {journal} {PLoS One}\ }\textbf {\bibinfo {volume} {5}}~(\bibinfo {number}
  {11}),\ \bibinfo {pages} {e13651}}\BibitemShut {NoStop}%
\bibitem [{\citenamefont {Melanie}(1993)}]{Mitchell}%
  \BibitemOpen
  \bibfield  {author} {\bibinfo {author} {\bibnamefont {Melanie}, \bibfnamefont
  {M.}}} (\bibinfo {year} {1993}),\ \href@noop {} {\bibinfo  {journal}
  {Complexity: Metaphors, Models, and Reality}\ }\BibitemShut {NoStop}%
\bibitem [{\citenamefont {Meunier}\ \emph {et~al.}(2010)\citenamefont
  {Meunier}, \citenamefont {Lambiotte},\ and\ \citenamefont
  {Bullmore}}]{Review-Bullmore}%
  \BibitemOpen
\bibfield  {journal} {  }\bibfield  {author} {\bibinfo {author} {\bibnamefont
  {Meunier}, \bibfnamefont {D.}}, \bibinfo {author} {\bibfnamefont
  {R.}~\bibnamefont {Lambiotte}}, \ and\ \bibinfo {author} {\bibfnamefont
  {E.~T.}\ \bibnamefont {Bullmore}}} (\bibinfo {year} {2010}),\ \href {\doibase
  10.3389/fnins.2010.00200} {\bibfield  {journal} {\bibinfo  {journal} {Front.
  Neurosci.}\ }\textbf {\bibinfo {volume} {4}},\ \bibinfo {pages}
  {200}}\BibitemShut {NoStop}%
\bibitem [{\citenamefont {Millman}\ \emph {et~al.}(2010)\citenamefont
  {Millman}, \citenamefont {Mihalas}, \citenamefont {Kirkwood},\ and\
  \citenamefont {Niebur}}]{Millman}%
  \BibitemOpen
  \bibfield  {author} {\bibinfo {author} {\bibnamefont {Millman}, \bibfnamefont
  {D.}}, \bibinfo {author} {\bibfnamefont {S.}~\bibnamefont {Mihalas}},
  \bibinfo {author} {\bibfnamefont {A.}~\bibnamefont {Kirkwood}}, \ and\
  \bibinfo {author} {\bibfnamefont {E.}~\bibnamefont {Niebur}}} (\bibinfo
  {year} {2010}),\ \href@noop {} {\bibfield  {journal} {\bibinfo  {journal}
  {Nat. Phys.}\ }\textbf {\bibinfo {volume} {6}}~(\bibinfo {number} {10}),\
  \bibinfo {pages} {801}}\BibitemShut {NoStop}%
\bibitem [{\citenamefont {Ming}\ and\ \citenamefont
  {Vit{\'a}nyi}(2014)}]{Kolmogorov}%
  \BibitemOpen
  \bibfield  {author} {\bibinfo {author} {\bibnamefont {Ming}, \bibfnamefont
  {L.}}, \ and\ \bibinfo {author} {\bibfnamefont {P.}~\bibnamefont
  {Vit{\'a}nyi}}} (\bibinfo {year} {2014}),\ \href@noop {} {\bibinfo  {journal}
  {Algorithms and Complexity}\ ,\ \bibinfo {pages} {187}}\BibitemShut {NoStop}%
\bibitem [{\citenamefont {Mitzenmacher}(2002)}]{Mitzenmacher}%
  \BibitemOpen
\bibfield  {journal} {  }\bibfield  {author} {\bibinfo {author} {\bibnamefont
  {Mitzenmacher}, \bibfnamefont {M.}}} (\bibinfo {year} {2002}),\ \href@noop {}
  {\bibfield  {journal} {\bibinfo  {journal} {Internet Mathematics}\ }\textbf
  {\bibinfo {volume} {1}},\ \bibinfo {pages} {226}}\BibitemShut {NoStop}%
\bibitem [{\citenamefont {Mora}\ and\ \citenamefont
  {Bialek}(2011)}]{Mora-Bialek}%
  \BibitemOpen
  \bibfield  {author} {\bibinfo {author} {\bibnamefont {Mora}, \bibfnamefont
  {T.}}, \ and\ \bibinfo {author} {\bibfnamefont {W.}~\bibnamefont {Bialek}}}
  (\bibinfo {year} {2011}),\ \href@noop {} {\bibfield  {journal} {\bibinfo
  {journal} {J. Stat. Phys.}\ }\textbf {\bibinfo {volume} {144}}~(\bibinfo
  {number} {2}),\ \bibinfo {pages} {268}}\BibitemShut {NoStop}%
\bibitem [{\citenamefont {Mora}\ \emph {et~al.}(2015)\citenamefont {Mora},
  \citenamefont {Deny},\ and\ \citenamefont {Marre}}]{Mora2015-retina}%
  \BibitemOpen
  \bibfield  {author} {\bibinfo {author} {\bibnamefont {Mora}, \bibfnamefont
  {T.}}, \bibinfo {author} {\bibfnamefont {S.}~\bibnamefont {Deny}}, \ and\
  \bibinfo {author} {\bibfnamefont {O.}~\bibnamefont {Marre}}} (\bibinfo {year}
  {2015}),\ \href@noop {} {\bibfield  {journal} {\bibinfo  {journal} {Phys.
  Rev. Lett.}\ }\textbf {\bibinfo {volume} {114}}~(\bibinfo {number} {7}),\
  \bibinfo {pages} {078105}}\BibitemShut {NoStop}%
\bibitem [{\citenamefont {Mora}\ \emph {et~al.}(2010)\citenamefont {Mora},
  \citenamefont {Walczak}, \citenamefont {Bialek},\ and\ \citenamefont
  {Callan~Jr}}]{Bialek-antibody}%
  \BibitemOpen
  \bibfield  {author} {\bibinfo {author} {\bibnamefont {Mora}, \bibfnamefont
  {T.}}, \bibinfo {author} {\bibfnamefont {A.}~\bibnamefont {Walczak}},
  \bibinfo {author} {\bibfnamefont {W.}~\bibnamefont {Bialek}}, \ and\ \bibinfo
  {author} {\bibfnamefont {C.}~\bibnamefont {Callan~Jr}}} (\bibinfo {year}
  {2010}),\ \href@noop {} {\bibfield  {journal} {\bibinfo  {journal} {Proc.
  Natl. Acad. Sci. USA.}\ }\textbf {\bibinfo {volume} {107}}~(\bibinfo {number}
  {12}),\ \bibinfo {pages} {5405}}\BibitemShut {NoStop}%
\bibitem [{\citenamefont {Mora}\ \emph {et~al.}(2016)\citenamefont {Mora},
  \citenamefont {Walczak}, \citenamefont {Del~Castello}, \citenamefont
  {Ginelli}, \citenamefont {Melillo}, \citenamefont {Parisi}, \citenamefont
  {Viale}, \citenamefont {Cavagna},\ and\ \citenamefont {Giardina}}]{Mora2016}%
  \BibitemOpen
  \bibfield  {author} {\bibinfo {author} {\bibnamefont {Mora}, \bibfnamefont
  {T.}}, \bibinfo {author} {\bibfnamefont {A.~M.}\ \bibnamefont {Walczak}},
  \bibinfo {author} {\bibfnamefont {L.}~\bibnamefont {Del~Castello}}, \bibinfo
  {author} {\bibfnamefont {F.}~\bibnamefont {Ginelli}}, \bibinfo {author}
  {\bibfnamefont {S.}~\bibnamefont {Melillo}}, \bibinfo {author} {\bibfnamefont
  {L.}~\bibnamefont {Parisi}}, \bibinfo {author} {\bibfnamefont
  {M.}~\bibnamefont {Viale}}, \bibinfo {author} {\bibfnamefont
  {A.}~\bibnamefont {Cavagna}}, \ and\ \bibinfo {author} {\bibfnamefont
  {I.}~\bibnamefont {Giardina}}} (\bibinfo {year} {2016}),\ \href@noop {}
  {\bibfield  {journal} {\bibinfo  {journal} {Nat. Phys.}\ }\textbf {\bibinfo
  {volume} {12}}~(\bibinfo {number} {12}),\ \bibinfo {pages}
  {1153}}\BibitemShut {NoStop}%
\bibitem [{\citenamefont {Moreau}\ and\ \citenamefont {Sontag}(2003)}]{Sontag}%
  \BibitemOpen
  \bibfield  {author} {\bibinfo {author} {\bibnamefont {Moreau}, \bibfnamefont
  {L.}}, \ and\ \bibinfo {author} {\bibfnamefont {E.}~\bibnamefont {Sontag}}}
  (\bibinfo {year} {2003}),\ \href@noop {} {\bibfield  {journal} {\bibinfo
  {journal} {Phys. Rev. E}\ }\textbf {\bibinfo {volume} {68}}~(\bibinfo
  {number} {2}),\ \bibinfo {pages} {020901}}\BibitemShut {NoStop}%
\bibitem [{\citenamefont {Moreira}\ and\ \citenamefont
  {Dickman}(1996)}]{Adriana}%
  \BibitemOpen
  \bibfield  {author} {\bibinfo {author} {\bibnamefont {Moreira}, \bibfnamefont
  {A.~G.}}, \ and\ \bibinfo {author} {\bibfnamefont {R.}~\bibnamefont
  {Dickman}}} (\bibinfo {year} {1996}),\ \href@noop {} {\bibfield  {journal}
  {\bibinfo  {journal} {Phys. Rev. E}\ }\textbf {\bibinfo {volume}
  {54}}~(\bibinfo {number} {4}),\ \bibinfo {pages} {R3090}}\BibitemShut
  {NoStop}%
\bibitem [{\citenamefont {Moretti}\ and\ \citenamefont
  {Mu{\~n}oz}(2013)}]{Moretti}%
  \BibitemOpen
  \bibfield  {author} {\bibinfo {author} {\bibnamefont {Moretti}, \bibfnamefont
  {P.}}, \ and\ \bibinfo {author} {\bibfnamefont {M.~A.}\ \bibnamefont
  {Mu{\~n}oz}}} (\bibinfo {year} {2013}),\ \href@noop {} {\bibfield  {journal}
  {\bibinfo  {journal} {Nature Comm.}\ }\textbf {\bibinfo {volume}
  {4}}}\BibitemShut {NoStop}%
\bibitem [{\citenamefont {Mu\~noz}\ \emph {et~al.}(2010)\citenamefont
  {Mu\~noz}, \citenamefont {Juh\'asz}, \citenamefont {Castellano},\ and\
  \citenamefont {\'Odor}}]{GPCN}%
  \BibitemOpen
  \bibfield  {author} {\bibinfo {author} {\bibnamefont {Mu\~noz}, \bibfnamefont
  {M.}}, \bibinfo {author} {\bibfnamefont {R.}~\bibnamefont {Juh\'asz}},
  \bibinfo {author} {\bibfnamefont {C.}~\bibnamefont {Castellano}}, \ and\
  \bibinfo {author} {\bibfnamefont {G.}~\bibnamefont {\'Odor}}} (\bibinfo
  {year} {2010}),\ \href {\doibase 10.1103/PhysRevLett.105.128701} {\bibfield
  {journal} {\bibinfo  {journal} {Phys. Rev. Lett.}\ }\textbf {\bibinfo
  {volume} {105}},\ \bibinfo {pages} {128701}}\BibitemShut {NoStop}%
\bibitem [{\citenamefont {Murphy}\ and\ \citenamefont {Miller}(2009)}]{Miller}%
  \BibitemOpen
  \bibfield  {author} {\bibinfo {author} {\bibnamefont {Murphy}, \bibfnamefont
  {B.~K.}}, \ and\ \bibinfo {author} {\bibfnamefont {K.~D.}\ \bibnamefont
  {Miller}}} (\bibinfo {year} {2009}),\ \href {\doibase
  10.1016/j.neuron.2009.02.005} {\bibfield  {journal} {\bibinfo  {journal}
  {Neuron}\ }\textbf {\bibinfo {volume} {61}}~(\bibinfo {number} {4}),\
  \bibinfo {pages} {635}}\BibitemShut {NoStop}%
\bibitem [{\citenamefont {Nadell}\ \emph {et~al.}(2013)\citenamefont {Nadell},
  \citenamefont {Bucci}, \citenamefont {Drescher}, \citenamefont {Levin},
  \citenamefont {Bassler},\ and\ \citenamefont {Xavier}}]{Nadell}%
  \BibitemOpen
  \bibfield  {author} {\bibinfo {author} {\bibnamefont {Nadell}, \bibfnamefont
  {C.~D.}}, \bibinfo {author} {\bibfnamefont {V.}~\bibnamefont {Bucci}},
  \bibinfo {author} {\bibfnamefont {K.}~\bibnamefont {Drescher}}, \bibinfo
  {author} {\bibfnamefont {S.~A.}\ \bibnamefont {Levin}}, \bibinfo {author}
  {\bibfnamefont {B.~L.}\ \bibnamefont {Bassler}}, \ and\ \bibinfo {author}
  {\bibfnamefont {J.~B.}\ \bibnamefont {Xavier}}} (\bibinfo {year} {2013}),\
  \href@noop {} {\bibfield  {journal} {\bibinfo  {journal} {Proc. R. Soc. B}\
  }\textbf {\bibinfo {volume} {280}}~(\bibinfo {number} {1755}),\ \bibinfo
  {pages} {20122770}}\BibitemShut {NoStop}%
\bibitem [{\citenamefont {Newman}(2003)}]{Newman-Review}%
  \BibitemOpen
  \bibfield  {author} {\bibinfo {author} {\bibnamefont {Newman}, \bibfnamefont
  {M.~E.~J.}}} (\bibinfo {year} {2003}),\ \href@noop {} {\bibfield  {journal}
  {\bibinfo  {journal} {SIAM Review}\ }\textbf {\bibinfo {volume}
  {45}}~(\bibinfo {number} {2}),\ \bibinfo {pages} {167}}\BibitemShut {NoStop}%
\bibitem [{\citenamefont {Newman}(2005)}]{Newman-powerlaws}%
  \BibitemOpen
  \bibfield  {author} {\bibinfo {author} {\bibnamefont {Newman}, \bibfnamefont
  {M.~E.~J.}}} (\bibinfo {year} {2005}),\ \href {\doibase
  10.1080/00107510500052444} {\bibfield  {journal} {\bibinfo  {journal}
  {Contemp Phys}\ }\textbf {\bibinfo {volume} {46}}~(\bibinfo {number} {5}),\
  \bibinfo {pages} {323}},\ \Eprint {http://arxiv.org/abs/cond-mat/0412004}
  {cond-mat/0412004} \BibitemShut {NoStop}%
\bibitem [{\citenamefont {Newman}(2010)}]{Newman-book}%
  \BibitemOpen
  \bibfield  {author} {\bibinfo {author} {\bibnamefont {Newman}, \bibfnamefont
  {M.~E.~J.}}} (\bibinfo {year} {2010}),\ \href@noop {} {\emph {\bibinfo
  {title} {Networks: An Introduction}}}\ (\bibinfo  {publisher} {Oxford
  University Press})\BibitemShut {NoStop}%
\bibitem [{\citenamefont {Newman}\ \emph {et~al.}(2004)\citenamefont {Newman},
  \citenamefont {Forgacs}, \citenamefont {Hinner}, \citenamefont {Maier},\ and\
  \citenamefont {Sackmann}}]{Forgacs2}%
  \BibitemOpen
  \bibfield  {author} {\bibinfo {author} {\bibnamefont {Newman}, \bibfnamefont
  {S.~A.}}, \bibinfo {author} {\bibfnamefont {G.}~\bibnamefont {Forgacs}},
  \bibinfo {author} {\bibfnamefont {B.}~\bibnamefont {Hinner}}, \bibinfo
  {author} {\bibfnamefont {C.~W.}\ \bibnamefont {Maier}}, \ and\ \bibinfo
  {author} {\bibfnamefont {E.}~\bibnamefont {Sackmann}}} (\bibinfo {year}
  {2004}),\ \href {http://stacks.iop.org/1478-3975/1/i=2/a=006} {\bibfield
  {journal} {\bibinfo  {journal} {Phys. Biol.}\ }\textbf {\bibinfo {volume}
  {1}}~(\bibinfo {number} {2}),\ \bibinfo {pages} {100}}\BibitemShut {NoStop}%
\bibitem [{\citenamefont {Nonnenmacher}\ \emph {et~al.}(2017)\citenamefont
  {Nonnenmacher}, \citenamefont {Behrens}, \citenamefont {Berens},
  \citenamefont {Bethge},\ and\ \citenamefont {Macke}}]{Macke2017}%
  \BibitemOpen
  \bibfield  {author} {\bibinfo {author} {\bibnamefont {Nonnenmacher},
  \bibfnamefont {M.}}, \bibinfo {author} {\bibfnamefont {C.}~\bibnamefont
  {Behrens}}, \bibinfo {author} {\bibfnamefont {P.}~\bibnamefont {Berens}},
  \bibinfo {author} {\bibfnamefont {M.}~\bibnamefont {Bethge}}, \ and\ \bibinfo
  {author} {\bibfnamefont {J.~H.}\ \bibnamefont {Macke}}} (\bibinfo {year}
  {2017}),\ \href@noop {} {\bibfield  {journal} {\bibinfo  {journal} {PLoS
  Comp. Biol.}\ }\textbf {\bibinfo {volume} {13}}~(\bibinfo {number} {10}),\
  \bibinfo {pages} {e1005718}}\BibitemShut {NoStop}%
\bibitem [{\citenamefont {Novikov}\ \emph {et~al.}(1997)\citenamefont
  {Novikov}, \citenamefont {Novikov}, \citenamefont {Shannahoff-Khalsa},
  \citenamefont {Schwartz},\ and\ \citenamefont {Wright}}]{Novikov}%
  \BibitemOpen
  \bibfield  {author} {\bibinfo {author} {\bibnamefont {Novikov}, \bibfnamefont
  {E.}}, \bibinfo {author} {\bibfnamefont {A.}~\bibnamefont {Novikov}},
  \bibinfo {author} {\bibfnamefont {D.}~\bibnamefont {Shannahoff-Khalsa}},
  \bibinfo {author} {\bibfnamefont {B.}~\bibnamefont {Schwartz}}, \ and\
  \bibinfo {author} {\bibfnamefont {J.}~\bibnamefont {Wright}}} (\bibinfo
  {year} {1997}),\ \href@noop {} {\bibfield  {journal} {\bibinfo  {journal}
  {Phys. Rev. E}\ }\textbf {\bibinfo {volume} {56}}~(\bibinfo {number} {3}),\
  \bibinfo {pages} {R2387}}\BibitemShut {NoStop}%
\bibitem [{\citenamefont {Nowak}(2006)}]{Nowak-book}%
  \BibitemOpen
  \bibfield  {author} {\bibinfo {author} {\bibnamefont {Nowak}, \bibfnamefont
  {M.~A.}}} (\bibinfo {year} {2006}),\ \href@noop {} {\emph {\bibinfo {title}
  {Evolutionary dynamics}}}\ (\bibinfo  {publisher} {Harvard University
  Press})\BibitemShut {NoStop}%
\bibitem [{\citenamefont {Nykter}\ \emph
  {et~al.}(2008{\natexlab{a}})\citenamefont {Nykter}, \citenamefont {Price},
  \citenamefont {Aldana}, \citenamefont {Ramsey}, \citenamefont {Kauffman},
  \citenamefont {Hood}, \citenamefont {Yli-Harja},\ and\ \citenamefont
  {Shmulevich}}]{Nykter1}%
  \BibitemOpen
  \bibfield  {author} {\bibinfo {author} {\bibnamefont {Nykter}, \bibfnamefont
  {M.}}, \bibinfo {author} {\bibfnamefont {N.~D.}\ \bibnamefont {Price}},
  \bibinfo {author} {\bibfnamefont {M.}~\bibnamefont {Aldana}}, \bibinfo
  {author} {\bibfnamefont {S.~A.}\ \bibnamefont {Ramsey}}, \bibinfo {author}
  {\bibfnamefont {S.~A.}\ \bibnamefont {Kauffman}}, \bibinfo {author}
  {\bibfnamefont {L.~E.}\ \bibnamefont {Hood}}, \bibinfo {author}
  {\bibfnamefont {O.}~\bibnamefont {Yli-Harja}}, \ and\ \bibinfo {author}
  {\bibfnamefont {I.}~\bibnamefont {Shmulevich}}} (\bibinfo {year}
  {2008}{\natexlab{a}}),\ \href@noop {} {\bibfield  {journal} {\bibinfo
  {journal} {Proc. Natl. Acad. Sci. USA.}\ }\textbf {\bibinfo {volume}
  {105}}~(\bibinfo {number} {6}),\ \bibinfo {pages} {1897}}\BibitemShut
  {NoStop}%
\bibitem [{\citenamefont {Nykter}\ \emph
  {et~al.}(2008{\natexlab{b}})\citenamefont {Nykter}, \citenamefont {Price},
  \citenamefont {Larjo}, \citenamefont {Aho}, \citenamefont {Kauffman},
  \citenamefont {Yli-Harja},\ and\ \citenamefont {Shmulevich}}]{Nykter2}%
  \BibitemOpen
  \bibfield  {author} {\bibinfo {author} {\bibnamefont {Nykter}, \bibfnamefont
  {M.}}, \bibinfo {author} {\bibfnamefont {N.~D.}\ \bibnamefont {Price}},
  \bibinfo {author} {\bibfnamefont {A.}~\bibnamefont {Larjo}}, \bibinfo
  {author} {\bibfnamefont {T.}~\bibnamefont {Aho}}, \bibinfo {author}
  {\bibfnamefont {S.~A.}\ \bibnamefont {Kauffman}}, \bibinfo {author}
  {\bibfnamefont {O.}~\bibnamefont {Yli-Harja}}, \ and\ \bibinfo {author}
  {\bibfnamefont {I.}~\bibnamefont {Shmulevich}}} (\bibinfo {year}
  {2008}{\natexlab{b}}),\ \href@noop {} {\bibfield  {journal} {\bibinfo
  {journal} {Phys. Rev. Lett}\ }\textbf {\bibinfo {volume} {100}}~(\bibinfo
  {number} {5}),\ \bibinfo {pages} {058702}}\BibitemShut {NoStop}%
\bibitem [{\citenamefont {{\'O}dor}(2008)}]{Odor}%
  \BibitemOpen
  \bibfield  {author} {\bibinfo {author} {\bibnamefont {{\'O}dor},
  \bibfnamefont {G.}}} (\bibinfo {year} {2008}),\ \href@noop {} {\emph
  {\bibinfo {title} {Universality in Nonequilibrium Lattice Systems:
  Theoretical Foundations}}}\ (\bibinfo  {publisher} {World Scientific},\
  \bibinfo {address} {Singapore})\BibitemShut {NoStop}%
\bibitem [{\citenamefont {Olami}\ \emph {et~al.}(1992)\citenamefont {Olami},
  \citenamefont {Feder},\ and\ \citenamefont {Christensen}}]{OFC}%
  \BibitemOpen
  \bibfield  {author} {\bibinfo {author} {\bibnamefont {Olami}, \bibfnamefont
  {Z.}}, \bibinfo {author} {\bibfnamefont {H.~J.~S.}\ \bibnamefont {Feder}}, \
  and\ \bibinfo {author} {\bibfnamefont {K.}~\bibnamefont {Christensen}}}
  (\bibinfo {year} {1992}),\ \href@noop {} {\bibfield  {journal} {\bibinfo
  {journal} {Phys. Rev. Lett.}\ }\textbf {\bibinfo {volume} {68}}~(\bibinfo
  {number} {8}),\ \bibinfo {pages} {1244}}\BibitemShut {NoStop}%
\bibitem [{\citenamefont {Oprisa}\ and\ \citenamefont
  {Toth}(2017{\natexlab{a}})}]{Toth1}%
  \BibitemOpen
  \bibfield  {author} {\bibinfo {author} {\bibnamefont {Oprisa}, \bibfnamefont
  {D.}}, \ and\ \bibinfo {author} {\bibfnamefont {P.}~\bibnamefont {Toth}}}
  (\bibinfo {year} {2017}{\natexlab{a}}),\ \href@noop {} {\bibinfo  {journal}
  {arXiv:1702.08039}\ }\BibitemShut {NoStop}%
\bibitem [{\citenamefont {Oprisa}\ and\ \citenamefont
  {Toth}(2017{\natexlab{b}})}]{Toth2}%
  \BibitemOpen
\bibfield  {journal} {  }\bibfield  {author} {\bibinfo {author} {\bibnamefont
  {Oprisa}, \bibfnamefont {D.}}, \ and\ \bibinfo {author} {\bibfnamefont
  {P.}~\bibnamefont {Toth}}} (\bibinfo {year} {2017}{\natexlab{b}}),\
  \href@noop {} {\bibinfo  {journal} {arXiv:1705.11023}\ }\BibitemShut
  {NoStop}%
\bibitem [{\citenamefont {Ospeck}\ \emph {et~al.}(2001)\citenamefont {Ospeck},
  \citenamefont {Egu{\'\i}luz},\ and\ \citenamefont {Magnasco}}]{Eguiluz2}%
  \BibitemOpen
\bibfield  {journal} {  }\bibfield  {author} {\bibinfo {author} {\bibnamefont
  {Ospeck}, \bibfnamefont {M.}}, \bibinfo {author} {\bibfnamefont {V.~M.}\
  \bibnamefont {Egu{\'\i}luz}}, \ and\ \bibinfo {author} {\bibfnamefont
  {M.~O.}\ \bibnamefont {Magnasco}}} (\bibinfo {year} {2001}),\ \href@noop {}
  {\bibfield  {journal} {\bibinfo  {journal} {Biophy. Jour.}\ }\textbf
  {\bibinfo {volume} {80}}~(\bibinfo {number} {6}),\ \bibinfo {pages}
  {2597}}\BibitemShut {NoStop}%
\bibitem [{\citenamefont {Packard}(1988)}]{Packard1988}%
  \BibitemOpen
  \bibfield  {author} {\bibinfo {author} {\bibnamefont {Packard}, \bibfnamefont
  {N.~H.}}} (\bibinfo {year} {1988}),\ \href@noop {} {\emph {\bibinfo {title}
  {Adaptation toward the edge of chaos}}}\ (\bibinfo  {publisher} {University
  of Illinois at Urbana-Champaign, Center for Complex Systems
  Research})\BibitemShut {NoStop}%
\bibitem [{\citenamefont {Pal}\ \emph {et~al.}(2014)\citenamefont {Pal},
  \citenamefont {Ghosh},\ and\ \citenamefont {Bose}}]{Bose1}%
  \BibitemOpen
  \bibfield  {author} {\bibinfo {author} {\bibnamefont {Pal}, \bibfnamefont
  {M.}}, \bibinfo {author} {\bibfnamefont {S.}~\bibnamefont {Ghosh}}, \ and\
  \bibinfo {author} {\bibfnamefont {I.}~\bibnamefont {Bose}}} (\bibinfo {year}
  {2014}),\ \href@noop {} {\bibfield  {journal} {\bibinfo  {journal} {Phys.
  Biol.}\ }\textbf {\bibinfo {volume} {12}}~(\bibinfo {number} {1}),\ \bibinfo
  {pages} {016001}}\BibitemShut {NoStop}%
\bibitem [{\citenamefont {Palmigiano}\ \emph {et~al.}(2017)\citenamefont
  {Palmigiano}, \citenamefont {Geisel}, \citenamefont {Wolf},\ and\
  \citenamefont {Battaglia}}]{Battaglia-synchro}%
  \BibitemOpen
  \bibfield  {author} {\bibinfo {author} {\bibnamefont {Palmigiano},
  \bibfnamefont {A.}}, \bibinfo {author} {\bibfnamefont {T.}~\bibnamefont
  {Geisel}}, \bibinfo {author} {\bibfnamefont {F.}~\bibnamefont {Wolf}}, \ and\
  \bibinfo {author} {\bibfnamefont {D.}~\bibnamefont {Battaglia}}} (\bibinfo
  {year} {2017}),\ \href@noop {} {\bibfield  {journal} {\bibinfo  {journal}
  {Nature Neurosci.}\ }\textbf {\bibinfo {volume} {20}}~(\bibinfo {number}
  {7}),\ \bibinfo {pages} {1014}}\BibitemShut {NoStop}%
\bibitem [{\citenamefont {Palva}\ \emph {et~al.}(2013)\citenamefont {Palva},
  \citenamefont {Zhigalov}, \citenamefont {Hirvonen}, \citenamefont {Korhonen},
  \citenamefont {Linkenkaer-Hansen},\ and\ \citenamefont {Palva}}]{Palva2013}%
  \BibitemOpen
  \bibfield  {author} {\bibinfo {author} {\bibnamefont {Palva}, \bibfnamefont
  {J.~M.}}, \bibinfo {author} {\bibfnamefont {A.}~\bibnamefont {Zhigalov}},
  \bibinfo {author} {\bibfnamefont {J.}~\bibnamefont {Hirvonen}}, \bibinfo
  {author} {\bibfnamefont {O.}~\bibnamefont {Korhonen}}, \bibinfo {author}
  {\bibfnamefont {K.}~\bibnamefont {Linkenkaer-Hansen}}, \ and\ \bibinfo
  {author} {\bibfnamefont {S.}~\bibnamefont {Palva}}} (\bibinfo {year}
  {2013}),\ \href@noop {} {\bibfield  {journal} {\bibinfo  {journal} {Proc.
  Nat. Acad. Sci. USA}\ }\textbf {\bibinfo {volume} {110}}~(\bibinfo {number}
  {9}),\ \bibinfo {pages} {3585}}\BibitemShut {NoStop}%
\bibitem [{\citenamefont {Parfitt}\ and\ \citenamefont {Shen}(2014)}]{Red2}%
  \BibitemOpen
  \bibfield  {author} {\bibinfo {author} {\bibnamefont {Parfitt}, \bibfnamefont
  {D.-E.}}, \ and\ \bibinfo {author} {\bibfnamefont {M.~M.}\ \bibnamefont
  {Shen}}} (\bibinfo {year} {2014}),\ \href@noop {} {\bibfield  {journal}
  {\bibinfo  {journal} {Phil. Trans. R. Soc. B}\ }\textbf {\bibinfo {volume}
  {369}}~(\bibinfo {number} {1657}),\ \bibinfo {pages} {20130542}}\BibitemShut
  {NoStop}%
\bibitem [{\citenamefont {Parga}\ and\ \citenamefont {Abbott}(2007)}]{Parga}%
  \BibitemOpen
  \bibfield  {author} {\bibinfo {author} {\bibnamefont {Parga}, \bibfnamefont
  {N.}}, \ and\ \bibinfo {author} {\bibfnamefont {L.~F.}\ \bibnamefont
  {Abbott}}} (\bibinfo {year} {2007}),\ \href@noop {} {\bibfield  {journal}
  {\bibinfo  {journal} {Front. Neurosci.}\ }\textbf {\bibinfo {volume} {1}},\
  \bibinfo {pages} {4}}\BibitemShut {NoStop}%
\bibitem [{\citenamefont {Parisi}(1993)}]{Parisi}%
  \BibitemOpen
  \bibfield  {author} {\bibinfo {author} {\bibnamefont {Parisi}, \bibfnamefont
  {G.}}} (\bibinfo {year} {1993}),\ \href
  {http://stacks.iop.org/2058-7058/6/i=9/a=35} {\bibfield  {journal} {\bibinfo
  {journal} {Physics World}\ }\textbf {\bibinfo {volume} {6}}~(\bibinfo
  {number} {9}),\ \bibinfo {pages} {42}}\BibitemShut {NoStop}%
\bibitem [{\citenamefont {Pasquale}\ \emph {et~al.}(2008)\citenamefont
  {Pasquale}, \citenamefont {Massobrio}, \citenamefont {Bologna}, \citenamefont
  {Chiappalone},\ and\ \citenamefont {Martinoia}}]{Pasquale2008}%
  \BibitemOpen
  \bibfield  {author} {\bibinfo {author} {\bibnamefont {Pasquale},
  \bibfnamefont {V.}}, \bibinfo {author} {\bibfnamefont {P.}~\bibnamefont
  {Massobrio}}, \bibinfo {author} {\bibfnamefont {L.}~\bibnamefont {Bologna}},
  \bibinfo {author} {\bibfnamefont {M.}~\bibnamefont {Chiappalone}}, \ and\
  \bibinfo {author} {\bibfnamefont {S.}~\bibnamefont {Martinoia}}} (\bibinfo
  {year} {2008}),\ \href@noop {} {\bibfield  {journal} {\bibinfo  {journal}
  {Neuroscience}\ }\textbf {\bibinfo {volume} {153}}~(\bibinfo {number} {4}),\
  \bibinfo {pages} {1354}}\BibitemShut {NoStop}%
\bibitem [{\citenamefont {Pastor-Satorras}\ \emph {et~al.}(2015)\citenamefont
  {Pastor-Satorras}, \citenamefont {Castellano}, \citenamefont {Van~Mieghem},\
  and\ \citenamefont {Vespignani}}]{Romu-Review}%
  \BibitemOpen
  \bibfield  {author} {\bibinfo {author} {\bibnamefont {Pastor-Satorras},
  \bibfnamefont {R.}}, \bibinfo {author} {\bibfnamefont {C.}~\bibnamefont
  {Castellano}}, \bibinfo {author} {\bibfnamefont {P.}~\bibnamefont
  {Van~Mieghem}}, \ and\ \bibinfo {author} {\bibfnamefont {A.}~\bibnamefont
  {Vespignani}}} (\bibinfo {year} {2015}),\ \href@noop {} {\bibfield  {journal}
  {\bibinfo  {journal} {Rev. Mod. Phys.}\ }\textbf {\bibinfo {volume}
  {87}}~(\bibinfo {number} {3}),\ \bibinfo {pages} {925}}\BibitemShut {NoStop}%
\bibitem [{\citenamefont {Pearlmutter}\ and\ \citenamefont
  {Houghton}(2009)}]{sleep}%
  \BibitemOpen
  \bibfield  {author} {\bibinfo {author} {\bibnamefont {Pearlmutter},
  \bibfnamefont {B.~A.}}, \ and\ \bibinfo {author} {\bibfnamefont {C.~J.}\
  \bibnamefont {Houghton}}} (\bibinfo {year} {2009}),\ \href@noop {} {\bibfield
   {journal} {\bibinfo  {journal} {Neural Comput.}\ }\textbf {\bibinfo {volume}
  {21}}~(\bibinfo {number} {6}),\ \bibinfo {pages} {1622}}\BibitemShut
  {NoStop}%
\bibitem [{\citenamefont {Perotti}\ \emph {et~al.}(2009)\citenamefont
  {Perotti}, \citenamefont {Billoni}, \citenamefont {Tamarit}, \citenamefont
  {Chialvo},\ and\ \citenamefont {Cannas}}]{Perotti}%
  \BibitemOpen
  \bibfield  {author} {\bibinfo {author} {\bibnamefont {Perotti}, \bibfnamefont
  {J.~I.}}, \bibinfo {author} {\bibfnamefont {O.~V.}\ \bibnamefont {Billoni}},
  \bibinfo {author} {\bibfnamefont {F.~A.}\ \bibnamefont {Tamarit}}, \bibinfo
  {author} {\bibfnamefont {D.~R.}\ \bibnamefont {Chialvo}}, \ and\ \bibinfo
  {author} {\bibfnamefont {S.~A.}\ \bibnamefont {Cannas}}} (\bibinfo {year}
  {2009}),\ \href@noop {} {\bibfield  {journal} {\bibinfo  {journal} {Phys.
  Rev. Lett.}\ }\textbf {\bibinfo {volume} {103}}~(\bibinfo {number} {10}),\
  \bibinfo {pages} {108701}}\BibitemShut {NoStop}%
\bibitem [{\citenamefont {Peruani}\ \emph {et~al.}(2012)\citenamefont
  {Peruani}, \citenamefont {Starru\ss{}}, \citenamefont {Jakovljevic},
  \citenamefont {S\o{}gaard-Andersen}, \citenamefont {Deutsch},\ and\
  \citenamefont {B\"ar}}]{Peruani}%
  \BibitemOpen
  \bibfield  {author} {\bibinfo {author} {\bibnamefont {Peruani}, \bibfnamefont
  {F.}}, \bibinfo {author} {\bibfnamefont {J.}~\bibnamefont {Starru\ss{}}},
  \bibinfo {author} {\bibfnamefont {V.}~\bibnamefont {Jakovljevic}}, \bibinfo
  {author} {\bibfnamefont {L.}~\bibnamefont {S\o{}gaard-Andersen}}, \bibinfo
  {author} {\bibfnamefont {A.}~\bibnamefont {Deutsch}}, \ and\ \bibinfo
  {author} {\bibfnamefont {M.}~\bibnamefont {B\"ar}}} (\bibinfo {year}
  {2012}),\ \href {\doibase 10.1103/PhysRevLett.108.098102} {\bibfield
  {journal} {\bibinfo  {journal} {Phys. Rev. Lett.}\ }\textbf {\bibinfo
  {volume} {108}},\ \bibinfo {pages} {098102}}\BibitemShut {NoStop}%
\bibitem [{\citenamefont {Petermann}\ \emph {et~al.}(2009)\citenamefont
  {Petermann}, \citenamefont {Thiagarajan}, \citenamefont {Lebedev},
  \citenamefont {Nicolelis}, \citenamefont {Chialvo},\ and\ \citenamefont
  {Plenz}}]{Peterman2009}%
  \BibitemOpen
  \bibfield  {author} {\bibinfo {author} {\bibnamefont {Petermann},
  \bibfnamefont {T.}}, \bibinfo {author} {\bibfnamefont {T.~A.}\ \bibnamefont
  {Thiagarajan}}, \bibinfo {author} {\bibfnamefont {M.}~\bibnamefont
  {Lebedev}}, \bibinfo {author} {\bibfnamefont {M.}~\bibnamefont {Nicolelis}},
  \bibinfo {author} {\bibfnamefont {D.~R.}\ \bibnamefont {Chialvo}}, \ and\
  \bibinfo {author} {\bibfnamefont {D.}~\bibnamefont {Plenz}}} (\bibinfo {year}
  {2009}),\ \href@noop {} {\bibfield  {journal} {\bibinfo  {journal} {Proc.
  Natl. Acad. Sci. USA}\ }\textbf {\bibinfo {volume} {106}}~(\bibinfo {number}
  {37}),\ \bibinfo {pages} {15921}}\BibitemShut {NoStop}%
\bibitem [{\citenamefont {Phillips}(2009)}]{Phillips2009}%
  \BibitemOpen
  \bibfield  {author} {\bibinfo {author} {\bibnamefont {Phillips},
  \bibfnamefont {J.}}} (\bibinfo {year} {2009}),\ \href@noop {} {\bibfield
  {journal} {\bibinfo  {journal} {Proc. Natl. Acad. Sci. USA.}\ }\textbf
  {\bibinfo {volume} {106}}~(\bibinfo {number} {9}),\ \bibinfo {pages}
  {3107}}\BibitemShut {NoStop}%
\bibitem [{\citenamefont {Pietronero}\ \emph {et~al.}(2001)\citenamefont
  {Pietronero}, \citenamefont {Tosatti}, \citenamefont {Tosatti},\ and\
  \citenamefont {Vespignani}}]{Pietronero}%
  \BibitemOpen
  \bibfield  {author} {\bibinfo {author} {\bibnamefont {Pietronero},
  \bibfnamefont {L.}}, \bibinfo {author} {\bibfnamefont {E.}~\bibnamefont
  {Tosatti}}, \bibinfo {author} {\bibfnamefont {V.}~\bibnamefont {Tosatti}}, \
  and\ \bibinfo {author} {\bibfnamefont {A.}~\bibnamefont {Vespignani}}}
  (\bibinfo {year} {2001}),\ \href@noop {} {\bibfield  {journal} {\bibinfo
  {journal} {Physica A}\ }\textbf {\bibinfo {volume} {293}}~(\bibinfo {number}
  {1}),\ \bibinfo {pages} {297}}\BibitemShut {NoStop}%
\bibitem [{\citenamefont {Pikovsky}\ \emph {et~al.}(2003)\citenamefont
  {Pikovsky}, \citenamefont {Rosenblum},\ and\ \citenamefont
  {Kurths}}]{synchro-book}%
  \BibitemOpen
  \bibfield  {author} {\bibinfo {author} {\bibnamefont {Pikovsky},
  \bibfnamefont {A.}}, \bibinfo {author} {\bibfnamefont {M.}~\bibnamefont
  {Rosenblum}}, \ and\ \bibinfo {author} {\bibfnamefont {J.}~\bibnamefont
  {Kurths}}} (\bibinfo {year} {2003}),\ \href@noop {} {\emph {\bibinfo {title}
  {Synchronization: a universal concept in nonlinear sciences}}},\
  Vol.~\bibinfo {volume} {12}\ (\bibinfo  {publisher} {Cambridge university
  press})\BibitemShut {NoStop}%
\bibitem [{\citenamefont {Pinto}\ and\ \citenamefont {Munoz}(2011)}]{Pinto}%
  \BibitemOpen
  \bibfield  {author} {\bibinfo {author} {\bibnamefont {Pinto}, \bibfnamefont
  {O.~A.}}, \ and\ \bibinfo {author} {\bibfnamefont {M.~A.}\ \bibnamefont
  {Munoz}}} (\bibinfo {year} {2011}),\ \href@noop {} {\bibfield  {journal}
  {\bibinfo  {journal} {PloS One}\ }\textbf {\bibinfo {volume} {6}}~(\bibinfo
  {number} {7}),\ \bibinfo {pages} {e21946}}\BibitemShut {NoStop}%
\bibitem [{\citenamefont {Plenz}(2013)}]{Plenz-Physics}%
  \BibitemOpen
  \bibfield  {author} {\bibinfo {author} {\bibnamefont {Plenz}, \bibfnamefont
  {D.}}} (\bibinfo {year} {2013}),\ \href@noop {} {\bibfield  {journal}
  {\bibinfo  {journal} {Physics}\ }\textbf {\bibinfo {volume} {6}},\ \bibinfo
  {pages} {47}}\BibitemShut {NoStop}%
\bibitem [{\citenamefont {Plenz}\ and\ \citenamefont
  {Niebur}(2014)}]{Schuster}%
  \BibitemOpen
  \bibfield  {author} {\bibinfo {author} {\bibnamefont {Plenz}, \bibfnamefont
  {D.}}, \ and\ \bibinfo {author} {\bibfnamefont {E.}~\bibnamefont {Niebur}}}
  (\bibinfo {year} {2014}),\ \href@noop {} {\emph {\bibinfo {title}
  {Criticality in neural systems}}}\ (\bibinfo  {publisher} {John Wiley \&
  Sons})\BibitemShut {NoStop}%
\bibitem [{\citenamefont {Plenz}\ and\ \citenamefont
  {Thiagarajan}(2007)}]{Plenz2007}%
  \BibitemOpen
  \bibfield  {author} {\bibinfo {author} {\bibnamefont {Plenz}, \bibfnamefont
  {D.}}, \ and\ \bibinfo {author} {\bibfnamefont {T.~C.}\ \bibnamefont
  {Thiagarajan}}} (\bibinfo {year} {2007}),\ \href@noop {} {\bibfield
  {journal} {\bibinfo  {journal} {Trends Neurosci}\ }\textbf {\bibinfo {volume}
  {30}}~(\bibinfo {number} {3}),\ \bibinfo {pages} {101 }}\BibitemShut
  {NoStop}%
\bibitem [{\citenamefont {Poblanno-Balp}\ and\ \citenamefont
  {Gershenson}(2011)}]{RBN-modular}%
  \BibitemOpen
  \bibfield  {author} {\bibinfo {author} {\bibnamefont {Poblanno-Balp},
  \bibfnamefont {R.}}, \ and\ \bibinfo {author} {\bibfnamefont
  {C.}~\bibnamefont {Gershenson}}} (\bibinfo {year} {2011}),\ \href@noop {}
  {\bibfield  {journal} {\bibinfo  {journal} {Artificial Life}\ }\textbf
  {\bibinfo {volume} {17}}~(\bibinfo {number} {4}),\ \bibinfo {pages}
  {331}}\BibitemShut {NoStop}%
\bibitem [{\citenamefont {Poil}\ \emph {et~al.}(2012)\citenamefont {Poil},
  \citenamefont {Hardstone}, \citenamefont {Mansvelder},\ and\ \citenamefont
  {Linkenkaer-Hansen}}]{Linken2012}%
  \BibitemOpen
  \bibfield  {author} {\bibinfo {author} {\bibnamefont {Poil}, \bibfnamefont
  {S.-S.}}, \bibinfo {author} {\bibfnamefont {R.}~\bibnamefont {Hardstone}},
  \bibinfo {author} {\bibfnamefont {H.~D.}\ \bibnamefont {Mansvelder}}, \ and\
  \bibinfo {author} {\bibfnamefont {K.}~\bibnamefont {Linkenkaer-Hansen}}}
  (\bibinfo {year} {2012}),\ \href@noop {} {\bibfield  {journal} {\bibinfo
  {journal} {J. Neurosci.}\ }\textbf {\bibinfo {volume} {32}}~(\bibinfo
  {number} {29}),\ \bibinfo {pages} {9817}}\BibitemShut {NoStop}%
\bibitem [{\citenamefont {Poland}\ and\ \citenamefont
  {Scheraga}(1970)}]{Poland}%
  \BibitemOpen
  \bibfield  {author} {\bibinfo {author} {\bibnamefont {Poland}, \bibfnamefont
  {D.}}, \ and\ \bibinfo {author} {\bibfnamefont {H.~A.}\ \bibnamefont
  {Scheraga}}} (\bibinfo {year} {1970}),\ \href@noop {} {\emph {\bibinfo
  {title} {Theory of helix-coil transitions in biopolymers}}}\ (\bibinfo
  {publisher} {Academic Press})\BibitemShut {NoStop}%
\bibitem [{\citenamefont {Pollack}\ and\ \citenamefont
  {Chin}(2008)}]{Pollack-book}%
  \BibitemOpen
  \bibfield  {author} {\bibinfo {author} {\bibnamefont {Pollack}, \bibfnamefont
  {G.~H.}}, \ and\ \bibinfo {author} {\bibfnamefont {W.-C.}\ \bibnamefont
  {Chin}}} (\bibinfo {year} {2008}),\ \href@noop {} {\emph {\bibinfo {title}
  {Phase Transitions in Cell Biology}}}\ (\bibinfo  {publisher}
  {Springer})\BibitemShut {NoStop}%
\bibitem [{\citenamefont {Priesemann}\ \emph {et~al.}(2009)\citenamefont
  {Priesemann}, \citenamefont {Munk},\ and\ \citenamefont {Wibral}}]{Viola-1}%
  \BibitemOpen
  \bibfield  {author} {\bibinfo {author} {\bibnamefont {Priesemann},
  \bibfnamefont {V.}}, \bibinfo {author} {\bibfnamefont {M.~H.}\ \bibnamefont
  {Munk}}, \ and\ \bibinfo {author} {\bibfnamefont {M.}~\bibnamefont {Wibral}}}
  (\bibinfo {year} {2009}),\ \href@noop {} {\bibfield  {journal} {\bibinfo
  {journal} {BMC Neurosci.}\ }\textbf {\bibinfo {volume} {10}}~(\bibinfo
  {number} {1}),\ \bibinfo {pages} {40}}\BibitemShut {NoStop}%
\bibitem [{\citenamefont {Priesemann}\ \emph {et~al.}(2013)\citenamefont
  {Priesemann}, \citenamefont {Valderrama}, \citenamefont {Wibral},\ and\
  \citenamefont {Le~Van~Quyen}}]{Viola-3}%
  \BibitemOpen
  \bibfield  {author} {\bibinfo {author} {\bibnamefont {Priesemann},
  \bibfnamefont {V.}}, \bibinfo {author} {\bibfnamefont {M.}~\bibnamefont
  {Valderrama}}, \bibinfo {author} {\bibfnamefont {M.}~\bibnamefont {Wibral}},
  \ and\ \bibinfo {author} {\bibfnamefont {M.}~\bibnamefont {Le~Van~Quyen}}}
  (\bibinfo {year} {2013}),\ \href@noop {} {\bibfield  {journal} {\bibinfo
  {journal} {PLoS Comput. Biol.}\ }\textbf {\bibinfo {volume} {9}}~(\bibinfo
  {number} {3}),\ \bibinfo {pages} {e1002985}}\BibitemShut {NoStop}%
\bibitem [{\citenamefont {Priesemann}\ \emph {et~al.}(2014)\citenamefont
  {Priesemann}, \citenamefont {Wibral}, \citenamefont {Valderrama},
  \citenamefont {Pr{\"o}pper}, \citenamefont {Le~Van~Quyen}, \citenamefont
  {Geisel}, \citenamefont {Triesch}, \citenamefont {Nikoli{\'c}},\ and\
  \citenamefont {Munk}}]{Viola-2}%
  \BibitemOpen
  \bibfield  {author} {\bibinfo {author} {\bibnamefont {Priesemann},
  \bibfnamefont {V.}}, \bibinfo {author} {\bibfnamefont {M.}~\bibnamefont
  {Wibral}}, \bibinfo {author} {\bibfnamefont {M.}~\bibnamefont {Valderrama}},
  \bibinfo {author} {\bibfnamefont {R.}~\bibnamefont {Pr{\"o}pper}}, \bibinfo
  {author} {\bibfnamefont {M.}~\bibnamefont {Le~Van~Quyen}}, \bibinfo {author}
  {\bibfnamefont {T.}~\bibnamefont {Geisel}}, \bibinfo {author} {\bibfnamefont
  {J.}~\bibnamefont {Triesch}}, \bibinfo {author} {\bibfnamefont
  {D.}~\bibnamefont {Nikoli{\'c}}}, \ and\ \bibinfo {author} {\bibfnamefont
  {M.~H.}\ \bibnamefont {Munk}}} (\bibinfo {year} {2014}),\ \href@noop {}
  {\bibfield  {journal} {\bibinfo  {journal} {Front. Syst. Neurosci.}\ }\textbf
  {\bibinfo {volume} {8}}}\BibitemShut {NoStop}%
\bibitem [{\citenamefont {Proekt}\ \emph {et~al.}(2012)\citenamefont {Proekt},
  \citenamefont {Banavar}, \citenamefont {Maritan},\ and\ \citenamefont
  {Pfaff}}]{Proekt}%
  \BibitemOpen
  \bibfield  {author} {\bibinfo {author} {\bibnamefont {Proekt}, \bibfnamefont
  {A.}}, \bibinfo {author} {\bibfnamefont {J.}~\bibnamefont {Banavar}},
  \bibinfo {author} {\bibfnamefont {A.}~\bibnamefont {Maritan}}, \ and\
  \bibinfo {author} {\bibfnamefont {D.}~\bibnamefont {Pfaff}}} (\bibinfo {year}
  {2012}),\ \href@noop {} {\bibfield  {journal} {\bibinfo  {journal} {Proc.
  Natl. Acad. Sci. USA.}\ }\textbf {\bibinfo {volume} {109}}~(\bibinfo {number}
  {26}),\ \bibinfo {pages} {10564}}\BibitemShut {NoStop}%
\bibitem [{\citenamefont {Prokopenko}(2013)}]{Proko}%
  \BibitemOpen
  \bibfield  {author} {\bibinfo {author} {\bibnamefont {Prokopenko},
  \bibfnamefont {M.}}} (\bibinfo {year} {2013}),\ in\ \href@noop {} {\emph
  {\bibinfo {booktitle} {ALIFE}}},\ pp.\ \bibinfo {pages}
  {140--144}\BibitemShut {NoStop}%
\bibitem [{\citenamefont {Pruessner}(2012)}]{Pruessner-book}%
  \BibitemOpen
  \bibfield  {author} {\bibinfo {author} {\bibnamefont {Pruessner},
  \bibfnamefont {G.}}} (\bibinfo {year} {2012}),\ \href@noop {} {\emph
  {\bibinfo {title} {Self-organised criticality: theory, models and
  characterisation}}}\ (\bibinfo  {publisher} {Cambridge University
  Press})\BibitemShut {NoStop}%
\bibitem [{\citenamefont {Puckett}\ and\ \citenamefont
  {Ouellette}(2014)}]{Ouellette2014}%
  \BibitemOpen
  \bibfield  {author} {\bibinfo {author} {\bibnamefont {Puckett}, \bibfnamefont
  {J.~G.}}, \ and\ \bibinfo {author} {\bibfnamefont {N.~T.}\ \bibnamefont
  {Ouellette}}} (\bibinfo {year} {2014}),\ \href@noop {} {\bibfield  {journal}
  {\bibinfo  {journal} {J. R. Soc. Interface}\ }\textbf {\bibinfo {volume}
  {11}}~(\bibinfo {number} {99}),\ \bibinfo {pages} {20140710}}\BibitemShut
  {NoStop}%
\bibitem [{\citenamefont {Rabinovich}\ \emph {et~al.}(2006)\citenamefont
  {Rabinovich}, \citenamefont {Varona}, \citenamefont {Selverston},\ and\
  \citenamefont {Abarbanel}}]{Rabinovich}%
  \BibitemOpen
  \bibfield  {author} {\bibinfo {author} {\bibnamefont {Rabinovich},
  \bibfnamefont {M.}}, \bibinfo {author} {\bibfnamefont {P.}~\bibnamefont
  {Varona}}, \bibinfo {author} {\bibfnamefont {A.}~\bibnamefont {Selverston}},
  \ and\ \bibinfo {author} {\bibfnamefont {H.}~\bibnamefont {Abarbanel}}}
  (\bibinfo {year} {2006}),\ \href@noop {} {\bibfield  {journal} {\bibinfo
  {journal} {Rev. Mod. Phys.}\ }\textbf {\bibinfo {volume} {78}},\ \bibinfo
  {pages} {1213}}\BibitemShut {NoStop}%
\bibitem [{\citenamefont {Raichle}(2011)}]{Raichle}%
  \BibitemOpen
  \bibfield  {author} {\bibinfo {author} {\bibnamefont {Raichle}, \bibfnamefont
  {M.~E.}}} (\bibinfo {year} {2011}),\ \href@noop {} {\bibfield  {journal}
  {\bibinfo  {journal} {Brain connectivity}\ }\textbf {\bibinfo {volume}
  {1}}~(\bibinfo {number} {1}),\ \bibinfo {pages} {3}}\BibitemShut {NoStop}%
\bibitem [{\citenamefont {Raichle}\ \emph {et~al.}(2001)\citenamefont
  {Raichle}, \citenamefont {MacLeod}, \citenamefont {Snyder}, \citenamefont
  {Powers}, \citenamefont {Gusnard},\ and\ \citenamefont {Shulman}}]{RSN2}%
  \BibitemOpen
  \bibfield  {author} {\bibinfo {author} {\bibnamefont {Raichle}, \bibfnamefont
  {M.~E.}}, \bibinfo {author} {\bibfnamefont {A.~M.}\ \bibnamefont {MacLeod}},
  \bibinfo {author} {\bibfnamefont {A.~Z.}\ \bibnamefont {Snyder}}, \bibinfo
  {author} {\bibfnamefont {W.~J.}\ \bibnamefont {Powers}}, \bibinfo {author}
  {\bibfnamefont {D.~A.}\ \bibnamefont {Gusnard}}, \ and\ \bibinfo {author}
  {\bibfnamefont {G.~L.}\ \bibnamefont {Shulman}}} (\bibinfo {year} {2001}),\
  \href@noop {} {\bibfield  {journal} {\bibinfo  {journal} {Proc. Natl. Acad.
  Sci. USA.}\ }\textbf {\bibinfo {volume} {98}}~(\bibinfo {number} {2}),\
  \bibinfo {pages} {676}}\BibitemShut {NoStop}%
\bibitem [{\citenamefont {Ramaswamy}(2010)}]{Ramaswamy}%
  \BibitemOpen
  \bibfield  {author} {\bibinfo {author} {\bibnamefont {Ramaswamy},
  \bibfnamefont {S.}}} (\bibinfo {year} {2010}),\ \href@noop {} {\bibinfo
  {journal} {Ann. Rev. Cond. Matt. Phys.}\ }\BibitemShut {NoStop}%
\bibitem [{\citenamefont {R{\"a}m{\"o}}\ \emph {et~al.}(2007)\citenamefont
  {R{\"a}m{\"o}}, \citenamefont {Kauffman}, \citenamefont {Kesseli},\ and\
  \citenamefont {Yli-Harja}}]{Ramo}%
  \BibitemOpen
\bibfield  {journal} {  }\bibfield  {author} {\bibinfo {author} {\bibnamefont
  {R{\"a}m{\"o}}, \bibfnamefont {P.}}, \bibinfo {author} {\bibfnamefont
  {S.}~\bibnamefont {Kauffman}}, \bibinfo {author} {\bibfnamefont
  {J.}~\bibnamefont {Kesseli}}, \ and\ \bibinfo {author} {\bibfnamefont
  {O.}~\bibnamefont {Yli-Harja}}} (\bibinfo {year} {2007}),\ \href@noop {}
  {\bibfield  {journal} {\bibinfo  {journal} {Physica D}\ }\textbf {\bibinfo
  {volume} {227}}~(\bibinfo {number} {1}),\ \bibinfo {pages} {100}}\BibitemShut
  {NoStop}%
\bibitem [{\citenamefont {R{\"a}m{\"o}}\ \emph {et~al.}(2006)\citenamefont
  {R{\"a}m{\"o}}, \citenamefont {Kesseli},\ and\ \citenamefont
  {Yli-Harja}}]{Ramo2}%
  \BibitemOpen
  \bibfield  {author} {\bibinfo {author} {\bibnamefont {R{\"a}m{\"o}},
  \bibfnamefont {P.}}, \bibinfo {author} {\bibfnamefont {J.}~\bibnamefont
  {Kesseli}}, \ and\ \bibinfo {author} {\bibfnamefont {O.}~\bibnamefont
  {Yli-Harja}}} (\bibinfo {year} {2006}),\ \href@noop {} {\bibfield  {journal}
  {\bibinfo  {journal} {J. Theor. Biol.}\ }\textbf {\bibinfo {volume}
  {242}}~(\bibinfo {number} {1}),\ \bibinfo {pages} {164}}\BibitemShut
  {NoStop}%
\bibitem [{\citenamefont {Ramos}\ \emph {et~al.}(2008)\citenamefont {Ramos},
  \citenamefont {L{\'o}pez}, \citenamefont {Hern{\'a}ndez-Garc{\'\i}a},\ and\
  \citenamefont {Munoz}}]{Bugs}%
  \BibitemOpen
  \bibfield  {author} {\bibinfo {author} {\bibnamefont {Ramos}, \bibfnamefont
  {F.}}, \bibinfo {author} {\bibfnamefont {C.}~\bibnamefont {L{\'o}pez}},
  \bibinfo {author} {\bibfnamefont {E.}~\bibnamefont
  {Hern{\'a}ndez-Garc{\'\i}a}}, \ and\ \bibinfo {author} {\bibfnamefont
  {M.~A.}\ \bibnamefont {Munoz}}} (\bibinfo {year} {2008}),\ \href@noop {}
  {\bibfield  {journal} {\bibinfo  {journal} {Phys. Rev. E}\ }\textbf {\bibinfo
  {volume} {77}}~(\bibinfo {number} {2}),\ \bibinfo {pages}
  {021102}}\BibitemShut {NoStop}%
\bibitem [{\citenamefont {Ravasz}\ \emph {et~al.}(2002)\citenamefont {Ravasz},
  \citenamefont {Somera}, \citenamefont {Mongru}, \citenamefont {Oltvai},\ and\
  \citenamefont {Barab{\'a}si}}]{Bara-meta}%
  \BibitemOpen
  \bibfield  {author} {\bibinfo {author} {\bibnamefont {Ravasz}, \bibfnamefont
  {E.}}, \bibinfo {author} {\bibfnamefont {A.~L.}\ \bibnamefont {Somera}},
  \bibinfo {author} {\bibfnamefont {D.~A.}\ \bibnamefont {Mongru}}, \bibinfo
  {author} {\bibfnamefont {Z.~N.}\ \bibnamefont {Oltvai}}, \ and\ \bibinfo
  {author} {\bibfnamefont {A.-L.}\ \bibnamefont {Barab{\'a}si}}} (\bibinfo
  {year} {2002}),\ \href@noop {} {\bibfield  {journal} {\bibinfo  {journal}
  {science}\ }\textbf {\bibinfo {volume} {297}}~(\bibinfo {number} {5586}),\
  \bibinfo {pages} {1551}}\BibitemShut {NoStop}%
\bibitem [{\citenamefont {Redner}(2001)}]{Redner-book}%
  \BibitemOpen
  \bibfield  {author} {\bibinfo {author} {\bibnamefont {Redner}, \bibfnamefont
  {S.}}} (\bibinfo {year} {2001}),\ \href@noop {} {\emph {\bibinfo {title} {A
  guide to first-passage processes}}}\ (\bibinfo  {publisher} {Cambridge
  University Press})\BibitemShut {NoStop}%
\bibitem [{\citenamefont {Reed}\ and\ \citenamefont {Hughes}(2002)}]{Reed}%
  \BibitemOpen
  \bibfield  {author} {\bibinfo {author} {\bibnamefont {Reed}, \bibfnamefont
  {W.~J.}}, \ and\ \bibinfo {author} {\bibfnamefont {B.~D.}\ \bibnamefont
  {Hughes}}} (\bibinfo {year} {2002}),\ \href@noop {} {\bibfield  {journal}
  {\bibinfo  {journal} {Phys. Rev. E}\ }\textbf {\bibinfo {volume}
  {66}}~(\bibinfo {number} {6}),\ \bibinfo {pages} {067103}}\BibitemShut
  {NoStop}%
\bibitem [{\citenamefont {Ribeiro}\ \emph {et~al.}(2008)\citenamefont
  {Ribeiro}, \citenamefont {Kauffman}, \citenamefont {Lloyd-Price},
  \citenamefont {Samuelsson},\ and\ \citenamefont {Socolar}}]{Ribeiro2008}%
  \BibitemOpen
  \bibfield  {author} {\bibinfo {author} {\bibnamefont {Ribeiro}, \bibfnamefont
  {A.~S.}}, \bibinfo {author} {\bibfnamefont {S.~A.}\ \bibnamefont {Kauffman}},
  \bibinfo {author} {\bibfnamefont {J.}~\bibnamefont {Lloyd-Price}}, \bibinfo
  {author} {\bibfnamefont {B.}~\bibnamefont {Samuelsson}}, \ and\ \bibinfo
  {author} {\bibfnamefont {J.~E.}\ \bibnamefont {Socolar}}} (\bibinfo {year}
  {2008}),\ \href@noop {} {\bibfield  {journal} {\bibinfo  {journal} {Phys.
  Rev. E}\ }\textbf {\bibinfo {volume} {77}}~(\bibinfo {number} {1}),\ \bibinfo
  {pages} {011901}}\BibitemShut {NoStop}%
\bibitem [{\citenamefont {Ribeiro}\ \emph {et~al.}(2010)\citenamefont
  {Ribeiro}, \citenamefont {Copelli}, \citenamefont {Caixeta}, \citenamefont
  {Belchior}, \citenamefont {Chialvo}, \citenamefont {Nicolelis},\ and\
  \citenamefont {Ribeiro}}]{Ribeiro2010}%
  \BibitemOpen
  \bibfield  {author} {\bibinfo {author} {\bibnamefont {Ribeiro}, \bibfnamefont
  {T.~L.}}, \bibinfo {author} {\bibfnamefont {M.}~\bibnamefont {Copelli}},
  \bibinfo {author} {\bibfnamefont {F.}~\bibnamefont {Caixeta}}, \bibinfo
  {author} {\bibfnamefont {H.}~\bibnamefont {Belchior}}, \bibinfo {author}
  {\bibfnamefont {D.~R.}\ \bibnamefont {Chialvo}}, \bibinfo {author}
  {\bibfnamefont {M.~A.~L.}\ \bibnamefont {Nicolelis}}, \ and\ \bibinfo
  {author} {\bibfnamefont {S.}~\bibnamefont {Ribeiro}}} (\bibinfo {year}
  {2010}),\ \href {\doibase 10.1371/journal.pone.0014129} {\bibfield  {journal}
  {\bibinfo  {journal} {PLoS One}\ }\textbf {\bibinfo {volume} {5}}~(\bibinfo
  {number} {11}),\ \bibinfo {pages} {e14129}}\BibitemShut {NoStop}%
\bibitem [{\citenamefont {Richmond}\ and\ \citenamefont
  {Solomon}(2001)}]{Solomon}%
  \BibitemOpen
  \bibfield  {author} {\bibinfo {author} {\bibnamefont {Richmond},
  \bibfnamefont {P.}}, \ and\ \bibinfo {author} {\bibfnamefont
  {S.}~\bibnamefont {Solomon}}} (\bibinfo {year} {2001}),\ \href@noop {}
  {\bibfield  {journal} {\bibinfo  {journal} {Int. J. Mod. Phys. C}\ }\textbf
  {\bibinfo {volume} {12}}~(\bibinfo {number} {03}),\ \bibinfo {pages}
  {333}}\BibitemShut {NoStop}%
\bibitem [{\citenamefont {Ridden}\ \emph {et~al.}(2015)\citenamefont {Ridden},
  \citenamefont {Chang}, \citenamefont {Zygalakis},\ and\ \citenamefont
  {MacArthur}}]{Pluripotency}%
  \BibitemOpen
  \bibfield  {author} {\bibinfo {author} {\bibnamefont {Ridden}, \bibfnamefont
  {S.~J.}}, \bibinfo {author} {\bibfnamefont {H.~H.}\ \bibnamefont {Chang}},
  \bibinfo {author} {\bibfnamefont {K.~C.}\ \bibnamefont {Zygalakis}}, \ and\
  \bibinfo {author} {\bibfnamefont {B.~D.}\ \bibnamefont {MacArthur}}}
  (\bibinfo {year} {2015}),\ \href {\doibase 10.1103/PhysRevLett.115.208103}
  {\bibfield  {journal} {\bibinfo  {journal} {Phys. Rev. Lett.}\ }\textbf
  {\bibinfo {volume} {115}},\ \bibinfo {pages} {208103}}\BibitemShut {NoStop}%
\bibitem [{\citenamefont {Rieke}\ \emph {et~al.}(1995)\citenamefont {Rieke},
  \citenamefont {Bodnar},\ and\ \citenamefont {Bialek}}]{Bialek1995}%
  \BibitemOpen
  \bibfield  {author} {\bibinfo {author} {\bibnamefont {Rieke}, \bibfnamefont
  {F.}}, \bibinfo {author} {\bibfnamefont {D.}~\bibnamefont {Bodnar}}, \ and\
  \bibinfo {author} {\bibfnamefont {W.}~\bibnamefont {Bialek}}} (\bibinfo
  {year} {1995}),\ \href@noop {} {\bibfield  {journal} {\bibinfo  {journal}
  {Proc. R. Soc. London, Ser. B}\ }\textbf {\bibinfo {volume} {262}}~(\bibinfo
  {number} {1365}),\ \bibinfo {pages} {259}}\BibitemShut {NoStop}%
\bibitem [{\citenamefont {Ringel}\ and\ \citenamefont {de~Bem}(2018)}]{Ringel}%
  \BibitemOpen
  \bibfield  {author} {\bibinfo {author} {\bibnamefont {Ringel}, \bibfnamefont
  {Z.}}, \ and\ \bibinfo {author} {\bibfnamefont {R.}~\bibnamefont {de~Bem}}}
  (\bibinfo {year} {2018}),\ \href@noop {} {\bibinfo  {journal}
  {arXiv:1802.02154}\ }\BibitemShut {NoStop}%
\bibitem [{\citenamefont {Rohlf}(2008)}]{Rohlf2008}%
  \BibitemOpen
\bibfield  {journal} {  }\bibfield  {author} {\bibinfo {author} {\bibnamefont
  {Rohlf}, \bibfnamefont {T.}}} (\bibinfo {year} {2008}),\ \href@noop {}
  {\bibfield  {journal} {\bibinfo  {journal} {EPL (Europhys. Lett.)}\ }\textbf
  {\bibinfo {volume} {84}}~(\bibinfo {number} {1}),\ \bibinfo {pages}
  {10004}}\BibitemShut {NoStop}%
\bibitem [{\citenamefont {Rohlf}\ and\ \citenamefont
  {Bornholdt}(2002)}]{Rohlf2002}%
  \BibitemOpen
  \bibfield  {author} {\bibinfo {author} {\bibnamefont {Rohlf}, \bibfnamefont
  {T.}}, \ and\ \bibinfo {author} {\bibfnamefont {S.}~\bibnamefont
  {Bornholdt}}} (\bibinfo {year} {2002}),\ \href@noop {} {\bibfield  {journal}
  {\bibinfo  {journal} {Physica A}\ }\textbf {\bibinfo {volume}
  {310}}~(\bibinfo {number} {1}),\ \bibinfo {pages} {245}}\BibitemShut
  {NoStop}%
\bibitem [{\citenamefont {Rohlf}\ \emph {et~al.}(2007)\citenamefont {Rohlf},
  \citenamefont {Gulbahce},\ and\ \citenamefont {Teuscher}}]{Rohlf2007}%
  \BibitemOpen
  \bibfield  {author} {\bibinfo {author} {\bibnamefont {Rohlf}, \bibfnamefont
  {T.}}, \bibinfo {author} {\bibfnamefont {N.}~\bibnamefont {Gulbahce}}, \ and\
  \bibinfo {author} {\bibfnamefont {C.}~\bibnamefont {Teuscher}}} (\bibinfo
  {year} {2007}),\ \href@noop {} {\bibfield  {journal} {\bibinfo  {journal}
  {Phys. Rev. Lett.}\ }\textbf {\bibinfo {volume} {99}}~(\bibinfo {number}
  {24}),\ \bibinfo {pages} {248701}}\BibitemShut {NoStop}%
\bibitem [{\citenamefont {Roli}\ \emph {et~al.}(2015)\citenamefont {Roli},
  \citenamefont {Villani}, \citenamefont {Filisetti},\ and\ \citenamefont
  {Serra}}]{Serra-review}%
  \BibitemOpen
  \bibfield  {author} {\bibinfo {author} {\bibnamefont {Roli}, \bibfnamefont
  {A.}}, \bibinfo {author} {\bibfnamefont {M.}~\bibnamefont {Villani}},
  \bibinfo {author} {\bibfnamefont {A.}~\bibnamefont {Filisetti}}, \ and\
  \bibinfo {author} {\bibfnamefont {R.}~\bibnamefont {Serra}}} (\bibinfo {year}
  {2015}),\ \href@noop {} {\bibinfo  {journal} {J. Systems Sciences and
  Complexity}\ ,\ \bibinfo {pages} {1}}\BibitemShut {NoStop}%
\bibitem [{\citenamefont {Rosenbaum}\ and\ \citenamefont
  {Doiron}(2014)}]{Doiron}%
  \BibitemOpen
\bibfield  {journal} {  }\bibfield  {author} {\bibinfo {author} {\bibnamefont
  {Rosenbaum}, \bibfnamefont {R.}}, \ and\ \bibinfo {author} {\bibfnamefont
  {B.}~\bibnamefont {Doiron}}} (\bibinfo {year} {2014}),\ \href@noop {}
  {\bibfield  {journal} {\bibinfo  {journal} {Phys. Rev. X}\ }\textbf {\bibinfo
  {volume} {4}}~(\bibinfo {number} {2}),\ \bibinfo {pages}
  {021039}}\BibitemShut {NoStop}%
\bibitem [{\citenamefont {Rosenfeld}(2013)}]{Rosenfeld-cancer}%
  \BibitemOpen
  \bibfield  {author} {\bibinfo {author} {\bibnamefont {Rosenfeld},
  \bibfnamefont {S.}}} (\bibinfo {year} {2013}),\ \href@noop {} {\bibfield
  {journal} {\bibinfo  {journal} {Gene Regulation and Systems Biology}\
  }\textbf {\bibinfo {volume} {7}},\ \bibinfo {pages} {23}}\BibitemShut
  {NoStop}%
\bibitem [{\citenamefont {R{\"o}ssert}\ \emph {et~al.}(2015)\citenamefont
  {R{\"o}ssert}, \citenamefont {Dean},\ and\ \citenamefont
  {Porrill}}]{cerebellum}%
  \BibitemOpen
  \bibfield  {author} {\bibinfo {author} {\bibnamefont {R{\"o}ssert},
  \bibfnamefont {C.}}, \bibinfo {author} {\bibfnamefont {P.}~\bibnamefont
  {Dean}}, \ and\ \bibinfo {author} {\bibfnamefont {J.}~\bibnamefont
  {Porrill}}} (\bibinfo {year} {2015}),\ \href@noop {} {\bibfield  {journal}
  {\bibinfo  {journal} {PLoS Comput. Biol.}\ }\textbf {\bibinfo {volume}
  {11}}~(\bibinfo {number} {10}),\ \bibinfo {pages} {e1004515}}\BibitemShut
  {NoStop}%
\bibitem [{\citenamefont {Rubinov}\ \emph {et~al.}(2011)\citenamefont
  {Rubinov}, \citenamefont {Sporns}, \citenamefont {Thivierge},\ and\
  \citenamefont {Breakspear}}]{Rubinov}%
  \BibitemOpen
  \bibfield  {author} {\bibinfo {author} {\bibnamefont {Rubinov}, \bibfnamefont
  {M.}}, \bibinfo {author} {\bibfnamefont {O.}~\bibnamefont {Sporns}}, \bibinfo
  {author} {\bibfnamefont {J.-P.}\ \bibnamefont {Thivierge}}, \ and\ \bibinfo
  {author} {\bibfnamefont {M.}~\bibnamefont {Breakspear}}} (\bibinfo {year}
  {2011}),\ \href {\doibase 10.1371/journal.pcbi.1002038} {\bibfield  {journal}
  {\bibinfo  {journal} {PLoS Comput. Biol.}\ }\textbf {\bibinfo {volume}
  {7}}~(\bibinfo {number} {6}),\ \bibinfo {pages} {e1002038}}\BibitemShut
  {NoStop}%
\bibitem [{\citenamefont {Rybarsch}\ and\ \citenamefont
  {Bornholdt}(2014)}]{Bornholdt2012}%
  \BibitemOpen
  \bibfield  {author} {\bibinfo {author} {\bibnamefont {Rybarsch},
  \bibfnamefont {M.}}, \ and\ \bibinfo {author} {\bibfnamefont
  {S.}~\bibnamefont {Bornholdt}}} (\bibinfo {year} {2014}),\ \enquote {\bibinfo
  {title} {Self-organized criticality in neural network models},}\ in\ \href
  {\doibase 10.1002/9783527651009.ch10} {\emph {\bibinfo {booktitle}
  {Criticality in Neural Systems}}}\ (\bibinfo  {publisher} {Wiley-VCH
  Verlag})\ pp.\ \bibinfo {pages} {227--254}\BibitemShut {NoStop}%
\bibitem [{\citenamefont {Sadilek}\ and\ \citenamefont
  {Thurner}(2015)}]{Thurner-Kuramoto}%
  \BibitemOpen
  \bibfield  {author} {\bibinfo {author} {\bibnamefont {Sadilek}, \bibfnamefont
  {M.}}, \ and\ \bibinfo {author} {\bibfnamefont {S.}~\bibnamefont {Thurner}}}
  (\bibinfo {year} {2015}),\ \href@noop {} {\bibfield  {journal} {\bibinfo
  {journal} {Sci. Rep.}\ }\textbf {\bibinfo {volume} {5}}}\BibitemShut
  {NoStop}%
\bibitem [{\citenamefont {Saito}\ and\ \citenamefont {Kikuchi}(2013)}]{Saito}%
  \BibitemOpen
  \bibfield  {author} {\bibinfo {author} {\bibnamefont {Saito}, \bibfnamefont
  {N.}}, \ and\ \bibinfo {author} {\bibfnamefont {M.}~\bibnamefont {Kikuchi}}}
  (\bibinfo {year} {2013}),\ \href@noop {} {\bibfield  {journal} {\bibinfo
  {journal} {New J. Phys.}\ }\textbf {\bibinfo {volume} {15}}~(\bibinfo
  {number} {5}),\ \bibinfo {pages} {053037}}\BibitemShut {NoStop}%
\bibitem [{\citenamefont {Sanchez-Vives}\ and\ \citenamefont
  {McCormick}(2000)}]{Vives-network}%
  \BibitemOpen
  \bibfield  {author} {\bibinfo {author} {\bibnamefont {Sanchez-Vives},
  \bibfnamefont {M.~V.}}, \ and\ \bibinfo {author} {\bibfnamefont {D.~A.}\
  \bibnamefont {McCormick}}} (\bibinfo {year} {2000}),\ \href@noop {}
  {\bibfield  {journal} {\bibinfo  {journal} {Nature Neurosci.}\ }\textbf
  {\bibinfo {volume} {3}}~(\bibinfo {number} {10}),\ \bibinfo {pages}
  {1027}}\BibitemShut {NoStop}%
\bibitem [{\citenamefont {di~Santo}\ \emph {et~al.}(2016)\citenamefont
  {di~Santo}, \citenamefont {Burioni}, \citenamefont {Vezzani},\ and\
  \citenamefont {Mu{\~n}oz}}]{SOB}%
  \BibitemOpen
  \bibfield  {author} {\bibinfo {author} {\bibnamefont {di~Santo},
  \bibfnamefont {S.}}, \bibinfo {author} {\bibfnamefont {R.}~\bibnamefont
  {Burioni}}, \bibinfo {author} {\bibfnamefont {A.}~\bibnamefont {Vezzani}}, \
  and\ \bibinfo {author} {\bibfnamefont {M.~A.}\ \bibnamefont {Mu{\~n}oz}}}
  (\bibinfo {year} {2016}),\ \href@noop {} {\bibfield  {journal} {\bibinfo
  {journal} {Phys. Rev. Lett.}\ }\textbf {\bibinfo {volume} {116}}~(\bibinfo
  {number} {24}),\ \bibinfo {pages} {240601}}\BibitemShut {NoStop}%
\bibitem [{\citenamefont {di~Santo}\ \emph
  {et~al.}(2017{\natexlab{a}})\citenamefont {di~Santo}, \citenamefont
  {Villegas}, \citenamefont {Burioni},\ and\ \citenamefont
  {Mu{\~n}oz}}]{Serena-LG}%
  \BibitemOpen
  \bibfield  {author} {\bibinfo {author} {\bibnamefont {di~Santo},
  \bibfnamefont {S.}}, \bibinfo {author} {\bibfnamefont {P.}~\bibnamefont
  {Villegas}}, \bibinfo {author} {\bibfnamefont {R.}~\bibnamefont {Burioni}}, \
  and\ \bibinfo {author} {\bibfnamefont {M.~A.}\ \bibnamefont {Mu{\~n}oz}}}
  (\bibinfo {year} {2017}{\natexlab{a}}),\ \href@noop {} {\bibinfo  {journal}
  {Submitted}\ }\BibitemShut {NoStop}%
\bibitem [{\citenamefont {di~Santo}\ \emph
  {et~al.}(2017{\natexlab{b}})\citenamefont {di~Santo}, \citenamefont
  {Villegas}, \citenamefont {Burioni},\ and\ \citenamefont
  {Mu{\~n}oz}}]{Serena-BP}%
  \BibitemOpen
\bibfield  {journal} {  }\bibfield  {author} {\bibinfo {author} {\bibnamefont
  {di~Santo}, \bibfnamefont {S.}}, \bibinfo {author} {\bibfnamefont
  {P.}~\bibnamefont {Villegas}}, \bibinfo {author} {\bibfnamefont
  {R.}~\bibnamefont {Burioni}}, \ and\ \bibinfo {author} {\bibfnamefont
  {M.~A.}\ \bibnamefont {Mu{\~n}oz}}} (\bibinfo {year} {2017}{\natexlab{b}}),\
  \href@noop {} {\bibfield  {journal} {\bibinfo  {journal} {Phys. Rev. E}\
  }\textbf {\bibinfo {volume} {95}}~(\bibinfo {number} {3}),\ \bibinfo {pages}
  {032115}}\BibitemShut {NoStop}%
\bibitem [{\citenamefont {di~Santo}\ \emph {et~al.}(2018)\citenamefont
  {di~Santo}, \citenamefont {Villegas}, \citenamefont {Burioni},\ and\
  \citenamefont {Mu{\~n}oz}}]{balanced}%
  \BibitemOpen
  \bibfield  {author} {\bibinfo {author} {\bibnamefont {di~Santo},
  \bibfnamefont {S.}}, \bibinfo {author} {\bibfnamefont {P.}~\bibnamefont
  {Villegas}}, \bibinfo {author} {\bibfnamefont {R.}~\bibnamefont {Burioni}}, \
  and\ \bibinfo {author} {\bibfnamefont {M.~A.}\ \bibnamefont {Mu{\~n}oz}}}
  (\bibinfo {year} {2018}),\ \href@noop {} {\bibinfo  {journal}
  {arXiv:1803.07858}\ }\BibitemShut {NoStop}%
\bibitem [{\citenamefont {Sauer}\ \emph {et~al.}(2007)\citenamefont {Sauer},
  \citenamefont {Heinemann},\ and\ \citenamefont {Zamboni}}]{Sauer}%
  \BibitemOpen
\bibfield  {journal} {  }\bibfield  {author} {\bibinfo {author} {\bibnamefont
  {Sauer}, \bibfnamefont {U.}}, \bibinfo {author} {\bibfnamefont
  {M.}~\bibnamefont {Heinemann}}, \ and\ \bibinfo {author} {\bibfnamefont
  {N.}~\bibnamefont {Zamboni}}} (\bibinfo {year} {2007}),\ \href@noop {}
  {\bibfield  {journal} {\bibinfo  {journal} {Science}\ }\textbf {\bibinfo
  {volume} {316}}~(\bibinfo {number} {5824}),\ \bibinfo {pages}
  {550}}\BibitemShut {NoStop}%
\bibitem [{\citenamefont {Schneidman}\ \emph {et~al.}(2006)\citenamefont
  {Schneidman}, \citenamefont {Berry}, \citenamefont {Segev},\ and\
  \citenamefont {Bialek}}]{Bialek-weak}%
  \BibitemOpen
  \bibfield  {author} {\bibinfo {author} {\bibnamefont {Schneidman},
  \bibfnamefont {E.}}, \bibinfo {author} {\bibfnamefont {M.~J.}\ \bibnamefont
  {Berry}}, \bibinfo {author} {\bibfnamefont {R.}~\bibnamefont {Segev}}, \ and\
  \bibinfo {author} {\bibfnamefont {W.}~\bibnamefont {Bialek}}} (\bibinfo
  {year} {2006}),\ \href@noop {} {\bibfield  {journal} {\bibinfo  {journal}
  {Nature}\ }\textbf {\bibinfo {volume} {440}}~(\bibinfo {number} {7087}),\
  \bibinfo {pages} {1007}}\BibitemShut {NoStop}%
\bibitem [{\citenamefont {Scholz}\ \emph {et~al.}(1987)\citenamefont {Scholz},
  \citenamefont {Kelso},\ and\ \citenamefont {Sch{\"o}ner}}]{Kelso1987}%
  \BibitemOpen
  \bibfield  {author} {\bibinfo {author} {\bibnamefont {Scholz}, \bibfnamefont
  {J.}}, \bibinfo {author} {\bibfnamefont {J.}~\bibnamefont {Kelso}}, \ and\
  \bibinfo {author} {\bibfnamefont {G.}~\bibnamefont {Sch{\"o}ner}}} (\bibinfo
  {year} {1987}),\ \href@noop {} {\bibfield  {journal} {\bibinfo  {journal}
  {Phys. Lett. A}\ }\textbf {\bibinfo {volume} {123}}~(\bibinfo {number} {8}),\
  \bibinfo {pages} {390}}\BibitemShut {NoStop}%
\bibitem [{\citenamefont {Schr{\"o}dinger}(1967)}]{Schrodinger}%
  \BibitemOpen
  \bibfield  {author} {\bibinfo {author} {\bibnamefont {Schr{\"o}dinger},
  \bibfnamefont {E.}}} (\bibinfo {year} {1967}),\ \href@noop {} {\emph
  {\bibinfo {title} {What is Life?}}}\ (\bibinfo  {publisher} {Cambridge
  University Press})\BibitemShut {NoStop}%
\bibitem [{\citenamefont {Schwab}\ \emph {et~al.}(2014)\citenamefont {Schwab},
  \citenamefont {Nemenman},\ and\ \citenamefont {Mehta}}]{Mehta}%
  \BibitemOpen
  \bibfield  {author} {\bibinfo {author} {\bibnamefont {Schwab}, \bibfnamefont
  {D.~J.}}, \bibinfo {author} {\bibfnamefont {I.}~\bibnamefont {Nemenman}}, \
  and\ \bibinfo {author} {\bibfnamefont {P.}~\bibnamefont {Mehta}}} (\bibinfo
  {year} {2014}),\ \href {\doibase 10.1103/PhysRevLett.113.068102} {\bibfield
  {journal} {\bibinfo  {journal} {Phys. Rev. Lett.}\ }\textbf {\bibinfo
  {volume} {113}},\ \bibinfo {pages} {068102}}\BibitemShut {NoStop}%
\bibitem [{\citenamefont {Segev}\ and\ \citenamefont
  {Ben-Jacob}(2001)}]{Segev1}%
  \BibitemOpen
  \bibfield  {author} {\bibinfo {author} {\bibnamefont {Segev}, \bibfnamefont
  {R.}}, \ and\ \bibinfo {author} {\bibfnamefont {E.}~\bibnamefont
  {Ben-Jacob}}} (\bibinfo {year} {2001}),\ \href@noop {} {\bibfield  {journal}
  {\bibinfo  {journal} {Physica A}\ }\textbf {\bibinfo {volume}
  {302}}~(\bibinfo {number} {1}),\ \bibinfo {pages} {64}}\BibitemShut {NoStop}%
\bibitem [{\citenamefont {Segev}\ \emph {et~al.}(2001)\citenamefont {Segev},
  \citenamefont {Shapira}, \citenamefont {Benveniste},\ and\ \citenamefont
  {Ben-Jacob}}]{Segev2}%
  \BibitemOpen
  \bibfield  {author} {\bibinfo {author} {\bibnamefont {Segev}, \bibfnamefont
  {R.}}, \bibinfo {author} {\bibfnamefont {Y.}~\bibnamefont {Shapira}},
  \bibinfo {author} {\bibfnamefont {M.}~\bibnamefont {Benveniste}}, \ and\
  \bibinfo {author} {\bibfnamefont {E.}~\bibnamefont {Ben-Jacob}}} (\bibinfo
  {year} {2001}),\ \href@noop {} {\bibfield  {journal} {\bibinfo  {journal}
  {Phys. Rev. E}\ }\textbf {\bibinfo {volume} {64}}~(\bibinfo {number} {1}),\
  \bibinfo {pages} {011920}}\BibitemShut {NoStop}%
\bibitem [{\citenamefont {Sejnowski}\ \emph {et~al.}(2014)\citenamefont
  {Sejnowski}, \citenamefont {Churchland},\ and\ \citenamefont
  {Movshon}}]{Techniques}%
  \BibitemOpen
  \bibfield  {author} {\bibinfo {author} {\bibnamefont {Sejnowski},
  \bibfnamefont {T.~J.}}, \bibinfo {author} {\bibfnamefont {P.~S.}\
  \bibnamefont {Churchland}}, \ and\ \bibinfo {author} {\bibfnamefont {J.~A.}\
  \bibnamefont {Movshon}}} (\bibinfo {year} {2014}),\ \href@noop {} {\bibfield
  {journal} {\bibinfo  {journal} {Nat. Neurosci.}\ }\textbf {\bibinfo {volume}
  {17}}~(\bibinfo {number} {11}),\ \bibinfo {pages} {1440}}\BibitemShut
  {NoStop}%
\bibitem [{\citenamefont {Seoane}\ and\ \citenamefont
  {Sol{\'e}}(2015)}]{Seoane}%
  \BibitemOpen
  \bibfield  {author} {\bibinfo {author} {\bibnamefont {Seoane}, \bibfnamefont
  {L.~F.}}, \ and\ \bibinfo {author} {\bibfnamefont {R.}~\bibnamefont
  {Sol{\'e}}}} (\bibinfo {year} {2015}),\ \href@noop {} {\bibinfo  {journal}
  {arXiv:1510.08697}\ }\BibitemShut {NoStop}%
\bibitem [{\citenamefont {Serra}\ \emph {et~al.}(2007)\citenamefont {Serra},
  \citenamefont {Villani}, \citenamefont {Graudenzi},\ and\ \citenamefont
  {Kauffman}}]{Serra2007}%
  \BibitemOpen
\bibfield  {journal} {  }\bibfield  {author} {\bibinfo {author} {\bibnamefont
  {Serra}, \bibfnamefont {R.}}, \bibinfo {author} {\bibfnamefont
  {M.}~\bibnamefont {Villani}}, \bibinfo {author} {\bibfnamefont
  {A.}~\bibnamefont {Graudenzi}}, \ and\ \bibinfo {author} {\bibfnamefont
  {S.}~\bibnamefont {Kauffman}}} (\bibinfo {year} {2007}),\ \href@noop {}
  {\bibfield  {journal} {\bibinfo  {journal} {J. Theor. Biol.}\ }\textbf
  {\bibinfo {volume} {246}}~(\bibinfo {number} {3}),\ \bibinfo {pages}
  {449}}\BibitemShut {NoStop}%
\bibitem [{\citenamefont {Serra}\ \emph {et~al.}(2004)\citenamefont {Serra},
  \citenamefont {Villani},\ and\ \citenamefont {Semeria}}]{Serra2004}%
  \BibitemOpen
  \bibfield  {author} {\bibinfo {author} {\bibnamefont {Serra}, \bibfnamefont
  {R.}}, \bibinfo {author} {\bibfnamefont {M.}~\bibnamefont {Villani}}, \ and\
  \bibinfo {author} {\bibfnamefont {A.}~\bibnamefont {Semeria}}} (\bibinfo
  {year} {2004}),\ \href@noop {} {\bibfield  {journal} {\bibinfo  {journal} {J.
  Theor. Biol.}\ }\textbf {\bibinfo {volume} {227}}~(\bibinfo {number} {1}),\
  \bibinfo {pages} {149}}\BibitemShut {NoStop}%
\bibitem [{\citenamefont {Sethna}(2006)}]{Sethna-book}%
  \BibitemOpen
  \bibfield  {author} {\bibinfo {author} {\bibnamefont {Sethna}, \bibfnamefont
  {J.}}} (\bibinfo {year} {2006}),\ \href@noop {} {\emph {\bibinfo {title}
  {Statistical mechanics: entropy, order parameters, and complexity}}},\
  Vol.~\bibinfo {volume} {14}\ (\bibinfo  {publisher} {Oxford Univ.
  Press})\BibitemShut {NoStop}%
\bibitem [{\citenamefont {Sethna}\ \emph {et~al.}(2001)\citenamefont {Sethna},
  \citenamefont {Dahmen},\ and\ \citenamefont {Myers}}]{Sethna}%
  \BibitemOpen
  \bibfield  {author} {\bibinfo {author} {\bibnamefont {Sethna}, \bibfnamefont
  {J.~P.}}, \bibinfo {author} {\bibfnamefont {K.~A.}\ \bibnamefont {Dahmen}}, \
  and\ \bibinfo {author} {\bibfnamefont {C.~R.}\ \bibnamefont {Myers}}}
  (\bibinfo {year} {2001}),\ \href@noop {} {\bibfield  {journal} {\bibinfo
  {journal} {Nature}\ }\textbf {\bibinfo {volume} {410}}~(\bibinfo {number}
  {6825}),\ \bibinfo {pages} {242}}\BibitemShut {NoStop}%
\bibitem [{\citenamefont {Shanahan}(2010)}]{Shanahan2010}%
  \BibitemOpen
  \bibfield  {author} {\bibinfo {author} {\bibnamefont {Shanahan},
  \bibfnamefont {M.}}} (\bibinfo {year} {2010}),\ \href@noop {} {\bibfield
  {journal} {\bibinfo  {journal} {Chaos}\ }\textbf {\bibinfo {volume}
  {20}}~(\bibinfo {number} {1}),\ \bibinfo {pages} {013108}}\BibitemShut
  {NoStop}%
\bibitem [{\citenamefont {Shew}\ \emph {et~al.}(2015)\citenamefont {Shew},
  \citenamefont {Clawson}, \citenamefont {Pobst}, \citenamefont {Karimipanah},
  \citenamefont {Wright},\ and\ \citenamefont {Wessel}}]{Shew2015b}%
  \BibitemOpen
  \bibfield  {author} {\bibinfo {author} {\bibnamefont {Shew}, \bibfnamefont
  {W.~L.}}, \bibinfo {author} {\bibfnamefont {W.~P.}\ \bibnamefont {Clawson}},
  \bibinfo {author} {\bibfnamefont {J.}~\bibnamefont {Pobst}}, \bibinfo
  {author} {\bibfnamefont {Y.}~\bibnamefont {Karimipanah}}, \bibinfo {author}
  {\bibfnamefont {N.~C.}\ \bibnamefont {Wright}}, \ and\ \bibinfo {author}
  {\bibfnamefont {R.}~\bibnamefont {Wessel}}} (\bibinfo {year} {2015}),\
  \href@noop {} {\bibinfo  {journal} {Nat. Phys.}\ ,\ \bibinfo {pages}
  {659}}\BibitemShut {NoStop}%
\bibitem [{\citenamefont {Shew}\ and\ \citenamefont
  {Plenz}(2013)}]{Plenz-Functional}%
  \BibitemOpen
\bibfield  {journal} {  }\bibfield  {author} {\bibinfo {author} {\bibnamefont
  {Shew}, \bibfnamefont {W.~L.}}, \ and\ \bibinfo {author} {\bibfnamefont
  {D.}~\bibnamefont {Plenz}}} (\bibinfo {year} {2013}),\ \href@noop {}
  {\bibfield  {journal} {\bibinfo  {journal} {The Neuroscientist}\ }\textbf
  {\bibinfo {volume} {19}}~(\bibinfo {number} {1}),\ \bibinfo {pages}
  {88}}\BibitemShut {NoStop}%
\bibitem [{\citenamefont {Shew}\ \emph {et~al.}(2009)\citenamefont {Shew},
  \citenamefont {Yang}, \citenamefont {Petermann}, \citenamefont {Roy},\ and\
  \citenamefont {Plenz}}]{Shew2009}%
  \BibitemOpen
  \bibfield  {author} {\bibinfo {author} {\bibnamefont {Shew}, \bibfnamefont
  {W.~L.}}, \bibinfo {author} {\bibfnamefont {H.}~\bibnamefont {Yang}},
  \bibinfo {author} {\bibfnamefont {T.}~\bibnamefont {Petermann}}, \bibinfo
  {author} {\bibfnamefont {R.}~\bibnamefont {Roy}}, \ and\ \bibinfo {author}
  {\bibfnamefont {D.}~\bibnamefont {Plenz}}} (\bibinfo {year} {2009}),\ \href
  {\doibase 10.1523/JNEUROSCI.3864-09.2009} {\bibfield  {journal} {\bibinfo
  {journal} {J. Neurosci.}\ }\textbf {\bibinfo {volume} {29}}~(\bibinfo
  {number} {49}),\ \bibinfo {pages} {15595}}\BibitemShut {NoStop}%
\bibitem [{\citenamefont {Shin}\ and\ \citenamefont {Kim}(2006)}]{Shin-Kim}%
  \BibitemOpen
  \bibfield  {author} {\bibinfo {author} {\bibnamefont {Shin}, \bibfnamefont
  {C.-W.}}, \ and\ \bibinfo {author} {\bibfnamefont {S.}~\bibnamefont {Kim}}}
  (\bibinfo {year} {2006}),\ \href@noop {} {\bibfield  {journal} {\bibinfo
  {journal} {Phys. Rev. E}\ }\textbf {\bibinfo {volume} {74}}~(\bibinfo
  {number} {4}),\ \bibinfo {pages} {045101}}\BibitemShut {NoStop}%
\bibitem [{\citenamefont {Shmulevich}\ and\ \citenamefont
  {Dougherty}(2010)}]{Ilya-review}%
  \BibitemOpen
  \bibfield  {author} {\bibinfo {author} {\bibnamefont {Shmulevich},
  \bibfnamefont {I.}}, \ and\ \bibinfo {author} {\bibfnamefont {E.~R.}\
  \bibnamefont {Dougherty}}} (\bibinfo {year} {2010}),\ \href@noop {} {\emph
  {\bibinfo {title} {Probabilistic Boolean networks: the modeling and control
  of gene regulatory networks}}}\ (\bibinfo  {publisher} {SIAM})\BibitemShut
  {NoStop}%
\bibitem [{\citenamefont {Shmulevich}\ \emph {et~al.}(2005)\citenamefont
  {Shmulevich}, \citenamefont {Kauffman},\ and\ \citenamefont
  {Aldana}}]{Shmu2005}%
  \BibitemOpen
  \bibfield  {author} {\bibinfo {author} {\bibnamefont {Shmulevich},
  \bibfnamefont {I.}}, \bibinfo {author} {\bibfnamefont {S.~A.}\ \bibnamefont
  {Kauffman}}, \ and\ \bibinfo {author} {\bibfnamefont {M.}~\bibnamefont
  {Aldana}}} (\bibinfo {year} {2005}),\ \href@noop {} {\bibfield  {journal}
  {\bibinfo  {journal} {Proc. Natl. Acad. Sci. USA.}\ }\textbf {\bibinfo
  {volume} {102}}~(\bibinfo {number} {38}),\ \bibinfo {pages}
  {13439}}\BibitemShut {NoStop}%
\bibitem [{\citenamefont {Shriki}\ \emph {et~al.}(2013)\citenamefont {Shriki},
  \citenamefont {Alstott}, \citenamefont {Carver}, \citenamefont {Holroyd},
  \citenamefont {Henson}, \citenamefont {Smith}, \citenamefont {Coppola},
  \citenamefont {Bullmore},\ and\ \citenamefont {Plenz}}]{Plenz-Shriki2013}%
  \BibitemOpen
  \bibfield  {author} {\bibinfo {author} {\bibnamefont {Shriki}, \bibfnamefont
  {O.}}, \bibinfo {author} {\bibfnamefont {J.}~\bibnamefont {Alstott}},
  \bibinfo {author} {\bibfnamefont {F.}~\bibnamefont {Carver}}, \bibinfo
  {author} {\bibfnamefont {T.}~\bibnamefont {Holroyd}}, \bibinfo {author}
  {\bibfnamefont {R.~N.}\ \bibnamefont {Henson}}, \bibinfo {author}
  {\bibfnamefont {M.~L.}\ \bibnamefont {Smith}}, \bibinfo {author}
  {\bibfnamefont {R.}~\bibnamefont {Coppola}}, \bibinfo {author} {\bibfnamefont
  {E.}~\bibnamefont {Bullmore}}, \ and\ \bibinfo {author} {\bibfnamefont
  {D.}~\bibnamefont {Plenz}}} (\bibinfo {year} {2013}),\ \href@noop {}
  {\bibfield  {journal} {\bibinfo  {journal} {J. Neurosci.}\ }\textbf {\bibinfo
  {volume} {33}}~(\bibinfo {number} {16}),\ \bibinfo {pages}
  {7079}}\BibitemShut {NoStop}%
\bibitem [{\citenamefont {Shriki}\ and\ \citenamefont
  {Yellin}(2016)}]{Shriki2016}%
  \BibitemOpen
  \bibfield  {author} {\bibinfo {author} {\bibnamefont {Shriki}, \bibfnamefont
  {O.}}, \ and\ \bibinfo {author} {\bibfnamefont {D.}~\bibnamefont {Yellin}}}
  (\bibinfo {year} {2016}),\ \href@noop {} {\bibfield  {journal} {\bibinfo
  {journal} {PLoS Comput. Biol.}\ }\textbf {\bibinfo {volume} {12}}~(\bibinfo
  {number} {2}),\ \bibinfo {pages} {e1004698}}\BibitemShut {NoStop}%
\bibitem [{\citenamefont {Simini}\ \emph {et~al.}(2010)\citenamefont {Simini},
  \citenamefont {Anfodillo}, \citenamefont {Carrer}, \citenamefont {Banavar},\
  and\ \citenamefont {Maritan}}]{Simini2010}%
  \BibitemOpen
  \bibfield  {author} {\bibinfo {author} {\bibnamefont {Simini}, \bibfnamefont
  {F.}}, \bibinfo {author} {\bibfnamefont {T.}~\bibnamefont {Anfodillo}},
  \bibinfo {author} {\bibfnamefont {M.}~\bibnamefont {Carrer}}, \bibinfo
  {author} {\bibfnamefont {J.~R.}\ \bibnamefont {Banavar}}, \ and\ \bibinfo
  {author} {\bibfnamefont {A.}~\bibnamefont {Maritan}}} (\bibinfo {year}
  {2010}),\ \href@noop {} {\bibfield  {journal} {\bibinfo  {journal} {Proc.
  Natl. Acad. Sci. USA.}\ }\textbf {\bibinfo {volume} {107}}~(\bibinfo {number}
  {17}),\ \bibinfo {pages} {7658}}\BibitemShut {NoStop}%
\bibitem [{\citenamefont {Simkin}\ and\ \citenamefont
  {Roychowdhury}(2011)}]{Willis}%
  \BibitemOpen
  \bibfield  {author} {\bibinfo {author} {\bibnamefont {Simkin}, \bibfnamefont
  {M.~V.}}, \ and\ \bibinfo {author} {\bibfnamefont {V.~P.}\ \bibnamefont
  {Roychowdhury}}} (\bibinfo {year} {2011}),\ \href@noop {} {\bibfield
  {journal} {\bibinfo  {journal} {Phys. Rep.}\ }\textbf {\bibinfo {volume}
  {502}}~(\bibinfo {number} {1}),\ \bibinfo {pages} {1}}\BibitemShut {NoStop}%
\bibitem [{\citenamefont {Simon}(1955)}]{Simon}%
  \BibitemOpen
  \bibfield  {author} {\bibinfo {author} {\bibnamefont {Simon}, \bibfnamefont
  {H.~A.}}} (\bibinfo {year} {1955}),\ \href@noop {} {\bibfield  {journal}
  {\bibinfo  {journal} {Biometrika}\ }\textbf {\bibinfo {volume}
  {42}}~(\bibinfo {number} {3/4}),\ \bibinfo {pages} {425}}\BibitemShut
  {NoStop}%
\bibitem [{\citenamefont {Sneppen}(2014)}]{Sneppen-book}%
  \BibitemOpen
  \bibfield  {author} {\bibinfo {author} {\bibnamefont {Sneppen}, \bibfnamefont
  {K.}}} (\bibinfo {year} {2014}),\ \href
  {https://books.google.es/books?id=BINxBAAAQBAJ} {\emph {\bibinfo {title}
  {Models of Life}}}\ (\bibinfo  {publisher} {Cambridge Univ.
  Press})\BibitemShut {NoStop}%
\bibitem [{\citenamefont {Sneppen}\ \emph {et~al.}(1995)\citenamefont
  {Sneppen}, \citenamefont {Bak}, \citenamefont {Flyvbjerg},\ and\
  \citenamefont {Jensen}}]{Sneppen1995}%
  \BibitemOpen
  \bibfield  {author} {\bibinfo {author} {\bibnamefont {Sneppen}, \bibfnamefont
  {K.}}, \bibinfo {author} {\bibfnamefont {P.}~\bibnamefont {Bak}}, \bibinfo
  {author} {\bibfnamefont {H.}~\bibnamefont {Flyvbjerg}}, \ and\ \bibinfo
  {author} {\bibfnamefont {M.~H.}\ \bibnamefont {Jensen}}} (\bibinfo {year}
  {1995}),\ \href@noop {} {\bibfield  {journal} {\bibinfo  {journal} {Proc.
  Natl. Acad. Sci. USA}\ }\textbf {\bibinfo {volume} {92}}~(\bibinfo {number}
  {11}),\ \bibinfo {pages} {5209}}\BibitemShut {NoStop}%
\bibitem [{\citenamefont {Sokolov}\ \emph {et~al.}(2007)\citenamefont
  {Sokolov}, \citenamefont {Aranson}, \citenamefont {Kessler},\ and\
  \citenamefont {Goldstein}}]{Goldstein-bacteria}%
  \BibitemOpen
  \bibfield  {author} {\bibinfo {author} {\bibnamefont {Sokolov}, \bibfnamefont
  {A.}}, \bibinfo {author} {\bibfnamefont {I.~S.}\ \bibnamefont {Aranson}},
  \bibinfo {author} {\bibfnamefont {J.~O.}\ \bibnamefont {Kessler}}, \ and\
  \bibinfo {author} {\bibfnamefont {R.~E.}\ \bibnamefont {Goldstein}}}
  (\bibinfo {year} {2007}),\ \href@noop {} {\bibfield  {journal} {\bibinfo
  {journal} {Phys. Rev. Lett.}\ }\textbf {\bibinfo {volume} {98}}~(\bibinfo
  {number} {15}),\ \bibinfo {pages} {158102}}\BibitemShut {NoStop}%
\bibitem [{\citenamefont {Sol\'e}(2011)}]{Sole-book}%
  \BibitemOpen
  \bibfield  {author} {\bibinfo {author} {\bibnamefont {Sol\'e}, \bibfnamefont
  {R.}}} (\bibinfo {year} {2011}),\ \href@noop {} {\emph {\bibinfo {title}
  {Phase transitions}}}\ (\bibinfo  {publisher} {Princeton University
  Press})\BibitemShut {NoStop}%
\bibitem [{\citenamefont {Sol{\'e}}\ and\ \citenamefont
  {Deisboeck}(2004)}]{Sole-cancer2}%
  \BibitemOpen
  \bibfield  {author} {\bibinfo {author} {\bibnamefont {Sol{\'e}},
  \bibfnamefont {R.}}, \ and\ \bibinfo {author} {\bibfnamefont
  {T.}~\bibnamefont {Deisboeck}}} (\bibinfo {year} {2004}),\ \href@noop {}
  {\bibfield  {journal} {\bibinfo  {journal} {J. Theor. Biol.}\ }\textbf
  {\bibinfo {volume} {228}}~(\bibinfo {number} {1}),\ \bibinfo {pages}
  {47}}\BibitemShut {NoStop}%
\bibitem [{\citenamefont {Sol{\'e}}(2003)}]{Sole-cancer1}%
  \BibitemOpen
  \bibfield  {author} {\bibinfo {author} {\bibnamefont {Sol{\'e}},
  \bibfnamefont {R.~V.}}} (\bibinfo {year} {2003}),\ \href@noop {} {\bibfield
  {journal} {\bibinfo  {journal} {The European Phys. Jour. B}\ }\textbf
  {\bibinfo {volume} {35}}~(\bibinfo {number} {1}),\ \bibinfo {pages}
  {117}}\BibitemShut {NoStop}%
\bibitem [{\citenamefont {Sol{\'e}}\ \emph
  {et~al.}(2002{\natexlab{a}})\citenamefont {Sol{\'e}}, \citenamefont
  {Alonso},\ and\ \citenamefont {McKane}}]{McKane}%
  \BibitemOpen
  \bibfield  {author} {\bibinfo {author} {\bibnamefont {Sol{\'e}},
  \bibfnamefont {R.~V.}}, \bibinfo {author} {\bibfnamefont {D.}~\bibnamefont
  {Alonso}}, \ and\ \bibinfo {author} {\bibfnamefont {A.}~\bibnamefont
  {McKane}}} (\bibinfo {year} {2002}{\natexlab{a}}),\ \href@noop {} {\bibfield
  {journal} {\bibinfo  {journal} {Philos. Trans. R. Soc. London, Ser. B}\
  }\textbf {\bibinfo {volume} {357}}~(\bibinfo {number} {1421}),\ \bibinfo
  {pages} {667}}\BibitemShut {NoStop}%
\bibitem [{\citenamefont {Sol{\'e}}\ \emph
  {et~al.}(2002{\natexlab{b}})\citenamefont {Sol{\'e}}, \citenamefont
  {Ferrer-Cancho}, \citenamefont {Montoya},\ and\ \citenamefont
  {Valverde}}]{Sole-adaptive}%
  \BibitemOpen
  \bibfield  {author} {\bibinfo {author} {\bibnamefont {Sol{\'e}},
  \bibfnamefont {R.~V.}}, \bibinfo {author} {\bibfnamefont {R.}~\bibnamefont
  {Ferrer-Cancho}}, \bibinfo {author} {\bibfnamefont {J.~M.}\ \bibnamefont
  {Montoya}}, \ and\ \bibinfo {author} {\bibfnamefont {S.}~\bibnamefont
  {Valverde}}} (\bibinfo {year} {2002}{\natexlab{b}}),\ \href@noop {}
  {\bibfield  {journal} {\bibinfo  {journal} {Complexity}\ }\textbf {\bibinfo
  {volume} {8}}~(\bibinfo {number} {1}),\ \bibinfo {pages} {20}}\BibitemShut
  {NoStop}%
\bibitem [{\citenamefont {Sol{\'e}}\ \emph {et~al.}(1999)\citenamefont
  {Sol{\'e}}, \citenamefont {Manrubia}, \citenamefont {Benton}, \citenamefont
  {Kauffman},\ and\ \citenamefont {Bak}}]{Sole-Manrubia}%
  \BibitemOpen
  \bibfield  {author} {\bibinfo {author} {\bibnamefont {Sol{\'e}},
  \bibfnamefont {R.~V.}}, \bibinfo {author} {\bibfnamefont {S.~C.}\
  \bibnamefont {Manrubia}}, \bibinfo {author} {\bibfnamefont {M.}~\bibnamefont
  {Benton}}, \bibinfo {author} {\bibfnamefont {S.}~\bibnamefont {Kauffman}}, \
  and\ \bibinfo {author} {\bibfnamefont {P.}~\bibnamefont {Bak}}} (\bibinfo
  {year} {1999}),\ \href@noop {} {\bibfield  {journal} {\bibinfo  {journal}
  {Trends in Ecology \& Evolution}\ }\textbf {\bibinfo {volume} {14}}~(\bibinfo
  {number} {4}),\ \bibinfo {pages} {156}}\BibitemShut {NoStop}%
\bibitem [{\citenamefont {Sol{\'e}}\ \emph {et~al.}(1996)\citenamefont
  {Sol{\'e}}, \citenamefont {Manrubia}, \citenamefont {Luque}, \citenamefont
  {Delgado},\ and\ \citenamefont {Bascompte}}]{Sole5}%
  \BibitemOpen
  \bibfield  {author} {\bibinfo {author} {\bibnamefont {Sol{\'e}},
  \bibfnamefont {R.~V.}}, \bibinfo {author} {\bibfnamefont {S.~C.}\
  \bibnamefont {Manrubia}}, \bibinfo {author} {\bibfnamefont {B.}~\bibnamefont
  {Luque}}, \bibinfo {author} {\bibfnamefont {J.}~\bibnamefont {Delgado}}, \
  and\ \bibinfo {author} {\bibfnamefont {J.}~\bibnamefont {Bascompte}}}
  (\bibinfo {year} {1996}),\ \href@noop {} {\bibfield  {journal} {\bibinfo
  {journal} {Complexity}\ }\textbf {\bibinfo {volume} {1}}~(\bibinfo {number}
  {4}),\ \bibinfo {pages} {13}}\BibitemShut {NoStop}%
\bibitem [{\citenamefont {Sol{\'e}}\ and\ \citenamefont
  {Miramontes}(1995)}]{Sole1995}%
  \BibitemOpen
  \bibfield  {author} {\bibinfo {author} {\bibnamefont {Sol{\'e}},
  \bibfnamefont {R.~V.}}, \ and\ \bibinfo {author} {\bibfnamefont
  {O.}~\bibnamefont {Miramontes}}} (\bibinfo {year} {1995}),\ \href@noop {}
  {\bibfield  {journal} {\bibinfo  {journal} {Physica D}\ }\textbf {\bibinfo
  {volume} {80}}~(\bibinfo {number} {1}),\ \bibinfo {pages} {171}}\BibitemShut
  {NoStop}%
\bibitem [{\citenamefont {Solovey}\ \emph {et~al.}(2012)\citenamefont
  {Solovey}, \citenamefont {Miller}, \citenamefont {Ojemann}, \citenamefont
  {Magnasco},\ and\ \citenamefont {Cecchi}}]{Magnasco1}%
  \BibitemOpen
  \bibfield  {author} {\bibinfo {author} {\bibnamefont {Solovey}, \bibfnamefont
  {G.}}, \bibinfo {author} {\bibfnamefont {K.~J.}\ \bibnamefont {Miller}},
  \bibinfo {author} {\bibfnamefont {J.~G.}\ \bibnamefont {Ojemann}}, \bibinfo
  {author} {\bibfnamefont {M.~O.}\ \bibnamefont {Magnasco}}, \ and\ \bibinfo
  {author} {\bibfnamefont {G.~A.}\ \bibnamefont {Cecchi}}} (\bibinfo {year}
  {2012}),\ \href@noop {} {\bibfield  {journal} {\bibinfo  {journal} {Front.
  Integrative Neurosci.}\ }\textbf {\bibinfo {volume} {6}}}\BibitemShut
  {NoStop}%
\bibitem [{\citenamefont {Song}\ \emph {et~al.}(2017)\citenamefont {Song},
  \citenamefont {Marsili},\ and\ \citenamefont {Jo}}]{Deep-Marsili}%
  \BibitemOpen
  \bibfield  {author} {\bibinfo {author} {\bibnamefont {Song}, \bibfnamefont
  {J.}}, \bibinfo {author} {\bibfnamefont {M.}~\bibnamefont {Marsili}}, \ and\
  \bibinfo {author} {\bibfnamefont {J.}~\bibnamefont {Jo}}} (\bibinfo {year}
  {2017}),\ \href@noop {} {\bibinfo  {journal} {arXiv:1710.11324}\
  }\BibitemShut {NoStop}%
\bibitem [{\citenamefont {Soriano}\ \emph {et~al.}(2006)\citenamefont
  {Soriano}, \citenamefont {Colombo},\ and\ \citenamefont {Ott}}]{Soriano2006}%
  \BibitemOpen
\bibfield  {journal} {  }\bibfield  {author} {\bibinfo {author} {\bibnamefont
  {Soriano}, \bibfnamefont {J.}}, \bibinfo {author} {\bibfnamefont
  {C.}~\bibnamefont {Colombo}}, \ and\ \bibinfo {author} {\bibfnamefont
  {A.}~\bibnamefont {Ott}}} (\bibinfo {year} {2006}),\ \href@noop {} {\bibfield
   {journal} {\bibinfo  {journal} {Phys. Rev. Lett.}\ }\textbf {\bibinfo
  {volume} {97}}~(\bibinfo {number} {25}),\ \bibinfo {pages}
  {258102}}\BibitemShut {NoStop}%
\bibitem [{\citenamefont {Soriano}\ \emph {et~al.}(2009)\citenamefont
  {Soriano}, \citenamefont {R{\"u}diger}, \citenamefont {Pullarkat},\ and\
  \citenamefont {Ott}}]{Soriano2009}%
  \BibitemOpen
  \bibfield  {author} {\bibinfo {author} {\bibnamefont {Soriano}, \bibfnamefont
  {J.}}, \bibinfo {author} {\bibfnamefont {S.}~\bibnamefont {R{\"u}diger}},
  \bibinfo {author} {\bibfnamefont {P.}~\bibnamefont {Pullarkat}}, \ and\
  \bibinfo {author} {\bibfnamefont {A.}~\bibnamefont {Ott}}} (\bibinfo {year}
  {2009}),\ \href@noop {} {\bibfield  {journal} {\bibinfo  {journal} {Biophys.
  Jour.}\ }\textbf {\bibinfo {volume} {96}}~(\bibinfo {number} {4}),\ \bibinfo
  {pages} {1649}}\BibitemShut {NoStop}%
\bibitem [{\citenamefont {Sornette}(1994)}]{Sornette-sweeping}%
  \BibitemOpen
  \bibfield  {author} {\bibinfo {author} {\bibnamefont {Sornette},
  \bibfnamefont {D.}}} (\bibinfo {year} {1994}),\ \href@noop {} {\bibfield
  {journal} {\bibinfo  {journal} {J. de Physique I}\ }\textbf {\bibinfo
  {volume} {4}}~(\bibinfo {number} {2}),\ \bibinfo {pages} {209}}\BibitemShut
  {NoStop}%
\bibitem [{\citenamefont {Sornette}(1998)}]{Sornette-multiplicative}%
  \BibitemOpen
  \bibfield  {author} {\bibinfo {author} {\bibnamefont {Sornette},
  \bibfnamefont {D.}}} (\bibinfo {year} {1998}),\ \href@noop {} {\bibfield
  {journal} {\bibinfo  {journal} {Phys. Rev. E}\ }\textbf {\bibinfo {volume}
  {57}}~(\bibinfo {number} {4}),\ \bibinfo {pages} {4811}}\BibitemShut
  {NoStop}%
\bibitem [{\citenamefont {Sornette}(2006)}]{Sornette-book}%
  \BibitemOpen
  \bibfield  {author} {\bibinfo {author} {\bibnamefont {Sornette},
  \bibfnamefont {D.}}} (\bibinfo {year} {2006}),\ \href@noop {} {\emph
  {\bibinfo {title} {Critical Phenomena in Natural Sciences}}}\ (\bibinfo
  {publisher} {Springer})\BibitemShut {NoStop}%
\bibitem [{\citenamefont {Sornette}(2009)}]{Sornette-powerlaws}%
  \BibitemOpen
  \bibfield  {author} {\bibinfo {author} {\bibnamefont {Sornette},
  \bibfnamefont {D.}}} (\bibinfo {year} {2009}),\ in\ \href@noop {} {\emph
  {\bibinfo {booktitle} {Encyclopedia of Complexity and Systems Science}}}\
  (\bibinfo  {publisher} {Springer})\ pp.\ \bibinfo {pages}
  {7009--7024}\BibitemShut {NoStop}%
\bibitem [{\citenamefont {Sornette}\ and\ \citenamefont
  {Cont}(1997)}]{Sornette+Cont}%
  \BibitemOpen
  \bibfield  {author} {\bibinfo {author} {\bibnamefont {Sornette},
  \bibfnamefont {D.}}, \ and\ \bibinfo {author} {\bibfnamefont
  {R.}~\bibnamefont {Cont}}} (\bibinfo {year} {1997}),\ \href@noop {}
  {\bibfield  {journal} {\bibinfo  {journal} {J. de Physique I}\ }\textbf
  {\bibinfo {volume} {7}}~(\bibinfo {number} {3}),\ \bibinfo {pages}
  {431}}\BibitemShut {NoStop}%
\bibitem [{\citenamefont {Sporns}(2010)}]{Sporns-book}%
  \BibitemOpen
  \bibfield  {author} {\bibinfo {author} {\bibnamefont {Sporns}, \bibfnamefont
  {O.}}} (\bibinfo {year} {2010}),\ \href@noop {} {\emph {\bibinfo {title}
  {{Networks of the Brain}}}}\ (\bibinfo  {publisher} {MIT Press,
  USA})\BibitemShut {NoStop}%
\bibitem [{\citenamefont {Sporns}\ \emph {et~al.}(2004)\citenamefont {Sporns},
  \citenamefont {Chialvo}, \citenamefont {Kaiser},\ and\ \citenamefont
  {Hilgetag}}]{SCKH}%
  \BibitemOpen
  \bibfield  {author} {\bibinfo {author} {\bibnamefont {Sporns}, \bibfnamefont
  {O.}}, \bibinfo {author} {\bibfnamefont {D.~R.}\ \bibnamefont {Chialvo}},
  \bibinfo {author} {\bibfnamefont {M.}~\bibnamefont {Kaiser}}, \ and\ \bibinfo
  {author} {\bibfnamefont {C.~C.}\ \bibnamefont {Hilgetag}}} (\bibinfo {year}
  {2004}),\ \href@noop {} {\bibfield  {journal} {\bibinfo  {journal} {Trends
  Cogn Sci}\ }\textbf {\bibinfo {volume} {8}}~(\bibinfo {number} {9}),\
  \bibinfo {pages} {418}}\BibitemShut {NoStop}%
\bibitem [{\citenamefont {Sporns}\ \emph {et~al.}(2005)\citenamefont {Sporns},
  \citenamefont {Tononi},\ and\ \citenamefont {K\"{o}tter}}]{Sporns2005}%
  \BibitemOpen
  \bibfield  {author} {\bibinfo {author} {\bibnamefont {Sporns}, \bibfnamefont
  {O.}}, \bibinfo {author} {\bibfnamefont {G.}~\bibnamefont {Tononi}}, \ and\
  \bibinfo {author} {\bibfnamefont {R.}~\bibnamefont {K\"{o}tter}}} (\bibinfo
  {year} {2005}),\ \href {\doibase 10.1371/journal.pcbi.0010042} {\bibfield
  {journal} {\bibinfo  {journal} {PLoS Comput. Biol.}\ }\textbf {\bibinfo
  {volume} {1}}~(\bibinfo {number} {4}),\ \bibinfo {pages} {e42}}\BibitemShut
  {NoStop}%
\bibitem [{\citenamefont {Stanley}(1987)}]{Stanley-book}%
  \BibitemOpen
  \bibfield  {author} {\bibinfo {author} {\bibnamefont {Stanley}, \bibfnamefont
  {H.~E.}}} (\bibinfo {year} {1987}),\ \href@noop {} {\emph {\bibinfo {title}
  {Introduction to phase transitions and critical phenomena}}}\ (\bibinfo
  {publisher} {Oxford Univ. Press})\BibitemShut {NoStop}%
\bibitem [{\citenamefont {Stassinopoulos}\ and\ \citenamefont
  {Bak}(1995)}]{Bak-brain}%
  \BibitemOpen
  \bibfield  {author} {\bibinfo {author} {\bibnamefont {Stassinopoulos},
  \bibfnamefont {D.}}, \ and\ \bibinfo {author} {\bibfnamefont
  {P.}~\bibnamefont {Bak}}} (\bibinfo {year} {1995}),\ \href@noop {} {\bibfield
   {journal} {\bibinfo  {journal} {Phys. Rev. E}\ }\textbf {\bibinfo {volume}
  {51}}~(\bibinfo {number} {5}),\ \bibinfo {pages} {5033}}\BibitemShut
  {NoStop}%
\bibitem [{\citenamefont {Stepp}\ \emph {et~al.}(2015)\citenamefont {Stepp},
  \citenamefont {Plenz},\ and\ \citenamefont {Srinivasa}}]{Plenz2015}%
  \BibitemOpen
  \bibfield  {author} {\bibinfo {author} {\bibnamefont {Stepp}, \bibfnamefont
  {N.}}, \bibinfo {author} {\bibfnamefont {D.}~\bibnamefont {Plenz}}, \ and\
  \bibinfo {author} {\bibfnamefont {N.}~\bibnamefont {Srinivasa}}} (\bibinfo
  {year} {2015}),\ \href@noop {} {\bibfield  {journal} {\bibinfo  {journal}
  {PLoS Comput. Biol.}\ }\textbf {\bibinfo {volume} {11}}~(\bibinfo {number}
  {1})}\BibitemShut {NoStop}%
\bibitem [{\citenamefont {Steriade}\ \emph {et~al.}(1996)\citenamefont
  {Steriade}, \citenamefont {Amzica},\ and\ \citenamefont
  {Contreras}}]{Steriade1996}%
  \BibitemOpen
  \bibfield  {author} {\bibinfo {author} {\bibnamefont {Steriade},
  \bibfnamefont {M.}}, \bibinfo {author} {\bibfnamefont {F.}~\bibnamefont
  {Amzica}}, \ and\ \bibinfo {author} {\bibfnamefont {D.}~\bibnamefont
  {Contreras}}} (\bibinfo {year} {1996}),\ \href@noop {} {\bibfield  {journal}
  {\bibinfo  {journal} {J. Neurosci.}\ }\textbf {\bibinfo {volume}
  {16}}~(\bibinfo {number} {1}),\ \bibinfo {pages} {392}}\BibitemShut {NoStop}%
\bibitem [{\citenamefont {Steriade}\ \emph {et~al.}(1993)\citenamefont
  {Steriade}, \citenamefont {Nunez},\ and\ \citenamefont
  {Amzica}}]{Steriade1993}%
  \BibitemOpen
  \bibfield  {author} {\bibinfo {author} {\bibnamefont {Steriade},
  \bibfnamefont {M.}}, \bibinfo {author} {\bibfnamefont {A.}~\bibnamefont
  {Nunez}}, \ and\ \bibinfo {author} {\bibfnamefont {F.}~\bibnamefont
  {Amzica}}} (\bibinfo {year} {1993}),\ \href@noop {} {\bibfield  {journal}
  {\bibinfo  {journal} {J. Neurosci.}\ }\textbf {\bibinfo {volume}
  {13}}~(\bibinfo {number} {8}),\ \bibinfo {pages} {3252}}\BibitemShut
  {NoStop}%
\bibitem [{\citenamefont {Stewart}\ and\ \citenamefont
  {Plenz}(2008)}]{development1}%
  \BibitemOpen
  \bibfield  {author} {\bibinfo {author} {\bibnamefont {Stewart}, \bibfnamefont
  {C.~V.}}, \ and\ \bibinfo {author} {\bibfnamefont {D.}~\bibnamefont {Plenz}}}
  (\bibinfo {year} {2008}),\ \href@noop {} {\bibfield  {journal} {\bibinfo
  {journal} {J. of Neurosci. Methods}\ }\textbf {\bibinfo {volume}
  {169}}~(\bibinfo {number} {2}),\ \bibinfo {pages} {405}}\BibitemShut
  {NoStop}%
\bibitem [{\citenamefont {Stoki{\'c}}\ \emph {et~al.}(2008)\citenamefont
  {Stoki{\'c}}, \citenamefont {Hanel},\ and\ \citenamefont
  {Thurner}}]{Thurner1}%
  \BibitemOpen
  \bibfield  {author} {\bibinfo {author} {\bibnamefont {Stoki{\'c}},
  \bibfnamefont {D.}}, \bibinfo {author} {\bibfnamefont {R.}~\bibnamefont
  {Hanel}}, \ and\ \bibinfo {author} {\bibfnamefont {S.}~\bibnamefont
  {Thurner}}} (\bibinfo {year} {2008}),\ \href@noop {} {\bibfield  {journal}
  {\bibinfo  {journal} {Phys. Rev. E}\ }\textbf {\bibinfo {volume}
  {77}}~(\bibinfo {number} {6}),\ \bibinfo {pages} {061917}}\BibitemShut
  {NoStop}%
\bibitem [{\citenamefont {Stoop}\ and\ \citenamefont
  {Gomez}(2016)}]{Stoop2016}%
  \BibitemOpen
  \bibfield  {author} {\bibinfo {author} {\bibnamefont {Stoop}, \bibfnamefont
  {R.}}, \ and\ \bibinfo {author} {\bibfnamefont {F.}~\bibnamefont {Gomez}}}
  (\bibinfo {year} {2016}),\ \href@noop {} {\bibfield  {journal} {\bibinfo
  {journal} {Phys. Rev. Lett.}\ }\textbf {\bibinfo {volume} {117}},\ \bibinfo
  {pages} {038102}}\BibitemShut {NoStop}%
\bibitem [{\citenamefont {Strogatz}(2014)}]{Strogatz-book}%
  \BibitemOpen
  \bibfield  {author} {\bibinfo {author} {\bibnamefont {Strogatz},
  \bibfnamefont {S.~H.}}} (\bibinfo {year} {2014}),\ \href@noop {} {\emph
  {\bibinfo {title} {Nonlinear dynamics and chaos: with applications to
  physics, biology, chemistry, and engineering}}}\ (\bibinfo  {publisher}
  {Westview press})\BibitemShut {NoStop}%
\bibitem [{\citenamefont {Stumpf}\ and\ \citenamefont {Porter}(2012)}]{Stumpf}%
  \BibitemOpen
  \bibfield  {author} {\bibinfo {author} {\bibnamefont {Stumpf}, \bibfnamefont
  {M.~P.}}, \ and\ \bibinfo {author} {\bibfnamefont {M.~A.}\ \bibnamefont
  {Porter}}} (\bibinfo {year} {2012}),\ \href@noop {} {\bibfield  {journal}
  {\bibinfo  {journal} {Science}\ }\textbf {\bibinfo {volume} {335}}~(\bibinfo
  {number} {6069}),\ \bibinfo {pages} {665}}\BibitemShut {NoStop}%
\bibitem [{\citenamefont {Summers}\ and\ \citenamefont
  {Litwin}(2006)}]{Summers}%
  \BibitemOpen
  \bibfield  {author} {\bibinfo {author} {\bibnamefont {Summers}, \bibfnamefont
  {J.}}, \ and\ \bibinfo {author} {\bibfnamefont {S.}~\bibnamefont {Litwin}}}
  (\bibinfo {year} {2006}),\ \href@noop {} {\bibfield  {journal} {\bibinfo
  {journal} {J. of Virology}\ }\textbf {\bibinfo {volume} {80}}~(\bibinfo
  {number} {1}),\ \bibinfo {pages} {20}}\BibitemShut {NoStop}%
\bibitem [{\citenamefont {Sumpter}(2010)}]{Collective1}%
  \BibitemOpen
  \bibfield  {author} {\bibinfo {author} {\bibnamefont {Sumpter}, \bibfnamefont
  {D.~J.}}} (\bibinfo {year} {2010}),\ \href@noop {} {\emph {\bibinfo {title}
  {Collective animal behavior}}}\ (\bibinfo  {publisher} {Princeton University
  Press})\BibitemShut {NoStop}%
\bibitem [{\citenamefont {Sussillo}\ \emph {et~al.}(2007)\citenamefont
  {Sussillo}, \citenamefont {Toyoizumi},\ and\ \citenamefont
  {Maass}}]{Sussillo}%
  \BibitemOpen
  \bibfield  {author} {\bibinfo {author} {\bibnamefont {Sussillo},
  \bibfnamefont {D.}}, \bibinfo {author} {\bibfnamefont {T.}~\bibnamefont
  {Toyoizumi}}, \ and\ \bibinfo {author} {\bibfnamefont {W.}~\bibnamefont
  {Maass}}} (\bibinfo {year} {2007}),\ \href@noop {} {\bibfield  {journal}
  {\bibinfo  {journal} {J. Neurophysiol.}\ }\textbf {\bibinfo {volume}
  {97}}~(\bibinfo {number} {6}),\ \bibinfo {pages} {4079}}\BibitemShut
  {NoStop}%
\bibitem [{\citenamefont {Suweis}\ \emph {et~al.}(2013)\citenamefont {Suweis},
  \citenamefont {Simini}, \citenamefont {Banavar},\ and\ \citenamefont
  {Maritan}}]{Suweis2013}%
  \BibitemOpen
  \bibfield  {author} {\bibinfo {author} {\bibnamefont {Suweis}, \bibfnamefont
  {S.}}, \bibinfo {author} {\bibfnamefont {F.}~\bibnamefont {Simini}}, \bibinfo
  {author} {\bibfnamefont {J.~R.}\ \bibnamefont {Banavar}}, \ and\ \bibinfo
  {author} {\bibfnamefont {A.}~\bibnamefont {Maritan}}} (\bibinfo {year}
  {2013}),\ \href@noop {} {\bibfield  {journal} {\bibinfo  {journal} {Nature}\
  }\textbf {\bibinfo {volume} {8500}},\ \bibinfo {pages} {449}}\BibitemShut
  {NoStop}%
\bibitem [{\citenamefont {Tabak}\ and\ \citenamefont
  {Latham}(2003)}]{Latham2003}%
  \BibitemOpen
  \bibfield  {author} {\bibinfo {author} {\bibnamefont {Tabak}, \bibfnamefont
  {J.}}, \ and\ \bibinfo {author} {\bibfnamefont {P.~E.}\ \bibnamefont
  {Latham}}} (\bibinfo {year} {2003}),\ \href@noop {} {\bibfield  {journal}
  {\bibinfo  {journal} {Neuroreport}\ }\textbf {\bibinfo {volume}
  {14}}~(\bibinfo {number} {11}),\ \bibinfo {pages} {1445}}\BibitemShut
  {NoStop}%
\bibitem [{\citenamefont {Tagliazucchi}\ \emph {et~al.}(2012)\citenamefont
  {Tagliazucchi}, \citenamefont {Balenzuela}, \citenamefont {Fraiman},\ and\
  \citenamefont {Chialvo}}]{Taglia}%
  \BibitemOpen
  \bibfield  {author} {\bibinfo {author} {\bibnamefont {Tagliazucchi},
  \bibfnamefont {E.}}, \bibinfo {author} {\bibfnamefont {P.}~\bibnamefont
  {Balenzuela}}, \bibinfo {author} {\bibfnamefont {D.}~\bibnamefont {Fraiman}},
  \ and\ \bibinfo {author} {\bibfnamefont {D.~R.}\ \bibnamefont {Chialvo}}}
  (\bibinfo {year} {2012}),\ \href@noop {} {\bibfield  {journal} {\bibinfo
  {journal} {Front. Physiol.}\ }\textbf {\bibinfo {volume} {3}},\ \bibinfo
  {pages} {15}}\BibitemShut {NoStop}%
\bibitem [{\citenamefont {Tagliazucchi}\ \emph {et~al.}(2016)\citenamefont
  {Tagliazucchi}, \citenamefont {Chialvo}, \citenamefont {Siniatchkin},
  \citenamefont {Amico}, \citenamefont {Brichant}, \citenamefont {Bonhomme},
  \citenamefont {Noirhomme}, \citenamefont {Laufs},\ and\ \citenamefont
  {Laureys}}]{Taglia-unconcious}%
  \BibitemOpen
  \bibfield  {author} {\bibinfo {author} {\bibnamefont {Tagliazucchi},
  \bibfnamefont {E.}}, \bibinfo {author} {\bibfnamefont {D.~R.}\ \bibnamefont
  {Chialvo}}, \bibinfo {author} {\bibfnamefont {M.}~\bibnamefont
  {Siniatchkin}}, \bibinfo {author} {\bibfnamefont {E.}~\bibnamefont {Amico}},
  \bibinfo {author} {\bibfnamefont {J.-F.}\ \bibnamefont {Brichant}}, \bibinfo
  {author} {\bibfnamefont {V.}~\bibnamefont {Bonhomme}}, \bibinfo {author}
  {\bibfnamefont {Q.}~\bibnamefont {Noirhomme}}, \bibinfo {author}
  {\bibfnamefont {H.}~\bibnamefont {Laufs}}, \ and\ \bibinfo {author}
  {\bibfnamefont {S.}~\bibnamefont {Laureys}}} (\bibinfo {year} {2016}),\
  \href@noop {} {\bibfield  {journal} {\bibinfo  {journal} {J. R. Soc.
  Interface}\ }\textbf {\bibinfo {volume} {13}}~(\bibinfo {number} {114}),\
  \bibinfo {pages} {20151027}}\BibitemShut {NoStop}%
\bibitem [{\citenamefont {Tagliazucchi}\ \emph {et~al.}(2013)\citenamefont
  {Tagliazucchi}, \citenamefont {von Wegner}, \citenamefont {Morzelewski},
  \citenamefont {Brodbeck}, \citenamefont {Jahnke},\ and\ \citenamefont
  {Laufs}}]{Taglia-breakdown}%
  \BibitemOpen
  \bibfield  {author} {\bibinfo {author} {\bibnamefont {Tagliazucchi},
  \bibfnamefont {E.}}, \bibinfo {author} {\bibfnamefont {F.}~\bibnamefont {von
  Wegner}}, \bibinfo {author} {\bibfnamefont {A.}~\bibnamefont {Morzelewski}},
  \bibinfo {author} {\bibfnamefont {V.}~\bibnamefont {Brodbeck}}, \bibinfo
  {author} {\bibfnamefont {K.}~\bibnamefont {Jahnke}}, \ and\ \bibinfo {author}
  {\bibfnamefont {H.}~\bibnamefont {Laufs}}} (\bibinfo {year} {2013}),\
  \href@noop {} {\bibfield  {journal} {\bibinfo  {journal} {Proc. Natl. Acad.
  Sci. USA.}\ }\textbf {\bibinfo {volume} {110}}~(\bibinfo {number} {38}),\
  \bibinfo {pages} {15419}}\BibitemShut {NoStop}%
\bibitem [{\citenamefont {Tang}\ \emph {et~al.}(2017)\citenamefont {Tang},
  \citenamefont {Zhang}, \citenamefont {Wang}, \citenamefont {Wang},\ and\
  \citenamefont {Chialvo}}]{Chialvo-proteins}%
  \BibitemOpen
  \bibfield  {author} {\bibinfo {author} {\bibnamefont {Tang}, \bibfnamefont
  {Q.-Y.}}, \bibinfo {author} {\bibfnamefont {Y.-Y.}\ \bibnamefont {Zhang}},
  \bibinfo {author} {\bibfnamefont {J.}~\bibnamefont {Wang}}, \bibinfo {author}
  {\bibfnamefont {W.}~\bibnamefont {Wang}}, \ and\ \bibinfo {author}
  {\bibfnamefont {D.~R.}\ \bibnamefont {Chialvo}}} (\bibinfo {year} {2017}),\
  \href {\doibase 10.1103/PhysRevLett.118.088102} {\bibfield  {journal}
  {\bibinfo  {journal} {Phys. Rev. Lett.}\ }\textbf {\bibinfo {volume} {118}},\
  \bibinfo {pages} {088102}}\BibitemShut {NoStop}%
\bibitem [{\citenamefont {T{\"a}uber}(2014)}]{Tauber}%
  \BibitemOpen
  \bibfield  {author} {\bibinfo {author} {\bibnamefont {T{\"a}uber},
  \bibfnamefont {U.~C.}}} (\bibinfo {year} {2014}),\ \href@noop {} {\emph
  {\bibinfo {title} {Critical dynamics: a field theory approach to equilibrium
  and non-equilibrium scaling behavior}}}\ (\bibinfo  {publisher} {Cambridge
  University Press})\BibitemShut {NoStop}%
\bibitem [{\citenamefont {T{\"a}uber}(2017)}]{Tauber2017}%
  \BibitemOpen
  \bibfield  {author} {\bibinfo {author} {\bibnamefont {T{\"a}uber},
  \bibfnamefont {U.~C.}}} (\bibinfo {year} {2017}),\ \href {\doibase
  10.1146/annurev-conmatphys-031016-025444} {\bibfield  {journal} {\bibinfo
  {journal} {Ann. Rev. Cond.Matt. Phys.}\ }\textbf {\bibinfo {volume}
  {8}}~(\bibinfo {number} {1}),\ \bibinfo {pages} {185}}\BibitemShut {NoStop}%
\bibitem [{\citenamefont {Taylor}(1961)}]{Taylor}%
  \BibitemOpen
  \bibfield  {author} {\bibinfo {author} {\bibnamefont {Taylor}, \bibfnamefont
  {L.}}} (\bibinfo {year} {1961}),\ \href@noop {} {\bibinfo  {journal}
  {Nature}\ ,\ \bibinfo {pages} {732}}\BibitemShut {NoStop}%
\bibitem [{\citenamefont {Tetzlaff}\ \emph {et~al.}(2010)\citenamefont
  {Tetzlaff}, \citenamefont {Okujeni}, \citenamefont {Egert}, \citenamefont
  {W{\"o}rg{\"o}tter},\ and\ \citenamefont {Butz}}]{development2}%
  \BibitemOpen
\bibfield  {journal} {  }\bibfield  {author} {\bibinfo {author} {\bibnamefont
  {Tetzlaff}, \bibfnamefont {C.}}, \bibinfo {author} {\bibfnamefont
  {S.}~\bibnamefont {Okujeni}}, \bibinfo {author} {\bibfnamefont
  {U.}~\bibnamefont {Egert}}, \bibinfo {author} {\bibfnamefont
  {F.}~\bibnamefont {W{\"o}rg{\"o}tter}}, \ and\ \bibinfo {author}
  {\bibfnamefont {M.}~\bibnamefont {Butz}}} (\bibinfo {year} {2010}),\
  \href@noop {} {\bibfield  {journal} {\bibinfo  {journal} {PLoS Comput.
  Biol.}\ }\textbf {\bibinfo {volume} {6}}~(\bibinfo {number} {12}),\ \bibinfo
  {pages} {e1001013}}\BibitemShut {NoStop}%
\bibitem [{\citenamefont {Tinker}\ and\ \citenamefont
  {Velazquez}(2014)}]{synchro-task}%
  \BibitemOpen
  \bibfield  {author} {\bibinfo {author} {\bibnamefont {Tinker}, \bibfnamefont
  {J.}}, \ and\ \bibinfo {author} {\bibfnamefont {J.~L.~P.}\ \bibnamefont
  {Velazquez}}} (\bibinfo {year} {2014}),\ \href@noop {} {\bibfield  {journal}
  {\bibinfo  {journal} {Front. Syst. Neurosci.}\ }\textbf {\bibinfo {volume}
  {8}},\ \bibinfo {pages} {73}}\BibitemShut {NoStop}%
\bibitem [{\citenamefont {Tka{\v{c}}ik}\ \emph {et~al.}(2014)\citenamefont
  {Tka{\v{c}}ik}, \citenamefont {Marre}, \citenamefont {Amodei}, \citenamefont
  {Schneidman}, \citenamefont {Bialek},\ and\ \citenamefont
  {Berry~II}}]{Bialek-searching}%
  \BibitemOpen
  \bibfield  {author} {\bibinfo {author} {\bibnamefont {Tka{\v{c}}ik},
  \bibfnamefont {G.}}, \bibinfo {author} {\bibfnamefont {O.}~\bibnamefont
  {Marre}}, \bibinfo {author} {\bibfnamefont {D.}~\bibnamefont {Amodei}},
  \bibinfo {author} {\bibfnamefont {E.}~\bibnamefont {Schneidman}}, \bibinfo
  {author} {\bibfnamefont {W.}~\bibnamefont {Bialek}}, \ and\ \bibinfo {author}
  {\bibfnamefont {M.~J.}\ \bibnamefont {Berry~II}}} (\bibinfo {year} {2014}),\
  \href@noop {} {\bibfield  {journal} {\bibinfo  {journal} {PLoS Comput.
  Biol.}\ }\textbf {\bibinfo {volume} {10}}~(\bibinfo {number} {1}),\ \bibinfo
  {pages} {e1003408}}\BibitemShut {NoStop}%
\bibitem [{\citenamefont {Tka{\v{c}}ik}\ \emph {et~al.}(2013)\citenamefont
  {Tka{\v{c}}ik}, \citenamefont {Marre}, \citenamefont {Mora}, \citenamefont
  {Amodei}, \citenamefont {Berry~II},\ and\ \citenamefont
  {Bialek}}]{Bialek-thermo1}%
  \BibitemOpen
  \bibfield  {author} {\bibinfo {author} {\bibnamefont {Tka{\v{c}}ik},
  \bibfnamefont {G.}}, \bibinfo {author} {\bibfnamefont {O.}~\bibnamefont
  {Marre}}, \bibinfo {author} {\bibfnamefont {T.}~\bibnamefont {Mora}},
  \bibinfo {author} {\bibfnamefont {D.}~\bibnamefont {Amodei}}, \bibinfo
  {author} {\bibfnamefont {M.~J.}\ \bibnamefont {Berry~II}}, \ and\ \bibinfo
  {author} {\bibfnamefont {W.}~\bibnamefont {Bialek}}} (\bibinfo {year}
  {2013}),\ \href@noop {} {\bibfield  {journal} {\bibinfo  {journal} {J. Stat.
  Mech.}\ }\textbf {\bibinfo {volume} {2013}}~(\bibinfo {number} {03}),\
  \bibinfo {pages} {P03011}}\BibitemShut {NoStop}%
\bibitem [{\citenamefont {Tka{\v{c}}ik}\ \emph {et~al.}(2015)\citenamefont
  {Tka{\v{c}}ik}, \citenamefont {Mora}, \citenamefont {Marre}, \citenamefont
  {Amodei}, \citenamefont {Palmer}, \citenamefont {Berry},\ and\ \citenamefont
  {Bialek}}]{Bialek-thermo2}%
  \BibitemOpen
  \bibfield  {author} {\bibinfo {author} {\bibnamefont {Tka{\v{c}}ik},
  \bibfnamefont {G.}}, \bibinfo {author} {\bibfnamefont {T.}~\bibnamefont
  {Mora}}, \bibinfo {author} {\bibfnamefont {O.}~\bibnamefont {Marre}},
  \bibinfo {author} {\bibfnamefont {D.}~\bibnamefont {Amodei}}, \bibinfo
  {author} {\bibfnamefont {S.~E.}\ \bibnamefont {Palmer}}, \bibinfo {author}
  {\bibfnamefont {M.~J.}\ \bibnamefont {Berry}}, \ and\ \bibinfo {author}
  {\bibfnamefont {W.}~\bibnamefont {Bialek}}} (\bibinfo {year} {2015}),\
  \href@noop {} {\bibfield  {journal} {\bibinfo  {journal} {Proc. Natl. Acad.
  Sci. USA.}\ }\textbf {\bibinfo {volume} {112}}~(\bibinfo {number} {37}),\
  \bibinfo {pages} {11508}}\BibitemShut {NoStop}%
\bibitem [{\citenamefont {Tkacik}\ \emph {et~al.}(2009)\citenamefont {Tkacik},
  \citenamefont {Schneidman}, \citenamefont {Berry}, \citenamefont {Michael},\
  and\ \citenamefont {Bialek}}]{spin-glass}%
  \BibitemOpen
  \bibfield  {author} {\bibinfo {author} {\bibnamefont {Tkacik}, \bibfnamefont
  {G.}}, \bibinfo {author} {\bibfnamefont {E.}~\bibnamefont {Schneidman}},
  \bibinfo {author} {\bibfnamefont {I.}~\bibnamefont {Berry}}, \bibinfo
  {author} {\bibfnamefont {J.}~\bibnamefont {Michael}}, \ and\ \bibinfo
  {author} {\bibfnamefont {W.}~\bibnamefont {Bialek}}} (\bibinfo {year}
  {2009}),\ \href@noop {} {\bibinfo  {journal} {arXiv preprint
  arXiv:0912.5409}\ }\BibitemShut {NoStop}%
\bibitem [{\citenamefont {Tomen}\ \emph {et~al.}(2014)\citenamefont {Tomen},
  \citenamefont {Rotermund},\ and\ \citenamefont {Ernst}}]{Tomen}%
  \BibitemOpen
\bibfield  {journal} {  }\bibfield  {author} {\bibinfo {author} {\bibnamefont
  {Tomen}, \bibfnamefont {N.}}, \bibinfo {author} {\bibfnamefont
  {D.}~\bibnamefont {Rotermund}}, \ and\ \bibinfo {author} {\bibfnamefont
  {U.}~\bibnamefont {Ernst}}} (\bibinfo {year} {2014}),\ \href@noop {}
  {\bibfield  {journal} {\bibinfo  {journal} {Front. Syst. Neurosci}\ }\textbf
  {\bibinfo {volume} {8}},\ \bibinfo {pages} {151}}\BibitemShut {NoStop}%
\bibitem [{\citenamefont {Toner}\ and\ \citenamefont {Tu}(1995)}]{Yuhai1995}%
  \BibitemOpen
  \bibfield  {author} {\bibinfo {author} {\bibnamefont {Toner}, \bibfnamefont
  {J.}}, \ and\ \bibinfo {author} {\bibfnamefont {Y.}~\bibnamefont {Tu}}}
  (\bibinfo {year} {1995}),\ \href@noop {} {\bibfield  {journal} {\bibinfo
  {journal} {Phys. Rev. Lett.}\ }\textbf {\bibinfo {volume} {75}}~(\bibinfo
  {number} {23}),\ \bibinfo {pages} {4326}}\BibitemShut {NoStop}%
\bibitem [{\citenamefont {Toner}\ \emph {et~al.}(2005)\citenamefont {Toner},
  \citenamefont {Tu},\ and\ \citenamefont {Ramaswamy}}]{Yuhai2005}%
  \BibitemOpen
  \bibfield  {author} {\bibinfo {author} {\bibnamefont {Toner}, \bibfnamefont
  {J.}}, \bibinfo {author} {\bibfnamefont {Y.}~\bibnamefont {Tu}}, \ and\
  \bibinfo {author} {\bibfnamefont {S.}~\bibnamefont {Ramaswamy}}} (\bibinfo
  {year} {2005}),\ \href@noop {} {\bibfield  {journal} {\bibinfo  {journal}
  {Ann. Phys.}\ }\textbf {\bibinfo {volume} {318}}~(\bibinfo {number} {1}),\
  \bibinfo {pages} {170}}\BibitemShut {NoStop}%
\bibitem [{\citenamefont {Tononi}\ \emph {et~al.}(1994)\citenamefont {Tononi},
  \citenamefont {Sporns},\ and\ \citenamefont {Edelman}}]{Tononi}%
  \BibitemOpen
  \bibfield  {author} {\bibinfo {author} {\bibnamefont {Tononi}, \bibfnamefont
  {G.}}, \bibinfo {author} {\bibfnamefont {O.}~\bibnamefont {Sporns}}, \ and\
  \bibinfo {author} {\bibfnamefont {G.~M.}\ \bibnamefont {Edelman}}} (\bibinfo
  {year} {1994}),\ \href@noop {} {\bibfield  {journal} {\bibinfo  {journal}
  {Proc. Natl. Acad. Sci. USA.}\ }\textbf {\bibinfo {volume} {91}}~(\bibinfo
  {number} {11}),\ \bibinfo {pages} {5033}}\BibitemShut {NoStop}%
\bibitem [{\citenamefont {Torres-Sosa}\ \emph {et~al.}(2012)\citenamefont
  {Torres-Sosa}, \citenamefont {Huang},\ and\ \citenamefont
  {Aldana}}]{Aldana2012}%
  \BibitemOpen
  \bibfield  {author} {\bibinfo {author} {\bibnamefont {Torres-Sosa},
  \bibfnamefont {C.}}, \bibinfo {author} {\bibfnamefont {S.}~\bibnamefont
  {Huang}}, \ and\ \bibinfo {author} {\bibfnamefont {M.}~\bibnamefont
  {Aldana}}} (\bibinfo {year} {2012}),\ \href@noop {} {\bibfield  {journal}
  {\bibinfo  {journal} {PLoS Comput. Biol.}\ }\textbf {\bibinfo {volume}
  {8}}~(\bibinfo {number} {9}),\ \bibinfo {pages} {e1002669}}\BibitemShut
  {NoStop}%
\bibitem [{\citenamefont {Touboul}\ and\ \citenamefont
  {Destexhe}(2010)}]{Touboul1}%
  \BibitemOpen
  \bibfield  {author} {\bibinfo {author} {\bibnamefont {Touboul}, \bibfnamefont
  {J.}}, \ and\ \bibinfo {author} {\bibfnamefont {A.}~\bibnamefont {Destexhe}}}
  (\bibinfo {year} {2010}),\ \href@noop {} {\bibfield  {journal} {\bibinfo
  {journal} {PloS One}\ }\textbf {\bibinfo {volume} {5}}~(\bibinfo {number}
  {2}),\ \bibinfo {pages} {e8982}}\BibitemShut {NoStop}%
\bibitem [{\citenamefont {Touboul}\ and\ \citenamefont
  {Destexhe}(2017)}]{Touboul2}%
  \BibitemOpen
  \bibfield  {author} {\bibinfo {author} {\bibnamefont {Touboul}, \bibfnamefont
  {J.}}, \ and\ \bibinfo {author} {\bibfnamefont {A.}~\bibnamefont {Destexhe}}}
  (\bibinfo {year} {2017}),\ \href@noop {} {\bibfield  {journal} {\bibinfo
  {journal} {Phys. Rev. E}\ }\textbf {\bibinfo {volume} {95}}~(\bibinfo
  {number} {1}),\ \bibinfo {pages} {012413}}\BibitemShut {NoStop}%
\bibitem [{\citenamefont {Toyoizumi}\ and\ \citenamefont
  {Abbott}(2011)}]{Abbott}%
  \BibitemOpen
  \bibfield  {author} {\bibinfo {author} {\bibnamefont {Toyoizumi},
  \bibfnamefont {T.}}, \ and\ \bibinfo {author} {\bibfnamefont {L.~F.}\
  \bibnamefont {Abbott}}} (\bibinfo {year} {2011}),\ \href@noop {} {\bibfield
  {journal} {\bibinfo  {journal} {Phys. Rev. E}\ }\textbf {\bibinfo {volume}
  {84}},\ \bibinfo {pages} {051908}}\BibitemShut {NoStop}%
\bibitem [{\citenamefont {Tsodyks}\ and\ \citenamefont {Markram}(1997)}]{TM97}%
  \BibitemOpen
  \bibfield  {author} {\bibinfo {author} {\bibnamefont {Tsodyks}, \bibfnamefont
  {M.}}, \ and\ \bibinfo {author} {\bibfnamefont {H.}~\bibnamefont {Markram}}}
  (\bibinfo {year} {1997}),\ \href@noop {} {\bibfield  {journal} {\bibinfo
  {journal} {Proc. Natl. Acad. Sci. USA}\ }\textbf {\bibinfo {volume} {94}},\
  \bibinfo {pages} {719}}\BibitemShut {NoStop}%
\bibitem [{\citenamefont {Tsuchiya}\ \emph {et~al.}(2015)\citenamefont
  {Tsuchiya}, \citenamefont {Giuliani}, \citenamefont {Hashimoto},
  \citenamefont {Erenpreisa},\ and\ \citenamefont {Yoshikawa}}]{Yoshikawa2}%
  \BibitemOpen
  \bibfield  {author} {\bibinfo {author} {\bibnamefont {Tsuchiya},
  \bibfnamefont {M.}}, \bibinfo {author} {\bibfnamefont {A.}~\bibnamefont
  {Giuliani}}, \bibinfo {author} {\bibfnamefont {M.}~\bibnamefont {Hashimoto}},
  \bibinfo {author} {\bibfnamefont {J.}~\bibnamefont {Erenpreisa}}, \ and\
  \bibinfo {author} {\bibfnamefont {K.}~\bibnamefont {Yoshikawa}}} (\bibinfo
  {year} {2015}),\ \href@noop {} {\bibfield  {journal} {\bibinfo  {journal}
  {PloS One}\ }\textbf {\bibinfo {volume} {10}}~(\bibinfo {number} {6}),\
  \bibinfo {pages} {e0128565}}\BibitemShut {NoStop}%
\bibitem [{\citenamefont {Tsuchiya}\ \emph {et~al.}(2016)\citenamefont
  {Tsuchiya}, \citenamefont {Giuliani}, \citenamefont {Hashimoto},
  \citenamefont {Erenpreisa},\ and\ \citenamefont {Yoshikawa}}]{Yoshikawa1}%
  \BibitemOpen
  \bibfield  {author} {\bibinfo {author} {\bibnamefont {Tsuchiya},
  \bibfnamefont {M.}}, \bibinfo {author} {\bibfnamefont {A.}~\bibnamefont
  {Giuliani}}, \bibinfo {author} {\bibfnamefont {M.}~\bibnamefont {Hashimoto}},
  \bibinfo {author} {\bibfnamefont {J.}~\bibnamefont {Erenpreisa}}, \ and\
  \bibinfo {author} {\bibfnamefont {K.}~\bibnamefont {Yoshikawa}}} (\bibinfo
  {year} {2016}),\ \href@noop {} {\bibfield  {journal} {\bibinfo  {journal}
  {PloS One}\ }\textbf {\bibinfo {volume} {11}}~(\bibinfo {number} {12}),\
  \bibinfo {pages} {e0167912}}\BibitemShut {NoStop}%
\bibitem [{\citenamefont {Turcotte}(1999)}]{Turcotte}%
  \BibitemOpen
  \bibfield  {author} {\bibinfo {author} {\bibnamefont {Turcotte},
  \bibfnamefont {D.}}} (\bibinfo {year} {1999}),\ \href
  {http://stacks.iop.org/0034-4885/62/i=10/a=201} {\bibfield  {journal}
  {\bibinfo  {journal} {Rep. Prog. Phys.}\ }\textbf {\bibinfo {volume}
  {62}}~(\bibinfo {number} {10}),\ \bibinfo {pages} {1377}}\BibitemShut
  {NoStop}%
\bibitem [{\citenamefont {Turing}(1950)}]{Turing1950}%
  \BibitemOpen
  \bibfield  {author} {\bibinfo {author} {\bibnamefont {Turing}, \bibfnamefont
  {A.~M.}}} (\bibinfo {year} {1950}),\ \href@noop {} {\bibfield  {journal}
  {\bibinfo  {journal} {Mind}\ }\textbf {\bibinfo {volume} {59}}~(\bibinfo
  {number} {236}),\ \bibinfo {pages} {433}}\BibitemShut {NoStop}%
\bibitem [{\citenamefont {Turing}(1952)}]{Turing}%
  \BibitemOpen
  \bibfield  {author} {\bibinfo {author} {\bibnamefont {Turing}, \bibfnamefont
  {A.~M.}}} (\bibinfo {year} {1952}),\ \href@noop {} {\bibfield  {journal}
  {\bibinfo  {journal} {Philos. Trans. R. Soc. London, Ser. B}\ }\textbf
  {\bibinfo {volume} {237}}~(\bibinfo {number} {641}),\ \bibinfo {pages}
  {37}}\BibitemShut {NoStop}%
\bibitem [{\citenamefont {Tyrcha}\ \emph {et~al.}(2013)\citenamefont {Tyrcha},
  \citenamefont {Roudi}, \citenamefont {Marsili},\ and\ \citenamefont
  {Hertz}}]{Marsili4}%
  \BibitemOpen
  \bibfield  {author} {\bibinfo {author} {\bibnamefont {Tyrcha}, \bibfnamefont
  {J.}}, \bibinfo {author} {\bibfnamefont {Y.}~\bibnamefont {Roudi}}, \bibinfo
  {author} {\bibfnamefont {M.}~\bibnamefont {Marsili}}, \ and\ \bibinfo
  {author} {\bibfnamefont {J.}~\bibnamefont {Hertz}}} (\bibinfo {year}
  {2013}),\ \href@noop {} {\bibfield  {journal} {\bibinfo  {journal} {Journal
  of Statistical Mechanics: Theory and Experiment}\ }\textbf {\bibinfo {volume}
  {2013}}~(\bibinfo {number} {03}),\ \bibinfo {pages} {P03005}}\BibitemShut
  {NoStop}%
\bibitem [{\citenamefont {Tyson}\ \emph {et~al.}(2003)\citenamefont {Tyson},
  \citenamefont {Chen},\ and\ \citenamefont {Novak}}]{Tyson}%
  \BibitemOpen
  \bibfield  {author} {\bibinfo {author} {\bibnamefont {Tyson}, \bibfnamefont
  {J.~J.}}, \bibinfo {author} {\bibfnamefont {K.~C.}\ \bibnamefont {Chen}}, \
  and\ \bibinfo {author} {\bibfnamefont {B.}~\bibnamefont {Novak}}} (\bibinfo
  {year} {2003}),\ \href@noop {} {\bibfield  {journal} {\bibinfo  {journal}
  {Current opinion in cell biology}\ }\textbf {\bibinfo {volume}
  {15}}~(\bibinfo {number} {2}),\ \bibinfo {pages} {221}}\BibitemShut {NoStop}%
\bibitem [{\citenamefont {Uhlig}\ \emph {et~al.}(2013)\citenamefont {Uhlig},
  \citenamefont {Levina}, \citenamefont {Geisel},\ and\ \citenamefont
  {Herrmann}}]{Levina2013}%
  \BibitemOpen
  \bibfield  {author} {\bibinfo {author} {\bibnamefont {Uhlig}, \bibfnamefont
  {M.}}, \bibinfo {author} {\bibfnamefont {A.}~\bibnamefont {Levina}}, \bibinfo
  {author} {\bibfnamefont {T.}~\bibnamefont {Geisel}}, \ and\ \bibinfo {author}
  {\bibfnamefont {M.}~\bibnamefont {Herrmann}}} (\bibinfo {year} {2013}),\
  \href@noop {} {\bibfield  {journal} {\bibinfo  {journal} {Front. Comput.
  Neurosci.}\ }\textbf {\bibinfo {volume} {7}}~(\bibinfo {number}
  {87})}\BibitemShut {NoStop}%
\bibitem [{\citenamefont {Van~Kampen}(1992)}]{vanKampen}%
  \BibitemOpen
  \bibfield  {author} {\bibinfo {author} {\bibnamefont {Van~Kampen},
  \bibfnamefont {N.~G.}}} (\bibinfo {year} {1992}),\ \href@noop {} {\emph
  {\bibinfo {title} {Stochastic processes in physics and chemistry}}},\
  Vol.~\bibinfo {volume} {1}\ (\bibinfo  {publisher} {Elsevier})\BibitemShut
  {NoStop}%
\bibitem [{\citenamefont {Vanni}\ \emph {et~al.}(2011)\citenamefont {Vanni},
  \citenamefont {Lukovi{\'c}},\ and\ \citenamefont {Grigolini}}]{Vanni}%
  \BibitemOpen
  \bibfield  {author} {\bibinfo {author} {\bibnamefont {Vanni}, \bibfnamefont
  {F.}}, \bibinfo {author} {\bibfnamefont {M.}~\bibnamefont {Lukovi{\'c}}}, \
  and\ \bibinfo {author} {\bibfnamefont {P.}~\bibnamefont {Grigolini}}}
  (\bibinfo {year} {2011}),\ \href@noop {} {\bibfield  {journal} {\bibinfo
  {journal} {Phys. Rev. Lett.}\ }\textbf {\bibinfo {volume} {107}}~(\bibinfo
  {number} {7}),\ \bibinfo {pages} {078103}}\BibitemShut {NoStop}%
\bibitem [{\citenamefont {Varela}\ \emph {et~al.}(2001)\citenamefont {Varela},
  \citenamefont {Lachaux}, \citenamefont {Rodriguez},\ and\ \citenamefont
  {Martinerie}}]{Varela}%
  \BibitemOpen
  \bibfield  {author} {\bibinfo {author} {\bibnamefont {Varela}, \bibfnamefont
  {F.}}, \bibinfo {author} {\bibfnamefont {J.-P.}\ \bibnamefont {Lachaux}},
  \bibinfo {author} {\bibfnamefont {E.}~\bibnamefont {Rodriguez}}, \ and\
  \bibinfo {author} {\bibfnamefont {J.}~\bibnamefont {Martinerie}}} (\bibinfo
  {year} {2001}),\ \href@noop {} {\bibfield  {journal} {\bibinfo  {journal}
  {Nature Rev. Neurosci.}\ }\textbf {\bibinfo {volume} {2}}~(\bibinfo {number}
  {4}),\ \bibinfo {pages} {229}}\BibitemShut {NoStop}%
\bibitem [{\citenamefont {Vattay}\ \emph {et~al.}(2015)\citenamefont {Vattay},
  \citenamefont {Salahub}, \citenamefont {Csabai}, \citenamefont {Nassimi},\
  and\ \citenamefont {Kaufmann}}]{quantum}%
  \BibitemOpen
  \bibfield  {author} {\bibinfo {author} {\bibnamefont {Vattay}, \bibfnamefont
  {G.}}, \bibinfo {author} {\bibfnamefont {D.}~\bibnamefont {Salahub}},
  \bibinfo {author} {\bibfnamefont {I.}~\bibnamefont {Csabai}}, \bibinfo
  {author} {\bibfnamefont {A.}~\bibnamefont {Nassimi}}, \ and\ \bibinfo
  {author} {\bibfnamefont {S.~A.}\ \bibnamefont {Kaufmann}}} (\bibinfo {year}
  {2015}),\ \href@noop {} {\bibfield  {journal} {\bibinfo  {journal} {J. of
  Phys.: Conference Series}\ }\textbf {\bibinfo {volume} {626}}~(\bibinfo
  {number} {1}),\ \bibinfo {pages} {012023}}\BibitemShut {NoStop}%
\bibitem [{\citenamefont {Veatch}\ \emph {et~al.}(2008)\citenamefont {Veatch},
  \citenamefont {Cicuta}, \citenamefont {Sengupta}, \citenamefont
  {Honerkamp-Smith}, \citenamefont {Holowka},\ and\ \citenamefont
  {Baird}}]{Cicuta2}%
  \BibitemOpen
  \bibfield  {author} {\bibinfo {author} {\bibnamefont {Veatch}, \bibfnamefont
  {S.~L.}}, \bibinfo {author} {\bibfnamefont {P.}~\bibnamefont {Cicuta}},
  \bibinfo {author} {\bibfnamefont {P.}~\bibnamefont {Sengupta}}, \bibinfo
  {author} {\bibfnamefont {A.}~\bibnamefont {Honerkamp-Smith}}, \bibinfo
  {author} {\bibfnamefont {D.}~\bibnamefont {Holowka}}, \ and\ \bibinfo
  {author} {\bibfnamefont {B.}~\bibnamefont {Baird}}} (\bibinfo {year}
  {2008}),\ \href@noop {} {\bibfield  {journal} {\bibinfo  {journal} {ACS Chem.
  Biol.}\ }\textbf {\bibinfo {volume} {3}}~(\bibinfo {number} {5}),\ \bibinfo
  {pages} {287}}\BibitemShut {NoStop}%
\bibitem [{\citenamefont {Veatch}\ \emph {et~al.}(2007)\citenamefont {Veatch},
  \citenamefont {Soubias}, \citenamefont {Keller},\ and\ \citenamefont
  {Gawrisch}}]{membranes1}%
  \BibitemOpen
  \bibfield  {author} {\bibinfo {author} {\bibnamefont {Veatch}, \bibfnamefont
  {S.~L.}}, \bibinfo {author} {\bibfnamefont {O.}~\bibnamefont {Soubias}},
  \bibinfo {author} {\bibfnamefont {S.~L.}\ \bibnamefont {Keller}}, \ and\
  \bibinfo {author} {\bibfnamefont {K.}~\bibnamefont {Gawrisch}}} (\bibinfo
  {year} {2007}),\ \href@noop {} {\bibfield  {journal} {\bibinfo  {journal}
  {Proc. Natl. Acad. Sci. USA.}\ }\textbf {\bibinfo {volume} {104}}~(\bibinfo
  {number} {45}),\ \bibinfo {pages} {17650}}\BibitemShut {NoStop}%
\bibitem [{\citenamefont {Veening}\ \emph {et~al.}(2008)\citenamefont
  {Veening}, \citenamefont {Smits},\ and\ \citenamefont
  {Kuipers}}]{Bet-hedging}%
  \BibitemOpen
  \bibfield  {author} {\bibinfo {author} {\bibnamefont {Veening}, \bibfnamefont
  {J.-W.}}, \bibinfo {author} {\bibfnamefont {W.~K.}\ \bibnamefont {Smits}}, \
  and\ \bibinfo {author} {\bibfnamefont {O.~P.}\ \bibnamefont {Kuipers}}}
  (\bibinfo {year} {2008}),\ \href@noop {} {\bibfield  {journal} {\bibinfo
  {journal} {Annu. Rev. Microbiol.}\ }\textbf {\bibinfo {volume} {62}},\
  \bibinfo {pages} {193}}\BibitemShut {NoStop}%
\bibitem [{\citenamefont {Vespignani}\ \emph {et~al.}(1998)\citenamefont
  {Vespignani}, \citenamefont {Dickman}, \citenamefont {Mu{\~n}oz},\ and\
  \citenamefont {Zapperi}}]{FES-PRL}%
  \BibitemOpen
  \bibfield  {author} {\bibinfo {author} {\bibnamefont {Vespignani},
  \bibfnamefont {A.}}, \bibinfo {author} {\bibfnamefont {R.}~\bibnamefont
  {Dickman}}, \bibinfo {author} {\bibfnamefont {M.~A.}\ \bibnamefont
  {Mu{\~n}oz}}, \ and\ \bibinfo {author} {\bibfnamefont {S.}~\bibnamefont
  {Zapperi}}} (\bibinfo {year} {1998}),\ \href@noop {} {\bibfield  {journal}
  {\bibinfo  {journal} {Phys. Rev. Lett.}\ }\textbf {\bibinfo {volume}
  {81}}~(\bibinfo {number} {25}),\ \bibinfo {pages} {5676}}\BibitemShut
  {NoStop}%
\bibitem [{\citenamefont {Vespignani}\ \emph {et~al.}(2000)\citenamefont
  {Vespignani}, \citenamefont {Dickman}, \citenamefont {Mu{\~n}oz},\ and\
  \citenamefont {Zapperi}}]{FES-PRE}%
  \BibitemOpen
  \bibfield  {author} {\bibinfo {author} {\bibnamefont {Vespignani},
  \bibfnamefont {A.}}, \bibinfo {author} {\bibfnamefont {R.}~\bibnamefont
  {Dickman}}, \bibinfo {author} {\bibfnamefont {M.~A.}\ \bibnamefont
  {Mu{\~n}oz}}, \ and\ \bibinfo {author} {\bibfnamefont {S.}~\bibnamefont
  {Zapperi}}} (\bibinfo {year} {2000}),\ \href@noop {} {\bibfield  {journal}
  {\bibinfo  {journal} {Phys. Rev. E}\ }\textbf {\bibinfo {volume}
  {62}}~(\bibinfo {number} {4}),\ \bibinfo {pages} {4564}}\BibitemShut
  {NoStop}%
\bibitem [{\citenamefont {Vicsek}\ \emph {et~al.}(1995)\citenamefont {Vicsek},
  \citenamefont {Czir{\'o}k}, \citenamefont {Ben-Jacob}, \citenamefont
  {Cohen},\ and\ \citenamefont {Shochet}}]{Vicsek}%
  \BibitemOpen
  \bibfield  {author} {\bibinfo {author} {\bibnamefont {Vicsek}, \bibfnamefont
  {T.}}, \bibinfo {author} {\bibfnamefont {A.}~\bibnamefont {Czir{\'o}k}},
  \bibinfo {author} {\bibfnamefont {E.}~\bibnamefont {Ben-Jacob}}, \bibinfo
  {author} {\bibfnamefont {I.}~\bibnamefont {Cohen}}, \ and\ \bibinfo {author}
  {\bibfnamefont {O.}~\bibnamefont {Shochet}}} (\bibinfo {year} {1995}),\
  \href@noop {} {\bibfield  {journal} {\bibinfo  {journal} {Phys. Rev. Lett.}\
  }\textbf {\bibinfo {volume} {75}}~(\bibinfo {number} {6}),\ \bibinfo {pages}
  {1226}}\BibitemShut {NoStop}%
\bibitem [{\citenamefont {Vicsek}\ and\ \citenamefont
  {Zafeiris}(2012)}]{Vicsek2012}%
  \BibitemOpen
  \bibfield  {author} {\bibinfo {author} {\bibnamefont {Vicsek}, \bibfnamefont
  {T.}}, \ and\ \bibinfo {author} {\bibfnamefont {A.}~\bibnamefont {Zafeiris}}}
  (\bibinfo {year} {2012}),\ \href@noop {} {\bibfield  {journal} {\bibinfo
  {journal} {Phys. Rep.}\ }\textbf {\bibinfo {volume} {517}}~(\bibinfo {number}
  {3}),\ \bibinfo {pages} {71}}\BibitemShut {NoStop}%
\bibitem [{\citenamefont {Villa~Mart{\'\i}n}\ \emph {et~al.}(2014)\citenamefont
  {Villa~Mart{\'\i}n}, \citenamefont {Bonachela},\ and\ \citenamefont
  {Mu{\~n}oz}}]{Paula1}%
  \BibitemOpen
  \bibfield  {author} {\bibinfo {author} {\bibnamefont {Villa~Mart{\'\i}n},
  \bibfnamefont {P.}}, \bibinfo {author} {\bibfnamefont {J.~A.}\ \bibnamefont
  {Bonachela}}, \ and\ \bibinfo {author} {\bibfnamefont {M.~A.}\ \bibnamefont
  {Mu{\~n}oz}}} (\bibinfo {year} {2014}),\ \href@noop {} {\bibfield  {journal}
  {\bibinfo  {journal} {Phys. Rev. E}\ }\textbf {\bibinfo {volume}
  {89}}~(\bibinfo {number} {1}),\ \bibinfo {pages} {012145}}\BibitemShut
  {NoStop}%
\bibitem [{\citenamefont {Villegas}\ \emph {et~al.}(2014)\citenamefont
  {Villegas}, \citenamefont {Moretti},\ and\ \citenamefont
  {Mu{\~n}oz}}]{Villegas1}%
  \BibitemOpen
  \bibfield  {author} {\bibinfo {author} {\bibnamefont {Villegas},
  \bibfnamefont {P.}}, \bibinfo {author} {\bibfnamefont {P.}~\bibnamefont
  {Moretti}}, \ and\ \bibinfo {author} {\bibfnamefont {M.~A.}\ \bibnamefont
  {Mu{\~n}oz}}} (\bibinfo {year} {2014}),\ \href@noop {} {\bibfield  {journal}
  {\bibinfo  {journal} {Sci. Rep.}\ }\textbf {\bibinfo {volume} {4}},\ \bibinfo
  {pages} {5990}}\BibitemShut {NoStop}%
\bibitem [{\citenamefont {Villegas}\ \emph {et~al.}(2016)\citenamefont
  {Villegas}, \citenamefont {Ruiz-Franco}, \citenamefont {Hidalgo},\ and\
  \citenamefont {Mu{\~n}oz}}]{Villegas-Boolean}%
  \BibitemOpen
  \bibfield  {author} {\bibinfo {author} {\bibnamefont {Villegas},
  \bibfnamefont {P.}}, \bibinfo {author} {\bibfnamefont {J.}~\bibnamefont
  {Ruiz-Franco}}, \bibinfo {author} {\bibfnamefont {J.}~\bibnamefont
  {Hidalgo}}, \ and\ \bibinfo {author} {\bibfnamefont {M.~A.}\ \bibnamefont
  {Mu{\~n}oz}}} (\bibinfo {year} {2016}),\ \href@noop {} {\bibfield  {journal}
  {\bibinfo  {journal} {Sci. Rep.}\ }\textbf {\bibinfo {volume} {6}},\ \bibinfo
  {pages} {34743}}\BibitemShut {NoStop}%
\bibitem [{\citenamefont {Visser}(2013)}]{Visser}%
  \BibitemOpen
  \bibfield  {author} {\bibinfo {author} {\bibnamefont {Visser}, \bibfnamefont
  {M.}}} (\bibinfo {year} {2013}),\ \href@noop {} {\bibfield  {journal}
  {\bibinfo  {journal} {New J. Phys.}\ }\textbf {\bibinfo {volume}
  {15}}~(\bibinfo {number} {4}),\ \bibinfo {pages} {043021}}\BibitemShut
  {NoStop}%
\bibitem [{\citenamefont {Vojta}(2006)}]{Vojta-Review}%
  \BibitemOpen
  \bibfield  {author} {\bibinfo {author} {\bibnamefont {Vojta}, \bibfnamefont
  {T.}}} (\bibinfo {year} {2006}),\ \href {\doibase
  10.1088/0305-4470/39/22/R01} {\bibfield  {journal} {\bibinfo  {journal} {J.
  Phys. A}\ }\textbf {\bibinfo {volume} {39}}~(\bibinfo {number} {22}),\
  \bibinfo {pages} {R143}}\BibitemShut {NoStop}%
\bibitem [{\citenamefont {van Vreeswijk}\ and\ \citenamefont
  {Sompolinsky}(1996)}]{Sompo}%
  \BibitemOpen
  \bibfield  {author} {\bibinfo {author} {\bibnamefont {van Vreeswijk},
  \bibfnamefont {C.}}, \ and\ \bibinfo {author} {\bibfnamefont
  {H.}~\bibnamefont {Sompolinsky}}} (\bibinfo {year} {1996}),\ \href@noop {}
  {\bibfield  {journal} {\bibinfo  {journal} {Science}\ }\textbf {\bibinfo
  {volume} {274}},\ \bibinfo {pages} {1724}}\BibitemShut {NoStop}%
\bibitem [{\citenamefont {Wagner}(2005)}]{Wagner}%
  \BibitemOpen
  \bibfield  {author} {\bibinfo {author} {\bibnamefont {Wagner}, \bibfnamefont
  {A.}}} (\bibinfo {year} {2005}),\ \href@noop {} {\emph {\bibinfo {title}
  {Robustness and evolvability in living systems}}}\ (\bibinfo  {publisher}
  {Princeton University Press Princeton})\BibitemShut {NoStop}%
\bibitem [{\citenamefont {Wang}\ \emph {et~al.}(2011)\citenamefont {Wang},
  \citenamefont {Lizier},\ and\ \citenamefont {Prokopenko}}]{FI}%
  \BibitemOpen
  \bibfield  {author} {\bibinfo {author} {\bibnamefont {Wang}, \bibfnamefont
  {X.~R.}}, \bibinfo {author} {\bibfnamefont {J.~T.}\ \bibnamefont {Lizier}}, \
  and\ \bibinfo {author} {\bibfnamefont {M.}~\bibnamefont {Prokopenko}}}
  (\bibinfo {year} {2011}),\ \href@noop {} {\bibfield  {journal} {\bibinfo
  {journal} {Artificial life}\ }\textbf {\bibinfo {volume} {17}}~(\bibinfo
  {number} {4}),\ \bibinfo {pages} {315}}\BibitemShut {NoStop}%
\bibitem [{\citenamefont {Watkins}\ \emph {et~al.}(2015)\citenamefont
  {Watkins}, \citenamefont {Pruessner}, \citenamefont {Chapman}, \citenamefont
  {Crosby},\ and\ \citenamefont {Jensen}}]{SOC-25}%
  \BibitemOpen
  \bibfield  {author} {\bibinfo {author} {\bibnamefont {Watkins}, \bibfnamefont
  {N.~W.}}, \bibinfo {author} {\bibfnamefont {G.}~\bibnamefont {Pruessner}},
  \bibinfo {author} {\bibfnamefont {S.~C.}\ \bibnamefont {Chapman}}, \bibinfo
  {author} {\bibfnamefont {N.~B.}\ \bibnamefont {Crosby}}, \ and\ \bibinfo
  {author} {\bibfnamefont {H.~J.}\ \bibnamefont {Jensen}}} (\bibinfo {year}
  {2015}),\ \href@noop {} {\bibinfo  {journal} {Space Sci. Rev.}\ ,\ \bibinfo
  {pages} {1}}\BibitemShut {NoStop}%
\bibitem [{\citenamefont {Watson}\ and\ \citenamefont {Galton}(1875)}]{Watson}%
  \BibitemOpen
\bibfield  {journal} {  }\bibfield  {author} {\bibinfo {author} {\bibnamefont
  {Watson}, \bibfnamefont {H.~W.}}, \ and\ \bibinfo {author} {\bibfnamefont
  {F.}~\bibnamefont {Galton}}} (\bibinfo {year} {1875}),\ \href@noop {}
  {\bibfield  {journal} {\bibinfo  {journal} {J. of the Anthropological Ins. of
  Great Britain and Ireland}\ }\textbf {\bibinfo {volume} {4}},\ \bibinfo
  {pages} {138}}\BibitemShut {NoStop}%
\bibitem [{\citenamefont {Watts}\ and\ \citenamefont {Strogatz}(1998)}]{SW}%
  \BibitemOpen
  \bibfield  {author} {\bibinfo {author} {\bibnamefont {Watts}, \bibfnamefont
  {D.~J.}}, \ and\ \bibinfo {author} {\bibfnamefont {S.~H.}\ \bibnamefont
  {Strogatz}}} (\bibinfo {year} {1998}),\ \href@noop {} {\bibfield  {journal}
  {\bibinfo  {journal} {Nature}\ }\textbf {\bibinfo {volume} {393}}~(\bibinfo
  {number} {6684}),\ \bibinfo {pages} {440}}\BibitemShut {NoStop}%
\bibitem [{\citenamefont {Werner}(2007)}]{Werner}%
  \BibitemOpen
  \bibfield  {author} {\bibinfo {author} {\bibnamefont {Werner}, \bibfnamefont
  {G.}}} (\bibinfo {year} {2007}),\ \href@noop {} {\bibfield  {journal}
  {\bibinfo  {journal} {Biosystems}\ }\textbf {\bibinfo {volume}
  {90}}~(\bibinfo {number} {2}),\ \bibinfo {pages} {496}}\BibitemShut {NoStop}%
\bibitem [{\citenamefont {West}(2010)}]{West-physio}%
  \BibitemOpen
  \bibfield  {author} {\bibinfo {author} {\bibnamefont {West}, \bibfnamefont
  {B.~J.}}} (\bibinfo {year} {2010}),\ \href@noop {} {\bibfield  {journal}
  {\bibinfo  {journal} {Front. Physiol.}\ }\textbf {\bibinfo {volume} {1}},\
  \bibinfo {pages} {12}}\BibitemShut {NoStop}%
\bibitem [{\citenamefont {West}(2017)}]{West-Scale}%
  \BibitemOpen
  \bibfield  {author} {\bibinfo {author} {\bibnamefont {West}, \bibfnamefont
  {G.}}} (\bibinfo {year} {2017}),\ \href
  {https://books.google.es/books?id=PZjCoAEACAAJ} {\emph {\bibinfo {title}
  {Scale}}}\ (\bibinfo  {publisher} {Orion Publishing Group,
  Limited})\BibitemShut {NoStop}%
\bibitem [{\citenamefont {West}\ \emph {et~al.}(1997)\citenamefont {West},
  \citenamefont {Brown},\ and\ \citenamefont {Enquist}}]{West}%
  \BibitemOpen
  \bibfield  {author} {\bibinfo {author} {\bibnamefont {West}, \bibfnamefont
  {G.~B.}}, \bibinfo {author} {\bibfnamefont {J.~H.}\ \bibnamefont {Brown}}, \
  and\ \bibinfo {author} {\bibfnamefont {B.~J.}\ \bibnamefont {Enquist}}}
  (\bibinfo {year} {1997}),\ \href@noop {} {\bibfield  {journal} {\bibinfo
  {journal} {Science}\ }\textbf {\bibinfo {volume} {276}}~(\bibinfo {number}
  {5309}),\ \bibinfo {pages} {122}}\BibitemShut {NoStop}%
\bibitem [{\citenamefont {Williams-Garcia}\ \emph {et~al.}(2017)\citenamefont
  {Williams-Garcia}, \citenamefont {Beggs},\ and\ \citenamefont
  {Ortiz}}]{Beggs-causal}%
  \BibitemOpen
  \bibfield  {author} {\bibinfo {author} {\bibnamefont {Williams-Garcia},
  \bibfnamefont {R.}}, \bibinfo {author} {\bibfnamefont {J.~M.}\ \bibnamefont
  {Beggs}}, \ and\ \bibinfo {author} {\bibfnamefont {G.}~\bibnamefont {Ortiz}}}
  (\bibinfo {year} {2017}),\ \href {\doibase 10.1209/0295-5075/119/18003}
  {\bibfield  {journal} {\bibinfo  {journal} {EPL}\ }\textbf {\bibinfo {volume}
  {119}}~(\bibinfo {number} {1}),\ \bibinfo {pages} {18003}}\BibitemShut
  {NoStop}%
\bibitem [{\citenamefont {Williams-Garc{\'\i}a}\ \emph
  {et~al.}(2014)\citenamefont {Williams-Garc{\'\i}a}, \citenamefont {Moore},
  \citenamefont {Beggs},\ and\ \citenamefont {Ortiz}}]{Rashid}%
  \BibitemOpen
  \bibfield  {author} {\bibinfo {author} {\bibnamefont {Williams-Garc{\'\i}a},
  \bibfnamefont {R.~V.}}, \bibinfo {author} {\bibfnamefont {M.}~\bibnamefont
  {Moore}}, \bibinfo {author} {\bibfnamefont {J.~M.}\ \bibnamefont {Beggs}}, \
  and\ \bibinfo {author} {\bibfnamefont {G.}~\bibnamefont {Ortiz}}} (\bibinfo
  {year} {2014}),\ \href@noop {} {\bibfield  {journal} {\bibinfo  {journal}
  {Phys. Rev. E}\ }\textbf {\bibinfo {volume} {90}}~(\bibinfo {number} {6}),\
  \bibinfo {pages} {062714}}\BibitemShut {NoStop}%
\bibitem [{\citenamefont {Willinger}\ \emph {et~al.}(2004)\citenamefont
  {Willinger}, \citenamefont {Alderson}, \citenamefont {Doyle},\ and\
  \citenamefont {Li}}]{More-normal}%
  \BibitemOpen
  \bibfield  {author} {\bibinfo {author} {\bibnamefont {Willinger},
  \bibfnamefont {W.}}, \bibinfo {author} {\bibfnamefont {D.}~\bibnamefont
  {Alderson}}, \bibinfo {author} {\bibfnamefont {J.~C.}\ \bibnamefont {Doyle}},
  \ and\ \bibinfo {author} {\bibfnamefont {L.}~\bibnamefont {Li}}} (\bibinfo
  {year} {2004}),\ in\ \href@noop {} {\emph {\bibinfo {booktitle} {Proc. Winter
  Simulation Conf.}}},\ Vol.~\bibinfo {volume} {1}\ (\bibinfo {organization}
  {IEEE})\BibitemShut {NoStop}%
\bibitem [{\citenamefont {Wilson}(1979)}]{Wilson1979}%
  \BibitemOpen
  \bibfield  {author} {\bibinfo {author} {\bibnamefont {Wilson}, \bibfnamefont
  {K.}}} (\bibinfo {year} {1979}),\ \href@noop {} {\bibfield  {journal}
  {\bibinfo  {journal} {Sci. Am.}\ }\textbf {\bibinfo {volume} {241}}~(\bibinfo
  {number} {2}),\ \bibinfo {pages} {158}}\BibitemShut {NoStop}%
\bibitem [{\citenamefont {Wilson}\ and\ \citenamefont {Kogut}(1974)}]{Wilson}%
  \BibitemOpen
  \bibfield  {author} {\bibinfo {author} {\bibnamefont {Wilson}, \bibfnamefont
  {K.}}, \ and\ \bibinfo {author} {\bibfnamefont {J.}~\bibnamefont {Kogut}}}
  (\bibinfo {year} {1974}),\ \href@noop {} {\bibfield  {journal} {\bibinfo
  {journal} {Phys. Rep.}\ }\textbf {\bibinfo {volume} {12}}~(\bibinfo {number}
  {2}),\ \bibinfo {pages} {75}}\BibitemShut {NoStop}%
\bibitem [{\citenamefont {Wolf}\ \emph {et~al.}(2005)\citenamefont {Wolf},
  \citenamefont {Vazirani},\ and\ \citenamefont {Arkin}}]{Bet-hedging2}%
  \BibitemOpen
  \bibfield  {author} {\bibinfo {author} {\bibnamefont {Wolf}, \bibfnamefont
  {D.~M.}}, \bibinfo {author} {\bibfnamefont {V.~V.}\ \bibnamefont {Vazirani}},
  \ and\ \bibinfo {author} {\bibfnamefont {A.~P.}\ \bibnamefont {Arkin}}}
  (\bibinfo {year} {2005}),\ \href@noop {} {\bibfield  {journal} {\bibinfo
  {journal} {J. Theor. Biol.}\ }\textbf {\bibinfo {volume} {234}}~(\bibinfo
  {number} {2}),\ \bibinfo {pages} {227}}\BibitemShut {NoStop}%
\bibitem [{\citenamefont {Wolfram}(2002)}]{Wolfram2002}%
  \BibitemOpen
  \bibfield  {author} {\bibinfo {author} {\bibnamefont {Wolfram}, \bibfnamefont
  {S.}}} (\bibinfo {year} {2002}),\ \href@noop {} {\emph {\bibinfo {title} {A
  new kind of science}}}\ (\bibinfo  {publisher} {Wolfram media Champaign,
  IL})\BibitemShut {NoStop}%
\bibitem [{\citenamefont {Yamaguchi}\ \emph {et~al.}(2017)\citenamefont
  {Yamaguchi}, \citenamefont {Kawaguchi},\ and\ \citenamefont
  {Sagawa}}]{Yamaguchi}%
  \BibitemOpen
  \bibfield  {author} {\bibinfo {author} {\bibnamefont {Yamaguchi},
  \bibfnamefont {H.}}, \bibinfo {author} {\bibfnamefont {K.}~\bibnamefont
  {Kawaguchi}}, \ and\ \bibinfo {author} {\bibfnamefont {T.}~\bibnamefont
  {Sagawa}}} (\bibinfo {year} {2017}),\ \href@noop {} {\bibfield  {journal}
  {\bibinfo  {journal} {Phys. Rev. E}\ }\textbf {\bibinfo {volume}
  {96}}~(\bibinfo {number} {1}),\ \bibinfo {pages} {012401}}\BibitemShut
  {NoStop}%
\bibitem [{\citenamefont {Yang}\ \emph {et~al.}(2012)\citenamefont {Yang},
  \citenamefont {Shew}, \citenamefont {Roy},\ and\ \citenamefont
  {Plenz}}]{Shew2012}%
  \BibitemOpen
  \bibfield  {author} {\bibinfo {author} {\bibnamefont {Yang}, \bibfnamefont
  {H.}}, \bibinfo {author} {\bibfnamefont {W.~L.}\ \bibnamefont {Shew}},
  \bibinfo {author} {\bibfnamefont {R.}~\bibnamefont {Roy}}, \ and\ \bibinfo
  {author} {\bibfnamefont {D.}~\bibnamefont {Plenz}}} (\bibinfo {year}
  {2012}),\ \href@noop {} {\bibfield  {journal} {\bibinfo  {journal} {J.
  Neurosci.}\ }\textbf {\bibinfo {volume} {32}}~(\bibinfo {number} {3}),\
  \bibinfo {pages} {1061}}\BibitemShut {NoStop}%
\bibitem [{\citenamefont {Yu}\ \emph {et~al.}(2014)\citenamefont {Yu},
  \citenamefont {Klaus}, \citenamefont {Yang},\ and\ \citenamefont
  {Plenz}}]{Yu2014}%
  \BibitemOpen
  \bibfield  {author} {\bibinfo {author} {\bibnamefont {Yu}, \bibfnamefont
  {S.}}, \bibinfo {author} {\bibfnamefont {A.}~\bibnamefont {Klaus}}, \bibinfo
  {author} {\bibfnamefont {H.}~\bibnamefont {Yang}}, \ and\ \bibinfo {author}
  {\bibfnamefont {D.}~\bibnamefont {Plenz}}} (\bibinfo {year} {2014}),\
  \href@noop {} {\bibfield  {journal} {\bibinfo  {journal} {PloS One}\ }\textbf
  {\bibinfo {volume} {9}}~(\bibinfo {number} {6}),\ \bibinfo {pages}
  {e99761}}\BibitemShut {NoStop}%
\bibitem [{\citenamefont {Yu}\ \emph {et~al.}(2011)\citenamefont {Yu},
  \citenamefont {Yang}, \citenamefont {Nakahara}, \citenamefont {Santos},
  \citenamefont {Nikoli{\'c}},\ and\ \citenamefont {Plenz}}]{Yu2011}%
  \BibitemOpen
  \bibfield  {author} {\bibinfo {author} {\bibnamefont {Yu}, \bibfnamefont
  {S.}}, \bibinfo {author} {\bibfnamefont {H.}~\bibnamefont {Yang}}, \bibinfo
  {author} {\bibfnamefont {H.}~\bibnamefont {Nakahara}}, \bibinfo {author}
  {\bibfnamefont {G.~S.}\ \bibnamefont {Santos}}, \bibinfo {author}
  {\bibfnamefont {D.}~\bibnamefont {Nikoli{\'c}}}, \ and\ \bibinfo {author}
  {\bibfnamefont {D.}~\bibnamefont {Plenz}}} (\bibinfo {year} {2011}),\
  \href@noop {} {\bibfield  {journal} {\bibinfo  {journal} {J. Neurosci.}\
  }\textbf {\bibinfo {volume} {31}}~(\bibinfo {number} {48}),\ \bibinfo {pages}
  {17514}}\BibitemShut {NoStop}%
\bibitem [{\citenamefont {Yule}(1925)}]{Yule}%
  \BibitemOpen
  \bibfield  {author} {\bibinfo {author} {\bibnamefont {Yule}, \bibfnamefont
  {G.~U.}}} (\bibinfo {year} {1925}),\ \href@noop {} {\bibfield  {journal}
  {\bibinfo  {journal} {Philos. Trans. R. Soc. London Ser.B}\ }\textbf
  {\bibinfo {volume} {213}},\ \bibinfo {pages} {21}}\BibitemShut {NoStop}%
\bibitem [{\citenamefont {Yuste}\ \emph {et~al.}(2005)\citenamefont {Yuste},
  \citenamefont {MacLean}, \citenamefont {Smith},\ and\ \citenamefont
  {Lansner}}]{Yuste}%
  \BibitemOpen
  \bibfield  {author} {\bibinfo {author} {\bibnamefont {Yuste}, \bibfnamefont
  {R.}}, \bibinfo {author} {\bibfnamefont {J.~N.}\ \bibnamefont {MacLean}},
  \bibinfo {author} {\bibfnamefont {J.}~\bibnamefont {Smith}}, \ and\ \bibinfo
  {author} {\bibfnamefont {A.}~\bibnamefont {Lansner}}} (\bibinfo {year}
  {2005}),\ \href@noop {} {\bibfield  {journal} {\bibinfo  {journal} {Nat. Rev.
  Neurosci.}\ }\textbf {\bibinfo {volume} {6}},\ \bibinfo {pages}
  {477}}\BibitemShut {NoStop}%
\bibitem [{\citenamefont {Zamponi}\ \emph {et~al.}(2018)\citenamefont
  {Zamponi}, \citenamefont {Zamponi}, \citenamefont {Cannas}, \citenamefont
  {Billoni}, \citenamefont {Helguera},\ and\ \citenamefont
  {Chialvo}}]{Chialvo-Mito}%
  \BibitemOpen
  \bibfield  {author} {\bibinfo {author} {\bibnamefont {Zamponi}, \bibfnamefont
  {N.}}, \bibinfo {author} {\bibfnamefont {E.}~\bibnamefont {Zamponi}},
  \bibinfo {author} {\bibfnamefont {S.~A.}\ \bibnamefont {Cannas}}, \bibinfo
  {author} {\bibfnamefont {O.~V.}\ \bibnamefont {Billoni}}, \bibinfo {author}
  {\bibfnamefont {P.~R.}\ \bibnamefont {Helguera}}, \ and\ \bibinfo {author}
  {\bibfnamefont {D.~R.}\ \bibnamefont {Chialvo}}} (\bibinfo {year} {2018}),\
  \href@noop {} {\bibfield  {journal} {\bibinfo  {journal} {Sci. Rep.}\
  }\textbf {\bibinfo {volume} {8}}~(\bibinfo {number} {1}),\ \bibinfo {pages}
  {363}}\BibitemShut {NoStop}%
\bibitem [{\citenamefont {Zapperi}\ \emph {et~al.}(1995)\citenamefont
  {Zapperi}, \citenamefont {Lauritsen},\ and\ \citenamefont {Stanley}}]{SOBP}%
  \BibitemOpen
  \bibfield  {author} {\bibinfo {author} {\bibnamefont {Zapperi}, \bibfnamefont
  {S.}}, \bibinfo {author} {\bibfnamefont {K.~B.}\ \bibnamefont {Lauritsen}}, \
  and\ \bibinfo {author} {\bibfnamefont {H.~E.}\ \bibnamefont {Stanley}}}
  (\bibinfo {year} {1995}),\ \href@noop {} {\bibfield  {journal} {\bibinfo
  {journal} {Phys. Rev. Lett.}\ }\textbf {\bibinfo {volume} {75}}~(\bibinfo
  {number} {22}),\ \bibinfo {pages} {4071}}\BibitemShut {NoStop}%
\bibitem [{\citenamefont {Zipf}(1949)}]{Zipf}%
  \BibitemOpen
  \bibfield  {author} {\bibinfo {author} {\bibnamefont {Zipf}, \bibfnamefont
  {G.~K.}}} (\bibinfo {year} {1949}),\ \href@noop {} {\emph {\bibinfo {title}
  {Human behavior and the principle of least effort}}}\ (\bibinfo  {publisher}
  {Addison-Wesley, Cambridge})\BibitemShut {NoStop}%
\end{thebibliography}%
%\begin{thebibliography}{113}

%\end{thebibliography}

\end{document}